\documentclass[preprint,13pt]{elsarticle}
\biboptions{sort&compress}

\usepackage{amsmath}
\usepackage{graphicx}
\usepackage{subfigure}
\usepackage[usenames,dvipsnames]{color}
\definecolor{darkblue}{RGB}{0,0,196}
\usepackage[colorlinks=true,linkcolor=darkblue,citecolor=darkblue,urlcolor=darkblue]{hyperref}
\usepackage{setspace}
\usepackage{multirow}
\usepackage{hyperref}
\usepackage{epstopdf}
\usepackage{xcolor}
\hypersetup{
  colorlinks   = true, %Colours links instead of ugly boxes
  urlcolor     = blue, %Colour for external hyperlinks
  linkcolor    = blue, %Colour of internal links
  citecolor   = blue %Colour of citations
}
\usepackage{graphicx}
\usepackage{amsmath,bbm}
\usepackage{amssymb,bm}
\usepackage{yfonts}
\usepackage{csquotes}
\usepackage{comment}
\usepackage[normalem]{ulem}
\newcommand{\pt}{$p_{\rm T}$}

\begin{document}
%\bibliographystyle{elsarticle-num}
%\bibliographystyle{apsrev4-2}
%\title{Hunting QGP-like properties in $pp$ collisions at LHC energies using event classifiers \\ - A review based on MC studies}

%\title{Recent Event Topology Studies to probe collective-like effects in small systems at the Large Hadron Collider \\}

\title{Event Topology Classifiers at the Large Hadron Collider \\}

\author[1]{Suraj Prasad}
\ead{Suraj.Prasad@cern.ch}
\author[2]{Sushanta Tripathy}
\ead{Sushanta.Tripathy@cern.ch}
\author[1]{Bhagyarathi Sahoo} 
\ead{Bhagyarathi.Sahoo@cern.ch}
\author[1]{Raghunath Sahoo\corref{cor1}}
\ead{Raghunath.Sahoo@cern.ch}
%\email[Corresponding author: ]{Raghunath.Sahoo@cern.ch}}
\affiliation[1]{organization={Department of Physics, Indian Institute of Technology Indore}, city={Simrol, Indore},  postcode={453552},
state={Madhya Pradesh},
country={India}}
\affiliation[2]{organization={Division of Particle and Nuclear Physics, Department of Physics, Lund University}, city={Lund},  postcode={22364},
country={Sweden}}
%\affiliation[2]{Experimental Physics division, CERN, 1211 Geneva, Switzerland}

\date{\today}% It is always \today, today,
             %  but any date may be explicitly specified
\cortext[cor1]{Corresponding author}

\begin{abstract}
%\linenumbers
Event classifiers are the most fundamental observables to probe the event topology of hadronic and nuclear collisions at relativistic energies. Over the last five decades, significant progress has been made to establish suitable event classifiers to probe different physics processes occurring in elementary $e^{+}e^{-}$ to heavy-ion collisions in a broad range of center of mass energies. One of the major motivations to revisit event classifiers at the Large Hadron Collider (LHC) originates from the recent measurements of high multiplicity proton-proton collisions, which have revealed that these small collision systems exhibit features similar to the formation of quark-gluon plasma (QGP), traditionally believed to be only achievable in heavy nucleus-nucleus collisions at ultra-relativistic energies. To pinpoint the origin of these QGP-like phenomena with substantially reduced autocorrelation and selection biases, and to bring all collision systems on equal footing, along with charged-particle multiplicity, lately several event topology classifiers such as transverse sphericity, transverse spherocity, relative transverse activity classifier, and charged-particle flattenicity have been used extensively in experiments as well as in the phenomenological front.  In addition, the infrared and collinear safety of event-shape observables makes them ideal for precision studies of jets and heavy-flavors at the LHC. In this review article, we summarise the motivation, scope, and practical use of these event-shape observables. The discussion integrates results and insights from all major LHC experiments, setting the stage for precision investigations for Run 3, Run 4, and future high luminosity upgrades of the LHC. 
In most cases, the event shape observables are found to be better probes in understanding the heavy-ion-like behavior seen at the LHC, while making a multi-differential study of multihadron production dynamics in hadronic and nuclear collisions.
\end{abstract}


\maketitle 
\tableofcontents
\newpage

\section{Introduction}
\label{sec:intro}
\begin{figure}[ht!]
\begin{center}
\includegraphics[scale = 0.7]{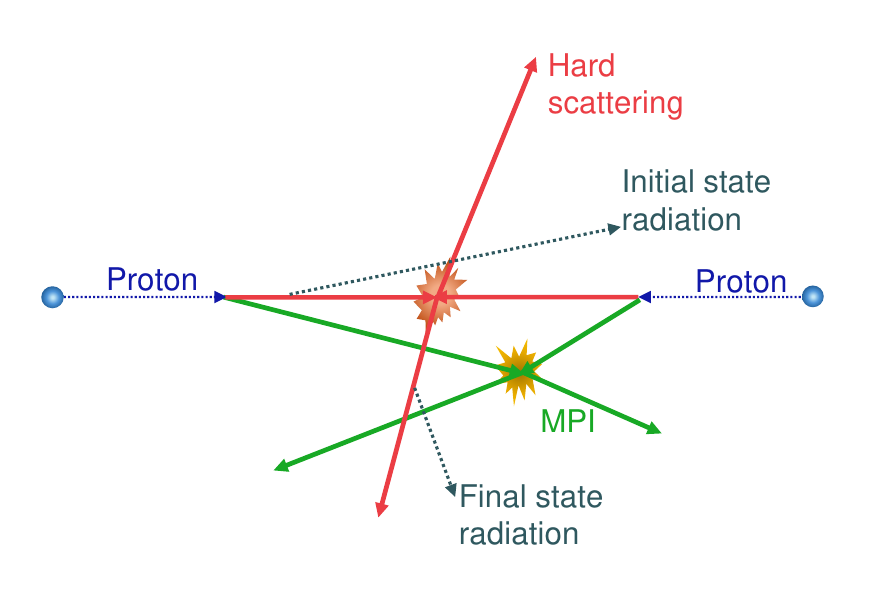}
\caption{Pictorial representation of multi-partonic interactions (MPI) in $pp$ collisions.}
\label{fig:NmpiCartoon}
\end{center}
\end{figure}

\begin{figure}[ht!]
\begin{center}
\includegraphics[scale = 0.5]{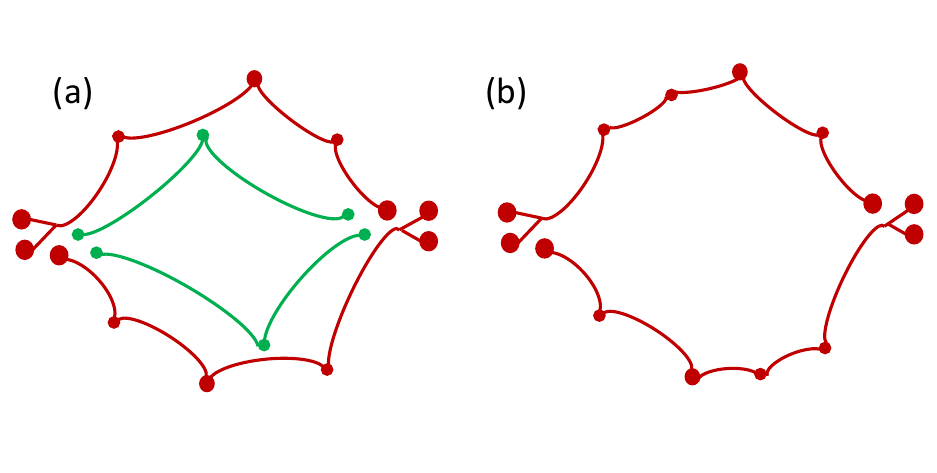}
\caption{Pictorial representation of Color Reconnection (CR) in the string fragmentation model: (a) two independent hard scatterings, (b) color reconnected string.}
\label{fig:CRCartoon}
\end{center}
\end{figure}

\noindent
%\linenumbers 
To understand the infancy of our Universe and its evolution, the little bangs of heavy-ions are made in collider experiments like the Large Hadron Collider (LHC) at CERN, Switzerland, and the Relativistic Heavy-Ion Collider (RHIC) at BNL, USA, which reproduces the conditions that happened during the fraction of a second after the Big Bang. The collision of two nuclei produces a hot and dense system of deconfined quarks and gluons for a short time (order of a few fm/c or 10$^{-23}$ sec). This deconfined and thermalized state of quarks and gluons is often referred to as quark-gluon plasma (QGP). Due to a very short timescale, it is impossible to probe QGP directly. However, several indirect signatures are proposed, such as the presence of strangeness enhancement, quarkonia suppression, jet quenching, collective flow, direct photon and dilepton production, which validate the presence of QGP. In proton-proton ($pp$) collisions, where the system size is significantly smaller than Pb-Pb, the main hard parton-parton scatterings produce high transverse momentum ($p_{\rm T}$) particles which can be described by the perturbative quantum chromodynamics (pQCD).  In the hard regime of QCD, jets are produced from parton scatterings at large $p_{\rm T}$. As in this regime, the coupling constant ($\alpha_{s}$) is significantly small, precise pQCD calculations can be performed. The semi-hard and soft parton scatterings, which are not a part of the main hard scatterings, form the underlying event (UE)~\cite{Diehl:2011yj, Field:2012kd}. As the soft-QCD calculations are cumbersome due to a larger value of $\alpha_{s}$, the description of the UE is modeled using different QCD-inspired phenomenological approaches~\cite{Christiansen:2024bhe}. At the LHC energies, due to high parton densities and the composite nature of hadrons, more than one binary interaction among partons is possible in a single $pp$ collision, which is often referred to as multi-parton interactions (MPI)~\cite{Sjostrand:1987su}. The MPI phenomenon has been widely supported by experimental data~\cite{ZEUS:2007xwd,D0:2009apj} and is one of the key ingredients in the Monte Carlo (MC) event generators.  Fig.~\ref{fig:NmpiCartoon} shows the pictorial representation of MPI in pp collisions.  In such MPI-based models, the final state in $pp$ collisions is sensitive to the modeling of MPI. They are also sensitive to non-perturbative effects in the final state, such as color reconnection (CR), where the strings connecting the partonic endpoints from each individual scattering can color reconnect (Fig. \ref{fig:CRCartoon}) and color Ropes, where overlapping strings can act coherently to form a color rope having a large effective string tension~\cite{Sjostrand:2014zea}.

 Recent measurements of high multiplicity $pp$ collisions at the LHC energies have revealed that these systems exhibit features similar to quark-gluon plasma, such as the presence of radial and elliptic flow~\cite{CMS:2016fnw,Grosse-Oetringhaus:2024bwr,ALICE:2019zfl,CMS:2010ifv,ATLAS:2015hzw,CMS:2015fgy,ATLAS:2016yzd,Busza:2018rrf, ALICE:2025bwp} and enhanced production of strange hadrons with respect to that of charged pions~\cite{ALICE:2017jyt,ALICE:2018pal,ALICE:2019avo}, traditionally believed to be only achievable in ultra-relativistic nucleus-nucleus collisions. These measurements may point towards a common underlying physics mechanism across collision systems. However, no jet quenching signatures have been reported in small collision systems~\cite{Nagle:2018nvi, ALICE:2023plt}. The event generators currently available with relevant physics models fail to simultaneously reproduce all the observed phenomena in small collisions. This makes the origin of QGP-like phenomena observed in small systems like $pp$ and p-Pb unclear. Out of several theoretical models, the explanations from the MPI-based models with final state non-pQCD phenomena like CR~\cite{Sjostrand:2014zea}, rope hadronisation~\cite{Bierlich:2015rha, Bierlich:2014xba}, and string shoving~\cite{Bierlich:2016vgw}, inherited in PYTHIA8, are able to explain many of the QGP-like signatures in $pp$ collisions. MPI is a phenomenological approach and can not be measured in experiments. Thus, many experimental measurements are performed as a function of charged particle multiplicity in the final state, which has a strong correlation with MPI in MC models. On the other hand, the measurements with charged particle multiplicity show a stronger than linear increase of high-$p_{\rm T}$ particle yields in high-multiplicity (HM) $pp$ collisions relative to that of minimum bias (MB) events~\cite{ALICE:2019dfi}. This signifies a selection bias towards local fluctuations of choosing only hard $pp$ collisions. Such biases can be removed when the event selection is performed with multiplicity measured in a different pseudorapidity interval with respect to the observable of interest. However, it is found that such measurements are still biased by the hard processes contributing to the high-$p_{\rm T}$ particles~\cite{ALICE:2022qxg}. Thus, in small collision systems, such selection biases in the measurements hinder the search for the origin of QGP-like behavior.

To pin down the origin of QGP-like phenomena with substantially diminished selection bias and to bring all the collision systems on an equal footing, along with charged-particle multiplicity ($N_{\rm{ch}}$), lately event shape observables such as transverse spherocity ($S_{0}$) and transverse sphericity ($S_{\rm T}$) have been inherited from the usage in $e^{+}e^{-}$ collisions since the late 1970s~\cite{MovillaFernandez:1997fr,  Donoghue:1979vi,Dasgupta:2003iq, Banfi:2004nk}. Additionally, there are other proposed event shape observables which are used in heavy-ion collisions and jet studies such as third-jet resolution parameter, $y_{23}$~\cite{Bethke:2009ehn,OPAL:1989fep,JADE:1982xwg};  aplanarity, $A$; transverse thrust, $\tau_{T}$; its minor component, $T_{\perp, \rm min}$ ~\cite{Brandt:1964sa,ATLAS:2012tch, CMS:2011usu}; $\mathcal{F}$-parameter~\cite{Banfi:2010xy}, reduced flow vector, $q_{n}$~\cite{STAR:2002hbo}, Fox-Wolfram moments~\cite{Fox:1978vu}, and event isotropy~\cite{Cesarotti:2020hwb, Cesarotti:2020ngq, ATLAS:2023mny}. Moreover, several new event shape observables have been recently constructed, for example, relative transverse activity classifiers such as $R_{\rm T}$, $R_{\rm T}^{\rm min}$, and $R_{\rm T}^{\rm max}$ developed by P.~Skands {\it et. al.} and charged-particle flattenicity ($\rho_{\rm ch}$) developed by A.~Ortiz {\it et. al.} to reduce the sensitivity to hard processes compared to the classifiers measured based only on charged-particle multiplicity~\cite{Martin:2016igp,Bencedi:2021tst,Ortiz:2017jaz,Weber:2018ddv, Ortiz:2024ndh}. In both experiments as well as in the phenomenological fronts, extensive explorations using the above-mentioned event-shape observables have been performed~\cite{ALICE:2019dfi, ALICE:2022qxg, MovillaFernandez:1997fr, Donoghue:1979vi, Dasgupta:2003iq, Banfi:2004nk, Bethke:2009ehn, OPAL:1989fep, JADE:1982xwg,   Brandt:1964sa, ATLAS:2012tch, Banfi:2010xy, STAR:2002hbo, Cesarotti:2020hwb,  Cesarotti:2020ngq, ATLAS:2023mny, Martin:2016igp, Bencedi:2021tst,Ortiz:2017jaz, Ortiz:2024ndh, Weber:2018ddv, ALICE:2012cor, ATLAS:2012uka, ALICE:2011ac, ALICE:2019bdw, ALICE:2019mmy, ALICE:2022fnb, ALICE:2023yuk,ALICE:2023csm,  ALICE:2023bga,ALICE:2023dei,ALICE:2024vaf, CMS:2025sws, CMS:2011usu}. Also, we have explored the particle production dynamics as a function of transverse spherocity for light, strange~\cite{Rath:2019izg,Tripathy:2019blo,Khuntia:2018qox, R:2023lku,Prasad:2024gqq} and heavy flavor~\cite{Deb:2020ige,Khatun:2019dml} sectors in proton-proton collisions. The possibility of thermalization in small collision systems has been explored using spherocity as an event-shape observable in Ref.~\cite{Deb:2019yjo,Deb:2020ezw}. The global observables, identified particle production, and anisotropic flow coefficients are studied as a function of transverse spherocity in Pb--Pb collisions in Refs.~\cite{Mallick:2020dzv, Tripathy:2025npe, Prasad:2021bdq, Mallick:2020ium, Mallick:2021hcs, Prasad:2022zbr, Prasad:2025ezg}. 

 Event shape observables describe the structure of hadronic events and properties of their energy flow. Event shape observables are among the first observables proposed to test QCD, and they have played a key role in the progress of understanding both perturbative and nonperturbative aspects of QCD~\cite{MovillaFernandez:1997fr,Donoghue:1979vi, Dasgupta:2003iq,  Banfi:2004nk}. Event shape in the hadronic collisions was investigated first at the Intersecting Storage Rings~\cite{AxialFieldSpectrometer:1983drg} and at the S$p\bar{p}$S~\cite{UA1:1987esv,UA2:1987hpe} at CERN to examine the emergence of jets. Later, event shape observables are explored at Tevatron to study the dependence of transverse energy of the leading jet and contribution from underlying events~\cite{CDF:2011yfm}. In $e^{+}e^{-}$ collisions, event shape variables are the popular observables to improve our understanding of QCD dynamics. In $e^{+}e^{-}$ and $ep$ deep-inelastic scattering experiments, the study of hadronic final states as a function of event shape has allowed testing the predictions of pQCD and extraction of the strong interaction coupling constant $\alpha_{s}$~\cite{MovillaFernandez:2001ed, ZEUS:2002tyf,L3:2002oql,ALEPH:2003obs, DELPHI:2004omy, OPAL:2004wof, H1:2005zsk}. In the late 1970s, event shape observables were instrumental in studying the nature of gluon bremsstrahlung. One of the key results that were confirmed from the usage of event shape observables in experiments is the deduction of gluons as vector particles, although the theoretical prediction during that time suggested that gluons are scalar in nature. Event shape observables have the capability to distinguish the collisions based on their geometrical shapes, and they can measure the extent to which energy flow departs from a dijet structure in an event. Event shape observables are required to possess the property of continuous globalness and are normally categorised in three types, i.e., directly global observables, global observables shared with exponentially suppressed forward terms, and shared with recoil terms (indirectly global observables)~\cite{Banfi:2004nk}. The discussed event shapes in this article fall into the directly global observables category. Event shape observables are expected to be instrumental in testing QCD as, by construction, they are collinear and infrared safe observables~\cite{Dasgupta:2003iq} (see next section for more details). Thus, they do not change their value if a parton is split into two collinear partons or an extra soft gluon is added. This is an important condition for the cancellation of divergences associated with such gluon emissions, which makes them ideal tools for making finite pQCD predictions. Thanks to these criteria of event shape variables, several studies related to jets and heavy-flavors are being performed in LHC experiments. In addition, these event-shape classifiers have shown a significant correlation with the number of MPI ($N_{\rm mpi}$), which makes them the ideal tool for the understanding of QGP-like effects. 

In this review article, we extensively discuss the purpose, coverage, and usage guidelines of these event shape classifiers, which can be crucial in view of the upcoming measurements in the precision era at the LHC. Since 1970, there have been several event classifiers proposed and studied. Over the past decade, event topology classifiers have gradually become part of the common variables of collider physics. At the ATLAS and CMS collaborations, variables such as transverse sphericity, thrust, and heavy-jet mass are routinely used to separate hard and soft components of particle production and to characterize the underlying event. In the ALICE collaboration, similar techniques have opened new perspectives on collectivity, strangeness production, and multi-partonic interactions in small and large systems. Although each experiment has its own detector geometry and acceptance, the underlying ideas remain the same, i.e., capturing the shape and complexity of event dynamics in a way that is independent of model assumptions and free from biases. Their growing use across experiments shows how these observables now bridge the traditional boundaries between high-energy and heavy-ion physics, providing a unified framework for studying QCD matter at the LHC. In this article, we will mostly highlight the event classifiers which are popular in the early 21$^{\rm{st}}$ century since the commencement of the LHC in view of the studies of small collision systems at the LHC. We will briefly discuss the most fundamental event classifier for the high-energy collisions, i.e., the charged-particle multiplicity measured both in mid and forward rapidities. Then the discussion will be followed by transverse sphero(i)city, which uses the information of charged-particle transverse momenta and the azimuthal angle for each particle. Then, we will discuss the event shape classifiers such as $R_{\rm T}$, $R_{\rm T}^{\rm min}$, and $R_{\rm T}^{\rm max}$, which take advantage of the underlying event and use the information of charged-particle multiplicity in the transverse region and azimuthal angle of the particles. Last but not least, the study will be focused on a relatively new event shape classifier, charged-particle flattenicity, which takes the information of final state particle multiplicity in different pseudorapidity and azimuthal angle ranges. Along with their novelty, the above-discussed event classifiers show significant correlation with $N_{\rm mpi}$, which will be extensively discussed in the next section. In most cases, the event shape observables are found to be better probes in understanding the heavy-ion-like behavior seen at the LHC. However, the coverage of event shape classifiers usually overlaps with each other. Thus, one needs to have a better understanding of these classifiers while using them. Keeping this in mind, we will summarise the findings from the experimental and phenomenological front, done so far, and we will provide the usage recommendations of these event classifiers. Additionally, to gain insight into the salient aspects of the pQCD domain, we discuss the event shape dependence of several hard probe observables associated with jets and heavy flavour production. By systematically investigating the event shape dependence of observables in both the soft and hard QCD sectors, we achieve a unified and comprehensive understanding of event shape dynamics in QCD. In this direction, we provide particular emphasis on jets and heavy flavour production, guided primarily by experimental data. Eventually, to bring all collision systems on equal footing, we will also discuss the extension of these event topology classifiers to heavy-ion collisions, where the particle density in phase space is higher than in $pp$ collisions. In this direction, the usefulness of machine learning is highlighted. 

Wherever available, this review discusses experimental results based on event classifiers across LHC experiments, complemented by relevant observations from the pre-LHC era. Given that the primary aim of this review is to summarise the use of event topology classifiers rather than to exhaustively study all theoretical or phenomenological models, representative simulations from two widely used event generators are included to illustrate the physical interpretation of these observables: PYTHIA 8 for $pp$ collisions and A Multi-Phase Transport (AMPT)~\cite{Lin:2004en, He:2017tla} for heavy-ion collisions. These examples are intended to provide a qualitative understanding of how event-shape observables relate to multihadron production dynamics, rather than to present new model studies. PYTHIA is a pQCD-inspired general-purpose Monte Carlo event generator used to simulate relativistic hadronic, leptonic, and heavy-ion collisions from RHIC to LHC energies. In PYTHIA8, models and theories are incorporated for a number of physics aspects, such as parton distributions, hard and soft interactions, initial- and final-state parton showers, multiparton interactions, fragmentation, partonic and hadronic decays~\cite{Corke:2010yf,Andersson:1983ia}. A detailed explanation of all physics processes involved in PYTHIA8 can be found in Ref.~\cite{Sjostrand:2007gs, manual}. The two key features of PYTHIA8, i.e., color reconnection and rope hadronisation, are explicitly discussed in the Appendix~\ref{appendixpythia} of the review. AMPT model, by contrast, is a hybrid model that combines both the initial partonic collisions and the final hadronic interactions effectively, as well as accounts for the transition between these two distinct phases of matter. It has four major components, namely, initial conditions, partonic interactions, conversion of partonic to hadronic matter through the hadronization mechanism, and hadron interactions. Appendix~\ref{sec:otherMCmodels}, highlights several additional MC models that are relevant to this review.

Moreover, in this review, the phenomena referred to as “QGP-like” in small collision systems are not treated as evidence for the formation of a distinct deconfined phase of partons. These observed phenomena emerge as a consequence of underlying multi-hadron production dynamics, which become prominent in high multiplicity environments. On the other hand, the event shape classifiers are used to select the sample with a specific final state topology, which is characterised by either high particle density in the final state or back-to-back jet topology. In addition, there are different physical processes that have distinct momentum space correlations, which can be captured by different event classifiers. These specific conditions, by the selection of specific event classifiers, can amplify the “QGP-like” signals irrespective of whether the origin is hydrodynamic, partonic, or purely hadronic in nature. Therefore, the observation of “QGP-like” signals in these selected events should be interpreted as the interplay among the manifestation of multihadron production dynamics, hadronization mechanisms, and global event geometry rather than referring only to the formation of QGP. Thus, in this review, the use of event shape classifiers is to classify and organize events based on the different multi-hadron production dynamics, while the question of ultimate physics interpretation of the observed mechanisms is open and model dependent.

The paper is organized as follows. We begin with a brief introduction and motivation about the event shape observables in Sec.~\ref{sec:intro}. In Sec.~\ref{sec:definitions}, we define each event classifier. Section~\ref{prelhcera} discusses the event shape classifiers from the pre-LHC era and provides a precursor for studies related to the LHC era. Section~\ref{sec:softprobes} discusses the soft probes with event classifiers, especially highlighting selection biases, correlation among the classifiers, and particle ratios obtained in $pp$ collisions. Section~\ref{hardprobes} briefly discusses the studies related to hard probes, i.e., jets and heavy-flavors with event shape classifiers. Further, we extend the discussion of event shape studies to heavy-ion collisions in Sec.~\ref{sec:HIC}. Section~\ref{machinelearning} provides a brief summary on the machine learning techniques to classify the events at LHC.  %The important findings of each event classifier are summarized in Sec.~\ref{sec:finalwords}. 
Section~\ref{sec:summary} presents the summary and provides a brief outlook. The event generation methodology and event/track selection criteria used according to experiments are discussed in the appendix Sec.~\ref{appendix}. To guide the readers, the summary of each section is mentioned in {\it italic} font at the end of the corresponding section.

\section{Event Classifiers}
\label{sec:definitions}

In this section, we define different event shape observables used at the LHC energies. In addition, the correlation of these event shape observables with  $N_{\rm mpi}$ is shown and discussed.
\noindent
\subsection{Different types of events: elastic and inelastic collisions}

\begin{figure}
    \centering
    \includegraphics[width=0.9\linewidth]{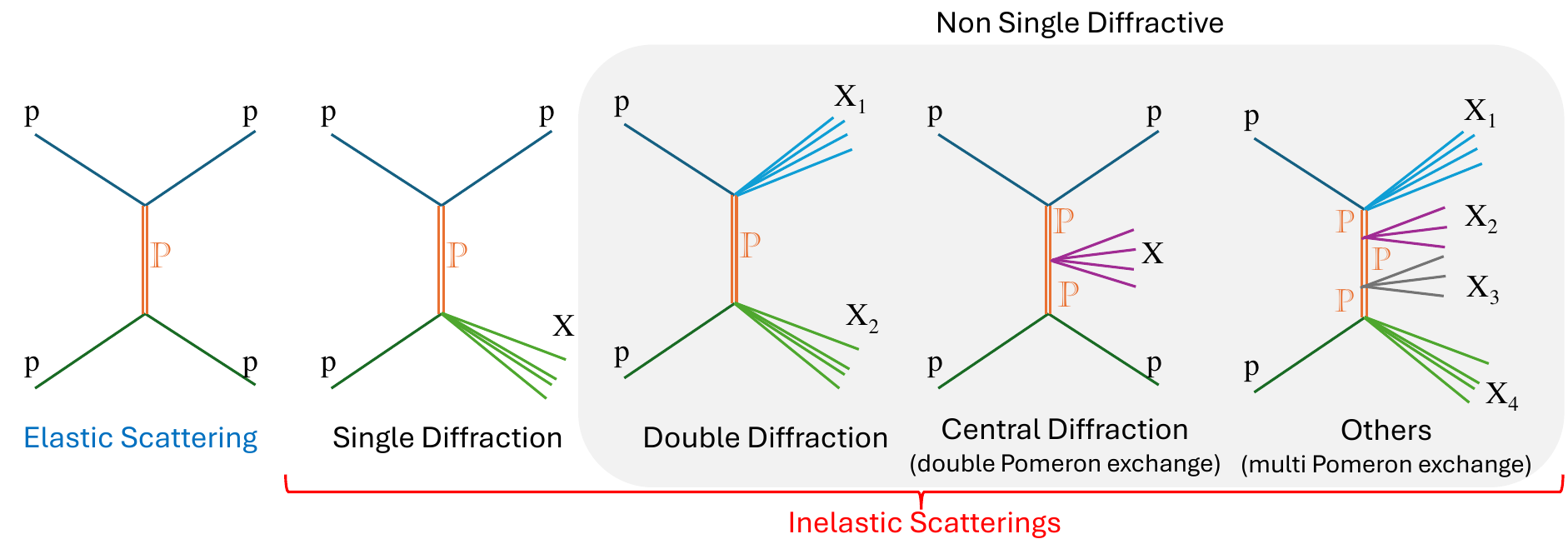}
    \caption{Schematic showing different types of interactions in $pp$ collisions.}
    \label{fig:elasticinelastic}
\end{figure}

The final state particle production in hadronic collisions, such as proton-on-proton, is a probe for the initial interactions among the colliding hadrons. An interaction can be classified as elastic or inelastic. In elastic collisions, the incoming and outgoing protons are the same, which takes place via exchange of Pomeron\footnote{Pomeron is a colour singlet particle that dominates the elastic scatterings at high energies and has the quantum numbers of the vacuum.}~\cite{ALICE:2012fjm}. In contrast, in inelastic collisions, the final state particles are not identical to the initial colliding protons. An inelastic event can be diffractive, which is signified by a small energy exchange between the protons. Sometimes, one or both protons interact with the Pomerons and dissociate into multiple final-state particles. The event is single diffractive (SD) if only one proton dissociates while the other one survives as it is ($p+p\rightarrow p+X$). Similarly, when both the protons dissociate to form multiple particles in the final state, it is called a double diffractive (DD) event ($p+p\rightarrow X_1 +X_2$). Sometimes, two Pomerons are exchanged, and both the protons survive along with the production of new particles, i.e., $p+p\rightarrow p+p+X$, it is called a central diffractive (CD) event. Additionally, when more than two Pomerons are exchanged, there is a possibility for both protons to dissociate and lead to the production of new particles. These events are usually accompanied by large particle multiplicity in the final state as compared to other diffractive events. Figure~\ref{fig:elasticinelastic} shows the pictorial representation of different types of possible interactions in $pp$ collisions. The inelastic events without SD are called Non-Single Diffractive (NSD). At the LHC, the measurements are usually performed for NSD or inelastic (INEL) events.

\subsection{Impact parameter and centrality}
\begin{figure}[ht!]
\centering
\includegraphics[width=0.65\linewidth]{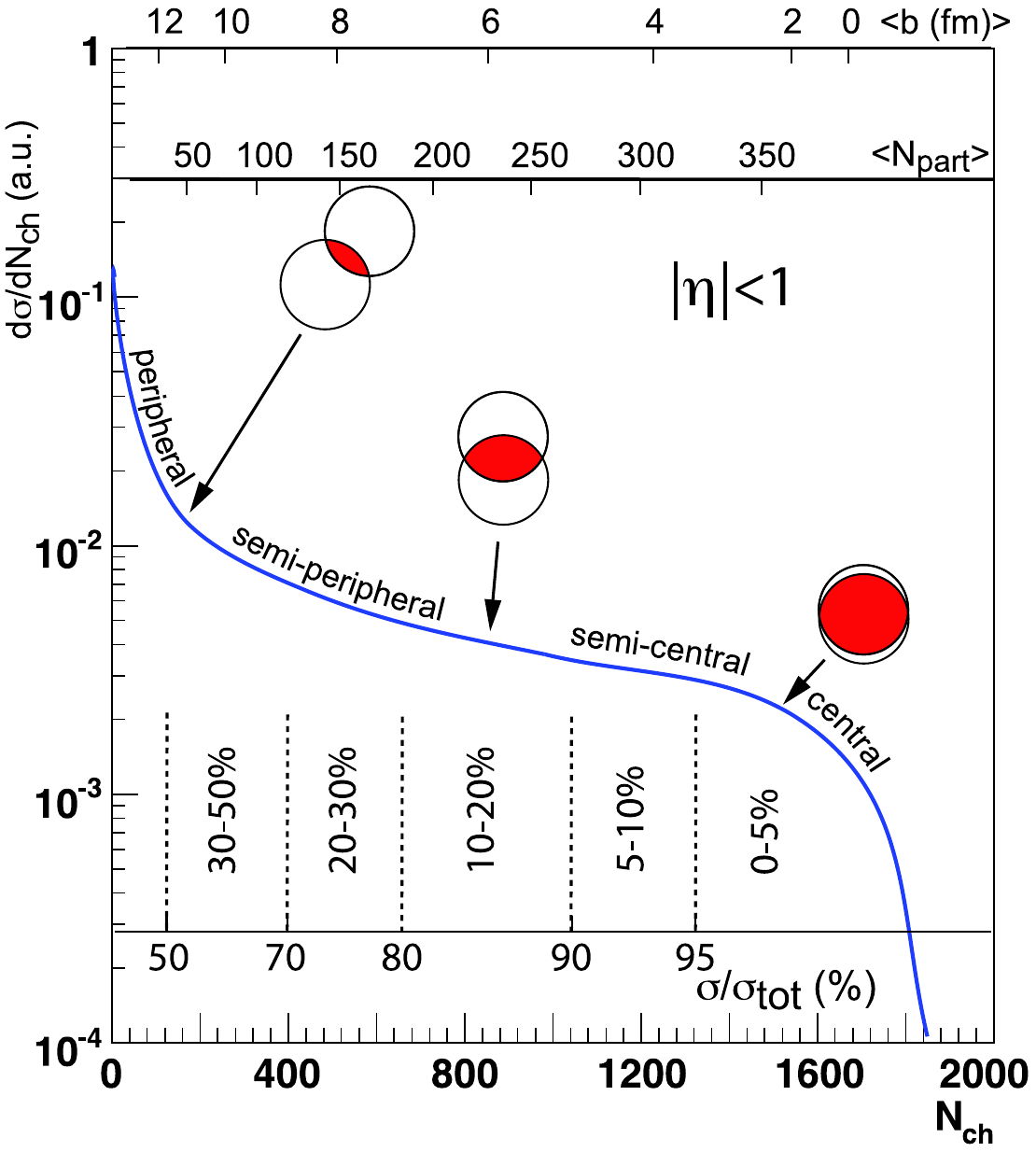}
\caption{A schematic representation of the correlation between the final state charged-particle multiplicity distribution ($N_{\rm ch}$) with the Glauber calculated quantities ($b$, and $N_{\rm part}$). The collision centrality classification in experiments is shown explicitly. Figure is taken from Ref.~\cite{Miller:2007ri}}.  
\label{fig:cocktail}
\end{figure}

The impact parameter ($b$) of a collision is one of the key parameters that substantially influence the particle production in the final state. However, its measurement is not trivial in experiments as the length scale of the impact parameter ranges in the level of a few fermis. Thus, experimental data are typically categorised by the centrality, defined as percentile of events obtained from the number of produced particles or by the number of participant nucleons, $N_{\rm part}$ that is registered in detectors through the estimation of number of spectator nucleons via $N_{\rm part} = 2A - N_{\rm spec}$, where $A$ is the mass number of the colliding nuclei (for symmetric collisions) and $N_{\rm spec}$ is the number of spectator nucleons. For example, in ALICE experiment, $N_{\rm spec}$ is given by $E_{\rm ZDC}/E_{\rm A}$, where, $E_{\rm ZDC}$ is the energy deposited in zero-degree calorimeters (ZDC) and $E_{\rm A}$ is beam energy per nucleon. Centrality allows for dividing events into different classes. However, phenomenologically, one needs to assign an impact parameter to a given centrality as depicted in Fig.~\ref{fig:cocktail}. Thus, theoretical techniques, using the Glauber formalism~\cite{Miller:2007ri,Loizides:2016,Glauber:1970,Wong:1994book} have been developed to allow the estimation of impact parameter along with the number of participant nucleons from experimental data, which considers multiple scattering of nucleons in nuclear targets. Here, we briefly explain how the total inelastic cross section, the number of binary collisions, and the number of participants depend on the impact parameter.

\begin{figure}
\centering
\includegraphics[width=0.8\linewidth]{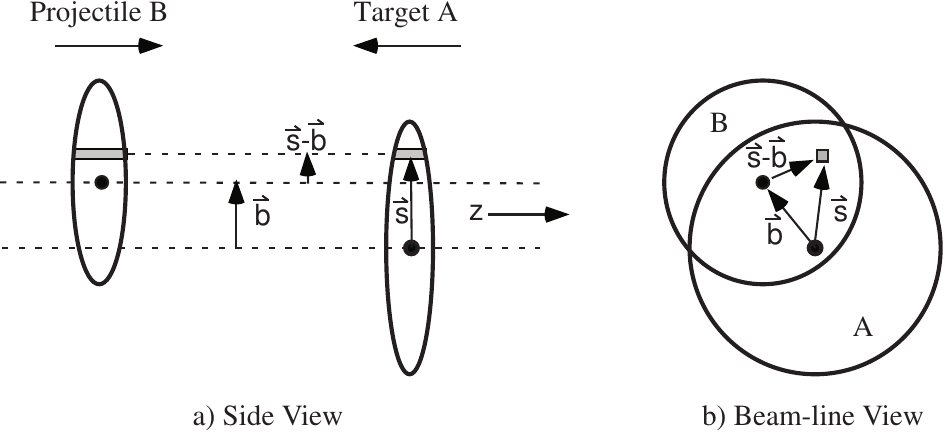}
\caption{
Schematic representation of the Optical Glauber Model geometry, with
transverse (a) and longitudinal (b) views. Figure is taken from Ref.~\cite{Miller:2007ri}} 
\label{fig:GlauberGeo}
\end{figure}
For a collision of two heavy-nuclei, $A$ and $B$ at relativistic speeds with impact parameter ${\bf b}$ as shown in Fig.~\ref{fig:GlauberGeo}, the inelastic cross section can be defined as

\begin{eqnarray}
\sigma^{\rm inel}_{AB}({\bf b})&=&\int 2\pi b \; d{\bf b}\left[1-\left(1-T_{AB}({\bf b})\sigma_{NN}^{\rm inel}\right)^{AB}\right]\\
&\simeq&\int 2\pi b \;\; d{\bf b}\left[1-\exp\left[-ABT_{AB}({\bf b})\sigma_{NN}^{\rm inel}\right]\right]\, .
\end{eqnarray}
Here, $T_{AB}({\bf b})$ is the nuclear overlap function and $\sigma_{\rm NN}^{\rm inel}$ denotes the nucleon-nucleon inelastic cross section.  For such nucleus-nucleus collisions, the total number of binary collisions is given as:
\begin{eqnarray}
N_{\rm coll}^{AB}({\bf b})&=&\sum_{n=1}^A n P(n,{\bf b})=ABT_{AB}({\bf b})\sigma_{\rm NN}^{\rm inel}\, ,
\end{eqnarray}
where $P(n,{\bf b})$ is the total probability of an interaction between nuclei $A$ and $B$. For a given impact parameter, {\bf b}, the number of participants  (or wounded nucleons) of nucleus $A$ is given by
\begin{eqnarray}
N_{\rm part}^A({\bf b})=B\int {{T}_B \left({\bf{s}-\bf{b} } \right)\left\{ {1 - \left[ {1 - T_A \left( {\bf{ s} } \right)\sigma^\mathrm{inel}_{\rm NN} } \right]^A }\right\}d^2 s}.
\end{eqnarray}
The number of participants in nucleus $A$ is proportional to the nuclear profile function at transverse positions ${\bf s}$, $T_{AB}({\bf s})$, weighted by the sum over the probability for a nucleon-nucleon collision at transverse position (${\bf {s-b}}$) in nucleus $B$. Thus at a given ${\bf b}$, the number of participants is given by
\begin{equation}
N_{\rm part}({\bf b})=N_{\rm part}^A({\bf b})+N_{\rm part}^B({\bf b})\, .
\end{equation}

In heavy-ion physics, theoretical calculations use $\bf b$ as an input to compare theoretical results to the experimental measurements. Both $N_{\rm part}({\bf b})$ and $N_{\rm coll}({\bf b})$ are calculated using Glauber model at a given $\bf b$, which are subsequently related to final state particle multiplicities~\cite{Kolb:2001}. 

In $pp$ collisions, one cannot define impact parameter and centrality as they are point-like collisions of two nucleons. In the Glauber model framework, which is used to describe heavy-ion collisions, treating the proton as point-like is an approximation. Thus, in experiments, the final state charged particle multiplicity is used to divide the event classes, which we discuss in more detail in the next subsection.

\subsection{Charged-particle multiplicity ($N_{\rm ch}$)}

\begin{figure}[ht!]
\begin{center}
\includegraphics[scale = 0.33]{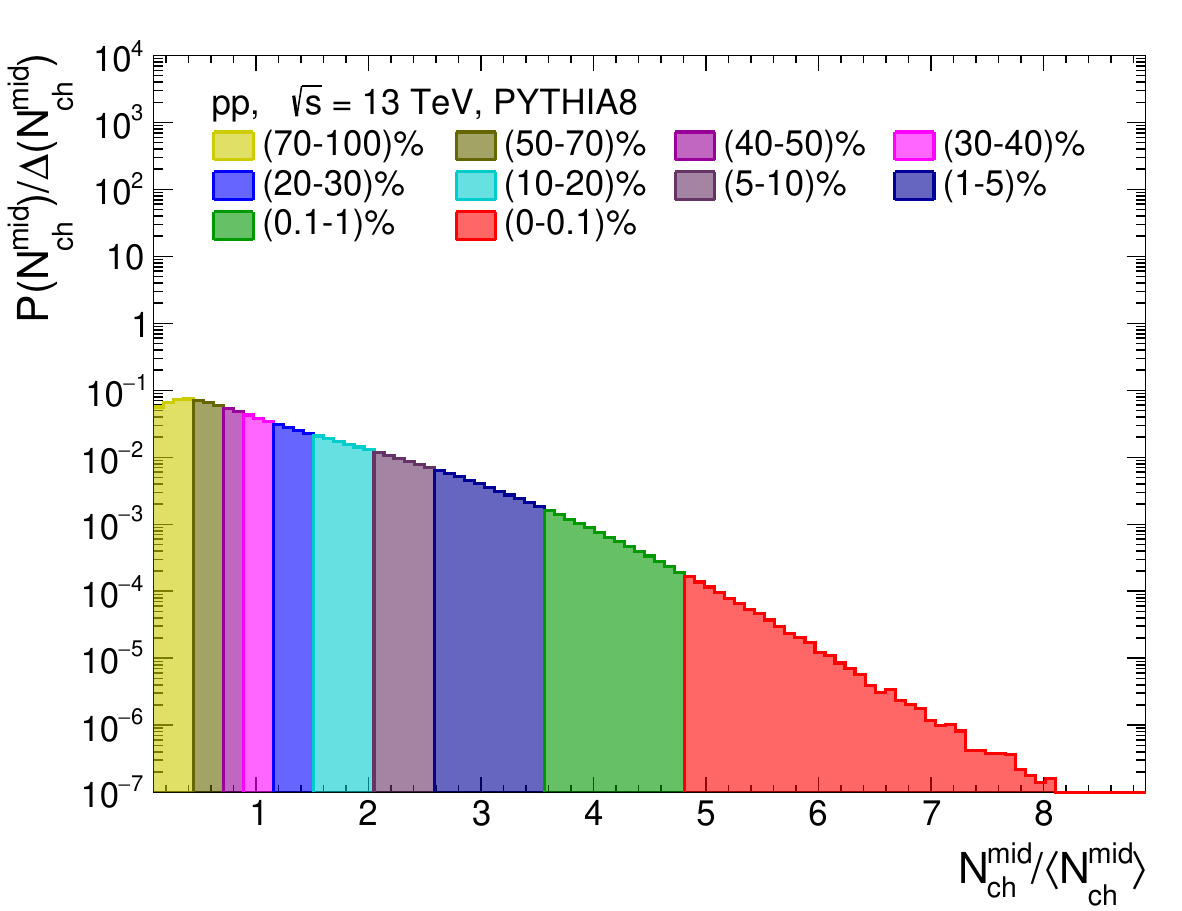}
\includegraphics[scale = 0.33]{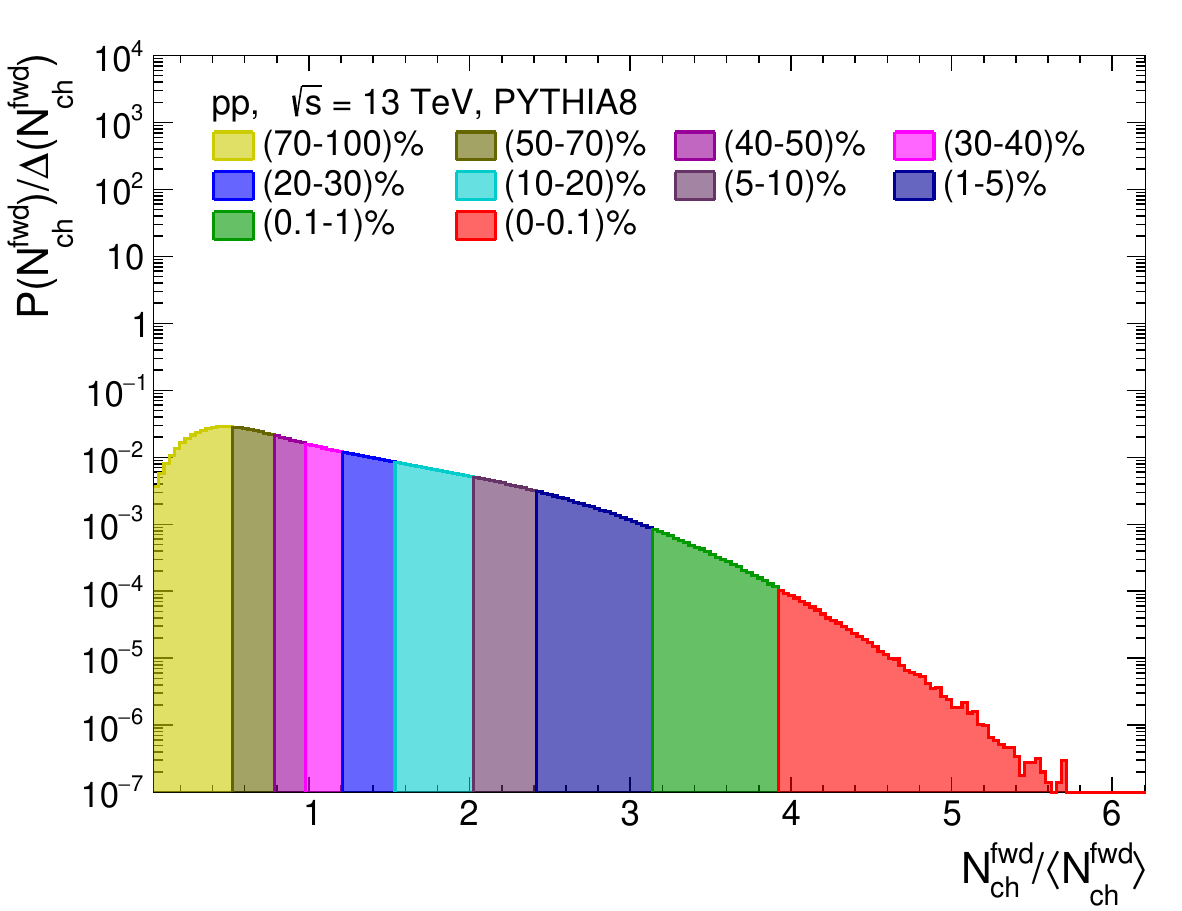}
\caption{Self-normalised charged-particle multiplicity distributions measured in mid (left panel) and forward (right panel) rapidities at $\sqrt{s}=13$ TeV using PYTHIA8.}
\label{fig:NchDist}
\end{center}
\end{figure}

One of the most fundamental observables to probe the system size of high-energy collisions is charged-particle multiplicity. In the ALICE experiment at the LHC, the charged-particle multiplicity is estimated at mid- and forward-pseudorapidity regions. The charged-particle multiplicity at the mid-pseudorapidity region (\textit{i.e.}, $|\eta| < 0.8$), denoted as $N_{\rm ch}^{\rm mid}$, is measured by the silicon pixel detector (SPD) and the time projection chamber (TPC). Contrary, the measurement of charged-particle multiplicity at the forward pseudorapidity region, denoted as $N_{\rm ch}^{\rm fwd}$, is performed by the V0 (V0A+V0C) detectors having the pseudorapidity coverage of $-3.7 < \eta < -1.7$ (V0C) and $2.8 < \eta < 5.1$ (V0A). The left and right panels of Fig.~\ref{fig:NchDist} show the self-normalised charged-particle multiplicity distributions obtained in mid and forward rapidity, respectively, at $\sqrt{s}=13$ TeV using PYTHIA8. The distributions are further divided into different percentiles and specified as multiplicity classes. 

Apart from the SPD, TPC, and V0 detectors discussed above, with the recent upgradation of ALICE in Run 3, the global detector named Fast Interaction Trigger (FIT) is used to estimate the collision centrality and the charged-particle multiplicity. FIT consists of three sub-detectors, namely, FT0, FV0, and FDD. The detailed description of the FIT detector can be found in Ref.~\cite{Bysiak:FIT}. The newly built Muon Forward Tracker (MFT) at forward rapidity ($-3.6 < \eta < -2.5$) can also be used to estimate the charged-particle multiplicity in Run 3 of ALICE~\cite{ALICE:MFTTDR}. Additionally, the CMS experiment at the LHC uses pairs of pixel clusters (known as tracklets) from two different layers (disks) of the silicon pixel detector to measure the charged-particle multiplicity produced in the range $|\eta| < 2.6$. These pairs have clusters with relatively small differences in $\eta$ and $\phi$. The correlation between $\eta$ and $\phi$ can be used to select tracklets corresponding to primary charged hadrons. The detailed description of tracklets and vertex reconstruction algorithms can be found in Ref.~\cite{ CMS:2024ykx, CMS:2011aqh, CMS:2010qvf}. The ATLAS experiment at the LHC uses tracking detectors such as the Inner Detector (ID) and the trigger detectors to measure the charged-particle multiplicity. The ID consists of a silicon pixel detector (Pixel), a silicon microstrip detector (SCT), and a transition radiation tracker (TRT) with almost full coverage in $\phi$, and $|\eta| < 2.5$. The detailed description of trigger selection and track reconstruction algorithms can be found in Ref.~\cite{ATLAS:2010jvh, ATLAS:2010zmc,ATLAS:2016zkp}. These three (ALICE, CMS, and ATLAS) are the major experiments at the LHC, and their measurement of final-state charged particle multiplicity as an event classifier is discussed.

\subsection{Number of multi-partonic interactions ($N_{\rm mpi}$)}

\begin{figure}[ht!]
\begin{center}
\includegraphics[scale = 0.4]{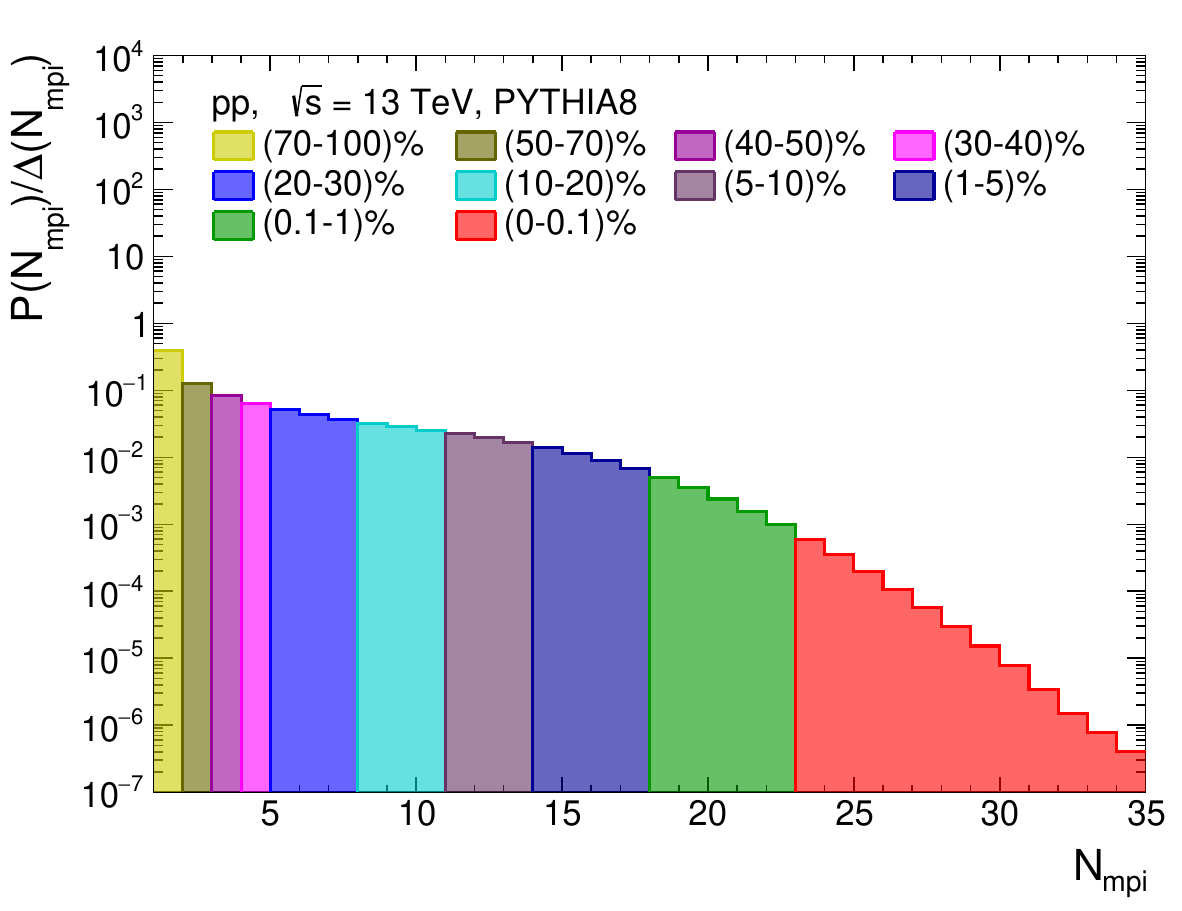}
\caption{Distribution of $N_{\rm mpi}$ in $pp$ collisions at $\sqrt{s}=13$ TeV using PYTHIA8.}
\label{fig:NmpiDist}
\end{center}
\end{figure}

Hadronic collisions at RHIC and LHC produce a large number of particles in the final state. The fundamental description of particle production in hadronic collisions, such as $pp$, can be understood using the QCD-inspired parton model.~\cite{Sjostrand:2004pf}. In this model, the hadrons are treated as a collection of elementary constituents -- quarks and gluons, collectively referred to as partons -- bound together by the strong interaction. In minimum-bias hadronic collisions, multiple partonic interactions are inevitable due to the composite structure of hadrons~\cite{Sjostrand:2004pf, Bartalini:2011jp}, with these MPIs typically being soft in nature. Additionally, in a hard $pp$ collision, which is used for testing the Standard Model (SM) to high accuracy or sometimes to hunt for physics beyond the SM, a hard scattering is accompanied by a number of additional soft interactions, known as underlying event activity (UE). The comprehension and accurate modelling of these additional soft interactions are thus crucial for properly interpreting these hard interactions. Due to the non-perturbative nature, large numbers, and interplay among these interactions, the current approximation of these additional scattering can only be modelled in the context of Monte Carlo (MC) simulation. Here, for an event with a fixed final state invariant mass, with the increase in the collision energy of the hadronic species, one can probe to a lower momentum fraction $x$, leading to a higher partonic flux and an increase in the multiple-partonic scattering cross section~\cite{Sjostrand:2004pf, Bartalini:2011jp}. Consequently, events with a lower invariant mass in the final state can have influence from larger multiple hard scatterings~\cite{Bartalini:2011jp}. In addition, most inelastic events in hadronic collisions contain several perturbatively calculable interactions when the pQCD theory is extended to the low-$p_{\rm T}$ regime with some distance above $\Lambda_{\rm QCD}$~\cite{Sjostrand:2004pf}. Such interactions in inelastic hadronic collisions, even though soft and contributing to MPIs abundantly, are of much significance as they can significantly affect the particle multiplicity in the final state~\cite{Sjostrand:2004pf}. Thus, with an increase in the soft event activity, one expects the corresponding $N_{\rm mpi}$ to rise. In other words, an event with a large value of $N_{\rm mpi}$ is most likely to be a softer event.

The authors in Ref.~\cite{Christiansen:2024bhe} provide an abstract discussion of MPI that can be applicable to most of the models. As MPIs are modeled as independent $2\rightarrow2$ partonic interactions, the corresponding pQCD cross section diverges and finally exceeds the total inelastic cross section for the limit $p_{\rm T}\rightarrow0$. Since the ratio of pQCD cross section to the total cross section is interpreted as the average number of MPIs per event, the divergence is not inherently problematic. Nevertheless, to avoid nonphysical values of cross sections at $p_{\rm T}\rightarrow0$, regularization is necessary. This regularization can be associated with color-screening effects, which limit the spatial resolution of individual partons as implemented in PYTHIA, or with initial-state saturation mechanisms as in EPOS.

In PYTHIA, a transverse momentum cut off scale, $p_{\rm T0}$ is introduced to regularize $2\rightarrow2$ partonic subprocesses, as follows:

\begin{equation}
\frac{{\rm d}\hat{\sigma}}{{\rm d}p_{\rm T}^{2}}\propto \frac{\alpha_{S}^{2}(p_{\rm T}^{2})}{p_{\rm T}^{4}}\rightarrow \frac{\alpha_{S}^{2}(p_{\rm T0}^{2}+p_{\rm T}^{2})}{(p_{\rm T0}^{2}+p_{\rm T}^{2})^{2}}.
\end{equation}
Here, $p_{\rm T0}$ depends upon the center of mass energy of the collision, given as:
\begin{equation}
p_{\rm T_{0}}(s)=p_{\rm T}^{\rm ref}\Bigg(\frac{\sqrt{s}}{\sqrt{s_0}}\Bigg)^{\epsilon}.
\end{equation}
	
Here, $p_{\rm T}^{\rm ref}$ and $\epsilon$ are free parameters that can be tuned to the experiments. $p_{\rm T0}$ increases with an increase in collision energy, which is due to the increase in the low-$x$ partons in the wavefunction of colliding protons. In PYTHIA~8, $N_{\rm{mpi}}$ additionally depends on the impact parameter, which is determined by the spatial overlap of the transverse profiles of the colliding protons. This impact-parameter dependence introduces a geometrical component to the MPI framework, which makes the description of MPI more realistic.

At the LHC, recent measurements of heavy-ion-like signatures for high multiplicity $pp$ collisions can be explained qualitatively with perturbative QCD-based event generators such as PYTHIA8~\cite{ALICE:2019avo}. These heavy-ion-like features are well reproduced when one probes high MPIs with color reconnection. However, the estimation of $N_{\rm mpi}$ in experiments is difficult. Thus, one looks for event shape observables that can be proportional to $N_{\rm mpi}$. This makes the studies of the correlation of MPI with event-shape observables crucial in MC simulations. Figure~\ref{fig:NmpiDist} shows the distribution of $N_{\rm mpi}$ obtained in $pp$ collisions at $\sqrt{s}=13$ TeV using PYTHIA8. Different colour representations in the figure denote different percentile cuts of the distribution, specified as different $N_{\rm mpi}$ classes. 

\begin{figure}[ht!]
\begin{center}
\includegraphics[scale = 0.33]{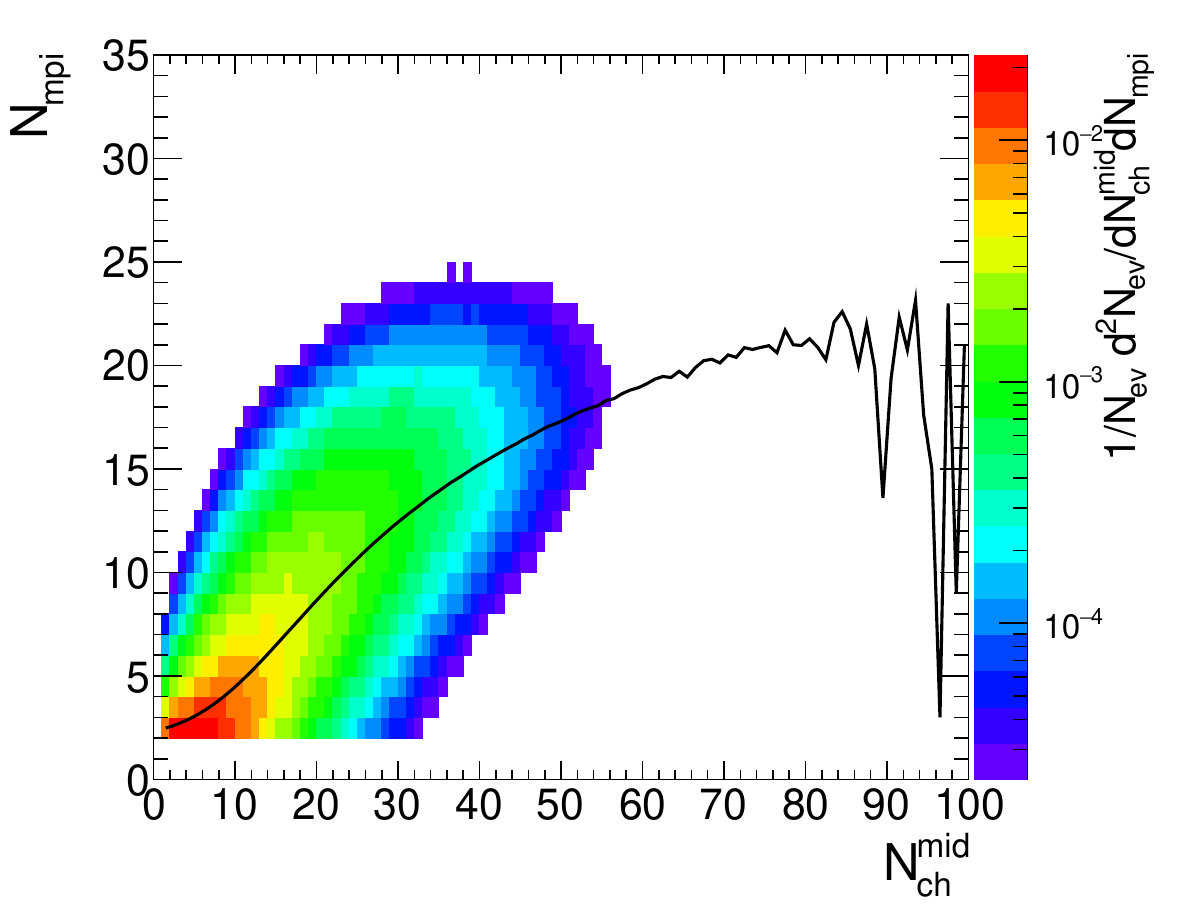}
\includegraphics[scale = 0.33]{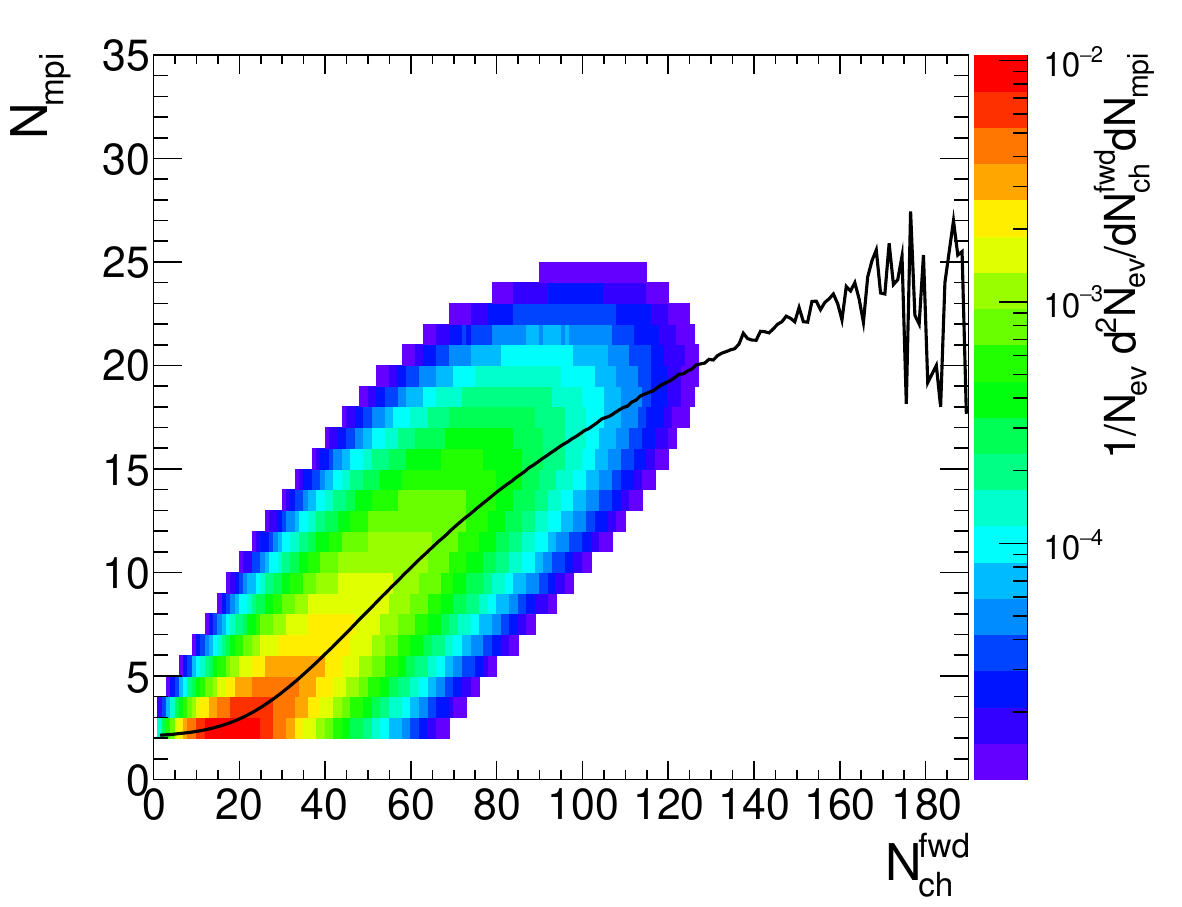}
\caption{Correlation of mid (left panel) and forward (right panel) rapidity charged-particle multiplicity with number of MPI in $pp$ collisions at $\sqrt{s}=13$ TeV using PYTHIA8. The solid line represents the average value of $y$-axis, \textit{i.e.}, $\langle N_{\rm mpi}\rangle$, for each bin in the $x$-axis, \textit{i.e.}, $N_{\rm ch}^{\rm mid}$ (left) and $N_{\rm ch}^{\rm fwd}$ (right).}
\label{fig:NchvsMPI}
\end{center}
\end{figure}

Figure~\ref{fig:NchvsMPI} shows the correlation of mid-rapidity charged-particle multiplicity ($N_{\rm ch}^{\rm mid}$) (left panel) and forward rapidity charged-particle multiplicity ($N_{\rm ch}^{\rm fwd}$) (right panel) with the number of $N_{\rm mpi}$ in $pp$ collisions at $\sqrt{s}=13$ TeV using PYTHIA8. The solid line shows the mean value of $N_{\rm mpi}$, i.e., $\langle N_{\rm mpi}\rangle$, for a given multiplicity value on the $x-$axis. As one notices, both mid- and forward-charged-particle multiplicity retain a fair degree of correlation with $N_{\rm mpi}$. From the figure, it is clear that irrespective of the pseudorapidity region, as the charged-particle multiplicity increases, the corresponding $\langle N_{\rm mpi}\rangle$ also increases, indicating that the events having higher value of $N_{\rm mpi}$ tend to produce a large number of particles at the final state and are softer as compared to the events having lower $N_{\rm mpi}$ value. However, a saturation behavior of the correlation of $N_{\rm mpi}$ with $N_{\rm ch}^{\rm mid}$ is seen for higher values of $N_{\rm ch}^{\rm mid}$. This indicates that beyond $N_{\rm ch}^{\rm mid}\simeq 50$, the increase in multiplicity in the mid-rapidity region probes the events with similar $N_{\rm mpi}$, $i.e,~\simeq$ 20.  In addition, one notices that using $N_{\rm ch}^{\rm fwd}$ one can reach a higher value of $\langle N_{\rm mpi}\rangle$ as compared to the usage of $N_{\rm ch}^{\rm mid}$.
%it is worth mentioning that the correlation of $N_{\rm ch}^{\rm fwd}$ with $N_{\rm mpi}$ is slightly larger as compared to the correlation of $N_{\rm ch}^{\rm mid}$ with $N_{\rm mpi}$.

\subsection{Transverse sphericity ($S_{\rm{T}}$)}
\begin{figure}[ht!]
\begin{center}
\includegraphics[scale = 0.4]{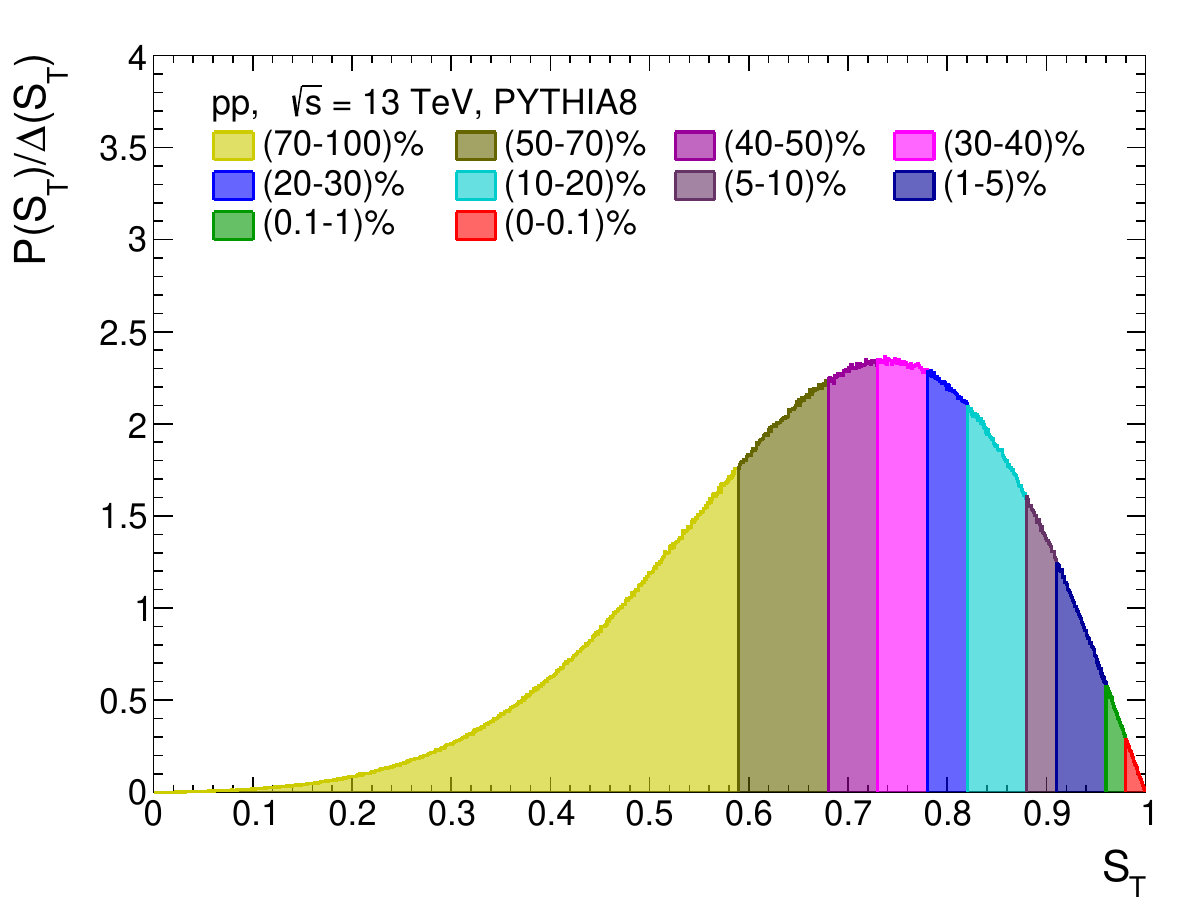}
\caption{Transverse sphericity distributions measured in $pp$ collisions at $\sqrt{s}=13$ TeV using PYTHIA8.}
\label{fig:STdist}
\end{center}
\end{figure}
Transverse sphericity, one of the oldest event shape observables, was used at the Stanford Linear Accelerator
Center (SLAC) to test the existence of jets in $e^{+}e^{-}$ processes at collision energies up to 7.4 GeV~\cite{Hanson:1975fe}.
Transverse sphericity is defined  in terms of the eigenvalues of the transverse momentum matrix ($S_{\rm xy}^{\rm Q}$), given as~\cite{ALICE:2012cor}:
\begin{equation}
    S_{\rm xy}^{\rm Q}=\frac{1}{\sum_{i}p_{{\rm T}_{i}}}\sum_{i}\begin{pmatrix}
p_{{\rm x}_{i}}^{2} & p_{{\rm x}_{i}}p_{{\rm y}_{i}} \\
p_{{\rm y}_{i}}p_{{\rm x}_{i}} & p_{{\rm y}_{i}}^{2}
\end{pmatrix} \text{[GeV/$c$].}
\label{SQxy}
\end{equation}
Here, $(p_{{\rm x}_{i}},p_{{\rm y}_{i}})$ are the projections of transverse momentum ($p_{\rm T_{i}}$) of $i$th particle in $(x,y)$ directions and $i$ runs over all the charged hadrons. Q denotes the quadrature property of the particle momenta, which makes the transverse momentum matrix in Eq.~\ref{SQxy}, $S_{\rm xy}^{\rm Q}$ is non-collinear safe. To make it collinear-safe, Eq.~\ref{SQxy} can be linearized as follows,
\begin{equation}
    S_{\rm xy}^{\rm L}=\frac{1}{\sum_{i}p_{{\rm T}_{i}}}\sum_{i}\frac{1}{p_{{\rm T}_{i}}}\begin{pmatrix}
p_{{\rm x}_{i}}^{2} & p_{{\rm x}_{i}}p_{{\rm y}_{i}} \\
p_{{\rm y}_{i}}p_{{\rm x}_{i}} & p_{{\rm y}_{i}}^{2}
\end{pmatrix} 
\label{SLxy}
\end{equation}
\href{}{}
\begin{figure}[ht!]
\begin{center}
\includegraphics[scale = 0.4]{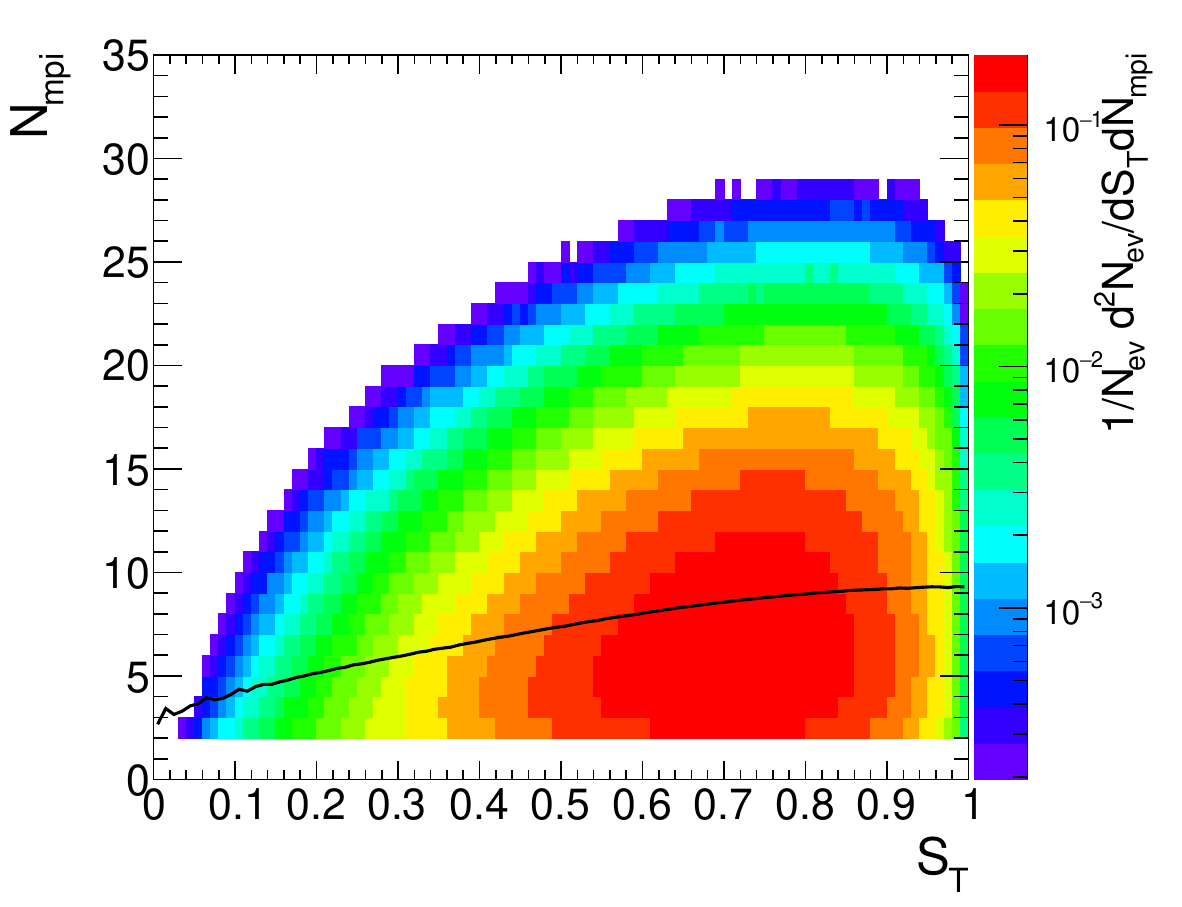}
\caption{Correlation of $S_{\rm T}$ with number of MPI. The black solid line represents $\langle N_{\rm mpi}\rangle$ as a function of $S_{\rm T}$.}
\label{fig:STvsMPI}
\end{center}
\end{figure}

Transverse sphericity can be defined in terms of the eigenvalues, $\lambda_{1}$, and $\lambda_{2}$ of the linearized transverse momentum matrix $S_{\rm xy}^{\rm L}$ such that $\lambda_1>\lambda_2$, and defined as follows.

\begin{eqnarray}
    S_{\rm T}=\frac{2\lambda_{2}}{\lambda_{1}+\lambda_{2}}
    \label{ST}
\end{eqnarray}

By construction, the value of $S_{\rm T}$ lies between 0 and 1. Here, the extreme limits, i.e., $S_{\rm T}\rightarrow 0$ selects the jetty or pencil-like events, while $S_{\rm T}\rightarrow 1$ indicates that the events are isotropic, which are dominated by the soft production of particles. In this work, $S_{\rm T}$ has been estimated with the charged particles in the mid-pseudorapidity region, i.e., $|\eta|<0.8$ having $p_{\rm T}>0.15$ GeV/$c$. Figure~\ref{fig:STdist} shows the distribution of transverse sphericity measured in $pp$ collisions at $\sqrt{s}=13$ TeV using PYTHIA8. In addition, different percentile slices are shown in the $S_{\rm T}$ distribution, referred to as sphericity classes, which are shown with different colours. Figure~\ref{fig:STvsMPI} shows the correlation of transverse sphericity with the number of multi-partonic interactions in $pp$ collisions at $\sqrt{s}=13$ TeV using PYTHIA8, where the solid line shows the mean value of $N_{\rm mpi}$ for each bin in $S_{\rm T}$. From Fig.~\ref{fig:STvsMPI}, it is clear that the transverse sphericity possesses a finite positive correlation with the number of multi-partonic interactions.

\subsection{Transverse spherocity ($S_{0}$)}
\label{sec:Spherodefn}
\begin{figure}[ht!]
\begin{center}
\includegraphics[scale = 0.4]{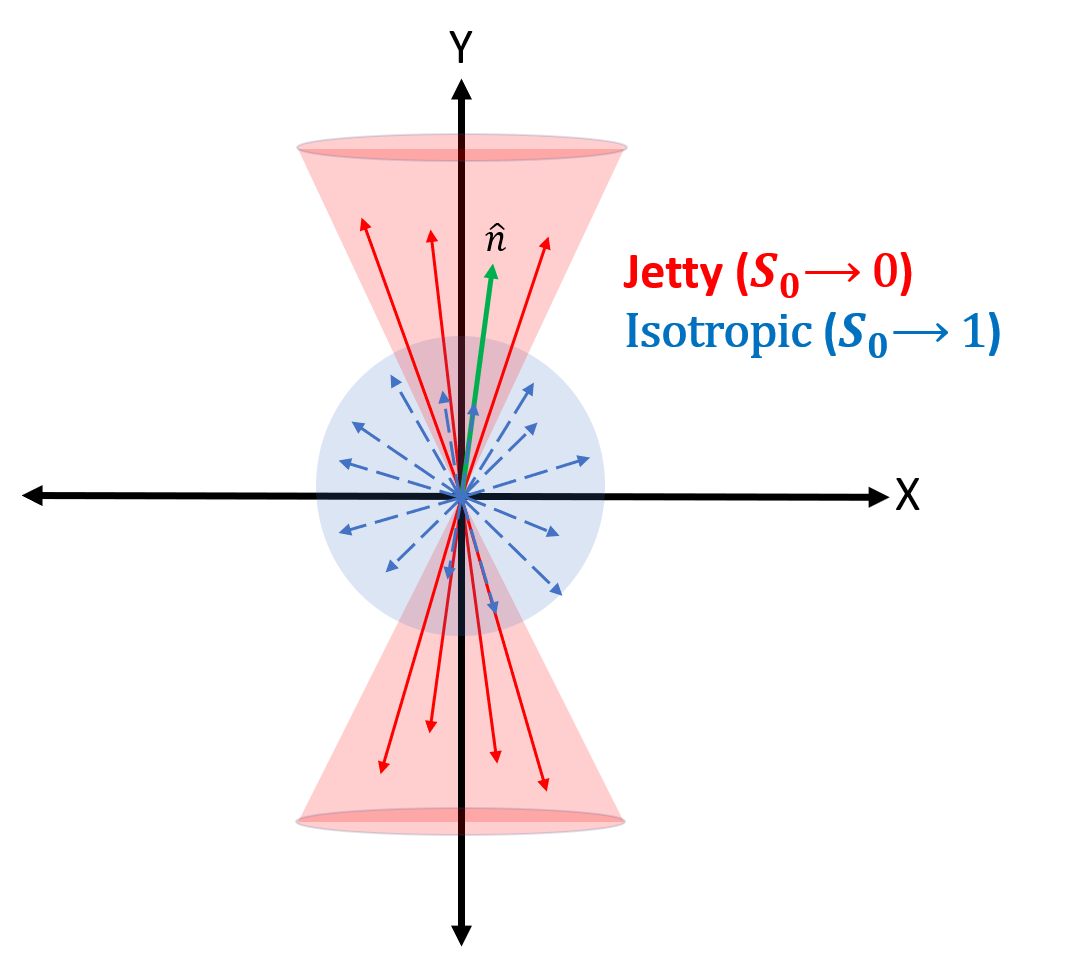}
\caption{Depiction of isotropic and jetty events based on the transverse spherocity selection~\cite{Khuntia:2018qox}.}
\label{fig:sphero}
\end{center}
\end{figure}

\begin{figure}[ht!]
\begin{center}
\includegraphics[scale = 0.33]{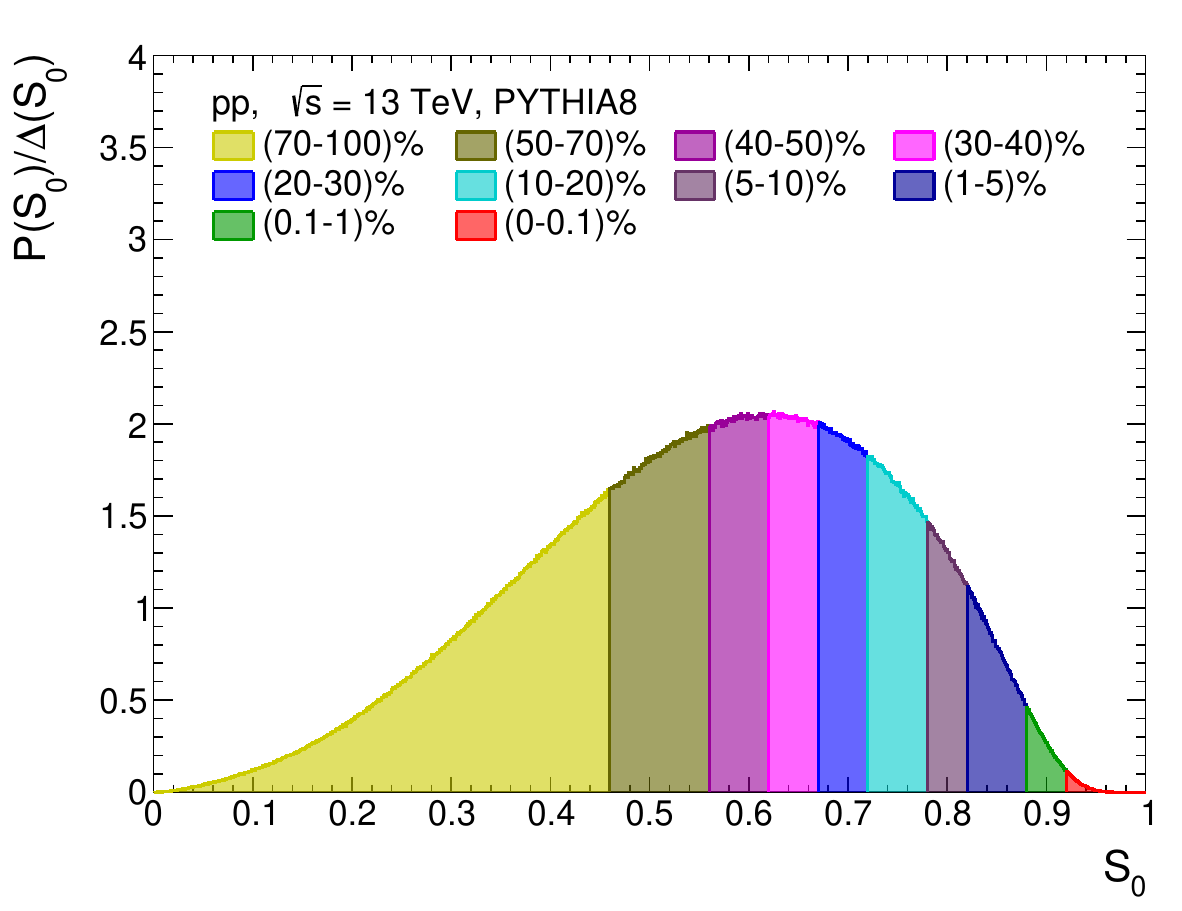}
\includegraphics[scale = 0.33]{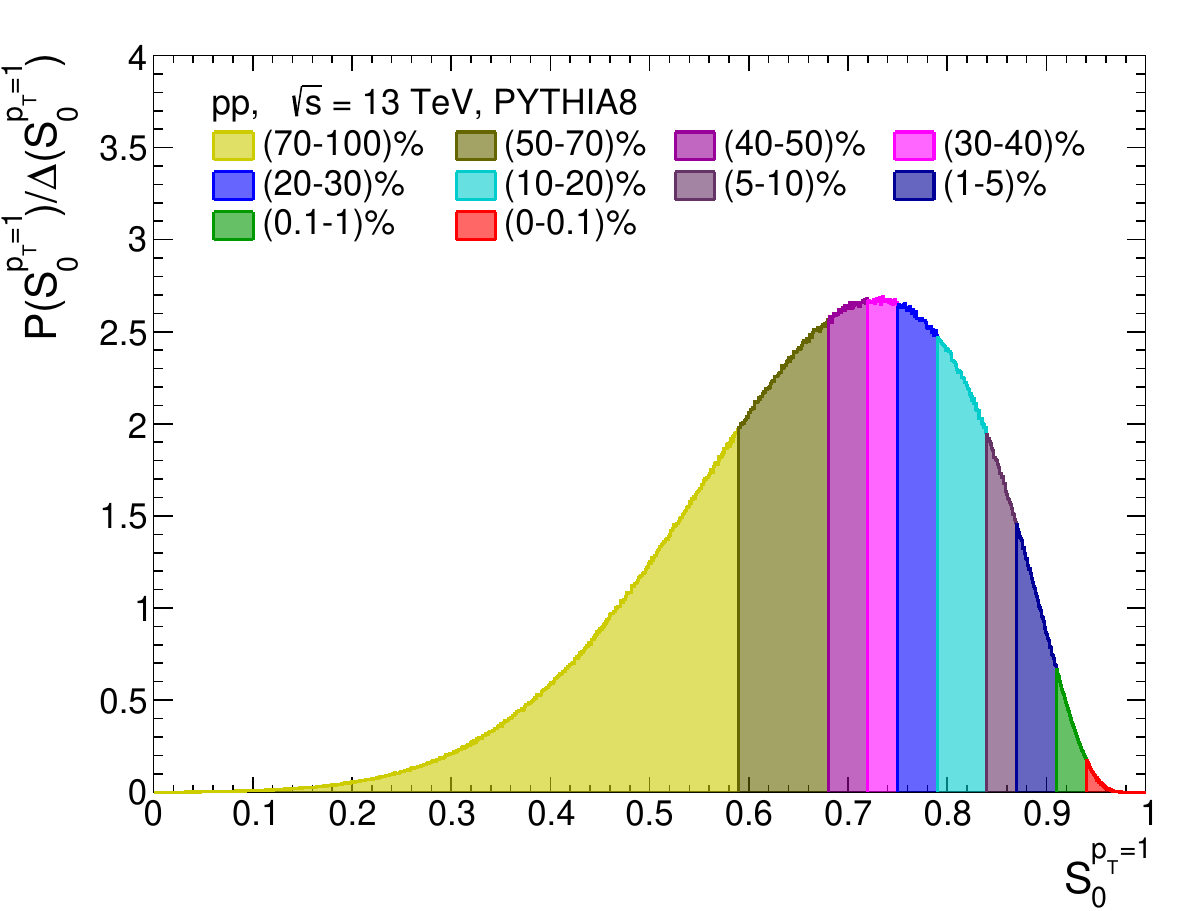}
\caption{Weighted and unweighted transverse spherocity distributions measured in $pp$ collisions at $\sqrt{s}=13$ TeV using PYTHIA8.}
\label{fig:S0dist}
\end{center}
\end{figure}

Similar to transverse sphericity, transverse spherocity is another event shape observable extensively used at the LHC energies to separate the soft versus hard QCD-dominated events. Transverse spherocity is defined for a unit vector $\hat{n} (n_{T},0)$ in the transverse plane, which follows a minimization criterion as shown in the following equation~\cite{Cuautle:2014yda, Cuautle:2015kra, Ortiz:2017jho, Khuntia:2018qox}
\begin{equation}
    S_{0} = \frac{\pi^{2}}{4}\min_{\hat{n}} \bigg(\frac{\sum_{i=1}^{N_{\rm had}}|\vec p_{{\rm T}_{i}}\times\hat{n}|}{\sum_{i=1}^{N_{\rm had}}~|\vec{p}_{{\rm T}_{i}}|}\bigg)^{2}.
\label{eq:sphero}
\end{equation}

$S_{0}$ is usually calculated using charged-particle tracks that lie in the mid-pseudorapidity ($|\eta|<0.8$) region and possess $p_{\rm T} > $ 0.15 GeV/$c$. To ensure a statistically meaningful concept of a topology, transverse spherocity is calculated for the events having more than 10 charged-particle tracks in the kinematic region mentioned above. These events are often referred to as $S_0$-integrated events. In Eq.~\eqref{eq:sphero}, $N_{\rm had}$ is the total number of charged hadrons in the event, and the multiplication of $\pi^2/4$ ensures $S_{0}$ to lie between 0 and 1. The two extreme limits of spherocity correspond to the two different configurations of event topology. Events with $S_{0} \rightarrow$ 0 consist of  single back-to-back jets, while events with $S_{0} \rightarrow$ 1 are  isotropic particle production dominated events. The different limits are depicted in Fig.~\ref{fig:sphero}. Usually in experimental measurements, the events located in the bottom 20\% of the $S_0$ distribution are referred to as jetty events, while the top 20\% of the $S_0$ distribution are referred to as isotropic events.

Historically, when event shape variables were introduced in $e^+e^-$ collisions, it was argued that infrared and collinear (IRC) safety of these variables is a key property when one relies on the calculation based on the perturbation series.  In other words, event shape variables are used to describe the parton structure in an event, which must be insensitive to the emission of soft and/or collinear radiation. These event shapes include thrust, squared jet-mass, heavy jet mass, jet broadening variables, sphericity, spherocity, aplanarity, etc., which are defined in Sec.~\ref{sec:definitions}. Later at the LHC energies, these event shape variables are being widely used for jet-related studies demanding the IRC safety of these quantities, which is a pQCD-based phenomenon. In the next subsections, we discuss the collinear and infrared safety for transverse spherocity as an example. However, it is important to mention that when dealing with the non-perturbative aspects of the collisions using the final state particles such as pions, kaons, etc, the IRC safety of event shape observables is not a key requirement.

\subsubsection{Infrared Safety}
In particle physics, infrared refers to the limit of the low-energy (large-wavelength) quanta, i.e., soft photons in QED and soft gluons in QCD. In high-energy collider experiments, soft gluons are frequently produced through parton radiation and showering. These gluons subsequently hadronise, giving rise to soft (low-$p_{\rm T}$) hadrons in the final state. The invariance of a quantity in the presence of these soft hadrons is called infrared safety. 

Let us consider a soft hadron with transverse momentum $\vec{p}_{{\rm T}_s}$ is emitted along with other hadrons such that $|\vec{p}_{{\rm T}_s}|=\varepsilon\rightarrow0$. Here, the subscript `$s$' stands for soft. New value of transverse spherocity modifies to:

\begin{equation}
    S^{\prime}_{0} = \frac{\pi^{2}}{4}\min_{\hat{n}} \bigg(\frac{\sum_{i=1}^{N_{\rm had}}|\vec p_{{\rm T}_{i}}\times\hat{n}|+|\vec{p}_{{\rm T}_s}\times\hat{n}|}{\sum_{i=1}^{N_{\rm had}}~|\vec{p}_{{\rm T}_{i}}|+|\vec{p}_{{\rm T}_s}|}\bigg)^{2}.
\end{equation}
If the angle between $\vec{p}_{{\rm T}_s}$ and $\hat{n}$ is $\theta_{s}$, then the above equation can be re-written as:
\begin{equation}
    S^{\prime}_{0} = \frac{\pi^{2}}{4}\min_{\hat{n}} \bigg(\frac{\sum_{i=1}^{N_{\rm had}}|\vec p_{{\rm T}_{i}}\times\hat{n}|+\varepsilon\sin{\theta_{s}}}{\sum_{i=1}^{N_{\rm had}}~|\vec{p}_{{\rm T}_{i}}|+\varepsilon}\bigg)^{2}.
    \label{eq:s0irstep2}
\end{equation}
In the infrared limit, i.e., $\varepsilon\rightarrow0$, Eq.~\eqref{eq:s0irstep2} modifies to:
\begin{equation}
    S^{\prime}_{0} = \frac{\pi^{2}}{4}\min_{\hat{n}} \bigg(\frac{\sum_{i=1}^{N_{\rm had}}|\vec p_{{\rm T}_{i}}\times\hat{n}|}{\sum_{i=1}^{N_{\rm had}}~|\vec{p}_{{\rm T}_{i}}|}\bigg)^{2}=S_{0}.
    \label{eq:s0irstep3}
\end{equation}

Thus, a soft (infrared) emission of particles renders transverse spherocity unchanged, making it infrared safe.
\subsubsection{Collinear Safety}
Let us consider a hadron with transverse momentum $p_{{\rm T}_{k}}$ splits into two collinear particles with transverse momenta $p_{{\rm T}_{k1}}$ and $p_{{\rm T}_{k2}}$, such that:
\begin{equation}
   p_{{\rm T}_{k1}}=zp_{{\rm T}_{k}}\;\;\;\; \text{and}\;\;\;\; p_{{\rm T}_{k2}}=(1-z)p_{{\rm T}_{k}}. 
\end{equation}
Here, $z$ and $(1-z)$ stand for the fractions of the momentum of the hadron in a collinear splitting and $0<z<1$ and $\vec{p}_{{\rm T}_{k}}$, $\vec{p}_{{\rm T}_{k1}}$, and $\vec{p}_{{\rm T}_{k2}}$ are parallel to each other. For any direction of $\hat{n}$,
\begin{equation}
|\vec p_{{\rm T}_{k1}}\times\hat{n}|+|\vec p_{{\rm T}_{k2}}\times\hat{n}|=z|\vec p_{{\rm T}_{k}}\times\hat{n}|+(1-z)|\vec p_{{\rm T}_{k}}\times\hat{n}|=|\vec p_{{\rm T}_{k}}\times\hat{n}|.
\end{equation}
Thus, a collinear splitting of a particle keeps the numerator of Eq.~\eqref{eq:sphero} unchanged.
Similarly, one can show for the denominator as follows.
\begin{equation}
|\vec{p}_{{\rm T}_{k1}}|+|\vec{p}_{{\rm T}_{k2}}|=z|\vec{p}_{{\rm T}_{k}}|+(1-z)|\vec{p}_{{\rm T}_{k}}|=|\vec{p}_{{\rm T}_{k}}|.
\end{equation}
The contribution of the original particle is the same as the sum of the collinear fragments in both the numerator and denominator of Eq.~\eqref{eq:sphero}. Hence, transverse spherocity remains unchanged for a collinear fragmentation of a particle, which makes it collinear safe.

\begin{figure}[ht!]
\begin{center}
\includegraphics[scale = 0.33]{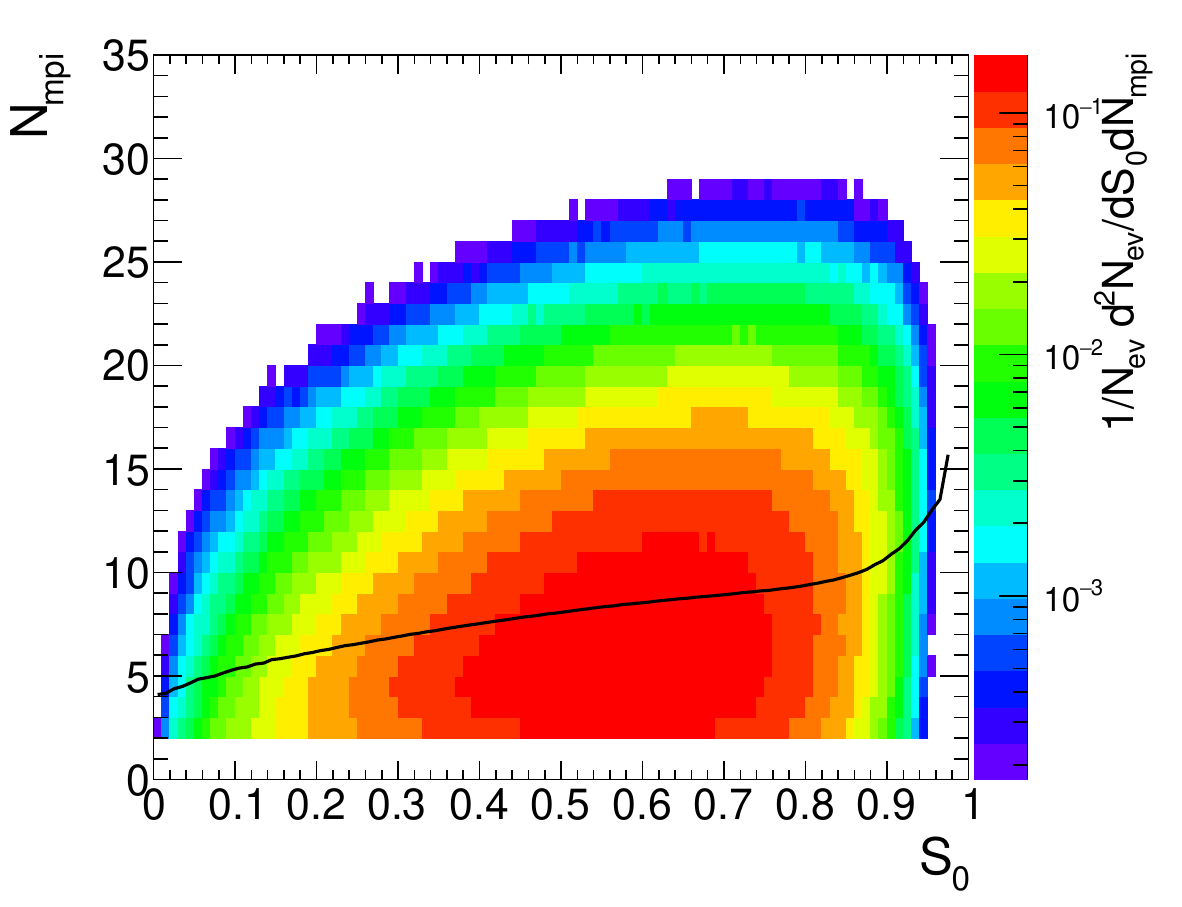}
\includegraphics[scale = 0.33]{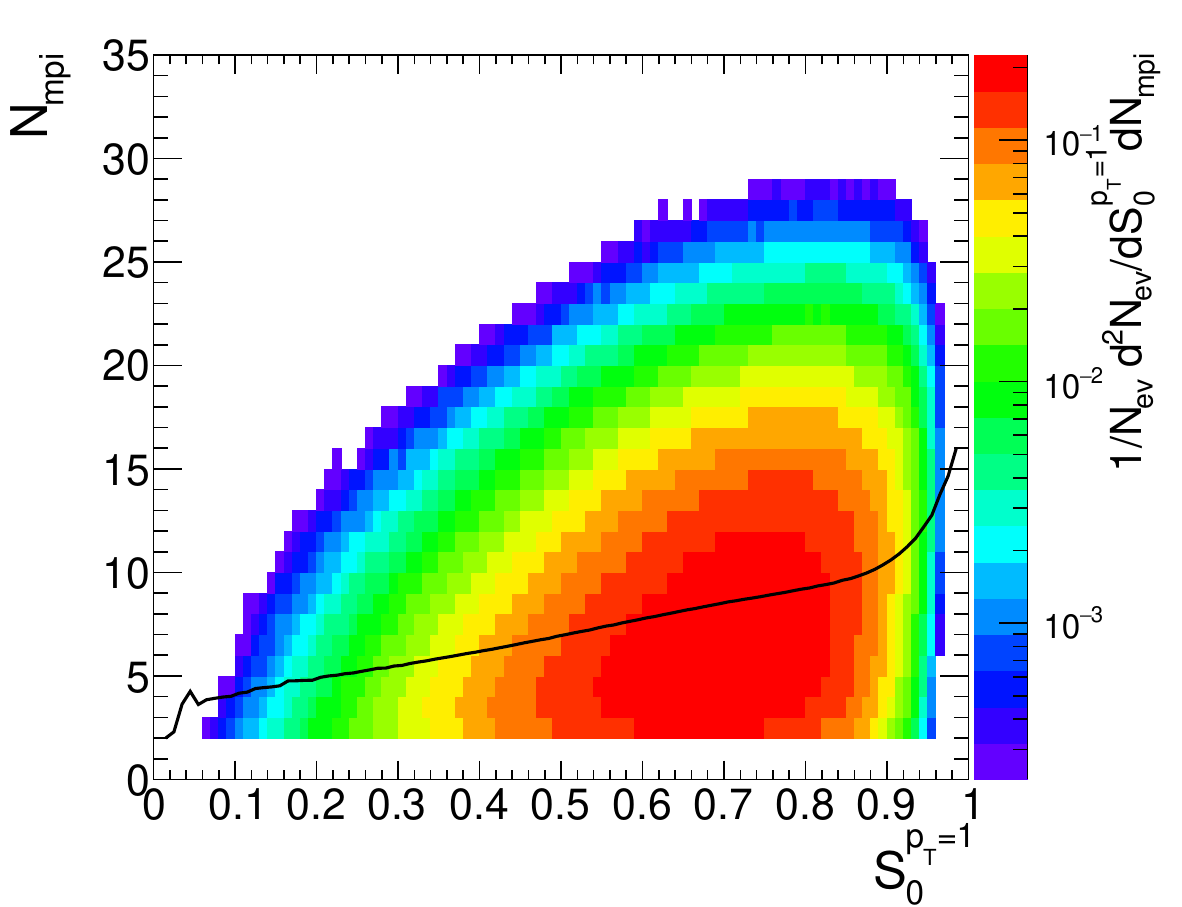}
\caption{Correlation of $S_{0}$ (left panel) and $S_{0}^{p_{\rm T} = 1}$ (right panel) with number of MPI in $pp$ collisions at $\sqrt{s}=13$ TeV using PYTHIA8. The black solid line represents $\langle N_{\rm mpi}\rangle$ as a function of $S_{\rm 0}$ (left) and $S_{0}^{p_{\rm T}=1}$ (right).}
\label{fig:S0vsMPI}
\end{center}
\end{figure}

It is worth emphasizing that Eq.~\eqref{eq:sphero} is $p_{\rm T}$ weighted and introduces neutral jet bias. This jet bias can be fixed by setting the $p_{\rm T}=1$ for all the charged tracks in Eq.~\eqref{eq:sphero}. Thus, one can write unweighted transverse spherocity ($S_{0}^{p_{\rm T} = 1}$) as follows.

\begin{equation}
    S_{0}^{p_{\rm T} = 1}=\frac{\pi^2}{4}\min_{\hat{n}}\Bigg(\frac{\sum_{i=1}^{N_{\rm had}}|\hat{p}_{{\rm T}_{i}} \times \hat{n}|}{N_{\rm had}}\Bigg)^{2}.
    \label{eq:spheropt1}
\end{equation}

\begin{figure}[ht!]
\begin{center}
\includegraphics[scale = 0.4]{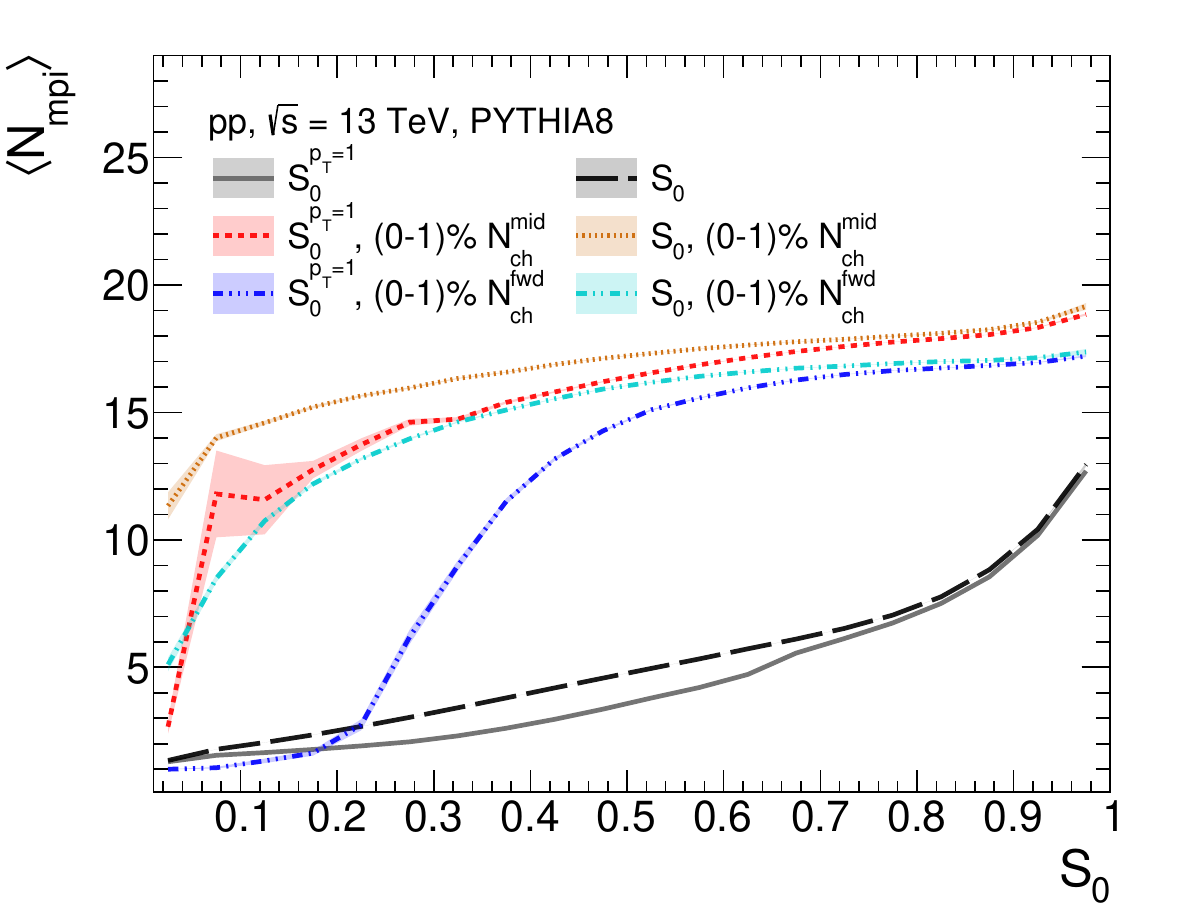}
\caption{Correlation of $S_{0}$ and $S_{0}^{p_{\rm T} = 1}$ with average number of MPI in minimum bias and high-multiplicity $pp$ collisions at $\sqrt{s}=13$ TeV using PYTHIA8.}
\label{fig:S0vsAvgMPI}
\end{center}
\end{figure}
As the biases towards high-$p_{\rm T}$ particles in Eq.~\eqref{eq:spheropt1} is removed by setting $p_{\rm T}=1$, the denominator in Eq.~\eqref{eq:sphero}, i.e., $\sum_{i}p_{{\rm T}_i}$ is now replaced by the number of charged hadrons, i.e., $N_{\rm had}$.

Figure~\ref{fig:S0dist} shows the distribution of weighted transverse spherocity ($S_{0}$) (left) and unweighted transverse spherocity ($S_{0}^{p_{\rm T}=1}$) (right) for minimum bias $pp$ collisions at $\sqrt{s}=13$ TeV using PYTHIA8. The transverse spherocity distributions are further divided into different percentiles and specified as different transverse spherocity classes. Figure~\ref{fig:S0vsMPI} shows the correlation of weighted ($S_{0}$) (left panel) and unweighted ($S_{0}^{p_{\rm T}=1}$) (right panel) transverse spherocity with $N_{\rm mpi}$  obtained in $pp$ collisions at $\sqrt{s}=13$ TeV using PYTHIA8. The solid line guides the mean values of $N_{\rm mpi}$, i.e., $\langle N_{\rm mpi}\rangle$, for each bin on the $x-$axis.  From Fig.~\ref{fig:S0vsMPI}, it is conclusive that both $S_{0}$, and $S_{0}^{p_{\rm T}=1}$ retains a fair degree of correlation with $N_{\rm mpi}$, where the higher values of $S_{0}$, and $S_{0}^{p_{\rm T}=1}$  corresponds to a large number of multi-partonic interactions, i.e., soft events. When compared to Fig.~\ref{fig:STvsMPI}, the correlation of $S_{\rm T}$ with $N_{\rm mpi}$ appears to be smaller in magnitude as compared to the correlation of transverse spherocity and charged-particle multiplicity with $N_{\rm mpi}$, especially for extreme isotropic events. 

%    MENTION WHICH HAS LARGE CORRELATION WITH MPI

Figure~\ref{fig:S0vsAvgMPI} shows the correlation of $S_{0}$ and $S_{0}^{p_{\rm T}=1}$ with $N_{\rm mpi}$. A similar correlation of $S_{0}$ and $S_{0}^{p_{\rm T}=1}$ with $N_{\rm mpi}$ is also shown for the highest (0-1)\% multiplicity class in the mid and forward pseudorapidity regions. Both $S_{0}$ and $S_{0}^{p_{\rm T}=1}$ possess a fair degree of positive correlation with $N_{\rm mpi}$, where events having spherocity value close to one have higher $\langle N_{\rm mpi}\rangle$ value. When no multiplicity selection is applied, $S_0$ probes higher values of $\langle N_{\rm mpi}\rangle$ as compared to $S_{0}^{p_{\rm T}=1}$ in the intermediate values of spherocity. The difference of $\langle N_{\rm mpi}\rangle$ for both weighted and unweighted transverse spherocity for the minimum bias case starts to vanish as one goes close to their extreme values, i.e., 0 and 1. In addition, by applying charged-particle event selection, the transverse spherocity achieves a higher value of $\langle N_{\rm mpi}\rangle$. For the top 1\% multiplicity class based on both $N_{\rm ch}^{\rm mid}$ and $N_{\rm ch}^{\rm fwd}$ we observe that $S_{0}$ probes a higher value of $N_{\rm mpi}$ than $S_{0}^{p_{\rm T}=1}$ towards lower values of spherocity. As the value of spherocity approaches `1', $\langle N_{\rm mpi}\rangle$ from both $S_{0}^{p_{\rm T}=1}$ and  $S_{0}$ approach each other and the difference in $\langle N_{\rm mpi}\rangle$ vanishes for $S_{0}\gtrsim0.8$.  Furthermore, both weighted and unweighted values of transverse spherocity for (0-1)\% $N_{\rm ch}^{\rm mid}$ class acquired a higher value of $\langle N_{\rm mpi}\rangle$ and a lower slope, relative to (0-1)\% $N_{\rm ch}^{\rm fwd}$ class. The lower slope of $S_{0}$ and $S_{0}^{p_{\rm T}=1}$ vs $\langle N_{\rm mpi}\rangle$ for the (0-1)\% class of $N_{\rm ch}^{\rm mid}$, may arise due to estimation of both spherocity and $N_{\rm ch}^{\rm mid}$ in the similar pseudorapidity region.

%fIG 4 TO BE REPLOTTED AND DISCUSSED

\begin{figure}[ht!]
    \centering
    \includegraphics[width=0.5\linewidth]{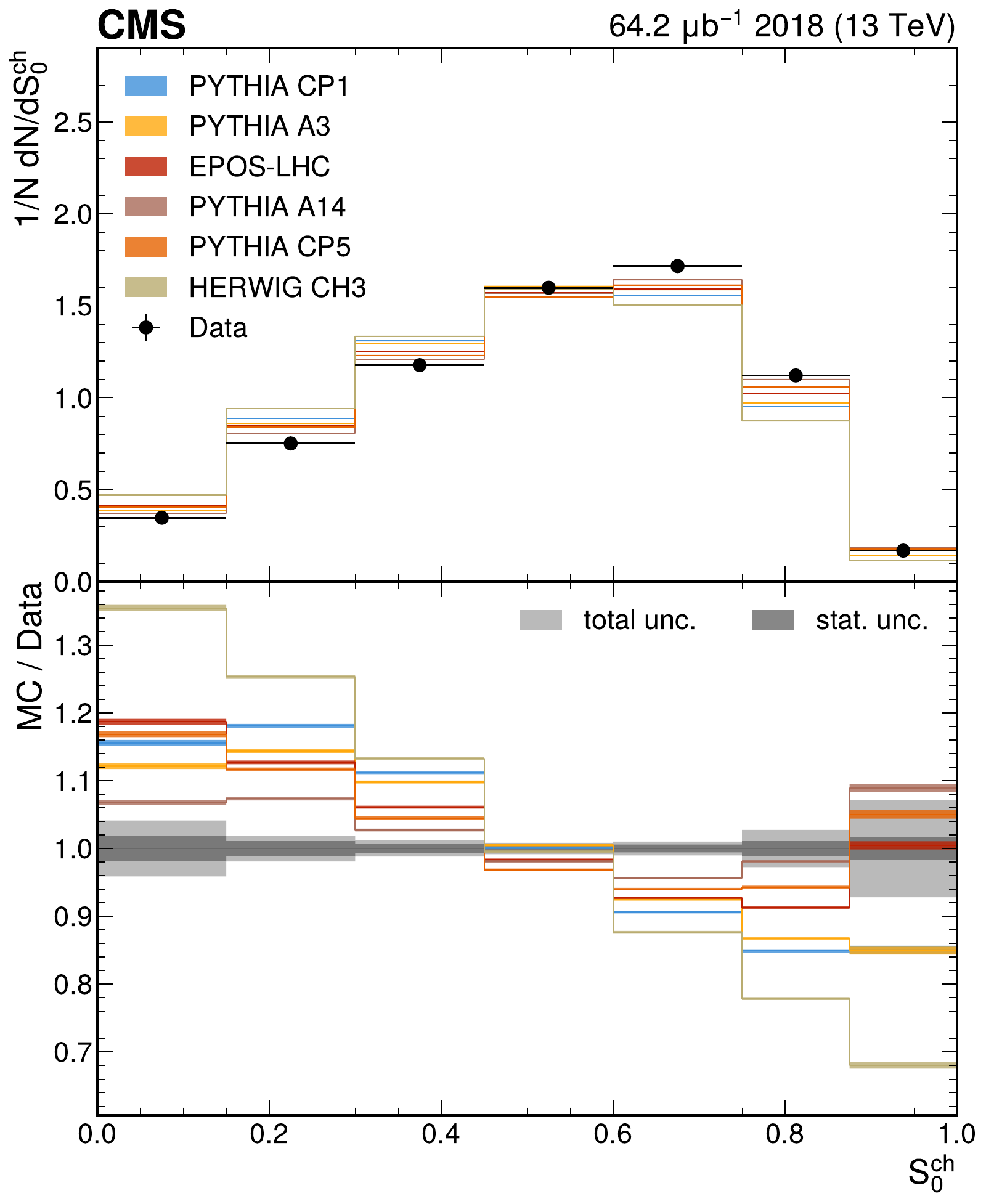}
    \caption{Unfolded distribution of $p_{\rm T}$-weighted transverse spherocity ($S_{0}^{\rm ch}$ or $S_{0}$) in $pp$ collisions at $\sqrt{s}=13$ TeV measured with CMS experiment and compared with predictions from PYTHIA CP1, CP5, A3, A14, tunes, the EPOS-LHC generator, and the HERWIG CH3 tune~\cite{CMS:2025sws}.}
    \label{fig:S0dataComp}
\end{figure}

Having discussed the correlation of transverse spherocity with $N_{\rm mpi}$, it is observed that the distribution of many event classifiers, including transverse spherocity, from MC event generators fails to reproduce the distribution of event shape observables measured in experiments. A sample is shown in Fig.~\ref{fig:S0dataComp}, where the unfolded distribution of transverse spherocity (denoted as $S_{0}^{\rm ch}$) from CMS experiment is compared with MC simulations using PYTHIA CP1~\cite{CMS:2019csb}, CP5~\cite{CMS:2019csb}, A3~\cite{ATLAS:2016puo}, A14~\cite{TheATLAScollaboration:2014rfk} tunes, the EPOS-LHC generator, and the HERWIG ~\cite{CMS:2020dqt} tune~\cite{CMS:2025sws}. The details about these specific tunes of PYTHIA and HERWIG event generators can be found in appendix~\ref{appendixpythia} and \ref{sec:otherMCmodels}, respectively. The unfolding procedure is used to correct the measured distributions of variables for detector effects, including but not limited to finite resolution and inefficiencies, to retrieve the true underlying physical distribution. This is because the raw data from detectors are usually affected by instrumental limitations. In contrast, unfolding is necessary to make meaningful comparisons with theoretical models or comparisons across experiments. In Fig.~\ref{fig:S0dataComp}, it can be observed that the distribution of experimentally measured transverse spherocity is more isotropic in nature than the predictions from different MC event generators. A similar behaviour is observed for other event classifiers, including sphericity, and thrust~\cite{CMS:2025sws}. The comparison of these event shape observables in different multiplicity regions is also made in Ref.~\cite{CMS:2025sws}, which reveals the fact that the mismatch or mis-modeling of event shapes in MC models is likely
not coming only from the distribution of the number of charged particles or poorly modeled longitudinal components. These results suggest a need to improve the modeling of $pp$ collisions and are crucial for understanding phenomena such as quark-gluon plasma and topological effects in non-Abelian gauge theories, such as QCD instantons~\cite{CMS:2025sws}.

\subsection{Relative transverse activity classifier ($R_{\rm T}$)}
\label{sec:RTdefn}
In relativistic collisions, the partonic interactions are characterized by pQCD or non-pQCD processes. Initial hard scattering in such collisions is a pQCD process due to the large energy and momentum transfers involved in such collisions. On the other hand, multi-partonic interactions, initial state radiations (ISR), final state radiations (FSR), and beam remnants are non-pQCD in nature. The partons involved in the multi-partonic interactions, initial and final state radiations, and the beam remnant collectively contribute to the underlying event. The particles contributing to UE do not originate from the fragmentation of partons produced in the hardest scattering. UE activity is one of the efficient tools theorised to affect particle production and can lead to observed QGP-like signatures in $pp$ collisions. 

\begin{figure}[ht!]
\begin{center}
\includegraphics[scale=0.35]{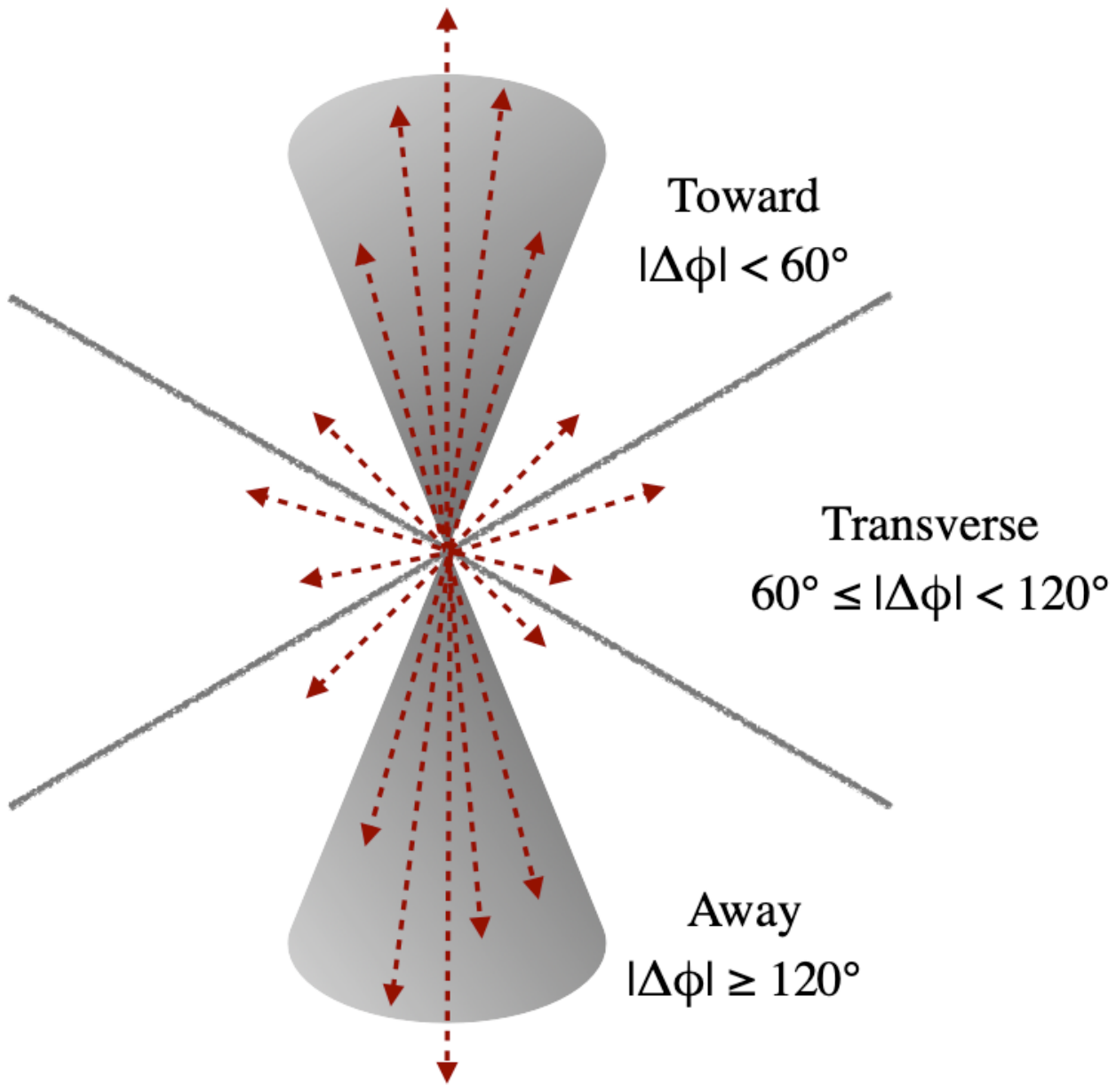}
\caption{Depiction of different topological regions with respect to the leading particle.}
\label{fig:myregions}
\end{center}
\end{figure}

\begin{figure}[ht!]
\begin{center}
\includegraphics[scale=0.4]{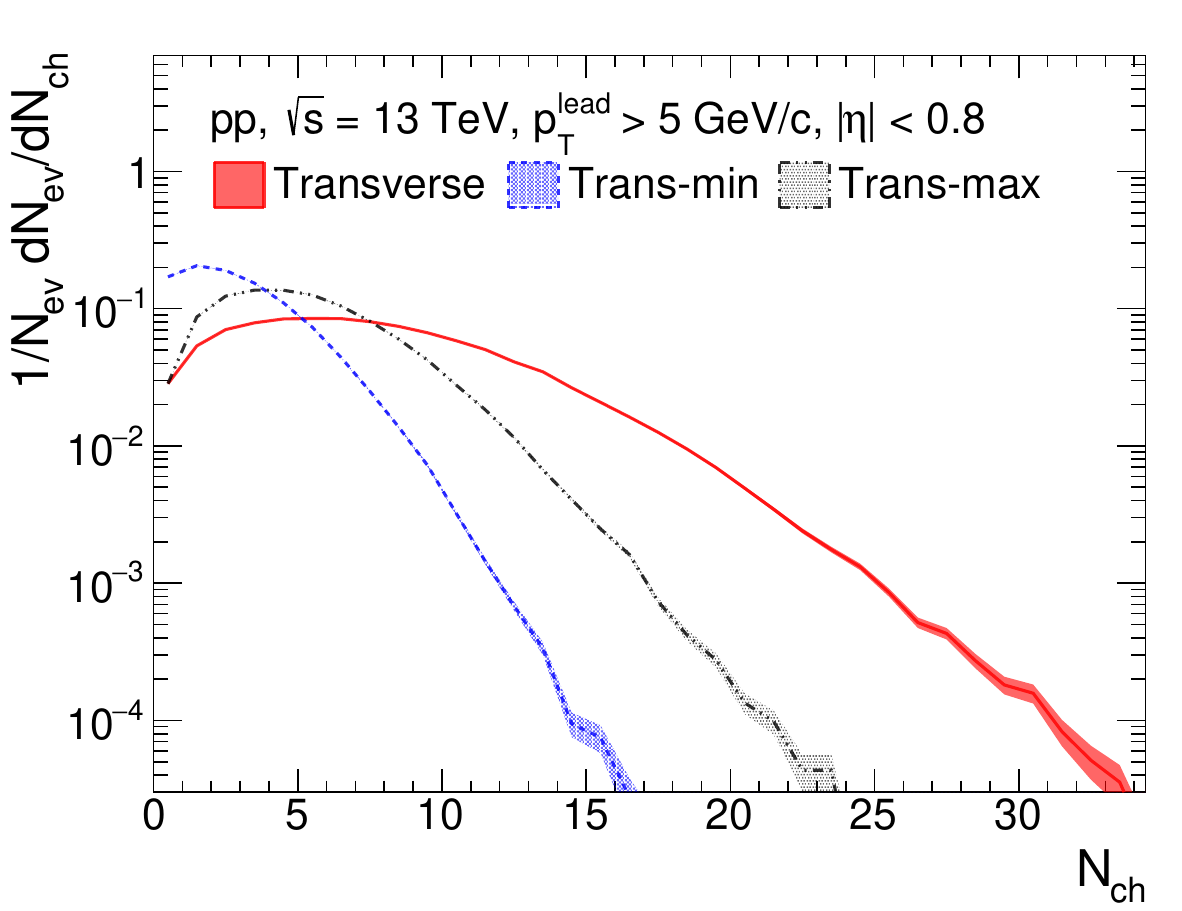}
\caption{$N_{\rm{ch}}$ distributions for transverse, trans-min and trans-max regions}
\label{fig:NchRTdist}
\end{center}
\end{figure}

\begin{figure*}[ht!]
\begin{center}
\includegraphics[width = 0.49\linewidth]{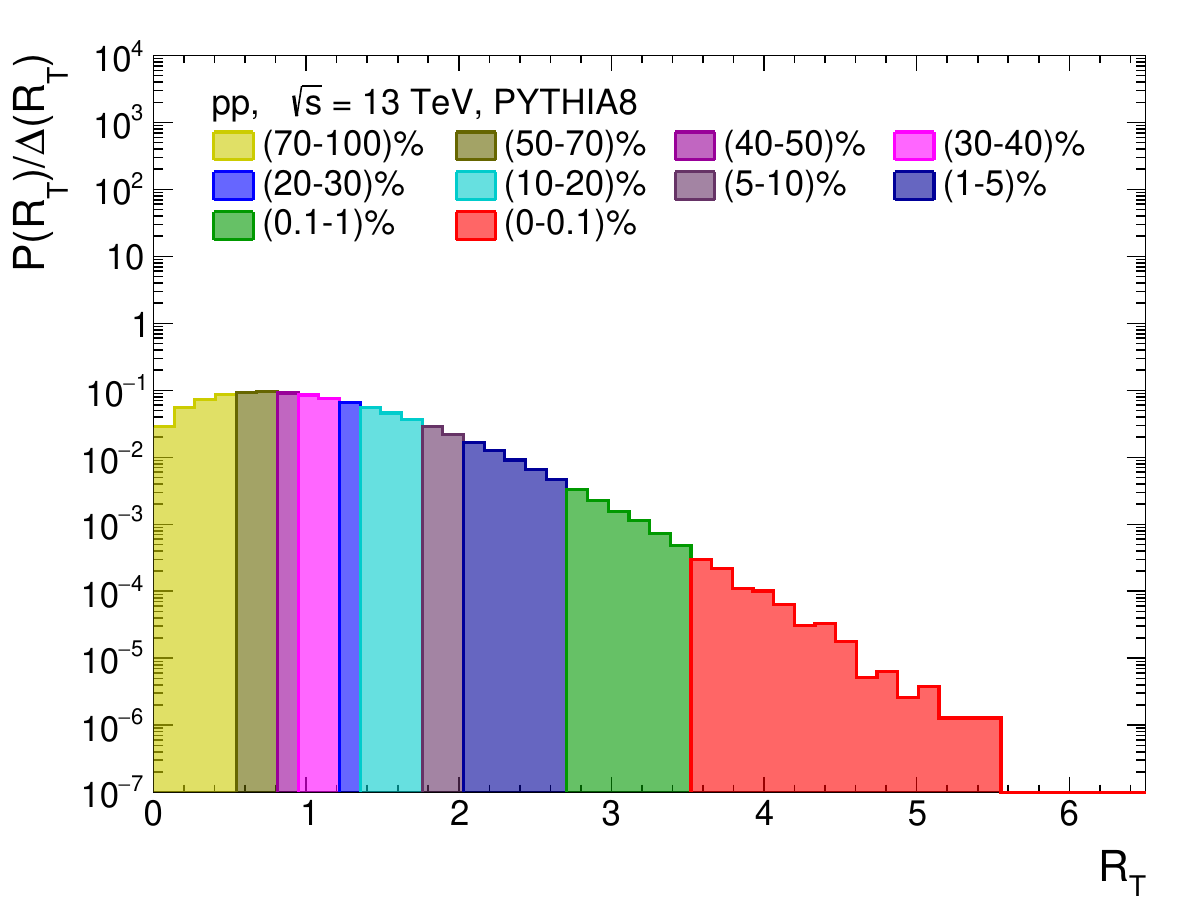}
\includegraphics[width = 0.49\linewidth]{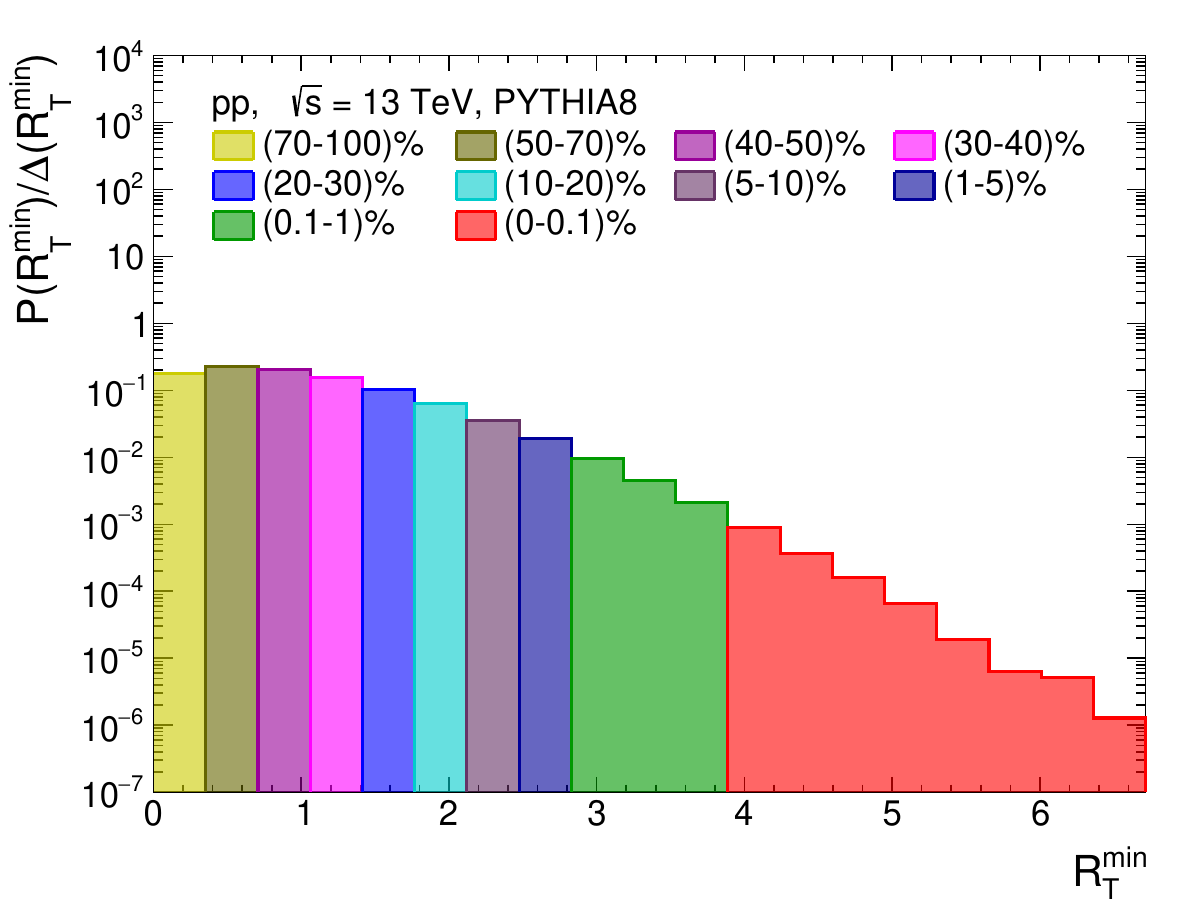}
\includegraphics[width = 0.49\linewidth]{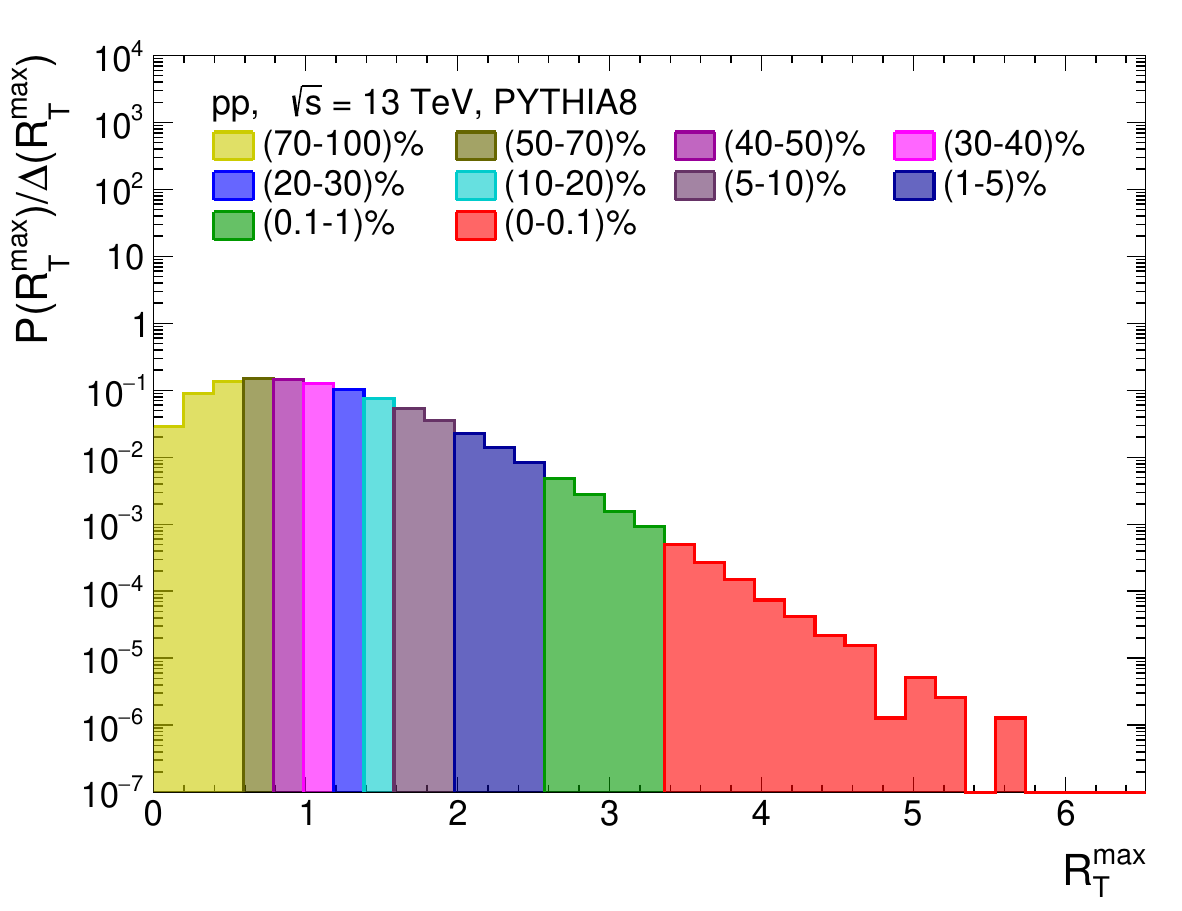}
\caption{Distribution of $R_{\rm{T}}$ measured in the transverse, trans-min and trans-max regions in $pp$ collisions at $\sqrt{s}=13$ TeV using PYTHIA8.}
\label{fig:RTdist}
\end{center}
\end{figure*}

\begin{figure*}[ht!]
\begin{center}
\includegraphics[width = 0.49\linewidth]{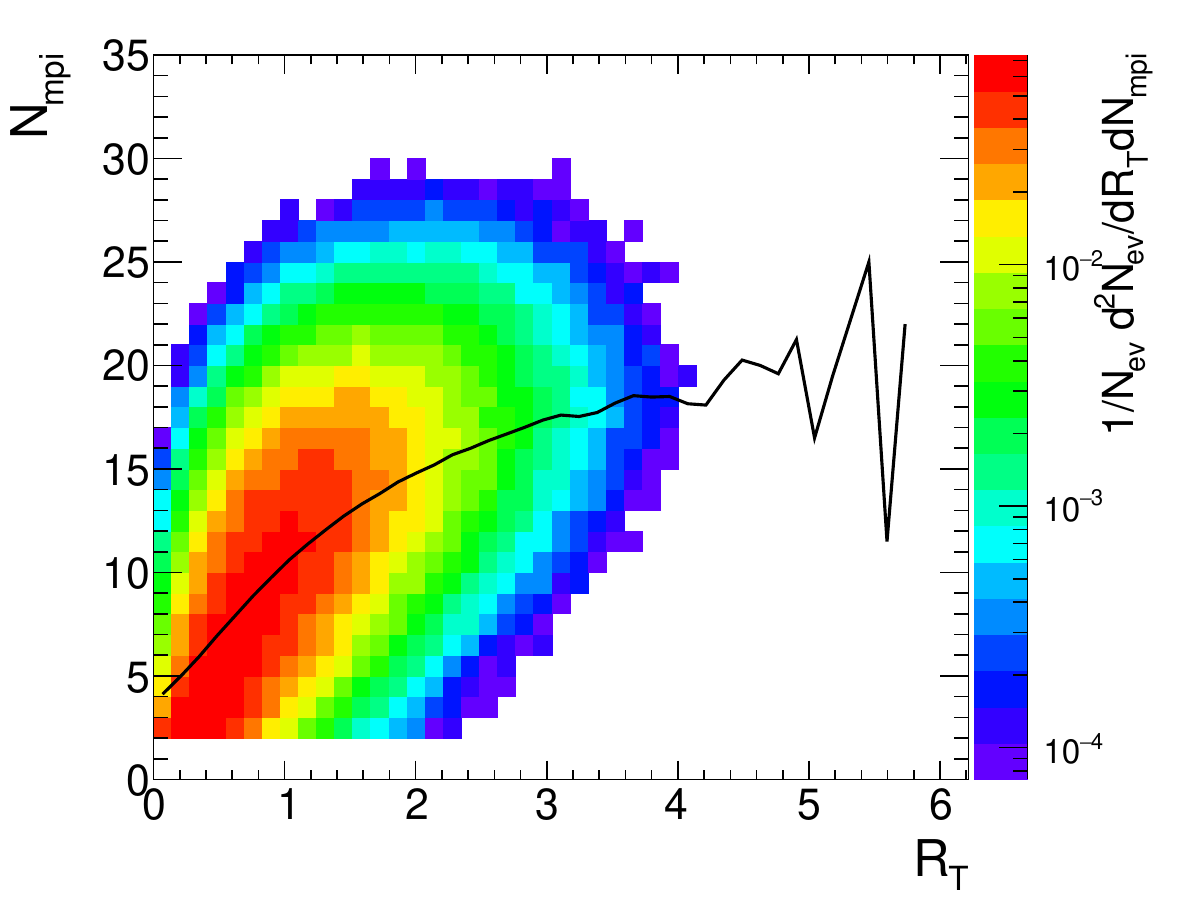}
\includegraphics[width = 0.49\linewidth]{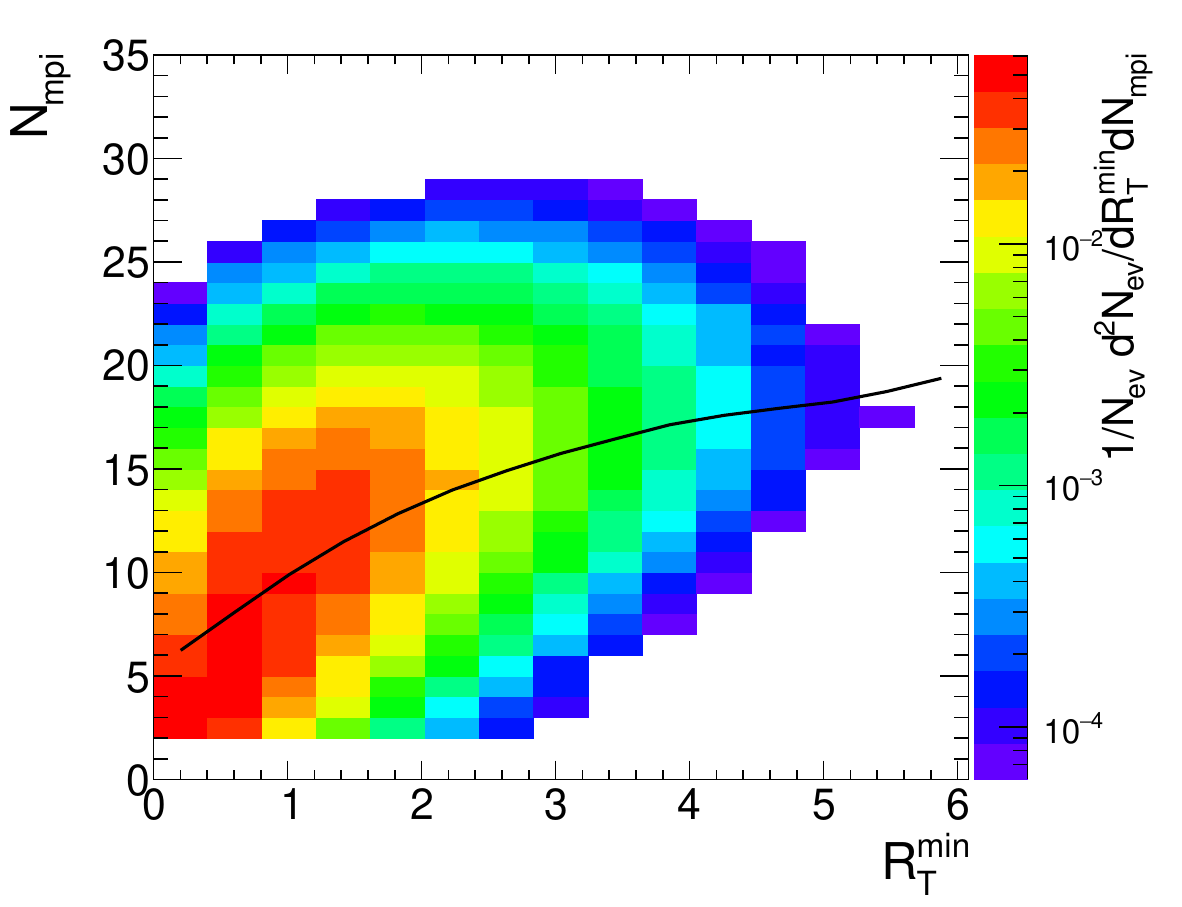}
\includegraphics[width = 0.49\linewidth]{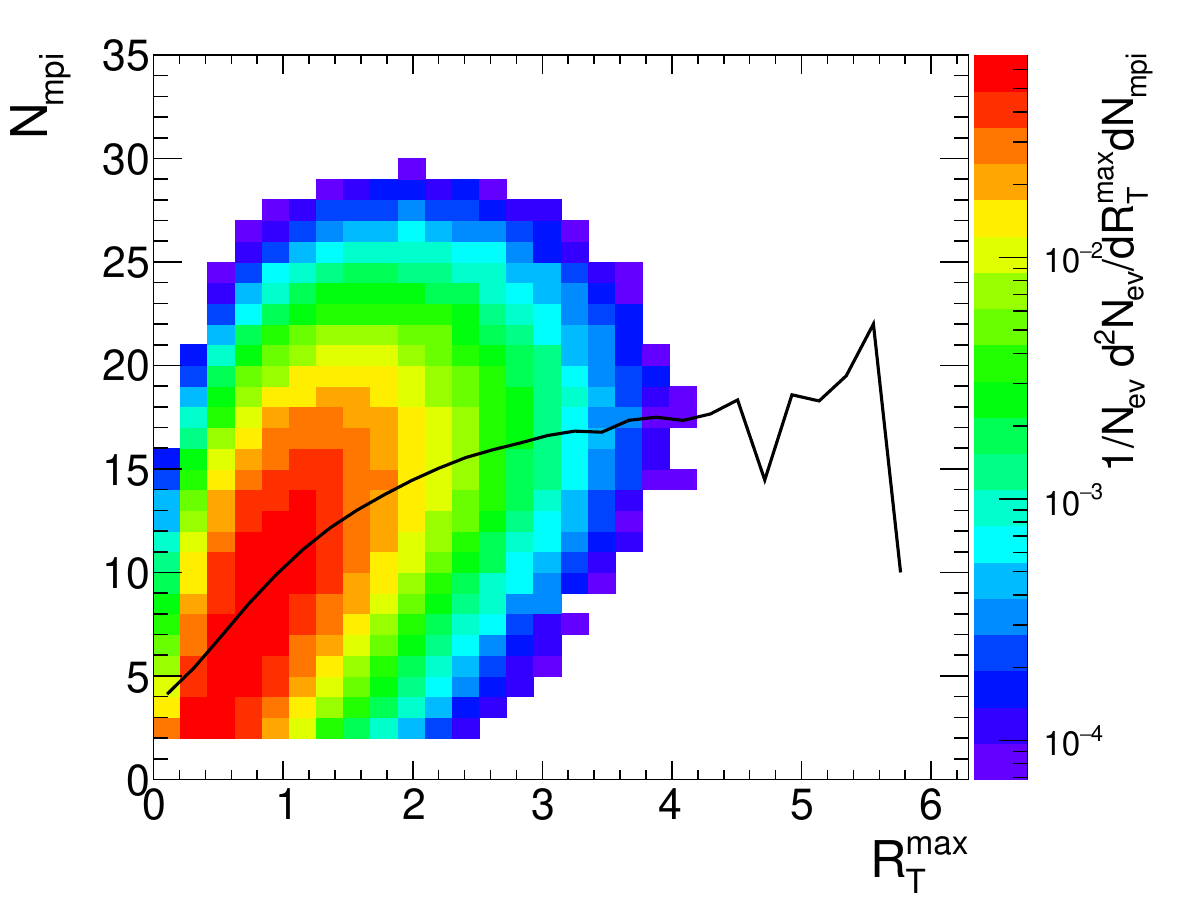}
\caption{Correlation of $R_{\rm{T}}$ measured in the transverse, trans-min and trans-max regions with the number of MPI in $pp$ collisions at $\sqrt{s}=13$ TeV using PYTHIA8. The black solid line represents $\langle N_{\rm mpi}\rangle$ as a function of $R_{\rm T}$ (upper left), $R_{\rm T}^{\rm min}$ (upper right), and $R_{\rm T}^{\rm max}$ (lower middle).}
\label{fig:RTvsNMPI}
\end{center}
\end{figure*}

To classify the events based on the contribution from the UE events, one can define different topological regions based on the contribution from the jet fragmentation and the UE. Figure~\ref{fig:myregions} depicts different topological regions, \textit{viz.,} toward, transverse and away, defined with respect to the highest $p_{\rm T}$ particle, also called the leading particle, and the corresponding momentum is $p_{\rm T}^{\rm leading}$. The toward region, as the name indicates, consists of the particles with an azimuthal angle less than $\pi/3$ relative to the leading particle, \textit{i.e.}, $|\Delta\phi|<\pi/3$, and has the largest contribution from the fragmentation of jets. The away region contains all the charged particles with $|\Delta\phi|\geq 2\pi/3$ while the transverse region is defined to be within $\pi/3\leq|\Delta\phi|<2\pi/3$. Since the transverse region is perpendicular to the leading jet axis, it is expected to have the least contribution from the jet and must be dominated by the underlying event activity.
% The following observables are expected to be sensitive to UE \cite{ATLASRTUE}.
% \begin{enumerate}
%     \item $\langle d^2 N_{\rm ch}/d\eta\;d\phi\rangle$: It shows the number of tracks per unit $\eta-\phi$.
%     \item $\langle d^2 \sum{p_{\rm T}}/d\eta\;d\phi\rangle$: It shows the scaler sum of track $p_{\rm T}$ per $\eta-\phi$.
% \end{enumerate}
% The charged-particle multiplicity of the transverse region is scaled with its event-average value to obtain $R_{\rm T}$, \textit{i.e.} $R_{\rm T}=N_{\rm ch}^{\rm T}/\langle N_{\rm ch}^{\rm T} \rangle$. 
To estimate the value of the relative transverse activity classifier ($R_{\rm T}$), the number of charged hadrons in the transverse regions is determined, denoted as $N_{\rm ch}^{\rm trans}$. Thus, one obtains the value of $R_{\rm T}$ using the following expression.

\begin{equation}
    R_{\rm T}=\frac{N_{\rm ch}^{\rm T}}{\langle N_{\rm ch}^{\rm T}\rangle}
    \label{eq:RT}
\end{equation}

Similarly, for $R_{\rm T}^{\rm min}$, and $R_{\rm T}^{\rm max}$, the transverse region is again subdivided into $\pi/3<\Delta\phi<2\pi/3$ and $-\pi/3>\Delta\phi>-2\pi/3$. The charged-particle multiplicity in these two regions is determined. For each event, the region with higher multiplicity (trans-max) contributes to $N_{\rm ch}^{\rm T-max}$ while the region with less multiplicity (trans-min) contributes to $N_{\rm ch}^{\rm T-min}$. Thus, one obtains, $R_{\rm T}^{\rm min}$ and $R_{\rm T}^{\rm max}$ as follows~\cite{Bencedi:2021tst}.

\begin{equation}
    R_{\rm T}^{\rm min}=\frac{N_{\rm ch}^{\rm T-min}}{\langle N_{\rm ch}^{\rm T-min}\rangle}
    \label{eq:RTmin}
\end{equation}

\begin{equation}
    R_{\rm T}^{\rm max}=\frac{N_{\rm ch}^{\rm T-max}}{\langle N_{\rm ch}^{\rm T-max}\rangle}
    \label{eq:RTmax}
\end{equation}

In Equations~\eqref{eq:RT}, \eqref{eq:RTmin}, and \eqref{eq:RTmax}, the angular brackets represent the event-average value of the observables. Figure~\ref{fig:NchRTdist} shows the distribution of charged-particle yields in the transverse ($N_{\rm ch}^{\rm T}$), trans-min ($N_{\rm ch}^{\rm T-min}$), and trans-max ($N_{\rm ch}^{\rm T-max}$) regions. The charged hadrons are measured in the mid-pseudorapidity region, i.e., $|\eta|<0.8$, with $p_{\rm T}^{\rm lead}>5$ GeV/$c$ in $pp$ collisions at $\sqrt{s}=13$ TeV using PYTHIA8. As expected, in Fig.~\ref{fig:NchRTdist}, the trans-min region has the lowest multiplicity yield, followed by trans-max. It is important to note that $N_{\rm ch}^{\rm T}=N_{\rm ch}^{\rm T-min}+N_{\rm ch}^{\rm T-max}$, which is also held by the distribution shown in Fig.~\ref{fig:NchRTdist}. Figure~\ref{fig:RTdist} shows the distribution of $R_{\rm T}$ (upper left), $R_{\rm T}^{\rm min}$ (upper right) and $R_{\rm T}^{\rm max}$ (lower) in $pp$ collisions at $\sqrt{s}=13$ TeV, and corresponding percentile slices.

In Fig.~\ref{fig:RTvsNMPI}, we show the correlation of $R_{\rm T}$ (upper left), $R_{\rm T}^{\rm min}$ (upper right), and $R_{\rm T}^{\rm max}$ (lower) with $N_{\rm mpi}$ obtained in $pp$ collisions at $\sqrt{s}=13$ TeV using PYTHIA8. Here, the solid line represents the mean values of $N_{\rm mpi}$ for the corresponding bin in the $x-$axis. All the classifiers, $R_{\rm T}$, $R_{\rm T}^{\rm max}$, and $R_{\rm T}^{\rm min}$ show a fair amount of correlations with $N_{\rm mpi}$. The correlation shown in Fig.~\ref{fig:RTvsNMPI} is significantly larger as compared to the correlation of transverse spherocity and transverse sphericity with $N_{\rm mpi}$. In other words, using the relative transverse activity classifiers, one can probe to higher values of $N_{\rm mpi}$ compared to $S_{0}$, $S_{0}^{p_{\rm T}=1}$, and $S_{\rm T}$.

\subsection{Charged-particle flattenicity ($\rho_{\rm{ch}}$)}

\begin{figure}[ht!]
\begin{center}
\includegraphics[scale = 0.4]{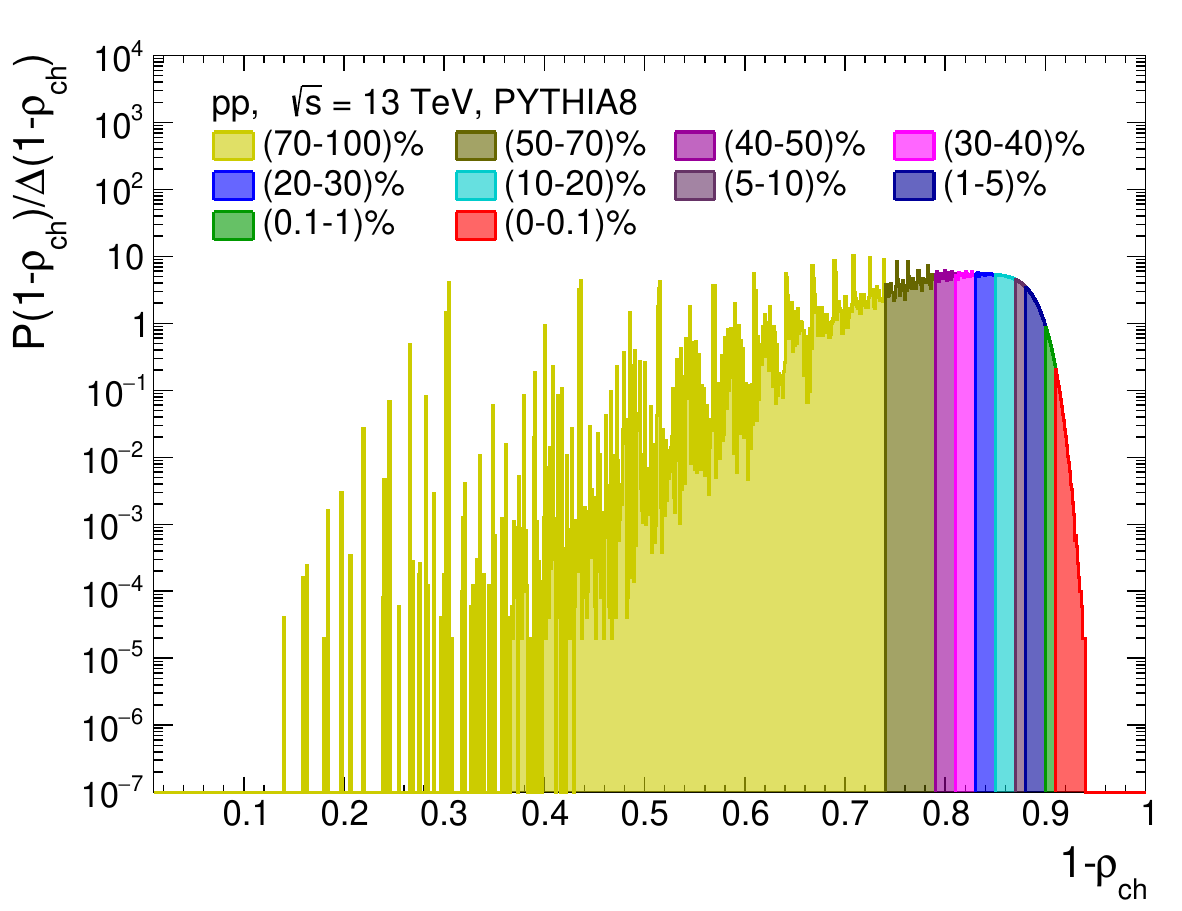}
\caption{Charged-particle flattenicity distributions measured in $pp$ collisions at $\sqrt{s}=13$ TeV using PYTHIA8.}
\label{fig:flatdist}
\end{center}
\end{figure}

Isotropic events are expected to have a uniform distribution of $p_{\rm T}$ throughout the  ($\eta-\phi$) region. To quantify the uniformity of $p_{\rm T}$ distribution event-by-event, the ($\eta-\phi$) region is divided into a $(10\times 10)$ grid. $p_{\rm T}$ in each cell ($p_{\rm T}^{\rm cell}$) can be measured, and corresponding mean ($\langle p_{\rm T}^{\rm cell}\rangle$) and standard deviation ($\sigma_{p_{\rm T}^{\rm cell}}$) can be estimated. Keeping the tracking capabilities of future upgrade for ALICE experiment (ALICE~3) in mind, the charged particles in $|\eta|<4.0$ and $p_{\rm T}>0.15$ GeV/c are considered for the estimation of $\langle p_{\rm T}^{\rm cell}\rangle$ and $\sigma_{p_{\rm T}^{\rm cell}}$. Thus, for an event, flattenicity ($\rho_{\rm ch}$) is defined as the relative standard deviation as follows~\cite{Ortiz:2022zqr},
\begin{equation}
    \rho_{\rm{ch}}=\frac{\sigma_{p_{\rm T}^{\rm cell}}}{\langle p_{\rm T}^{\rm cell}\rangle}
    \label{eq:flatpt}
\end{equation}

\begin{figure}[ht!]
\begin{center}
\includegraphics[scale=0.4]{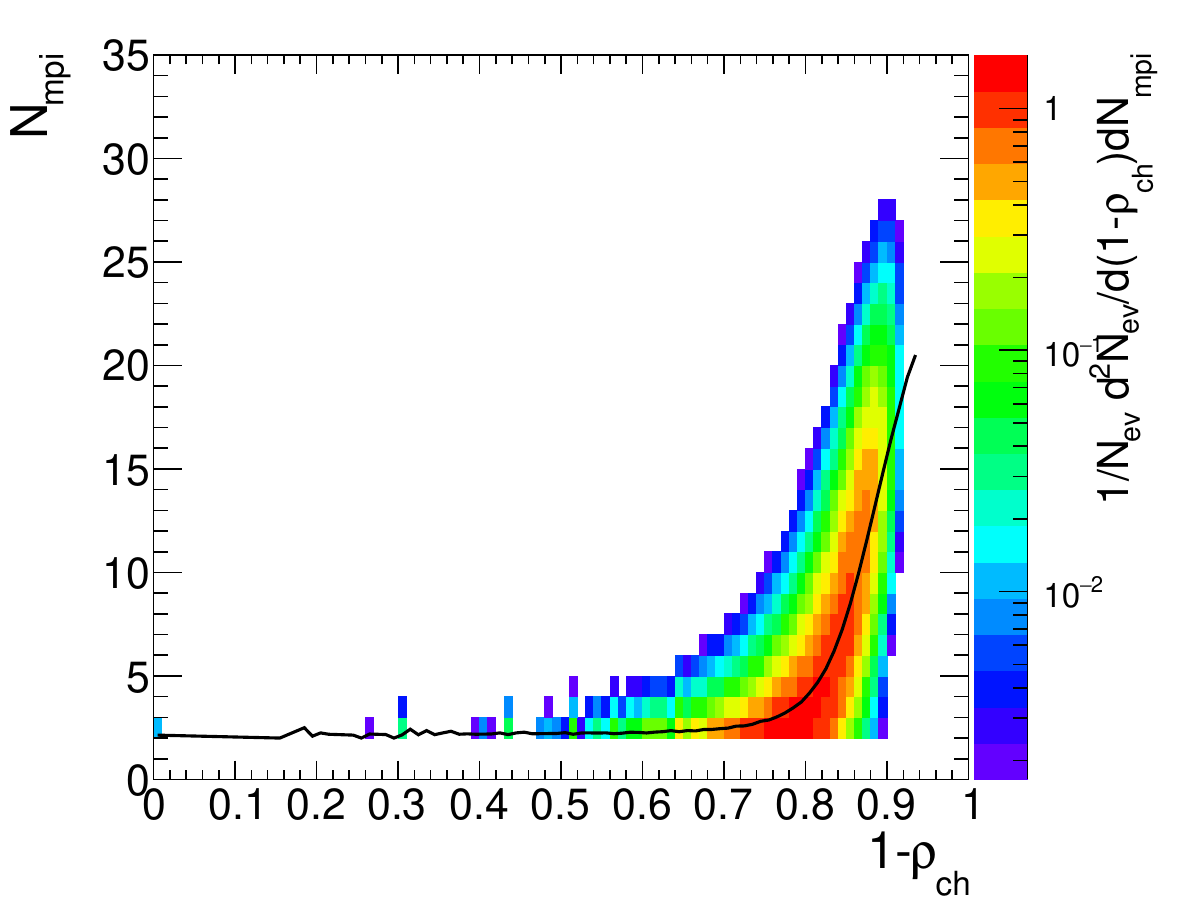}
\caption{Correlation of flattenicity with number of MPI. The black solid line represents $\langle N_{\rm mpi}\rangle$ as a function of $1-\rho_{\rm ch}$.}
\label{fig:FlatvsMPI}
\end{center}
\end{figure}

Events with both jet signals and underlying event activity are expected to have a larger spread, and correspondingly, $\sigma_{p_{\rm T}^{\rm cell}}$ will be higher. In contrast, events having only soft production of particles will have a narrower distribution of $p_{\rm T}^{\rm cell}$, and therefore corresponding $\sigma_{p_{\rm T}^{\rm cell}}$ will be smaller. It is important to note that the presence of jet signals in the jetty events can also enhance $\langle p_{\rm T}^{\rm cell}\rangle$ in the denominator of Eq.~\eqref{eq:flatpt}; however, the increase in $\langle p_{\rm T}^{\rm cell}\rangle$ will be overshadowed by the corresponding rise in $\sigma_{p_{\rm T}^{\rm cell}}$ in the numerator. Thus, the value of $\rho_{\rm ch}$ is expected to be smaller for isotropic events, while a large value of $\sigma_{p_{\rm T}^{\rm cell}}$ corresponds to a jetty event.

However, the current PID and tracking capabilities at ALICE (LHC) and the STAR experiment (RHIC) are limited to $|\eta|<1.0$. As a result, using the above definition, one can not estimate the value of $\rho_{\rm ch}$ at these experiments. Moreover, the measurement of both the event shape and the particle of interest in the same pseudorapidity region induces a bias. The bias can be mitigated by choosing the event shape estimator in the forward rapidity region, which ensures it is in a different pseudorapidity region compared to the particle of interest. Most of the present detectors can measure particle multiplicity in the forward rapidity. Thus, a redefinition of flattenicity is necessary to account for the issues mentioned above, which would consider the charged-particle multiplicity in each cell at the forward rapidity instead of $p_{\rm T}^{\rm cell}$. Flattenicity ($\rho_{\rm ch}$) is determined in the forward-rapidity region, where the ($\eta-\phi$) space is divided into ($8\times 8$) cells and charged particles in each cell `$i$' ($N_{\rm ch}^{\rm{cell}, i}$) is estimated. Thus, $\rho_{\rm ch}$ can be estimated using the following equation~\cite{Ortiz:2022mfv}.
\begin{equation}
    \rho_{\rm ch} = \frac{\sqrt{\sum_{i}(N_{\rm ch}^{\rm{cell}, i}-\langle N_{\rm ch}^{\rm{cell}}\rangle)^2/N^{2}_{\rm{cell}}}}{\langle N_{\rm ch}^{\rm{cell}}\rangle}
    \label{eq:flatNch}
\end{equation}
Here, $\langle N_{\rm ch}^{\rm{cell}}\rangle$ is the mean number of charged particles in the cells. By construction, $\rho_{\rm ch}$ ranges from 0 to 1, where the lower limit, $\rho_{\rm ch}\rightarrow0$ indicates isotropic events and $\rho_{\rm ch}\rightarrow1$ indicates jetty events. To be consistent with other event-shape observables, here, throughout the manuscript, we use $(1-\rho_{\rm ch})$ instead of $\rho_{\rm ch}$, where events with $(1-\rho_{\rm ch})\rightarrow 1$ are likely to be isotropic while $(1-\rho_{\rm ch})\rightarrow 0$ indicates to a jetty event. This is shown in Fig. \ref{fig:flatdist}.

Figure~\ref{fig:FlatvsMPI} shows the correlation of charged-particle flattenicity, ($1-\rho_{\rm ch}$), with $N_{\rm mpi}$. $\rho_{\rm ch}$ is measured using the charged particles having $p_{\rm T}>0$ the V0 acceptance region of ALICE, i.e., $-3.7 < \eta < -1.7$ and $2.8 < \eta < 5.1$, where ($\eta-\phi$) space is divided into a ($8\times8$) grid. The correlation of $(1-\rho_{\rm ch})$ with $N_{\rm mpi}$ remains almost flat at 0 till $(1-\rho_{\rm ch})<0.6$, then starts to rise rapidly, where most of the events lie.

It is noteworthy to mention that the charged-particle flattenicity distribution depends on the detector acceptance range from which the charged-particle multiplicity is measured. In this review, we discuss the flattenicity distribution that is obtained by using the multiplicity from the V0M acceptance. In Run 3, a preliminary study has been performed using the FV0A detector to study the flattenicity dependence of strangeness production~\cite{QM2025:flatPoster}. However, in the future, it would be interesting to investigate charged-particle flattenicity distribution and the studied observables as a function of charged-particle flattenicity, considering the charged-particle multiplicity from the newly built FT0M and MFT forward detectors of ALICE in Run 3. It is worth mentioning that, to make the studies coherent and detector/experiment independent, one needs to employ an unfolding procedure to retrieve a true flattenicity distribution from the folded distributions with the detector effects.     
%\begin{figure}[ht!]
%\includegraphics[scale=0.4]{RTdist.pdf}
%\includegraphics[scale=0.4]{NTalldist.pdf}
%\includegraphics[scale=0.4]{RTalldist.pdf}
%\caption{}
%\label{}
%\end{figure}

%\subsection{Other event topology classifiers}
%In addition to the event shape classifiers discussed in the previous sections, there are also other observables that can make event classifications based on different properties of events. They are discussed below.
\subsection{Flow vectors ($q_{n}$)}
In heavy-ion collisions, the initial configuration of the collision overlap region can affect the final state azimuthal distribution of charged particles. In non-central collisions, the collision overlap region is elliptic in the transverse plane, and thus, the final state azimuthal distributions of particles are elliptic, leading to a larger value of elliptic flow in the mid-central collisions as compared to the most-central collisions. Further, the density fluctuations can lead to observations of higher-order flow coefficients. Interestingly, one can make classifications of events based on the azimuthal distribution of particles in the final state using the reduced flow vectors ($q_{n}$), defined as follows~\cite{STAR:2002hbo, ATLAS:2015qwl}:
\begin{equation}
q_{\rm n}=\frac{|Q_n|}{\sqrt{M}},
\end{equation}
where
\begin{equation}
   Q_{n} =\sum_{j=1}^{M}e^{in\phi_{j}}.    
\end{equation}
Here, $M$ is the total number of charged hadrons, and $\phi_{j}$ is the azimuthal angle of the $j$th hadron. $n$ denotes the order of the reduced flow vectors. Here, $q_2$ can separate events based on the elliptic distribution of the hadrons in the transverse plane. Similarly, $q_3$ isolates events based on the triangular distribution of the hadrons in the transverse plane. For $M\rightarrow\infty$, the reduced flow vectors are proportional to anisotropic flow coefficients of similar order~\cite{ATLAS:2015qwl}. However, this proportionality becomes weaker as $M$ is reduced~\cite{ATLAS:2015qwl}. Thus, $q_{n}$ are not good observables for the low multiplicity events~\cite{ATLAS:2015qwl}.

\subsection{Aplanarity ($A$)}

Aplanarity is another interesting event shape observable that characterizes the shape of the underlying event (UE) based on the momentum of the charged hadrons. Aplanarity is estimated from the linearized sphericity tensor $S^{\rm \mu \nu}$, and defined as ~\cite{Hanson:1975fe, CMS:2018mdd, ALICE:2012cor};

\begin{equation}
    S^{\rm \mu \nu}= \frac{\sum_{i =1}^{N_{\rm ch}} p^{{\rm \mu}}_{i} p^{{\rm \nu}}_{i}/|p_{i}| }{\sum_{i =1}^{N_{\rm ch}} |p_{i}|}.
    \label{Smunu}
\end{equation}
Here, the index $i$ runs over all the charged hadrons in a given event. $\mu$ and $\nu$ indices refer to one of the (x, y, z) components of the momentum of a particle. The individual eigenvalues $\lambda_1, \lambda_2,  \lambda_3$ of the momentum tensor $S^{\rm \mu \nu}$ are normalized and ordered such that $\lambda_1 > \lambda_2 > \lambda_3$ and by definition $\lambda_1+\lambda_2+\lambda_3$ =1. Using these eigenvalues, one can construct various event shape observables such that;

\begin{itemize}
    \item Aplanarity: $A$ = $\frac{3}{2}$$ \lambda_3$ measures the amount of transverse momentum in or out of the plane formed by the two leading order eigenvectors, $\lambda_1$ and  $\lambda_2$. Theoretically, $A$ lies between $0 \leq A < 1/2$; however, the typical measured values fall between $0 \leq A < 0.3$.  The value of $A$ close to zero indicates a relatively planar event, while $A\rightarrow 1/2$ indicates the isotropic events.

    \item Sphericity: S = $\frac{3}{2} (\lambda_2 + \lambda_3) $ quantifies the total transverse momentum with respect to the sphericity axis defined by the four momenta used in event shape measurement (usually, the first eigenvector). The theoretical values fall between $0 \leq S < 1$. However, the inclusion of the smallest eigenvalue $ \lambda_3$, the typical maximum value achieved in the experiment is close to S $\simeq 0.8$. S$\rightarrow0$ indicates a dijet kind of event, while S$\rightarrow1$ indicates the isotropic events.

    \item $Y = \frac{\sqrt{3}}{2}$ ($\lambda_{2} - \lambda_{3}$). Events with both $S$ and $Y$ values are small are cigar-shaped, and collinear, while events with $Y = \frac{1}{\sqrt{3}} S$ (that implies $\lambda_{3} = 0$) are disc-shaped and coplanar.

    \item $C$ = 3($\lambda_1 \lambda_2 + \lambda_1 \lambda_3 + \lambda_2 \lambda_3$)
    measures the events with three jets. C$\rightarrow0$ indicates dijet events.

    \item $D$ = 27$\lambda_1 \lambda_2  \lambda_3 $ measures the events with four jets. D$\rightarrow0$ for dijet or trijet events.

\end{itemize}

\subsection{Event isotropy}
\label{sec:eventisotropy}
Recently, a novel event shape observable called event isotropy ($\mathcal{I}$) has been designed to identify uniform radiation patterns. Event isotropy is defined as the dimensionless distance of a collider event $\mathcal{E}$ and isotropic radiation pattern $\mathcal{U}$ of the same energy~\cite{Cesarotti:2020hwb, Cesarotti:2020ngq, ATLAS:2023mny}. This distance is evaluated by solving optimal transport problems, using the ‘Energy-Mover’s Distance’ (EMD). It is designed for new physics signals that are far from QCD-like.

\begin{equation}
    \mathcal{I} \equiv \rm EMD(\mathcal{E}, \mathcal{U})
\end{equation}

The definition of event isotropy depends on the isotropic reference geometries such as spherical, cylindrical, circular (or Rings), Rings with Dipole geometry, and corresponding ground measure~\cite{Cesarotti:2020hwb, Cesarotti:2020ngq, ATLAS:2023mny}. By construction, the event isotropy $\mathcal{I}$ is bounded on $\mathcal{I}$ $\in$ [0, 1].  Perfectly isotropic events take a value of $\mathcal{I}$ = 0, meaning there is zero distance between radiation patterns. Nearly least isotropic (jetty type) events take values approaching $\mathcal{I}$ = 1. To follow the similar
historical convention with all other event shape observables,  events that the least isotropic (‘dijet-like’) topology is near values of 0 and
the most isotropic topology is near values of 1, it is always preferred to use 1- $\mathcal{I}$, instead of $\mathcal{I}$. It is found that event isotropies can differentiate between quasi-isotropic events better than the existing observables $C$ and $D$ parameter~\cite{Cesarotti:2020hwb, Cesarotti:2020ngq, ATLAS:2023mny}.

\subsection{Fox-Wolfram moments}
\label{sec:Fox-moments}
In $e^{+}e^{-}$ collisions, a set of unique event shape observables named as Fox-Wolfram moments (FWMs) is introduced to discuss the geometrical shape of the entire jets in an event~\cite{Fox:1978vu}.  Later, FWMs are used in collider physics, particularly in high-energy physics, including studies of Higgs and top physics~\cite{Bernaciak:2012nh} and hadron collisions~\cite{Spiller:2015axa}. At LHC energies, FWMs are used to study the medium modifications of jet shapes in relativistic heavy-ion collisions~\cite{Kong:2024wdk}. FWMs, which measure the angular distribution of the energy flow, are characterized by the superposition of spherical harmonics $Y_{l}^{m}$. It is defined as:

\begin{align}
    \mathcal{H}_{l} & \equiv  \left( \frac{4\pi}{2l + 1}\right) \sum_{m = -l}^{+l} \left|\sum_{i} Y_l^m (\Omega_{i}) \frac{|\vec{p_i}|}{\sqrt{s}} \right|^{2} \nonumber \\& 
    = \sum_{i,j}  \frac{|\vec{p_i}| |\vec{p_j}|}{s} P_l (cos\phi_{ij})
\end{align}

where the index $i$ and $j$ run over all the hadrons in an event, $\phi_{ij}$ is the angle between particle $i$ and $j$. The energy momentum  conservation equation demands $\mathcal{H}_{0} = 0$ and  $\mathcal{H}_{1} = 1$. In practice, $\mathcal{H}_{2}$ and  $\mathcal{H}_{3}$ are used to describe the geometrical shape of the jets in an event. These moments are sensitive to the number, angular correlation, and energy distribution of jets.

\subsection{Thrust, major and minor, oblateness}
\label{sec:thrust}
Apart from the directly global observables discussed so far, there are other sets of global observables that exist in the literature, such as transverse thrust ($\tau_{\perp}$), and are defined in terms of the so-called thrust axis for each event. This thrust axis is defined in such a way that the total transverse momentum of the jets used in the measurement is minimized. Mathematically,   

\begin{equation}
    T =  \max_{\hat{n}_{\perp}} \frac{\sum_{i}|\vec{p}_{i} \cdot \hat{n}_{\perp}|}{\sum_{i} |\vec{p_{i}}|}
    \label{T}
\end{equation}

\begin{equation}
    T_{\perp} = \max_{\hat{n}_{\perp}} \frac{|\vec{p}_{{\rm T}_i} \cdot \hat{n}_{\perp}|}{\sum_{i} |p_{{\rm T}_i}|}
    \label{Tper}
\end{equation}
\begin{equation}
   \tau_{\perp} = 1 - T_{\perp} 
    \label{thrust}
\end{equation}

where the sum runs over all the hadrons produced in the final state with total momentum $p_{\rm i}$, transverse momentum $p_{\rm Ti}$, and the unit vector $ \hat{n}_{\perp}$ is used to maximize the sum. The variable $\tau_{\perp}$ is defined in terms global transverse thrust $T_{\perp}$ in order to keep the common-event shape definition. The allowed values for $\tau_{\perp}$ lies in between the ranges $0 < \tau_{\perp} < 1/3$. This is due to the fact that the allowed values of $T_{\perp}$ fall within the range $0 < T_{\perp} < 2/3$. The smaller values of $\tau_{\perp}$ denote extreme two-jet ({\it i.e}, cigar-like) events, whereas the higher values denote spherical events. Here, $T$ is the global event-shape variable thrust. The value of $T$ can vary between 0.5 for spherical events to 1.0 for narrow jet events. Since the thrust variable is linear in momentum, it is more sensitive to low-momentum particles than sphericity. 

\begin{figure}[ht!]
\begin{center}
\includegraphics[scale=0.35]{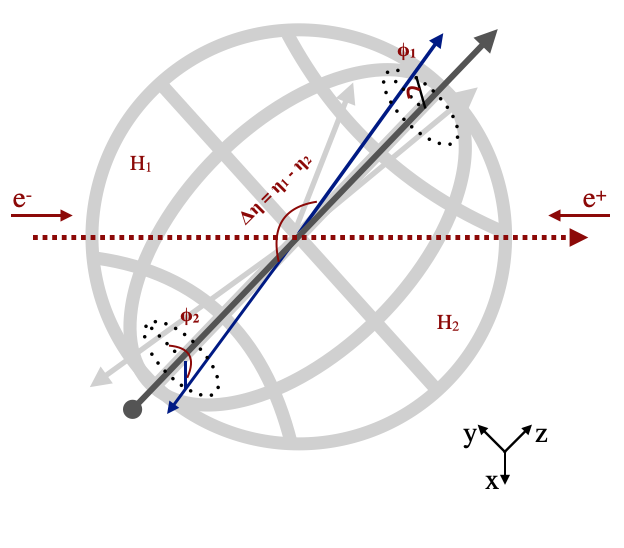}
\caption{Depiction of thrust axis used as reference to study the final state particles.}
\label{fig:thrustcartoon}
\end{center}
\end{figure}

A plane through the origin and perpendicular to the thrust axis divides the event into two hemispheres $H_1$ and $H_2$ as shown in Fig.~\ref{fig:thrustcartoon}. To obtain the thrust major $T_{\rm maj}$, the maximization of Eq.~\ref{T} is performed within that plane~\cite{Barber:1979yr}. Furthermore, the thrust minor $T_{\rm min}$ is given by the argument of Eq.~\ref{T} and estimated in the direction perpendicular to both thrust major and thrust axes. It is defined as;

\begin{equation}
     T_{\rm min} = \frac{\sum_{i}|\vec{p}_{i} \times \hat{n}_{\perp}|}{\sum_{i}  |p_{i}|}
    \label{minor}
\end{equation}
The observable  $T_{\rm min}$ is called the directly global thrust minor, which measures the out-of-event-plane energy flow.

 Using the thrust major  $T_{\rm maj}$ and thrust minor $T_{\rm min}$ one can define oblatness ($O$) as~\cite{Barber:1979yr}
 \begin{equation}
     O = T_{\rm maj} - T_{\rm min}
\end{equation}

\begin{figure}[ht!]
\begin{center}
\includegraphics[scale=0.27]{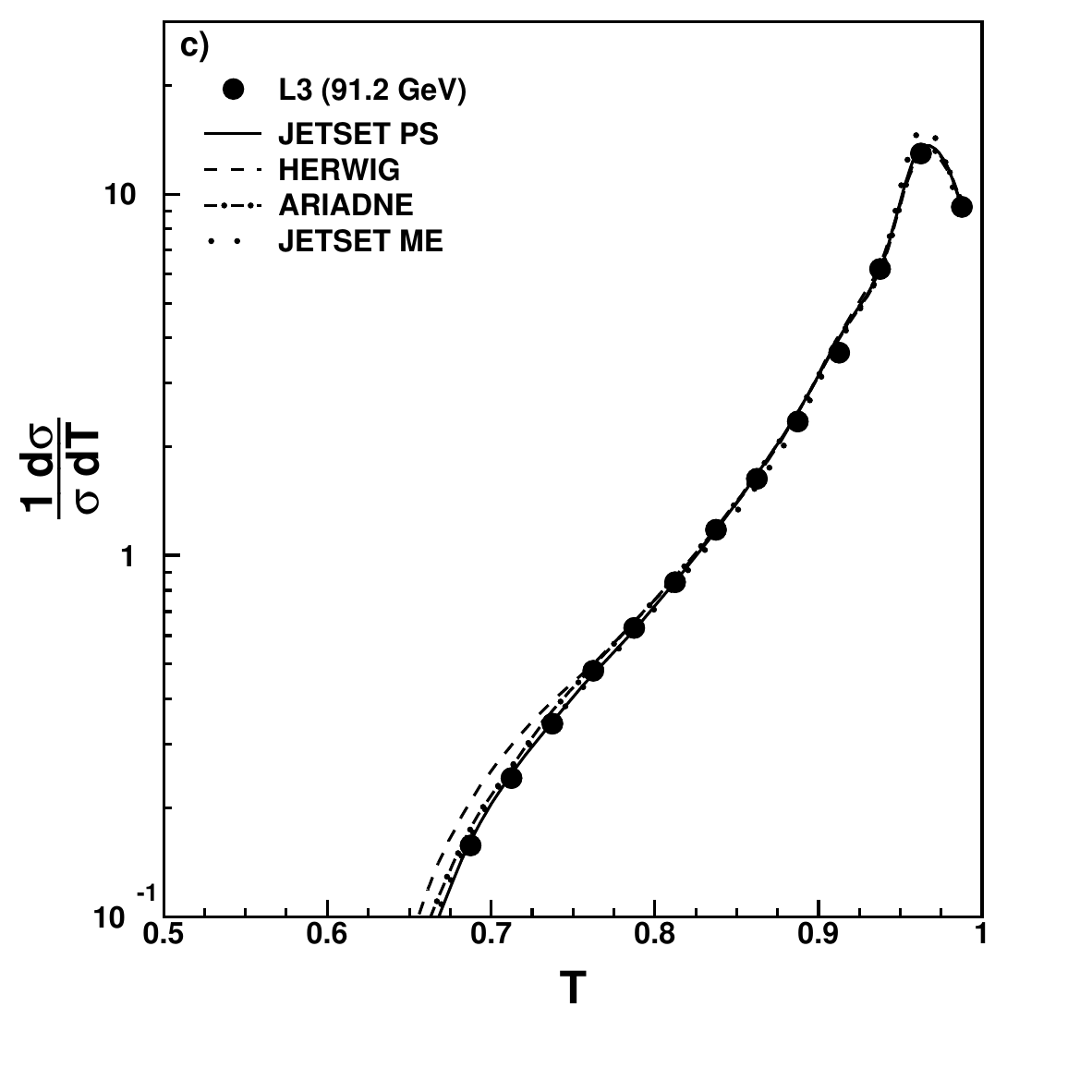}
\includegraphics[scale=0.27]{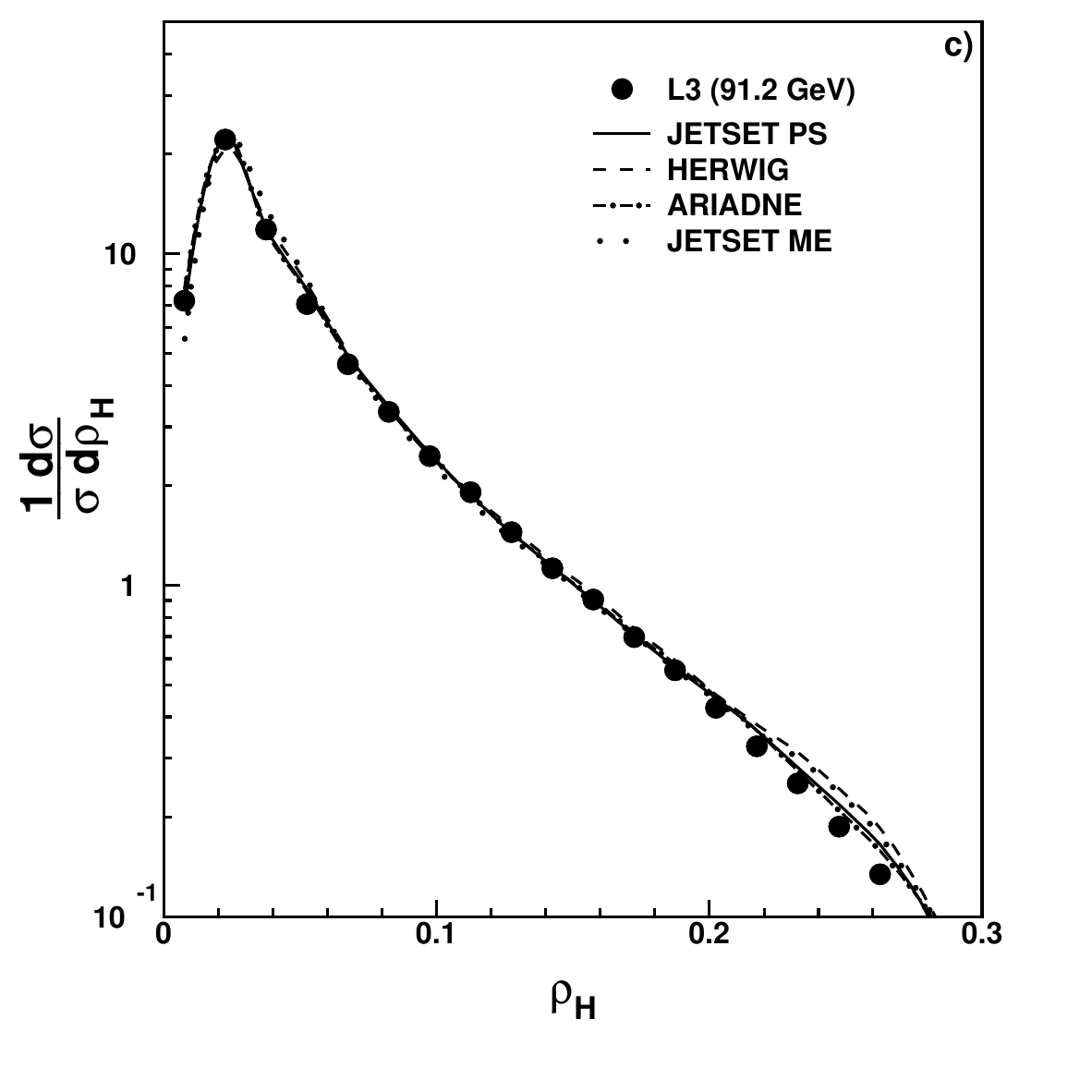}
\includegraphics[scale=0.27]{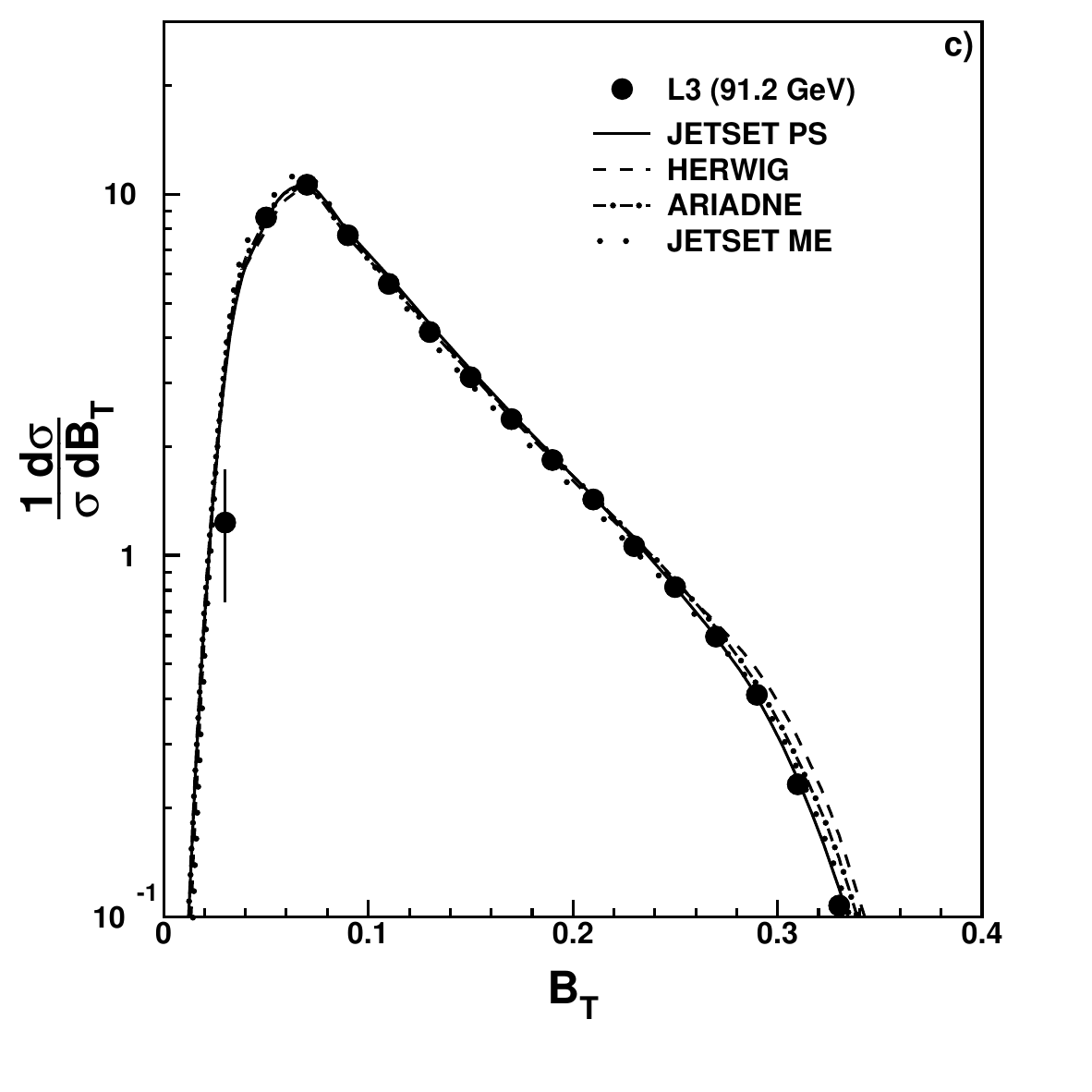}
\includegraphics[scale=0.27]{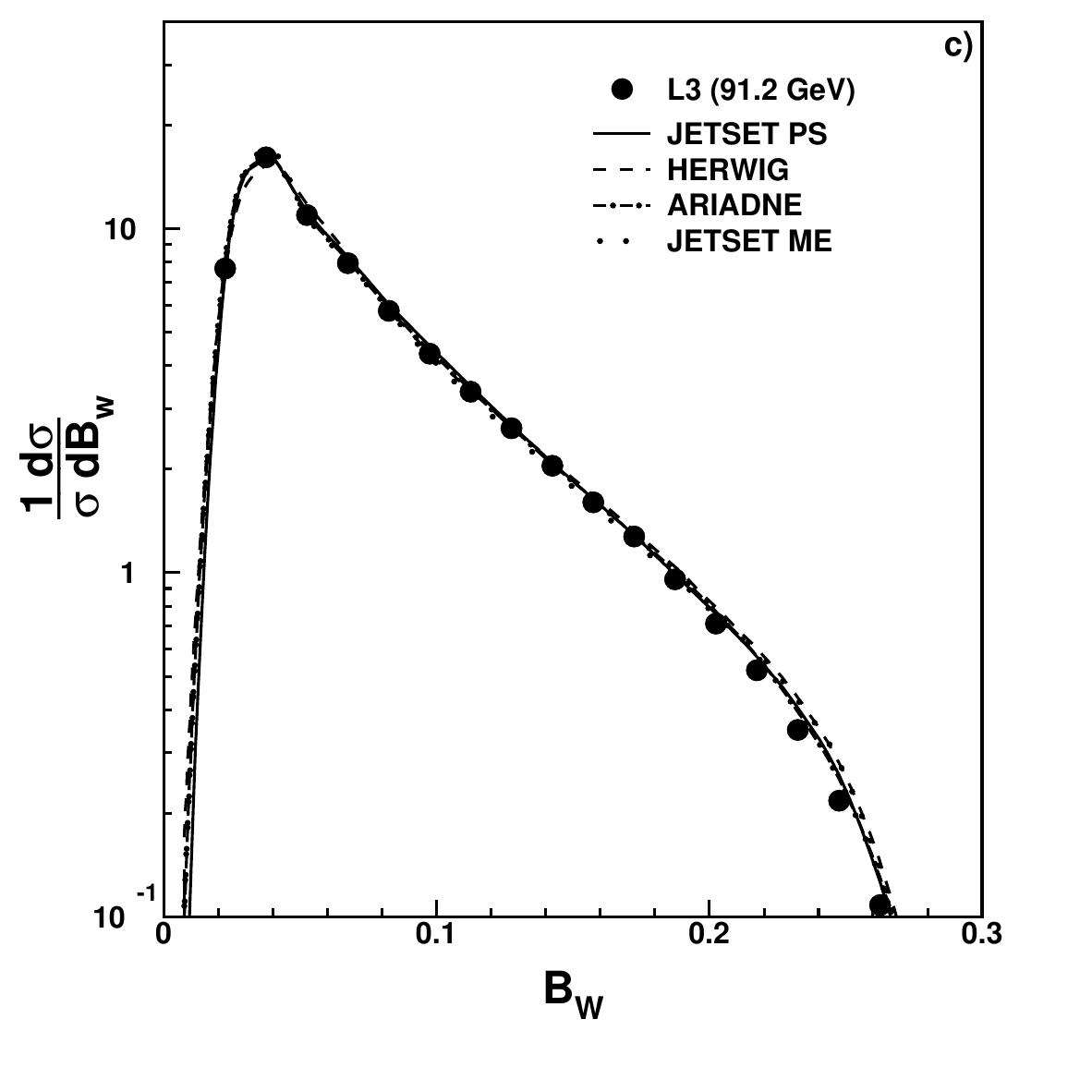}
\includegraphics[scale=0.27]{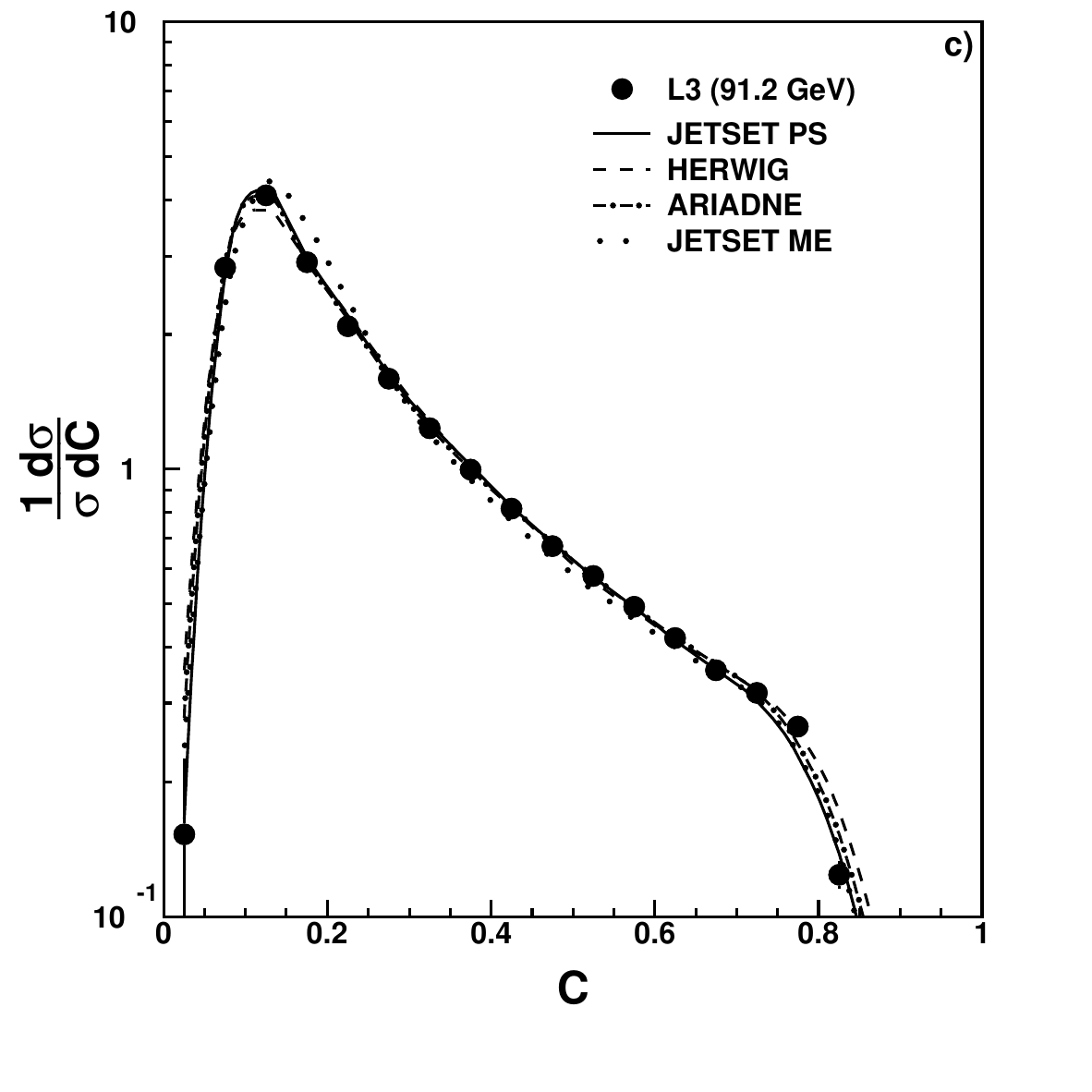}
\includegraphics[scale=0.27]{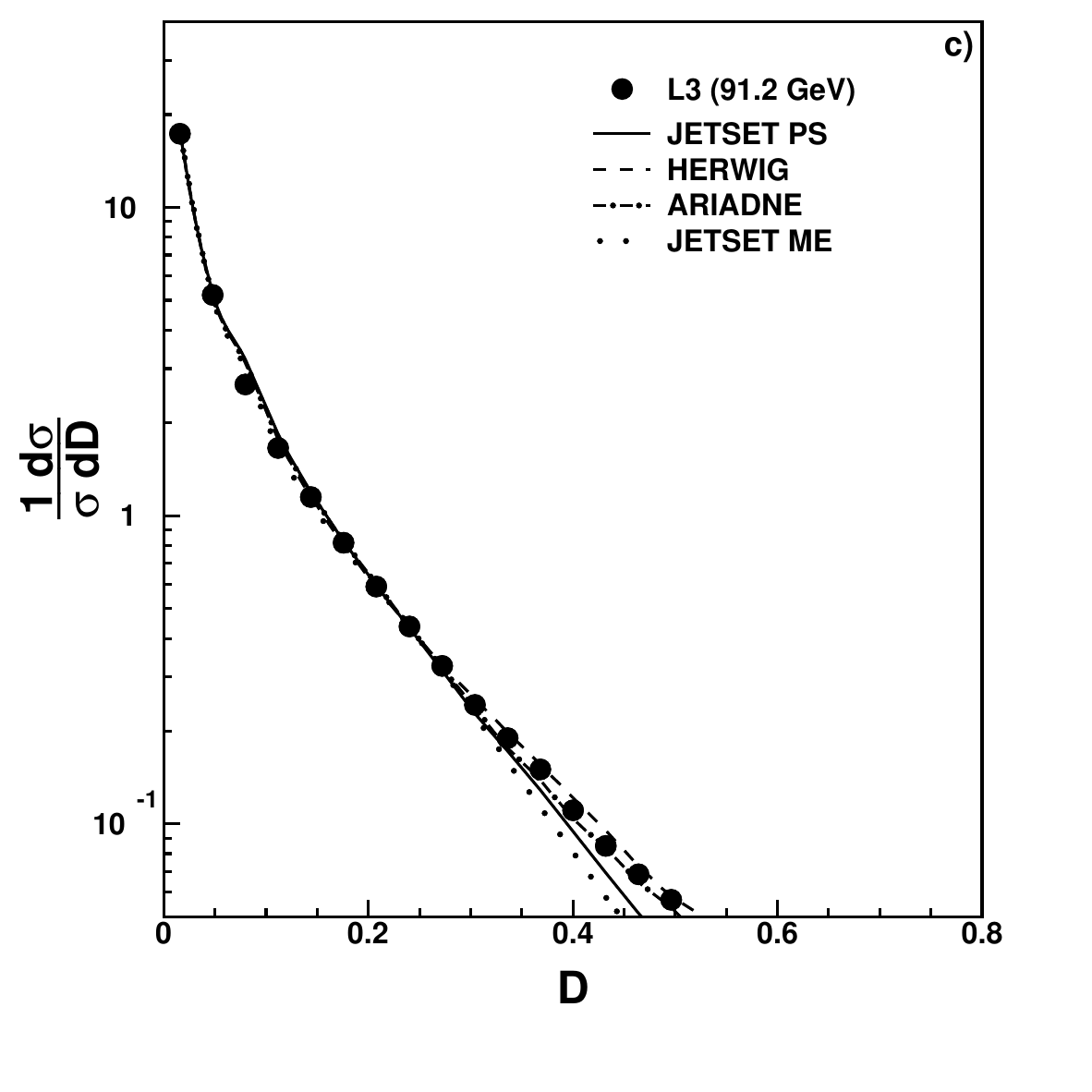}
\caption{The distribution of thrust $T$, heavy jet mass $\rho$, total and wide-jet broadenings $B_{T}$ and $B_{W}$, $C$ , and $D$ parameters at $\sqrt{s}$ =  91.2 GeV in comparison with various QCD model prediction~\cite{L3:1992nwf,L3:1992btq, L3:1995eyy,L3:1997bxr,L3:1997dkv,L3:1998ubl,L3:2000shd,L3:2002oql,L3:2004cdh}}
\label{fig:thrust}
\end{center}
\end{figure}

\begin{figure}[ht!]
\begin{center}
\includegraphics[scale=0.85]{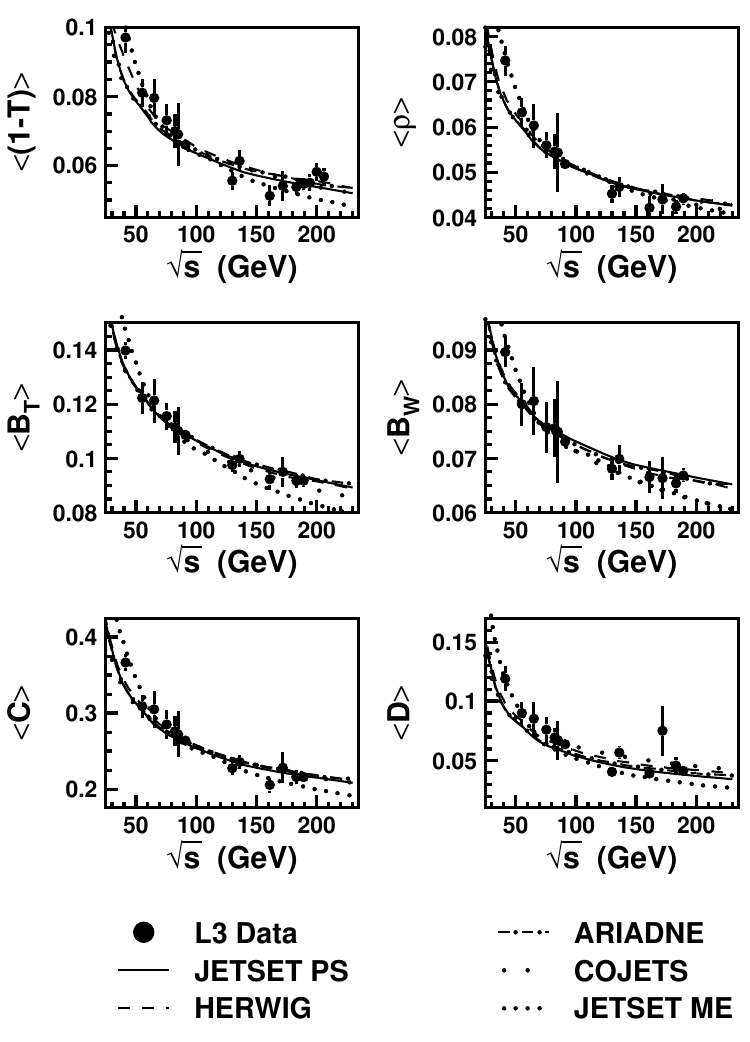}
\caption{The first moment of the event shape variables thrust $T$, heavy jet mass $\rho$, total and wide-jet broadenings $B_{T}$ and $B_{W}$, $C$, and $D$ parameters as a function of  center of mass energy $\sqrt{s}$, compared with several QCD model prediction~\cite{L3:1992nwf,L3:1992btq, L3:1995eyy,L3:1997bxr,L3:1997dkv,L3:1998ubl,L3:2000shd,L3:2002oql,L3:2004cdh}}
\label{fig:thrustvsenergy}
\end{center}
\end{figure}
\subsection{Heavy-jet mass ($M_{\rm H}$)}
 The invariant mass of the particle in each hemisphere, ${\rm H}_1$ and ${\rm H}_2$, is calculated, and the heavy-jet mass $M_{\rm H}$ is defined as the larger of the two~\cite{Clavelli:1981yh}. For a given center of mass energy $\sqrt{s}$, one can define the scaled heavy jet mass $\rho_{\rm H}$ as follows;
 \begin{equation}
     \rho_{\rm H} =  \frac{M_{\rm H}^{2}}{s}
 \end{equation}

The experiments at LHC incorporate another variant of the jet mass variable which uses the invariant mass ($M_{{\rm H}_{k}}$) of the jet constituents in the hemisphere, ${\rm H}_{k}$ and the scalar sum of the momenta of both sides of the hemisphere, $P$~\cite{CMS:2014tkl, CMS:2018svp}.
Thus, the new definition of the jet mass is given as follows:
\begin{equation}
    \rho_{{\rm H}_{k}}=\frac{M_{{\rm H}_{k}}^{2}}{P^{2}},
\end{equation}
where $k=1$ and 2 stand for either of the hemispheres. One can define the total jet mass of an event as follows:
\begin{equation}
    \rho_{\rm tot}=\rho_{{\rm H}_{1}}+\rho_{{\rm H}_{2}}.
\end{equation}
A transverse component of the total jet mass can also be calculated in a similar manner, which is denoted as $\rho_{\rm tot}^{\rm T}$.

It is important to note that several studies incorporate jet grooming techniques to selectively remove soft wide-angle radiation within jets that is less likely to have originated from the hard parton that initiated the jet~\cite{Larkoski:2017jix, Kogler:2018hem}. These grooming techniques thereby enable a more meaningful heavy jet mass. Some of the popular grooming techniques include trimming~\cite{Krohn:2009th}, pruning~\cite{Ellis:2009su, Ellis:2009me}, filtering or mass drop tagger (MDT)~\cite{Butterworth:2008iy}, modified MDT~\cite{Dasgupta:2013ihk}, and soft drop~\cite{Larkoski:2014wba}.

\subsection{Jet broadening variables $B_{\rm T}$ and $B_{\rm W}$ }
Jet broadening variables measure the fraction of energy that is perpendicular to the thrust axis. For each of the two hemispheres ${\rm H}_k$, one can define~\cite{Kittel:2005fu}
\begin{equation}
    B_{k} =  \frac{\sum_{i \in {\rm H}_{k}}|\vec{p}_{i} \times \hat{n}_{\perp}|}{2 \sum_{i}  |\vec{p}_{i}|}.
\end{equation}
Using the definition of $ B_{k}$, two variables are constructed, such that~\cite{Catani:1992jc}
\begin{equation}
    B_{\rm T} = B_{1} + B_{2} \;\;\;\; \text{and} \;\;\;\; B_{\rm W} = \max(B_{1}, B_{2})
\end{equation}
where $B_{\rm T}$ is the total jet broadening and $B_{\rm W}$ is the wide-jet broadening.

At LHC, the calculation of the jet broadening parameter involves the transverse momentum weighted pseudorapidity ($\eta_{{\rm H}_k}$) and azimuthal angle ($\phi_{{\rm H}_k}$) in each hemisphere $H_{k}$ (for $k=1$ and 2). $\eta_{{\rm H}_k}$ and $\phi_{{\rm H}_k}$ are calculated as follows~\cite{CMS:2018svp}:
\begin{equation}
    \eta_{{\rm H}_k}=\frac{\sum_{i\in {\rm H}_k}p_{{\rm T}_i}\eta_{i}}{\sum_{i\in {\rm H}_k}p_{{\rm T}_i}}\;\;\;\;\text{and}\;\;\;\;
    \phi_{{\rm H}_k}=\frac{\sum_{i\in {\rm H}_k}p_{{\rm T}_i}\phi_{i}}{\sum_{i\in {\rm H}_k}p_{{\rm T}_i}}.
\end{equation}
Subsequently, the jet broadening variable in each of the two hemispheres is redefined as follows~\cite{CMS:2018svp}:
\begin{equation}
    B_{{\rm H}_k}=\frac{\sum_{i\in {\rm H}_{k}}p_{{\rm T}_{i}}\sqrt{(\eta_{i}-\eta_{{\rm H}_k})^{2}+(\phi_{i}-\phi_{{\rm H}_k})^{2}}}{2\sum_{i}p_{{\rm T}_{i}}}.
\end{equation}

These observables $\tau_{\perp}$, $\rho_{\rm H}$ or $\rho $,  $B_{\rm T}$, and $B_{\rm W}$ are most commonly used in QCD studies in $e^{+}e^{-}$ collisions because QCD calculations including the resummation of leading and next-to-leading logarithm (NLLA calculations) terms to all order exist for these observables~\cite{Catani:1992jc, Catani:1992ua}. The distribution of thrust variable $T$, scaled heavy jet mass $\rho_{\rm H}$, total jet broadenings $B_{T}$, and wide-jet broadenings $B_{W}$, $C$, and $D$ parameters in $e^{+}e^{-}$ collisions at center of mass energy $\sqrt{s}$ =  91.2 GeV is shown in Fig.~\ref{fig:thrust}. The experimental data are compared with QCD-based models such as JETSET PS, HERWIG, ARIADNE, and JETSET ME~\cite{L3:1992nwf,L3:1992btq, L3:1995eyy,L3:1997bxr,L3:1997dkv,L3:1998ubl,L3:2000shd,L3:2002oql,L3:2004cdh}. A qualitative agreement between experimental data and model prediction is observed. Furthermore, Fig.~\ref{fig:thrustvsenergy} shows the first-order moment of these same event shape variables as a function of the center of mass energy dependence in $e^{+}e^{-}$ collisions.  Figure~\ref{fig:thrustvsenergy} shows that these event shape variables are dominant at the lower center of mass energies and may have a negligible contribution towards the higher center of mass energy, which is the main focus of this review.

{\it We compile the definitions and practical choices behind the event classifiers used at hadron colliders, starting from charged-particle multiplicity to transverse sphericity/spherocity, $R_{\rm T}$, flattenicity, thrust-family variables, aplanarity, heavy-jet mass, broadenings, and flow vectors. Emphasis is given on infrared/collinear safety, acceptance/region definitions, and on how these observables correlate with multiparton interactions and underlying-event activity. This section highlights the developments of event classifiers and sets the conventions used in the LHC era applications. Let us start from the pre-LHC era history of these classifiers in the next section.}
 
 %The detailed discussion about these observables are outside the scope of this manuscript. 
\section{Pre-LHC era}
\label{prelhcera}
Decades before LHC, event shape observables were designed from a simple idea of the geometry of an event carrying the fingerprints of the underlying QCD dynamics. For a precise quantification of the geometry, several event shape observables that have been discussed in the previous section, such as thrust, sphericity, oblateness, Fox–Wolfram moments, heavy-jet mass, etc., have been developed since the 1970s. The definitions and perturbative properties of the event shape observables were highlighted in early theoretical work, which essentially introduced a means that collider experiments could share across machines and energies~\cite{Ellis:1980wv}. 

At Positron–Electron Tandem Ring Accelerator (PETRA) in the German national laboratory DESY in Hamburg, Germany, in the late 1970s, $e^{+}e^{-}$ collisions were among the first stages for the usage of event shape classifiers. Three-jet topologies were observed and quantified by event-shape measures (e.g., oblateness), giving the first unambiguous evidence for hard gluon radiation and, shortly after, confirmation that the gluon is a spin-1 vector boson~\cite{Barber:1979yr}. Those results were a decisive, data-driven validation of QCD’s gauge structure, and they cemented the role of event topology as a discovery and characterization tool.

Studies during the Large Electron–Positron Collider (LEP) era turned the usage of event classifiers in discovery into precision. With large datasets at $\sqrt{s} \simeq$ 91–209 GeV, LEP experiments measured differential distributions and moments of thrust, C-parameter, and then extracted $\alpha_{s}$ using fixed-order predictions matched to resummation. The LEP QCD program made event shapes a precision instrument which demonstrates the running of $\alpha_{s}$ and tests the hadronization models~\cite{Soldner-Rembold:2004lzl}. 

In deep-inelastic scattering at Hadron–Electron Ring Accelerator (HERA) at DESY in Hamburg, event shapes found a complementary laboratory. The H1 and ZEUS experiments measured thrust, jet broadening, jet mass, and the C-parameter over a wide range of $Q^{2}$ using a Breit frame where the jet from the scattered quark is kinematically separated from the proton remnant~\cite{H1:2005zsk, ZEUS:2006vwm}. These data were compared with QCD predictions that include resummation and simple power-correction terms. Two clear lessons were learned from this. First, event shapes show how perturbative and non-perturbative physics mix, and this balance changes with the energy scale (roughly like $1/Q$). Second, practical choices, such as working in the right frame and hemisphere, matter for keeping observables stable against soft/collinear radiation (IRC safety) and for limiting contamination from soft background activity for hard QCD measurements. 

Hadron colliders before the LHC added the missing piece on how to use event geometry when the initial state carries color and the underlying event (UE) is unavoidable. The Tevatron, a circular particle accelerator at Fermilab, pioneered two pillars that are still central today. First, the Collider Detector at Fermilab (CDF) experiment formalized the UE methodology (dividing in toward, away, and transverse regions) for separating hard recoil from UE activity, which is conceptually similar to more recent observable $R_{\rm T}$, self-normalized activity classifiers at the LHC. Second, CDF performed the first unfolded event-shape measurements for transverse thrust and thrust minor to quantify where calorimeter-level shapes agree (or do not) with modern generators. Together with theory adapting global shapes to hadron colliders, those studies defined how we think about ``jetty vs. isotropic'' events at hadron colliders~\cite{CDF:2011yfm, Field:2001aok, Banfi:2010zza}.

Finally, it is worth noting a parallel study at the B-factories: event shapes like thrust and Fox–Wolfram moments became standard handles to separate spherical $B$ events from jetty continuum, which provided an early pragmatic bridge between physics-motivated shapes and multivariate classifiers~\cite{Belle:2004yyi}. That practice had shown an early forecast of the modern LHC convergence between interpretable event shapes and data-trained event-level discriminants via machine learning, about which we will discuss more in Sec.~\ref{machinelearning}.

{\it Across PETRA, LEP, HERA, the Tevatron, and the B-factories, a consistent picture emerges:
\begin{itemize}
\item  IRC-safe, global event shapes encode robust information about partonic geometry
\item Careful definitions and frames suppress non-perturbative and UE effects for the study in pQCD sectors
\item Precision comparisons to theory calibrate both perturbative predictions and hadronization models
\item Combining event shape information extends sensitivity without discarding interpretability
\item UE methodology that was developed in CDF laid the foundation for several UE-based classifiers in the LHC era
\end{itemize}

These are precisely the pillars we build on in the LHC era. So, let us now move to the LHC era with a focus on soft and hard probes in LHC with event shape classifiers and their physics implications.}

\section{Soft QCD Probes with event shape classifiers}
\label{sec:softprobes}

As discussed in Sec.~\ref{sec:intro}, the soft-QCD calculations are usually cumbersome due to a larger value of $\alpha_{s}$. Thus, the description of the soft-QCD processes or the underlying event (UE) is modeled using different QCD-inspired phenomenological approaches. Unlike in the pre-LHC era, event shape classifiers are playing an instrumental role at the LHC for the precision studies of soft QCD processes in small systems. In addition, they provide important tools for studying the QGP-like behavior discovered for small collision systems. In this section, we discuss the role of event classifiers in understanding soft QCD processes in terms of multi-differential studies with several experimental observables.

\subsection{Integrated yield and transverse momentum spectra}

\begin{figure}[ht!]
\begin{center}
\includegraphics[scale=0.4]{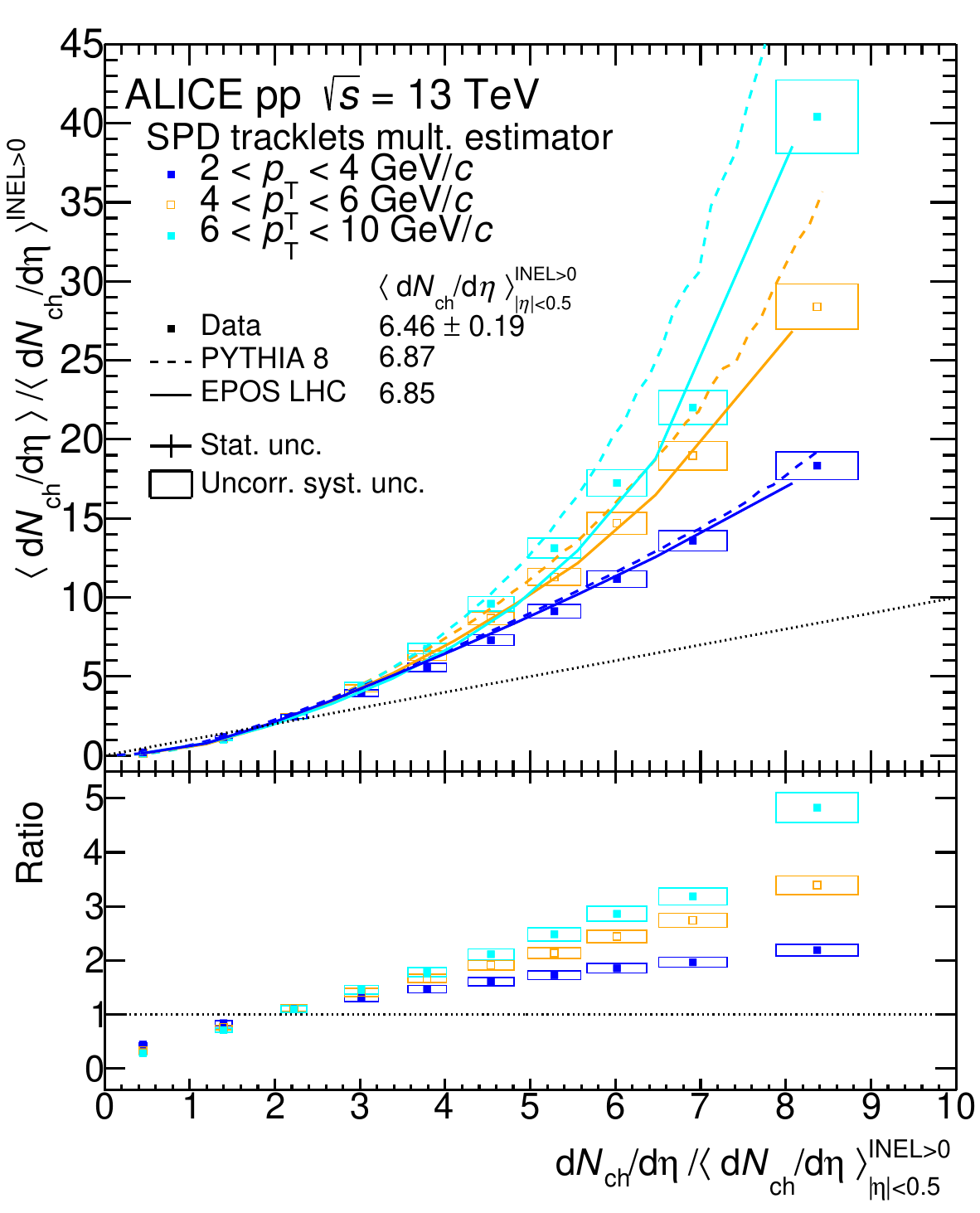}

\caption{Self-normalied yield of charged particles integrated over different $p_{\rm T}$-intervals: $2<p_{\rm T}<4~{\rm GeV/c}$ (blue), $4<p_{\rm T}<6~{\rm GeV/c}$ (yellow) and $6<p_{\rm T}<10~{\rm GeV/c}$ (cyan) as a function of $p_{\rm T}$-integrated self-normalised charged-particle multiplicity density~\cite{ALICE:2019dfi}. ALICE data are compared with PYTHIA~8 and EPOS LHC estimations.}
\label{fig:selfnormyield}
\end{center}
\end{figure}

\begin{figure*}[ht!]
\begin{center}
\includegraphics[width = 0.49\linewidth]{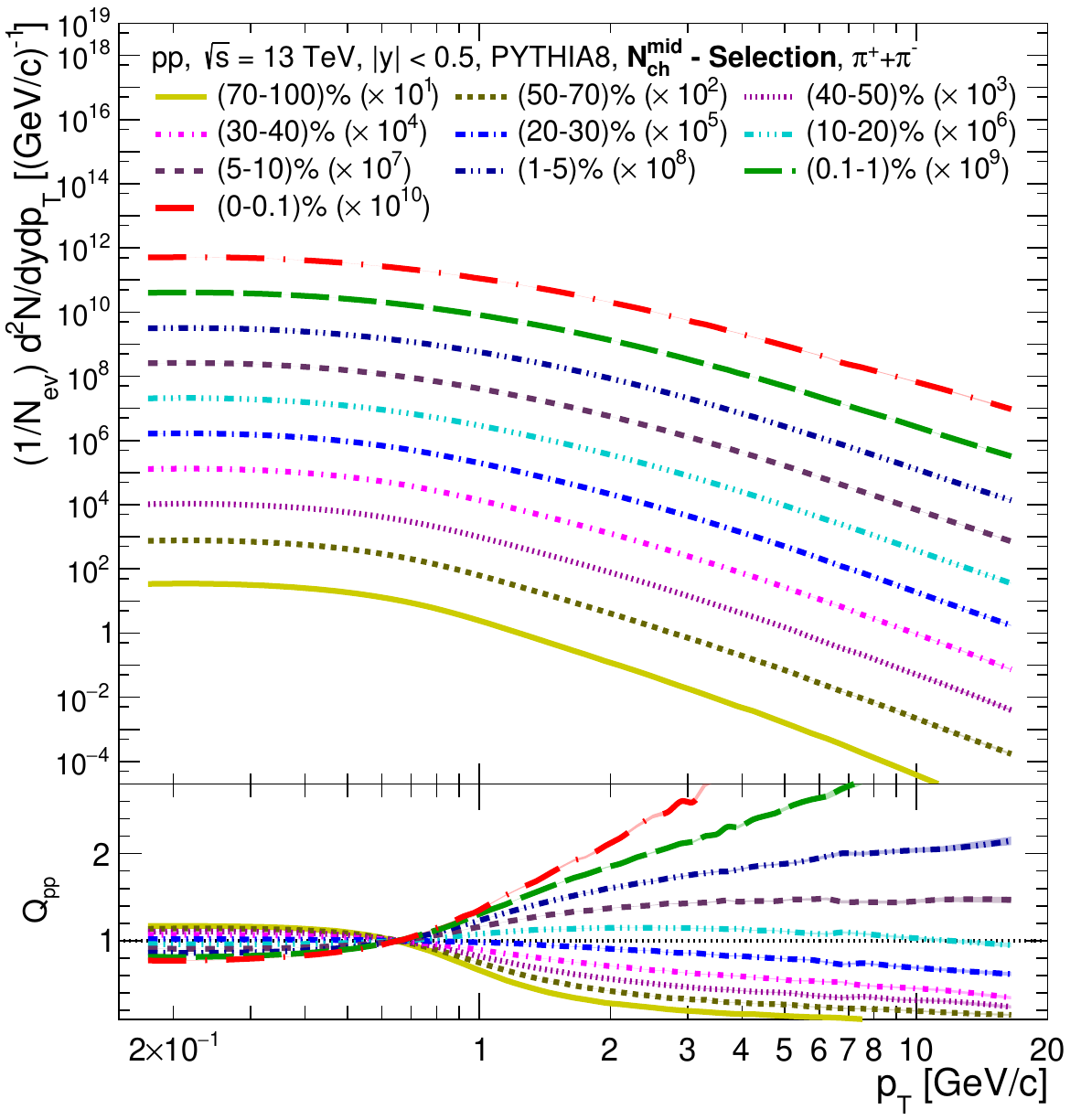}
\includegraphics[width = 0.49\linewidth]{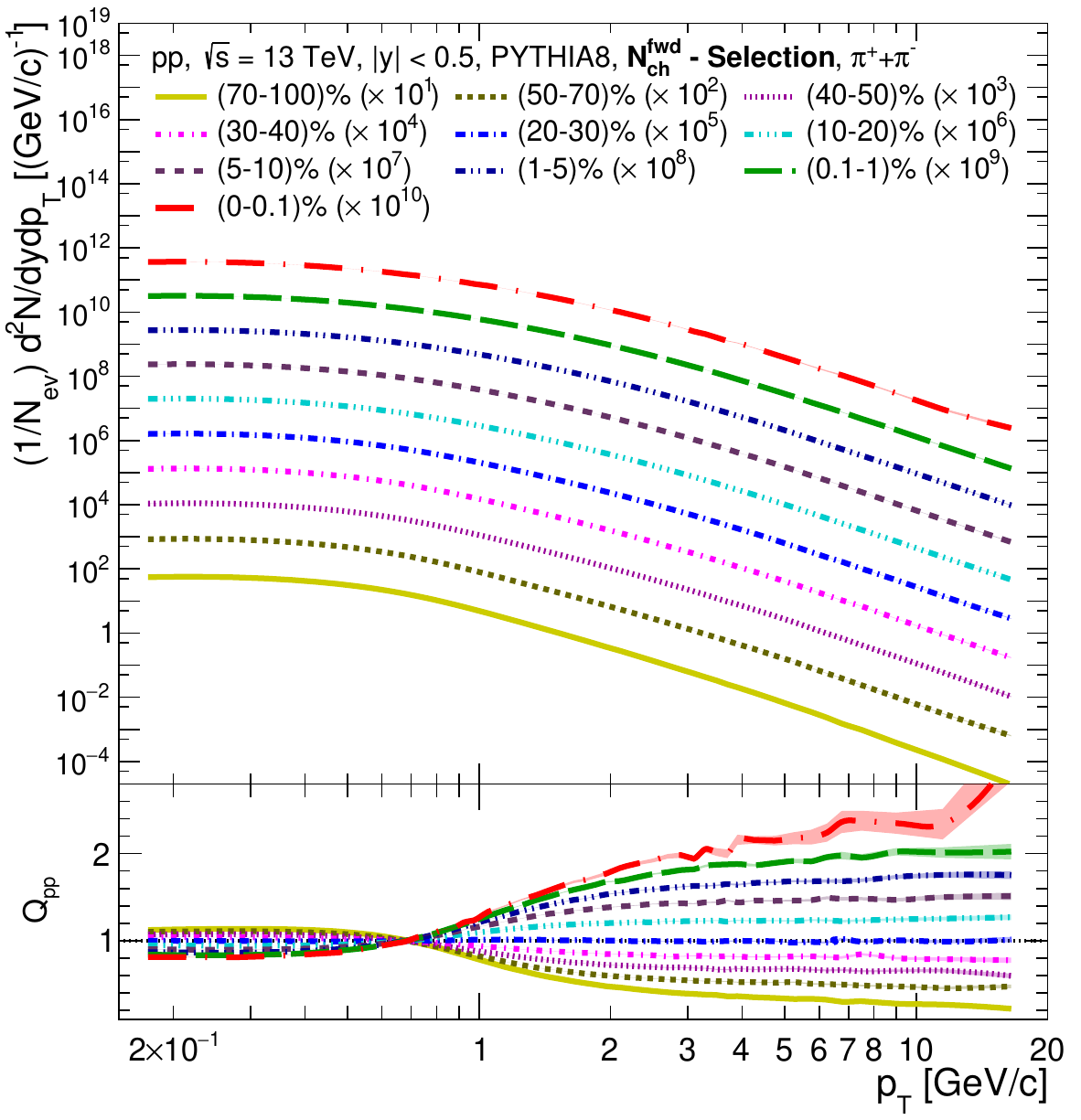}
\caption{ Transverse momentum distribution of charged pions for different slices of midrapidity (left panel) and forward rapidity (right panel) charged-particle multiplicity in $pp$ collisions at $\sqrt{s}=13$ TeV using PYTHIA8. The bottom panels show the ratio with respect to the minimum bias (MB) events.}
\label{fig:pionspectraNch}
\end{center}
\end{figure*}

Let us now resume our discussion about selection and autocorrelation biases, which we started in Sec.~\ref{sec:intro} with the results of transverse momentum spectra. As discussed, to have a proxy for MPI in experiments, several measurements have been performed as a function of charged-particle multiplicity measured in mid- and forward-pseudorapidity regions, which, in MC studies, show significant correlations with $N_{\rm mpi}$. However, measurements by event selections based only on mid-rapidity multiplicity have shown a stronger than linear increase of high-$p_{\rm T}$ particles in high multiplicity (HM) collisions relative to the yield in minimum-bias (MB) $pp$ collisions~\cite{ALICE:2017jyt, ALICE:2019dfi}, as illustrated in Fig.~\ref{fig:selfnormyield}. This is also highlighted in the left panel of Fig.~\ref{fig:pionspectraNch}, where the transverse momentum distribution of charged pions for different slices of midrapidity charged-particle multiplicity in $pp$ collisions at $\sqrt{s}=13$ TeV is shown from PYTHIA8 simulations. The $p_{\rm T}$-dependence of selection bias can be understood from the ratio panels with $Q_{\rm pp}$, defined as follows~\cite{ALICE:2024vaf, R:2023lku, Ortiz:2020rwg}.

\begin{equation}
Q_{\rm pp} = \frac{d^2N_{\rm ch}^{\rm ES}/\langle N_{\rm ch}^{\rm ES} \rangle d\eta dp_{\rm T}} {d^2N_{\rm ch}^{\text{ES-int}}/\langle N_{\rm ch}^{\text{ES-int}} \rangle d\eta dp_{\rm T}}
\label{eq:Qpp}
\end{equation}

Here, $N_{\rm ch}^{\rm ES}$ and $N_{\rm ch}^{\text{ES-int}}$ denote the number of charged particles in a given percentile of event shape and the event shape integrated events. In the ratio plot with respect to the minimum bias collisions, one can clearly see that with increasing multiplicity, the particle production at high-$p_{\rm T}$ is biased towards the hard processes, which gives rise to a stronger than linear increase. This indicates a selection bias towards local fluctuations of choosing only hard $pp$ collisions. Such biases can be diminished when the event selection is performed with the charged particle multiplicity estimated in a different pseudorapidity interval than the observable of interest. The right panel of Fig.~\ref{fig:pionspectraNch} shows the transverse momentum distribution of charged pions for different slices of forward charged-particle multiplicity in $pp$ collisions at $\sqrt{s}=13$ TeV from PYTHIA8 simulations.  It is found that these measurements are affected by the hard processes at high-$p_{\rm T}$~\cite{ALICE:2022qxg}, which is evident from the $\rm Q_{\rm pp}$ in the bottom panel of Fig.~\ref{fig:pionspectraNch}. Thus, such selection biases in measurements hinder the search for the origin of QGP-like behavior in small collision systems.

\begin{figure}[ht!]
\begin{center}
\includegraphics[scale=0.40]{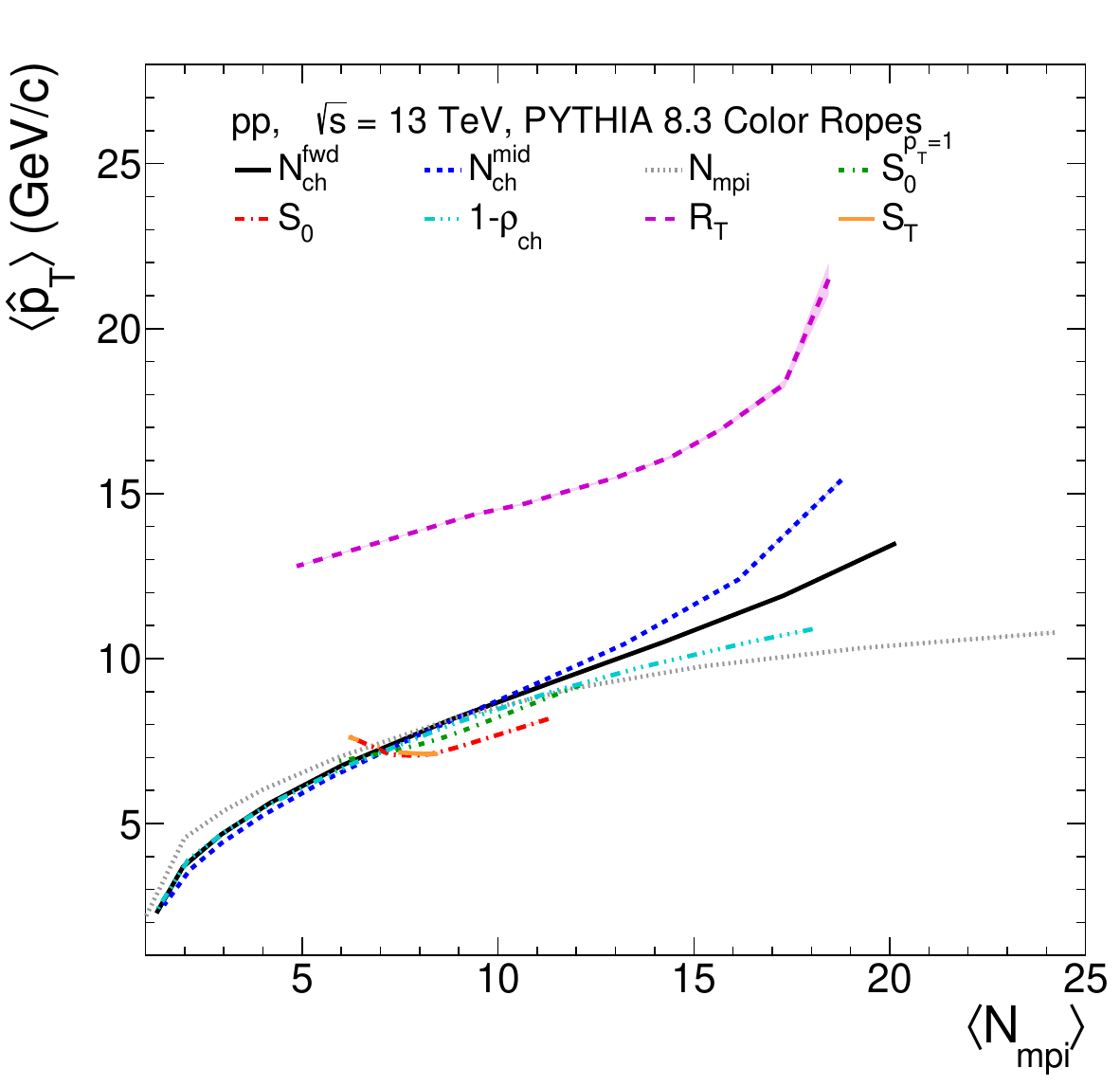}
\caption{Correlation between $\langle N_{\rm mpi}\rangle$ and $\langle \hat{p}_{\rm T}\rangle$ as a function of different percentiles of event classifiers. Comparisons are made for the Color Ropes tune of PYTHIA~8.}
\label{fig:MPIvspThat}
\end{center}
\end{figure}
\begin{figure}[ht!]
\begin{center}
\includegraphics[scale=0.35]{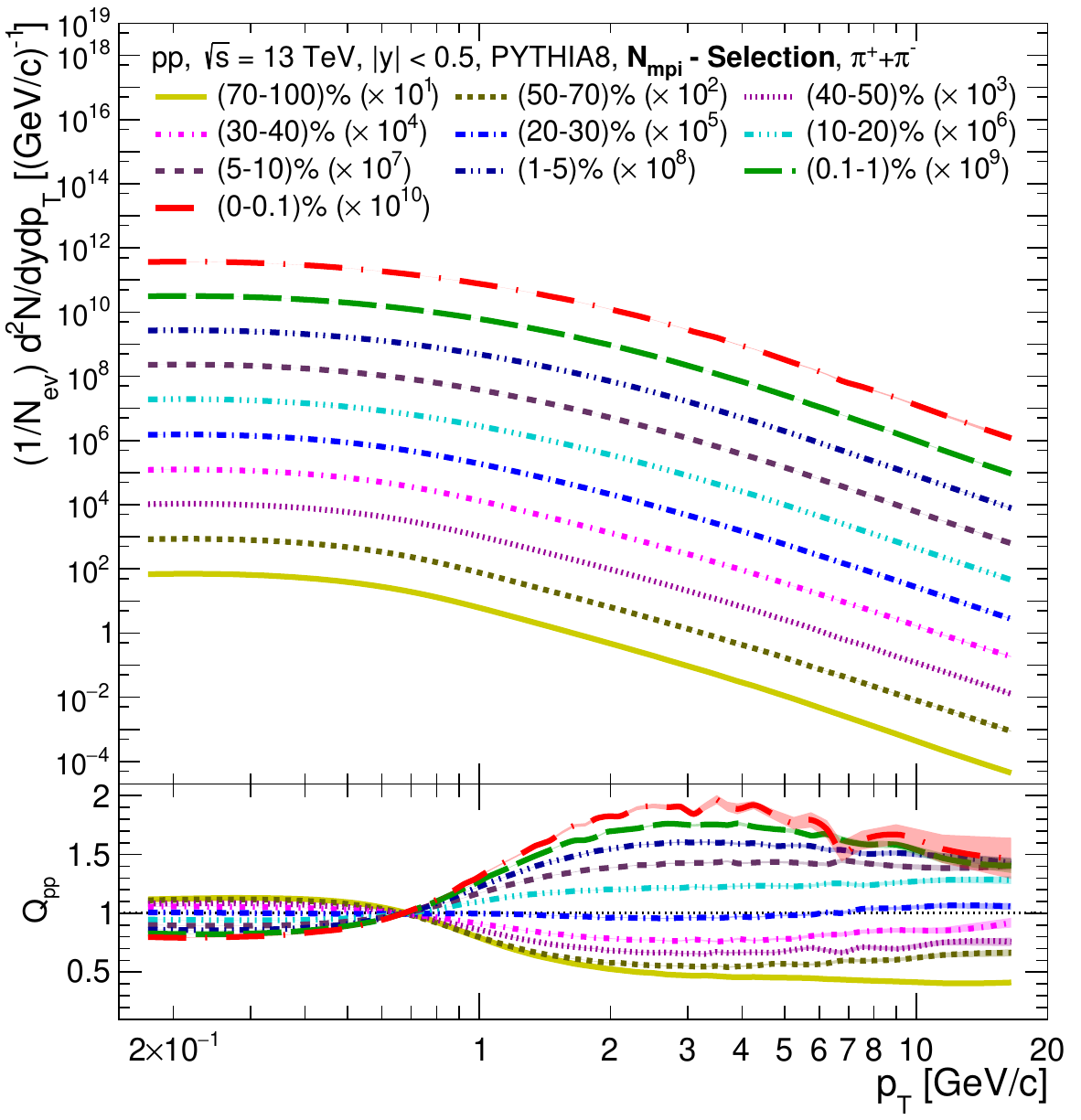}
\caption{Top panel shows the transverse momentum distribution of charged pions for different slices of $N_{\rm mpi}$ in $pp$ collisions at $\sqrt{s}=13$ TeV using PYTHIA8. The bottom panel shows the ratio with respect to the minimum bias (MB) events.}
\label{fig:pionspectraNmpi}
\end{center}
\end{figure}

\begin{figure}[ht!]
\begin{center}
\includegraphics[scale=0.35]{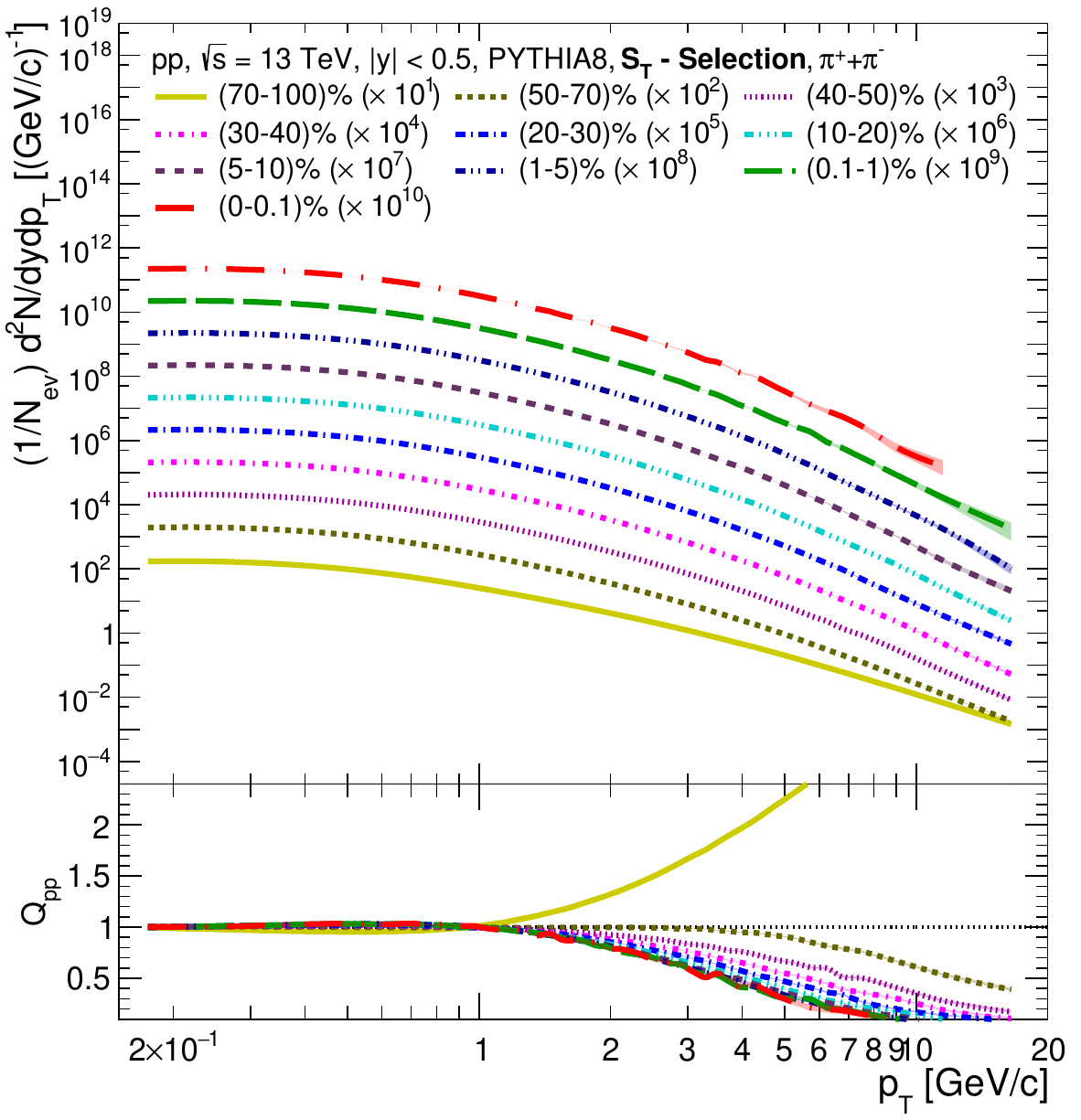}
\caption{Top panel shows the transverse momentum distribution of charged pions for different slices of transverse sphericity ($S_{\rm T}$) in $pp$ collisions at $\sqrt{s}=13$ TeV using PYTHIA8. The bottom panel shows the ratio with respect to the minimum bias (MB) events.}
\label{fig:pionspectraST}
\end{center}
\end{figure}

Let us now move our discussion to the event shape classifiers. Since this section mostly focuses on the soft-QCD sector to find out QGP-like effects, one needs to explore event classifiers that are sensitive to MPI with a significant reduction of selection bias. Event shape observables are expected to be sensitive to MPI. Figure~\ref{fig:MPIvspThat} shows a correlation between the average number of multi-partonic interactions ($\langle N_{\rm mpi}\rangle$) and the average transverse momentum transfer of the hardest
parton-parton interaction ($\langle \hat{p}_{\rm T}\rangle$) as a function of different percentiles of event classifiers.  It can be clearly seen that all the discussed event classifiers are sensitive to MPI. Among those, $R_{\rm T}$ probes higher $p_{\rm T}$ ranges due to high-$p_{\rm T}$ trigger selection. As there is an inherent multiplicity cut of 10 particles, the MPI-coverage for spherocity and sphericity selections is limited. 

Figure~\ref{fig:pionspectraNmpi} shows the transverse momentum distribution of charged pions for different slices of $N_{\rm mpi}$ in $pp$ collisions at $\sqrt{s}=13$ TeV using PYTHIA8. This figure provides a benchmark for the expected behavior of $p_{\rm T}$ for other event classifiers discussed in this review for the studies in the soft QCD sector.

\begin{figure*}[ht!]
\begin{center}
\includegraphics[width = 0.49\linewidth]{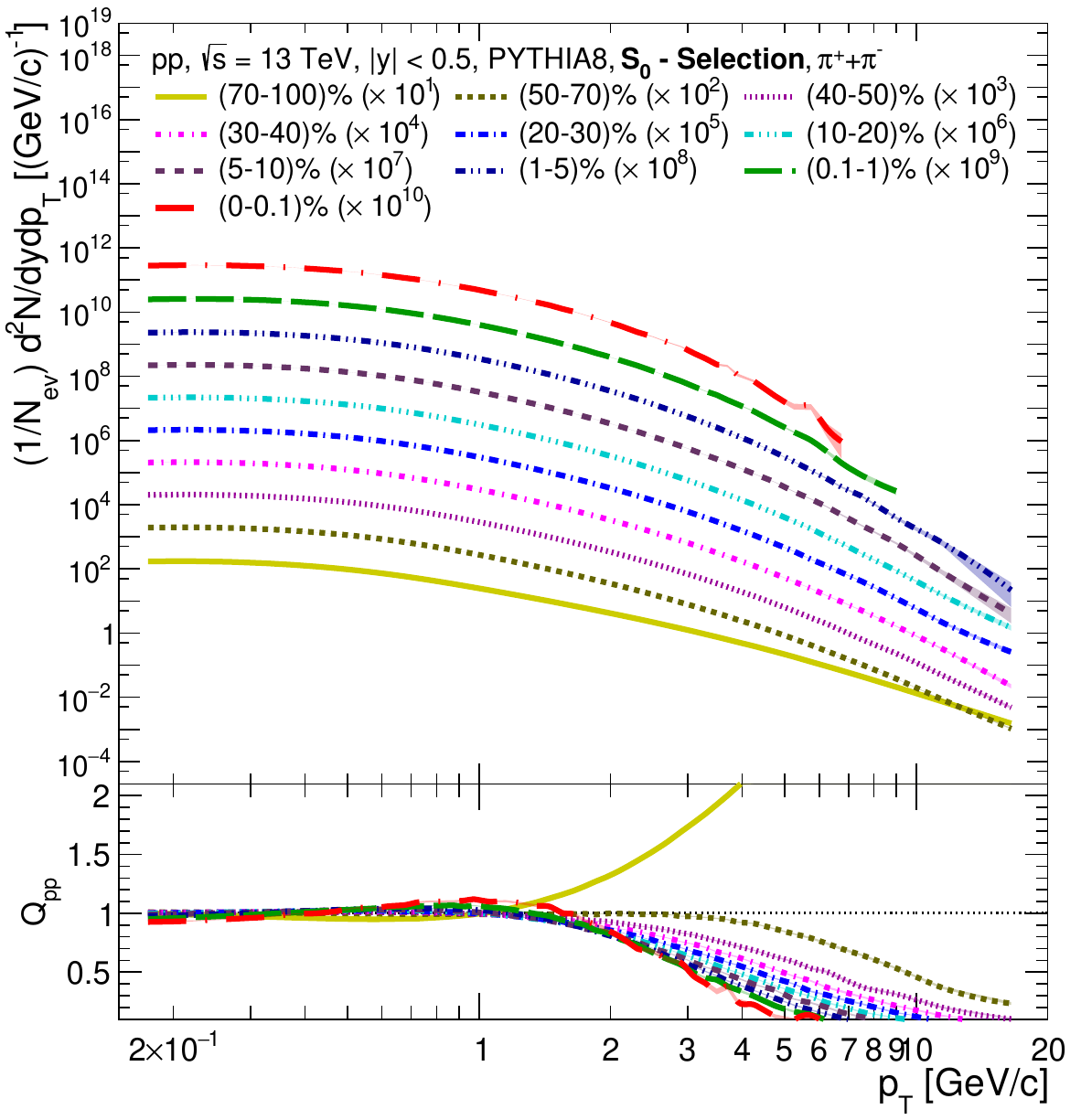}
\includegraphics[width = 0.49\linewidth]{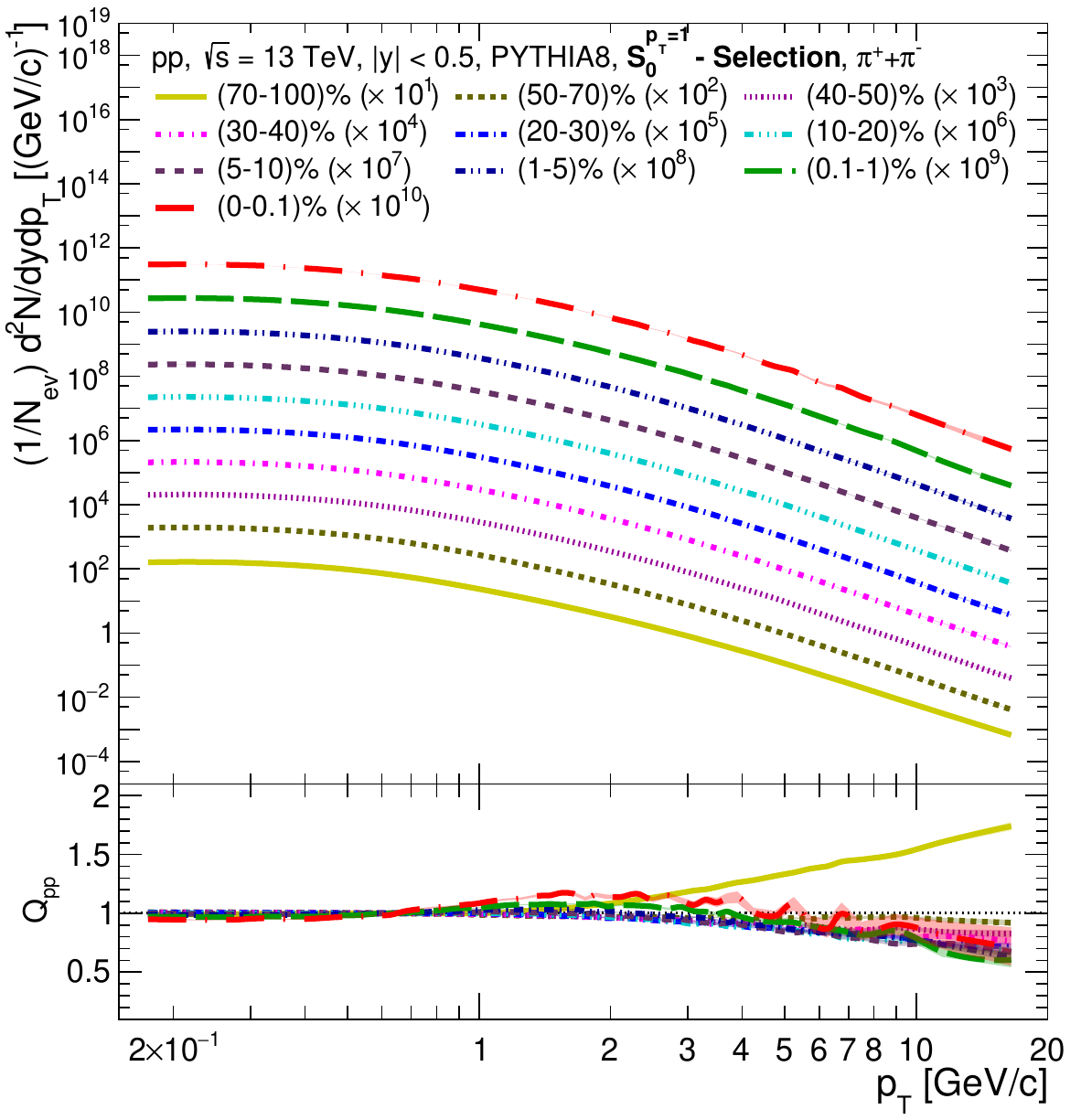}
\caption{Top panel shows the transverse momentum distribution of charged pions for different slices of $S_{0}$ (left) and $S_{0}^{p_{\rm T}=1}$ (right) in $pp$ collisions at $\sqrt{s}=13$ TeV using PYTHIA8. The bottom panel shows the ratio with respect to the minimum bias (MB) events.}
\label{fig:pionspectraS0}
\end{center}
\end{figure*}

%\begin{figure*}[ht!]
%\begin{center}
%\includegraphics[width = 0.49\linewidth]{RTFigures/pTspectra_RTpion.pdf}
%\includegraphics[width = 0.49\linewidth]{RTFigures/pTspectra_RTminpion.pdf}
%\includegraphics[width = 0.49\linewidth]{RTFigures/pTspectra_RTmaxpion.pdf}
%\caption{Top panel shows the transverse momentum distribution of charged pions for different slices of $R_{\rm T}$~(left), $R_{\rm T}^{\rm min}$~(upper right), and $R_{\rm T}^{\rm max}$~(right) in $pp$ collisions at $\sqrt{s}=13$ TeV using PYTHIA8. The bottom panel shows the ratio with respect to the minimum bias (MB) events.}
%\label{fig:pionspectraRT}
%\end{center}
%\end{figure*}

\begin{figure*}[ht!]
\begin{center}
\includegraphics[scale=0.65]{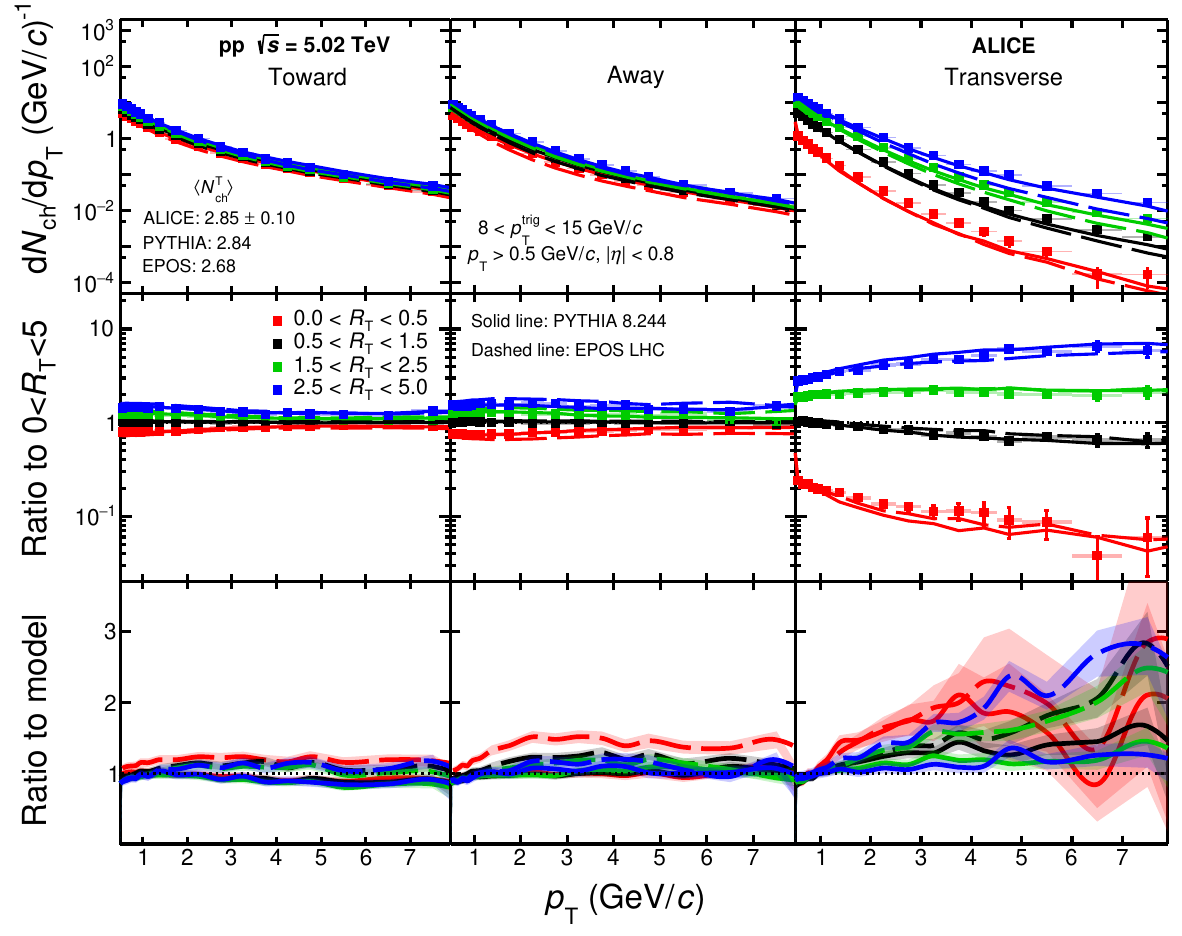}
\caption{Transverse momentum spectra of charged particles in toward, away, and transverse regions, from left to right for different events selected based on $R_{\rm T}$ in $pp$ collisions at $\sqrt{s}=5.02$ TeV with ALICE~\cite{ALICE:2023csm}. The middle panel shows the ratio of charged-particle spectra for different regions of $R_{\rm T}$ to the charged-particle spectra measured for the $R_{\rm T}\geq 0$ case~\cite{ALICE:2023csm}. The bottom panel shows the ratio of data to models like PYTHIA~8 and EPOS LHC.}
\label{fig:ChargedspectraRTppALICE}
\end{center}
\end{figure*}

\begin{figure*}[ht!]
\begin{center}
\includegraphics[scale=0.65]{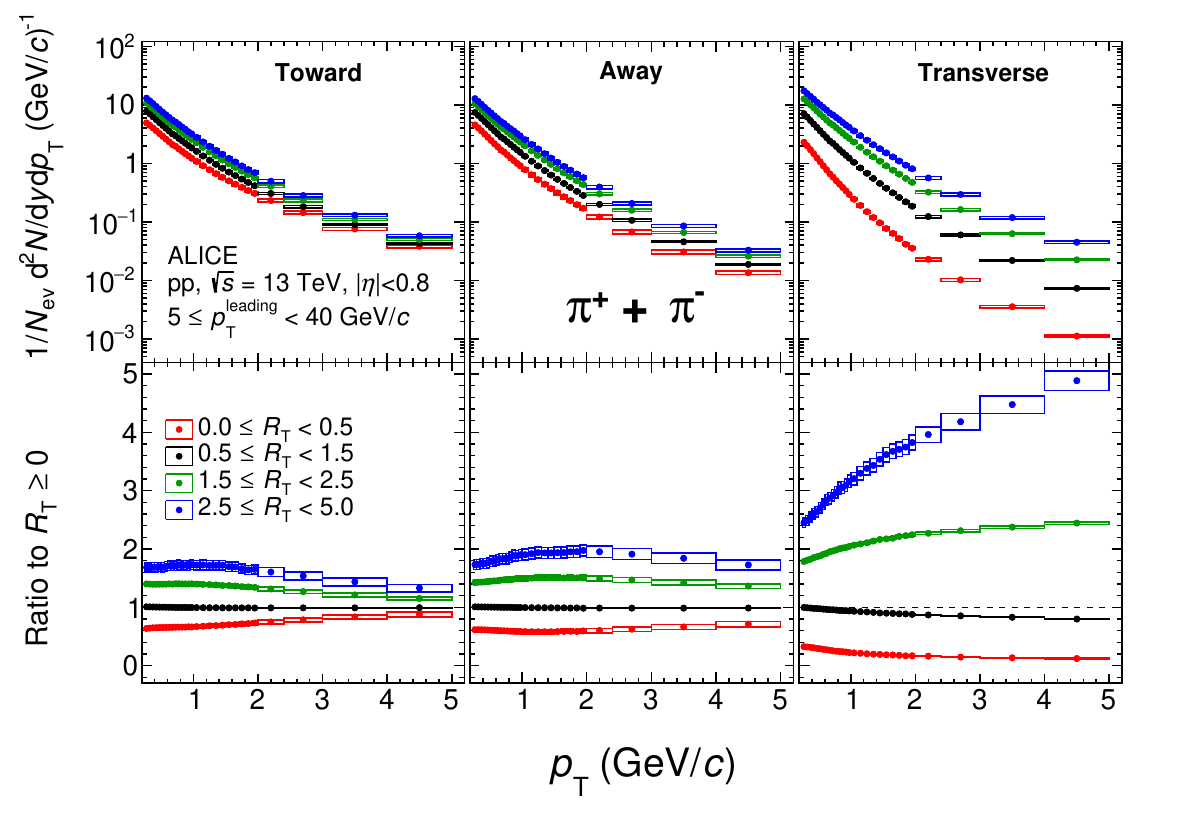}
\caption{Transverse momentum spectra of pions in toward, away, and transverse regions, from left to right for different events selected based on $R_{\rm T}$ in $pp$ collisions at $\sqrt{s}=13$ TeV with ALICE~\cite{ALICE:2023yuk}. The lower panel shows the ratio of pion spectra for different regions of $R_{\rm T}$ to the pion spectra measured for the $R_{\rm T}\geq 0$ case~\cite{ALICE:2023yuk}.}
\label{fig:pionspectraRTALICE}
\end{center}
\end{figure*}

When studied as a function of sphericity and (un)weighted spherocity in Figs.~\ref{fig:pionspectraST} and~\ref{fig:pionspectraS0}, respectively, one observes a hint of selection bias similar to Fig.~\ref{fig:pionspectraNch} for all these event shape selections. In particular, Figs.~\ref{fig:pionspectraST} and ~\ref{fig:pionspectraS0} show that low-$p_{\rm T}$ particles are insensitive to sphericity and spherocity selections. At high-$p_{\rm T}$, the behavior of $Q_{\rm pp}$ is opposite to that of $N_{\rm mpi}$ selection. As discussed above, this could be an artifact of the inherent multiplicity cut of 10 particles. Further, the events with smaller values of spherocity and sphericity are biased towards jets, leading to an enhancement structure in high-$p_{\rm T}$ regions, which is otherwise absent. Similar features of $Q_{\rm pp}$ for spherocity selection are also shown in Ref.~\cite{R:2023lku}. This can also be confirmed from Fig.~\ref{fig:MPIvspThat}, where event selections with $S_0$, $S_{0}^{p_{\rm T}=1}$ and $S_{\rm T}$ show a small rise in $\langle\hat{p}_{\rm T}\rangle$ towards lower $\langle N_{\rm mpi}\rangle$. 

Experimentally the effect of the auto-correlation bias in $R_{\rm T}$ is explicitly studied by ALICE via the measurement of charged particles' transverse momentum spectra in $pp$ collisions for toward, away, and transverse regions as a function of $R_{\rm T}$ as shown in Fig.~\ref{fig:ChargedspectraRTppALICE}~\cite{ALICE:2023csm}. Also, the measurement of pion spectra as a function of $R_{\rm T}$ in $pp$ collisions at $\sqrt{s}=13$ TeV with ALICE, shown in Fig.~\ref{fig:pionspectraRTALICE}~\cite{ALICE:2023yuk}. Here, the middle panel of Fig.~\ref{fig:ChargedspectraRTppALICE} and the lower panel of Fig.~\ref{fig:pionspectraRTALICE} show the ratio with respect to the $R_{\rm T}\geq 0$ case. Interestingly, for both charged-particle and pion spectra in the toward and away regions, with an increase in $p_{\rm T}$, the spectra measured in different $R_{\rm T}$ values approach to $R_{\rm T}\geq 0$ case, as can be seen in the lower panel. However, for the case of the transverse region, a deviation from the $R_{\rm T}\geq 0$ case is observed as one approaches a higher value of $p_{\rm T}$. As $R_{\rm T}$ is measured using the charged-particle multiplicity in the transverse regions, the deviation from $R_{\rm T}\geq 0$ for the transverse case can be attributed to the autocorrelation in the measurement of charged pions, where the event shape, $R_{\rm T}$ is defined. On the other hand, for the toward and away regions, where this auto-correlation bias is absent and due to a large correlation of $R_{\rm T}$ with $N_{\rm mpi}$, the pion spectra approach the $R_{\rm T}\geq 0$ towards the higher $p_{\rm T}$ regions.

%\begin{figure}[ht!]
%\begin{center}
%\includegraphics[scale=0.4]{FlattenicityFigures/pTspectra_flatpion.pdf}
%\caption{Top panel shows the transverse momentum distribution of charged pions for different slices of charged-particle flattenicity ($\rho_{\rm ch}$) in $pp$ collisions at $\sqrt{s}=13$ TeV using PYTHIA8. The bottom panel shows the ratio with respect to the minimum bias (MB) events.}
%\label{fig:pionspectraflat}
%\end{center}
%\end{figure}

\begin{figure*}[ht!]
\begin{center}
\includegraphics[scale=0.7]{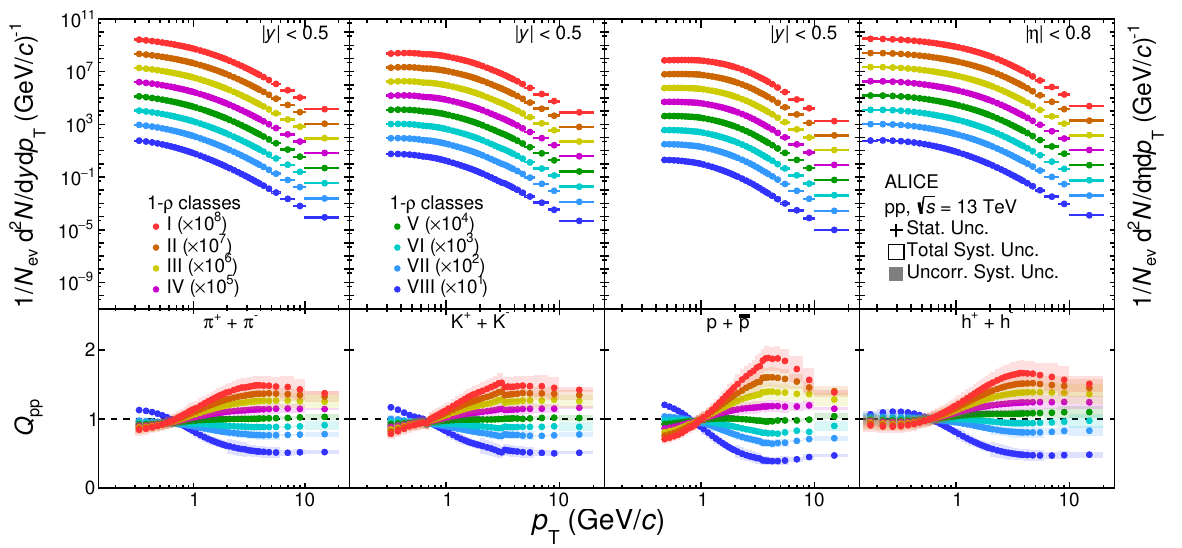}
\caption{The upper panel shows the transverse momentum distribution of charged pions, kaons, protons, and charged hadrons, from left to right, for different charged-particle flattenicity classes in $pp$ collisions at $\sqrt{s}=13$ TeV with ALICE. The lower panel shows $Q_{\rm pp}$ versus $p_{\rm T}$ for the corresponding flattenicity event class. The figure is taken from Ref.~\cite{ALICE:2024vaf}.}
\label{fig:aliceflat}
\end{center}
\end{figure*}

The closest to the trend of MPI selections (grey line in Figure~\ref{fig:MPIvspThat}) is the charged-particle flattenicity, which indicates that flattenicity is the robust observable against the selection bias of choosing high-$p_{\rm T}$ particles. This is also evident from the measurement from ALICE in Fig.~\ref{fig:aliceflat}. The measured $Q_{\rm pp}$ versus $p_{\rm T}$ for different flattenicity event classes shows a similar behavior as shown in Fig.~\ref{fig:pionspectraNmpi}, where the event classes are defined based on the number of multiple partonic interactions.\\

\begin{itshape}
The key takeaway messages of this section are summarized below:

\begin{itemize}
\item The correlation between MPI and different event shape observables highlights the reduction of selection and auto-correlation biases that are present in the study of particle production with charged particle multiplicity.

\item Unlike other event shape observables such as spherocity, sphericity, and relative transverse activity classifier, the charged-particle flattenicity shows the least selection bias of choosing high-$p_{\rm T}$ particles, which can be evident from the variation of $Q_{\rm pp}$ with $p_{\rm T}$.
\end{itemize}
\end{itshape}

%\subsection{Correlation among event classifiers}

 %Figure~\ref{fig:spherovsNchmid} shows the correlation of weighted (top) and unweighted (bottom) transverse spherocity with the charged-particle multiplicity measured at the midrapidity region.

%\begin{figure}[ht!]
%\begin{center}
%\includegraphics[scale=0.4]{SpherocityFigures/S0vsNchcorrel.pdf}
%\includegraphics[scale=0.4]{SpherocityFigures/S0pt1vsNchcorrel.pdf}
%\caption{Weighted (top) and unweighted (bottom) transverse spherocity versus charged-particle multiplicity measured at the midrapidity region in $pp$ collisions at $\sqrt{s}=13$ TeV using PYTHIA8.}
%\label{fig:spherovsNchmid}
%\end{center}
%\end{figure}

%\begin{figure}[ht!]
%\begin{center}
%\includegraphics[scale=0.4]{SpherocityFigures/S0vsV0Mcorrel.pdf}
%\includegraphics[scale=0.4]{SpherocityFigures/S0pt1vsV0Mcorrel.pdf}
%\caption{Weighted (top) and unweighted (bottom) transverse spherocity versus charged-particle multiplicity measured at the forward rapidity region in $pp$ collisions at $\sqrt{s}=13$ TeV using PYTHIA8.}
%\label{fig:spherovsNchfwd}
%\end{center}
%\end{figure}

% \begin{figure}[ht!]
% \begin{center}
% \includegraphics[scale=0.4]{SpherocityFigures/S0vsmeanpTcorrel.pdf}
% \includegraphics[scale=0.4]{SpherocityFigures/S0pt1vsmeanpTcorrel.pdf}
% \end{center}
% \caption{}
% \label{}
% \end{figure}

%\subsection {NMPI vs event shape}

\subsection {Average transverse momentum}
\label{sec:avgpT}
\begin{figure}[ht!]
\begin{center}
\includegraphics[scale=0.36]{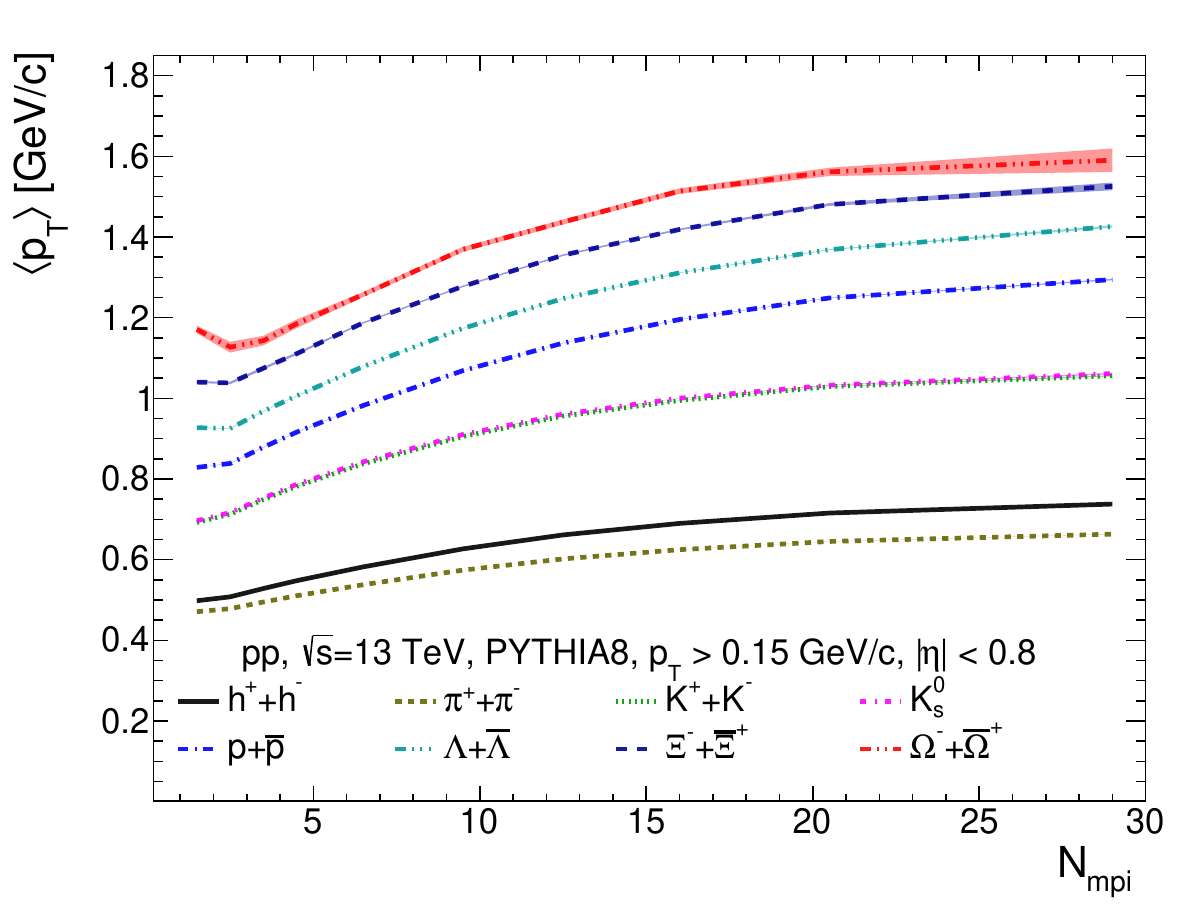}
\end{center}
\caption{Mean transverse momentum of all charged and identified hadrons as a function of $N_{\rm mpi}$ in $pp$ collisions at $\sqrt{s}=13$ TeV using PYTHIA8.}
\label{fig:meanpTNmpi}
\end{figure}

\begin{figure*}[ht!]
\begin{center}
\includegraphics[width = 0.49\linewidth]{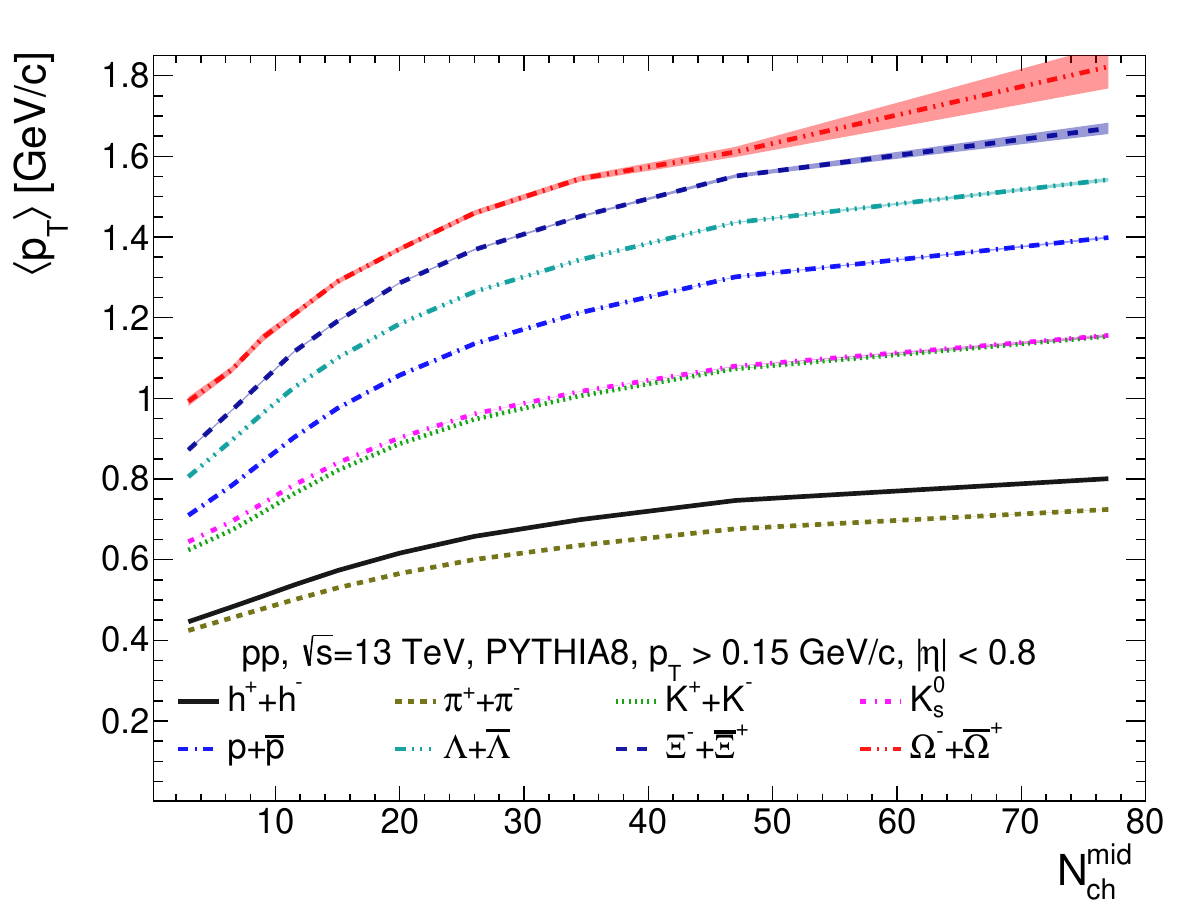}
\includegraphics[width = 0.49\linewidth]{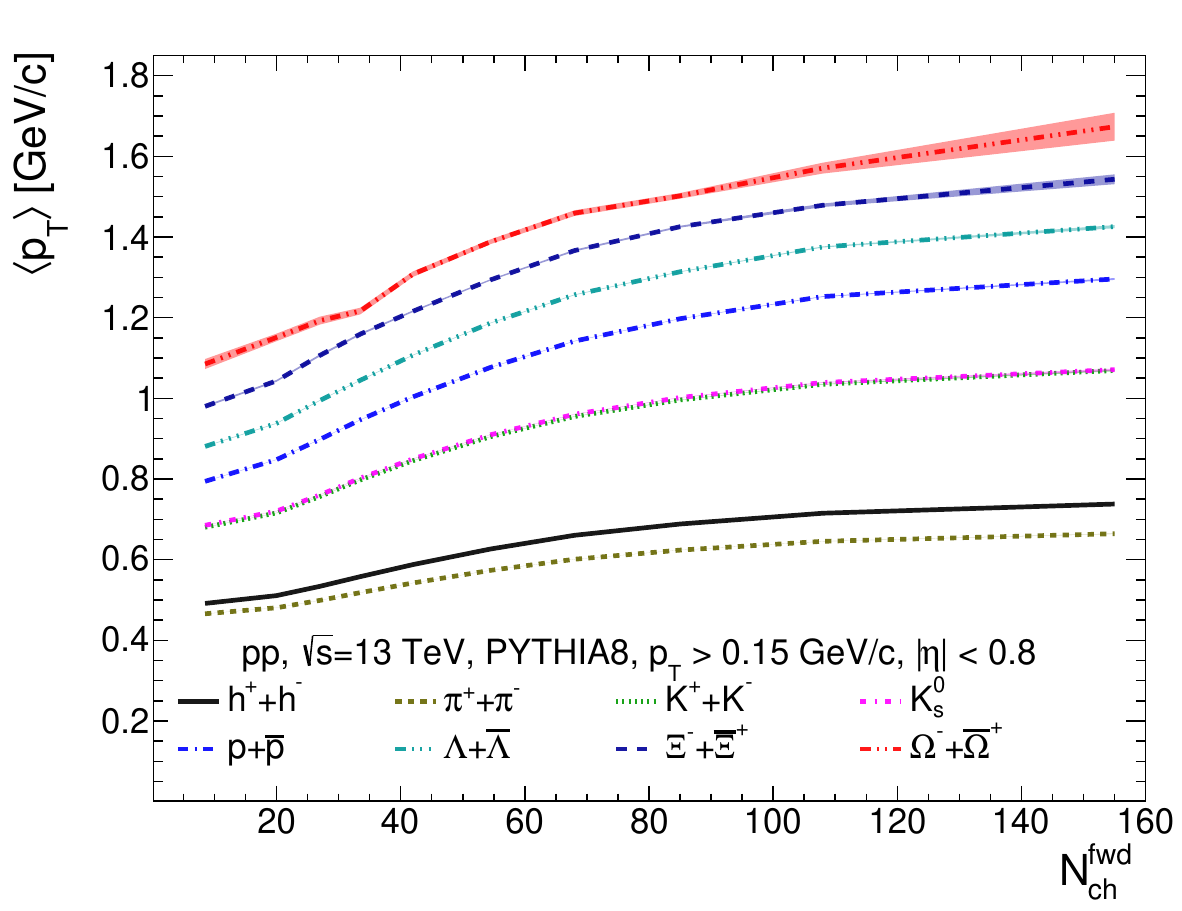}
\end{center}
\caption{Mean transverse momentum of all charged and identified hadrons as a function of charged-particle multiplicity in mid-pseudorapidity (left) and forward pseudorapidity (right) in $pp$ collisions at $\sqrt{s}=13$ TeV using PYTHIA8.}
\label{fig:meanpTNch}
\end{figure*}

In a hydrodynamic expanding system, the presence of high energy density and outward pressure gives a radial boost to all the produced hadrons. The presence of this outward radial boost, or radial flow, gives an additional boost to the produced hadrons in addition to their thermal expansion. This results in an increase in the transverse momentum of the particles. Here, in the presence of such a radial boost, the hadrons with a larger mass are significantly affected as compared to the hadrons with a lower mass. This boost effect is usually reflected in the mean transverse momentum ($\langle p_{\rm T}\rangle$) of the produced hadrons, where the presence of a higher radially boosted medium is reflected in the enhanced value of $\langle p_{\rm T}\rangle$ of the hadrons. Recently, such radial flow-like features are observed in $pp$ collisions~\cite{ALICE:2020nkc, ALICE:2022wpn}, which were traditionally considered for baseline measurements for the signatures of quark-gluon plasma.

Interestingly, PYTHIA8 with color reconnection is able to qualitatively explain such radial flow-like features observed in $pp$ collisions in experiments without taking the hydrodynamic evolution of partons into account~\cite{OrtizVelasquez:2013ofg}. In PYTHIA8, when a string connecting two partons moves, it gives a common boost to the string fragments. In the absence of CR, if a parton is emitted in the midrapidity, the other end of the string will be part of the proton moving forward, leading to a small boost~\cite{OrtizVelasquez:2013ofg}. However, in the presence of CR, the partons from two independent scatterings can color reconnect and lead to an enhanced boost~\cite{OrtizVelasquez:2013ofg}. In addition, this boosting effect is expected to be largely enhanced in the case of a large number of multi-partonic interactions~\cite{OrtizVelasquez:2013ofg}. Figure~\ref{fig:meanpTNmpi} shows $\langle p_{\rm T}\rangle$ of charged, identified, and strange hadrons measured in midrapidity as a function of $N_{\rm mpi}$ in $pp$ collisions using PYTHIA8. For any particle species, $\langle p_{\rm T}\rangle$ is observed to rise to higher values with an increase in $N_{\rm mpi}$ till $N_{\rm mpi}\lesssim 20$. This is because, with an increase in $N_{\rm mpi}$, the partons can color reconnect easily, thus leading to a larger boost and an enhanced value of $\langle p_{\rm T}\rangle$. However, since the total available energy for particle production is limited, $N_{\rm mpi}$ increases at the cost of the average value of the partonic momentum. Thus, for the $N_{\rm mpi}\gtrsim 20$ region, CR suffers from an increase in $N_{\rm mpi}$ and a drop in partonic momentum, leading to a saturation behavior in hadron $\langle p_{\rm T}\rangle$. Another remarkable observation can be made by looking at the rise of $\langle p_{\rm T}\rangle$ with $N_{\rm mpi}$ for different species of hadrons. The increase of $\langle p_{\rm T}\rangle$ with $N_{\rm mpi}$ becomes steeper when the particle mass increases, indicating a larger radial boost for particles with a higher mass. In addition, a small rise in the low $N_{\rm mpi}$ region is also observed, which becomes more distinct with an increase in particle mass. This is expected to arise from the production of heavier particles from the initial hard interactions. As in PYTHIA8, CR with high $N_{\rm mpi}$ gives rise to enhanced radial flow-like effects, we try to probe the effects of $N_{\rm mpi}$ selection on $\langle p_{\rm T}\rangle$ with different event shape observables.

Figure~\ref{fig:meanpTNch} shows $\langle p_{\rm T}\rangle$ of particles as a function of charged-particle multiplicity measured in the midrapidity ($N_{\rm ch}^{\rm mid}$) (left) and forward rapidity ($N_{\rm ch}^{\rm fwd}$) (right) in $pp$ collisions at $\sqrt{s}=13$ TeV using PYTHIA8. The $\langle p_{\rm T}\rangle$ of particles as a function of both $N_{\rm ch}^{\rm mid}$ and $N_{\rm ch}^{\rm fwd}$ have a similar behavior going from low multiplicity to high multiplicity, as compared to $\langle p_{\rm T}\rangle$ vs $N_{\rm mpi}$ shown in Fig.~\ref{fig:meanpTNmpi}. However, we do not see the saturation behavior of $\langle p_{\rm T}\rangle$ with the increase in multiplicity in the midrapidity and forward rapidity regions. Although $N_{\rm ch}^{\rm mid}$ and $N_{\rm ch}^{\rm fwd}$ have significant correlations with $N_{\rm mpi}$, as shown in Fig.~\ref{fig:NchvsMPI},  due to a significant contribution of jet topologies in the high-multiplicity regions a continuous rise in $\langle p_{\rm T}\rangle$ with multiplicity is observed. The rise of $\langle p_{\rm T}\rangle$ with multiplicity shown in Fig ~\ref{fig:meanpTNch} is qualitatively consistent with experimental findings~\cite{ALICE:2019avo, ALICE:2020nkc}. Also, one observes a significant rise in the steepness of $\langle p_{\rm T}\rangle$ vs $N_{\rm ch}$ with increasing particle mass. The steepness further increases when one considers the $N_{\rm ch}^{\rm mid}$ for the event selection compared to $N_{\rm ch}^{\rm fwd}$.

\begin{figure}[ht!]
\begin{center}
\includegraphics[scale=0.4]{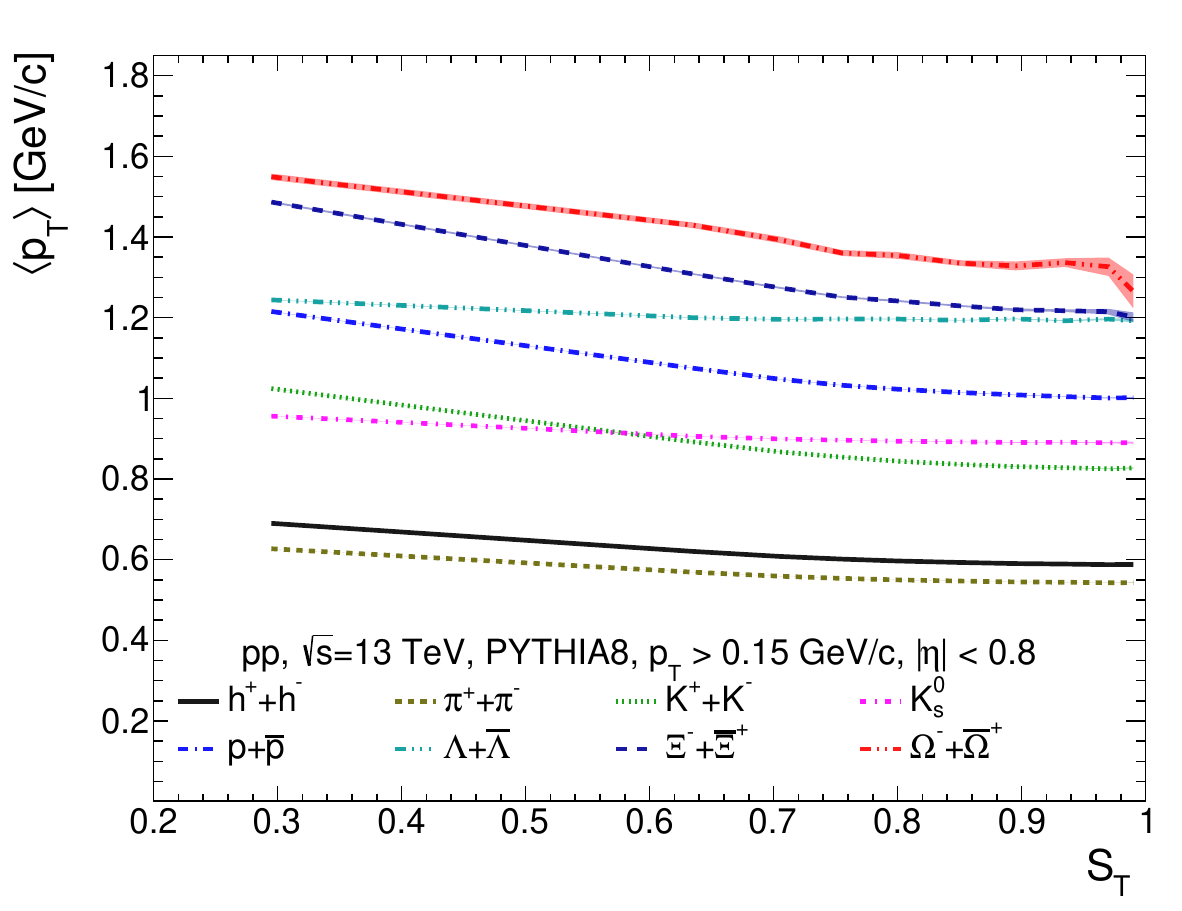}
\end{center}
\caption{Mean transverse momentum of all charged and identified hadrons as a function of transverse sphericity ($S_{\rm T}$) in $pp$ collisions at $\sqrt{s}=13$ TeV using PYTHIA8.}
\label{fig:meanpTST}
\end{figure}

\begin{figure}[ht!]
\begin{center}
\includegraphics[scale=0.6]{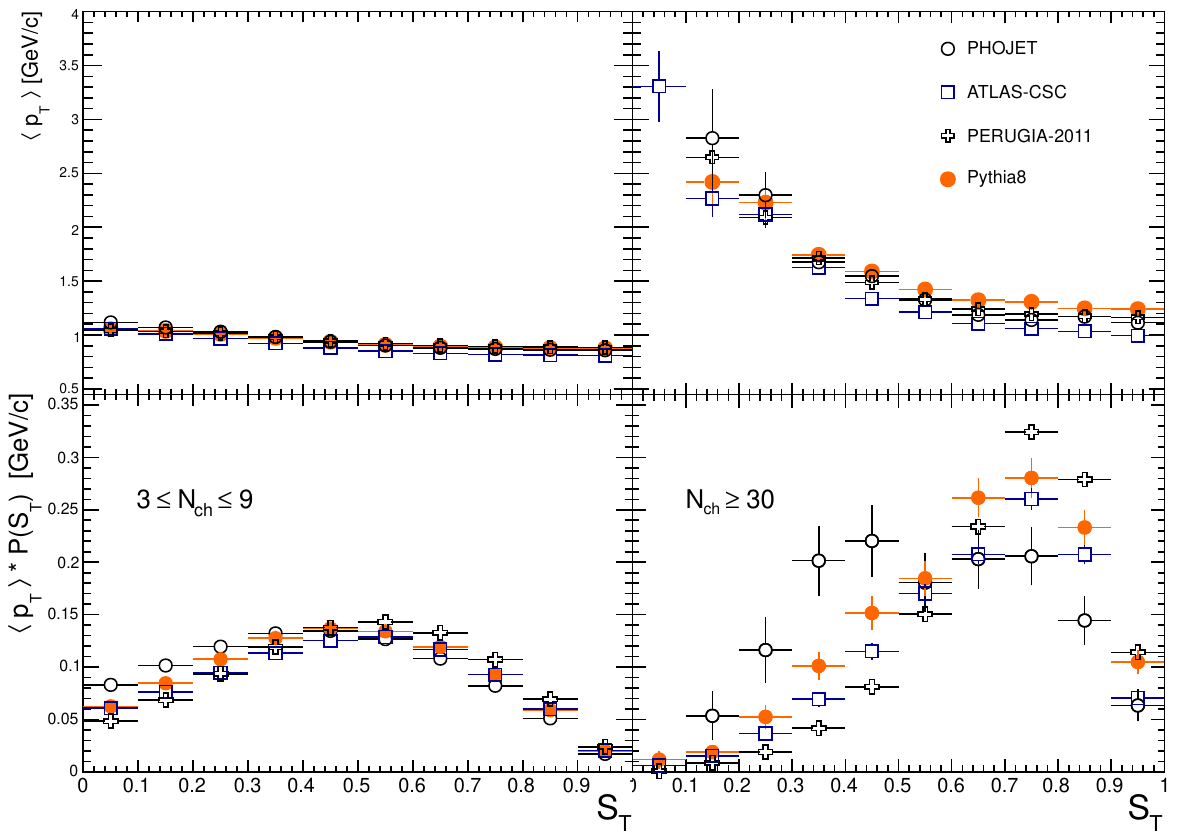}
\end{center}
\caption{Mean transverse momentum of all charged particles as a function of transverse sphericity ($S_{\rm T}$) for two multiplicity bins 3 $\leq$  N$_{\rm ch}$ $\leq$ 9 (left) and  N$_{\rm ch}$ $\geq$ 30 in minimum bias $pp$ collisions at $\sqrt{s}$ = 7 TeV obtained using four different MC generator: PHOJET, ATLAS-CSC, PERUGIA-2011, and PYTHIA8 is shown in the top panel. The bottom panel shows the contribution of
each sphericity bin to the final $\langle p_{\rm T}\rangle$~\cite{ALICE:2012cor}}
\label{fig:meanpTSTALICE}
\end{figure}

Figure~\ref{fig:meanpTST} shows $\langle p_{\rm T}\rangle$ of particles as a function of $S_{\rm T}$ in $pp$ collisions at $\sqrt{s}=13$ TeV using PYTHIA8. Here, the intrinsic mass dependence of  $\langle p_{\rm T}\rangle$ is observed. However, unlike $N_{\rm mpi}$, $N_{\rm ch}^{\rm mid}$ or $N_{\rm ch}^{\rm fwd}$ the steepness of $\langle p_{\rm T}\rangle$ towards the isotropic events is not observed. Furthermore, the event selection based on $S_{\rm T}$ is found to have a bias towards the charged particles. As can be seen in Fig.~\ref{fig:meanpTST}, the variation of $\langle p_{\rm T}\rangle$ with $S_{\rm T}$ is negligibly small for the neutral particles, such as $\Lambda+\bar{\Lambda}$ and $K^{0}_{\rm S}$. In addition, $\langle p_{\rm T}\rangle$ is observed to maintain a consistent value and slightly decreases when going towards the isotropic events. Due to the weak correlation of $S_{\rm T}$ with $N_{\rm mpi}$, one observes little to no significant effects of event selection based on $S_{\rm T}$ in $\langle p_{\rm T}\rangle$. Similarly, the $\langle p_{\rm T}\rangle$ of all charged particles as a function of $S_{\rm T}$ in minimum bias $pp$ collisions at $\sqrt{s}$ = 7 TeV is shown in the top panel of Fig.~\ref{fig:meanpTSTALICE} for two multiplicity bins, i.e 3 $\leq$  N$_{\rm ch}$ $\leq$ 9 (left), and  N$_{\rm ch}$ $\geq$ 30 (right) obtained using four different MC generators: PHOJET, ATLAS-CSC, PERUGIA-2011, and PYTHIA8. From Fig.~\ref{fig:meanpTSTALICE} it is observed that a large dependence of $\langle p_{\rm T}\rangle$ on sphericity event-shape is found at higher multiplicity regions. The bottom panel shows the contribution of each sphericity bin to the final $\langle p_{\rm T}\rangle$, i.e., $\langle p_{\rm T}\rangle$ weighted by the value P($S_{\rm T}$). The contribution to $\langle p_{\rm T}\rangle$ in the multiplicity bin is twice as large for PHOJET compared to ATLAS-CSC, although these two MC models have the same contribution to $\langle p_{\rm T}\rangle$~\cite{ALICE:2012cor}.

\begin{figure*}[ht!]
\begin{center}
\includegraphics[width = 0.49\linewidth]{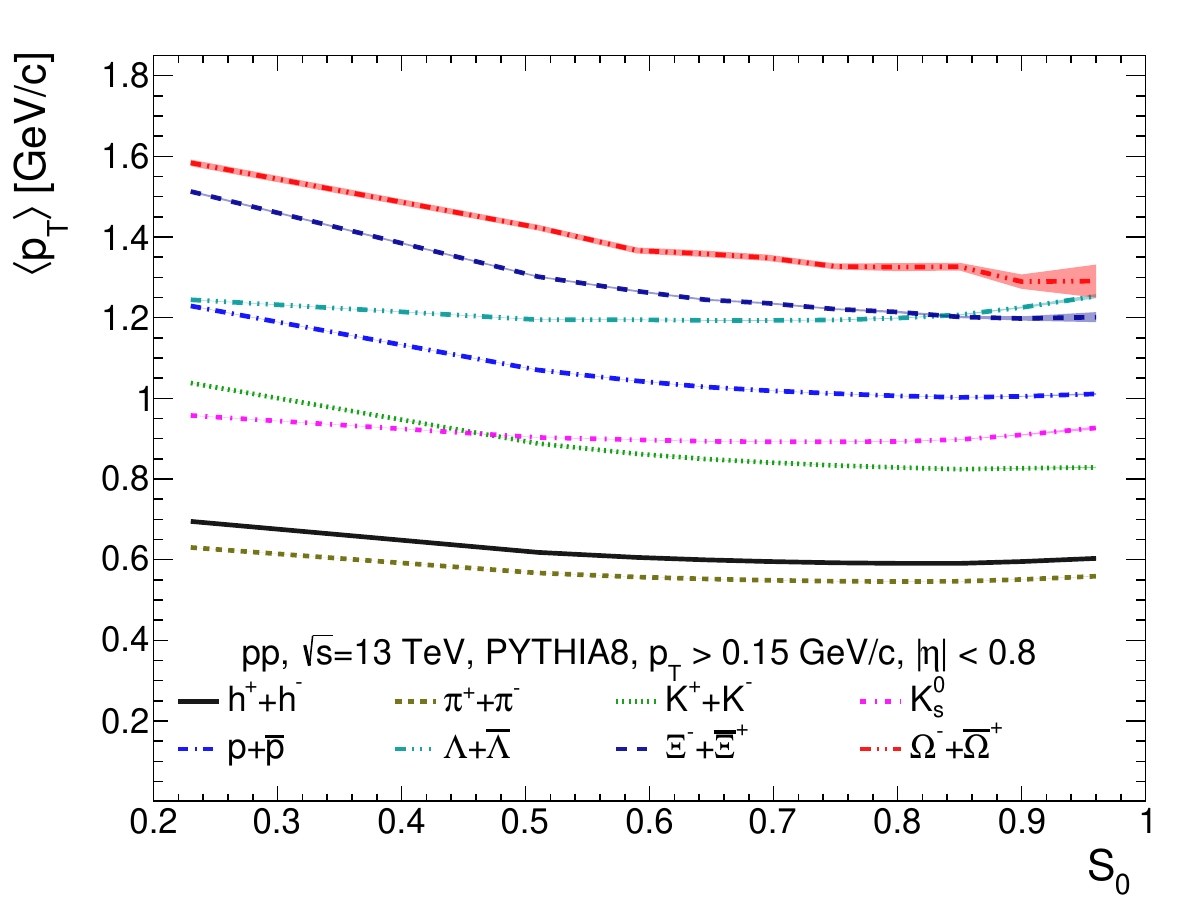}
\includegraphics[width = 0.49\linewidth]{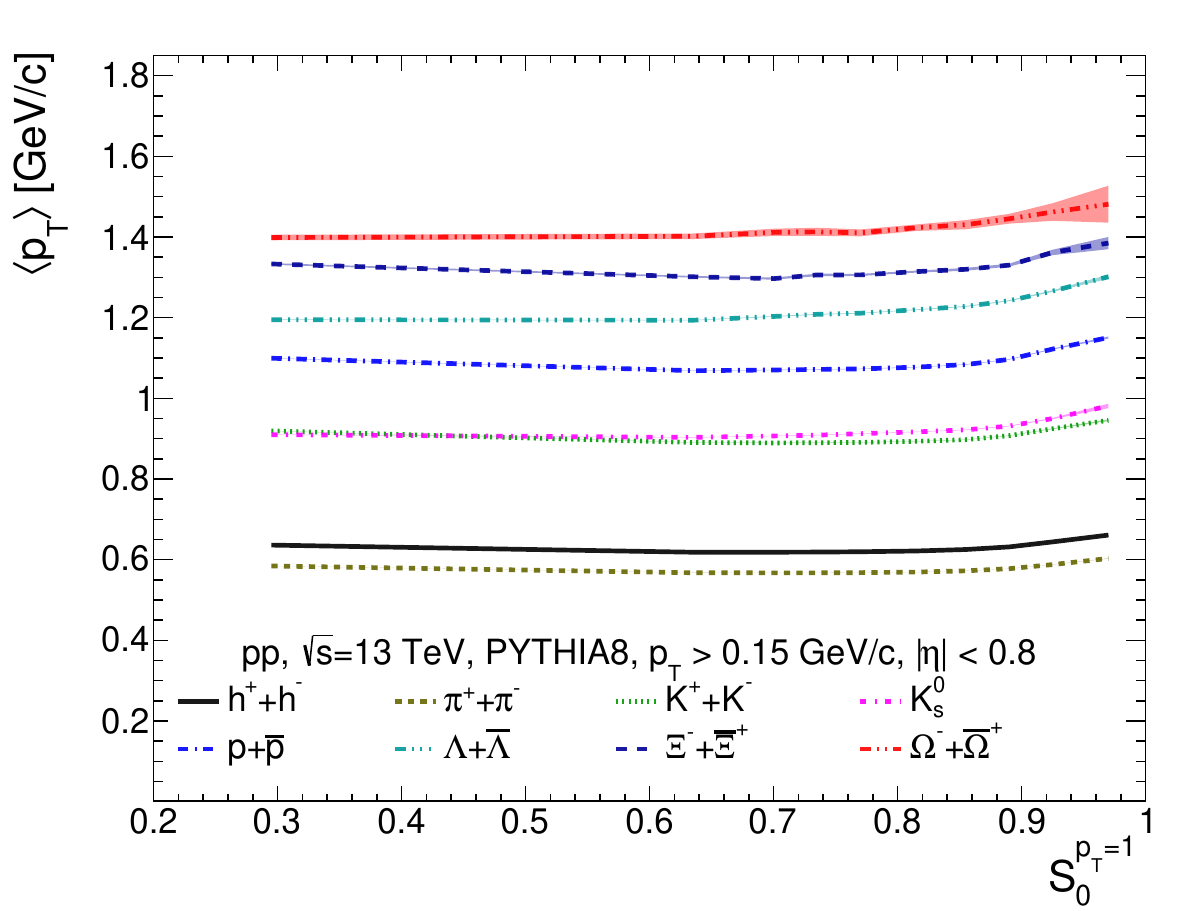}
\end{center}
\caption{Mean transverse momentum of all charged and identified hadrons as a function of $p_{\rm T}-$weighted transverse spherocity ($S_{0}$) (left) and unweighted transverse spherocity ($S_{0}^{p_{\rm T}=1}$) (right)  in $pp$ collisions at $\sqrt{s}=13$ TeV using PYTHIA8.}
\label{fig:meanpTS0}
\end{figure*}

\begin{figure}[ht!]
\begin{center}
\includegraphics[scale=0.4]{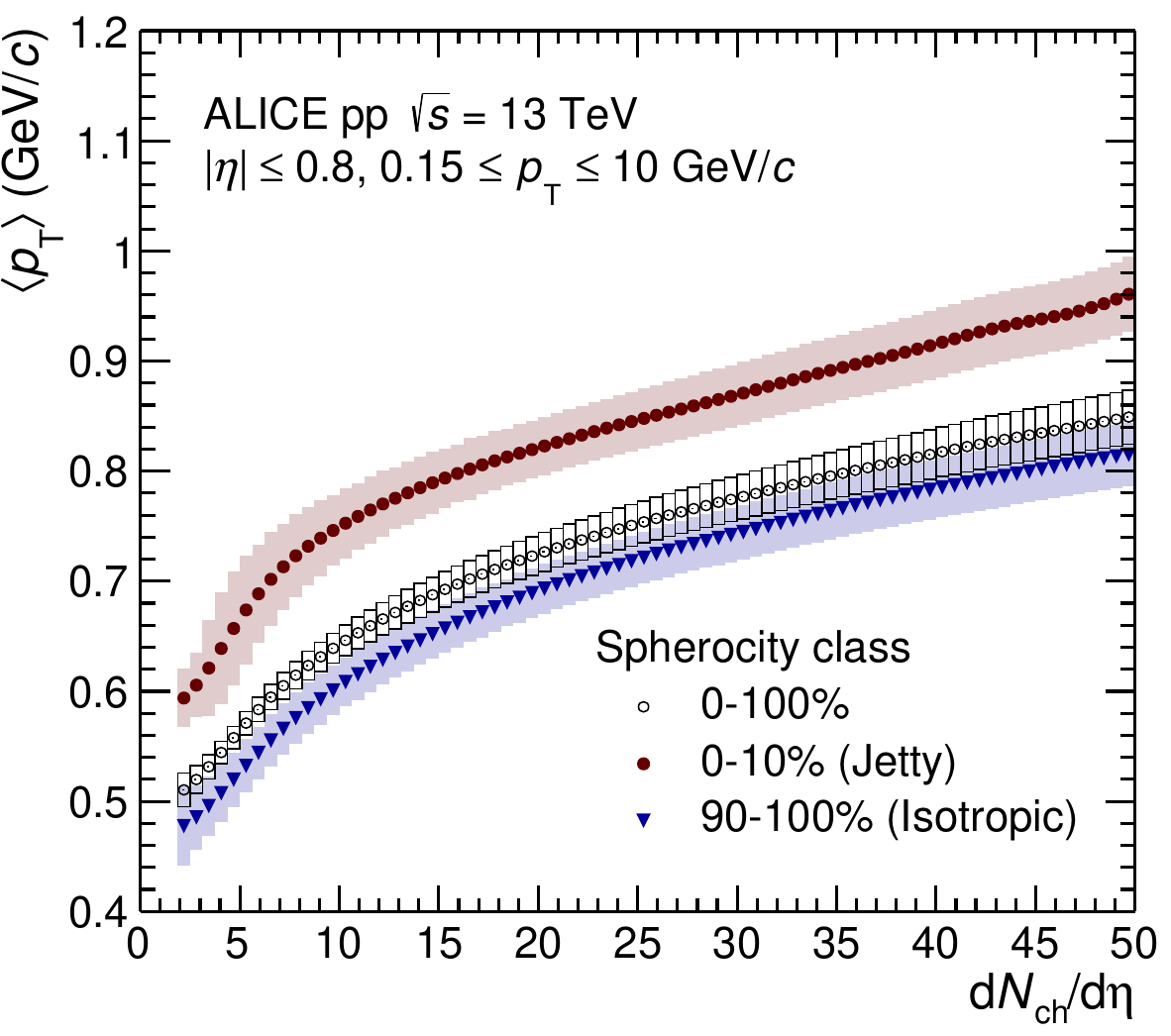}
\end{center}
\caption{Average transverse momentum ($\langle p_{\rm T}\rangle$) as a function of event multiplicity for $S_0$-integrated (0-100\%), jetty (0-10\%) and isotropic (90-100\%) events in $pp$ collisions at $\sqrt{s}=13$ TeV with ALICE~\cite{ALICE:2019dfi}.}
\label{fig:meanpTS0Nch}
\end{figure}
Figure~\ref{fig:meanpTS0} shows $\langle p_{\rm T}\rangle$ of particles measured in $|\eta|<0.8$ and $p_{\rm T}>0.15$ GeV/$c$ as a function of transverse spherocity in $pp$ collisions at $\sqrt{s}=13$ TeV using PYTHIA8. The left panel shows $S_{0}$ versus $\langle p_{\rm T}\rangle$ and the right panel represents $S_{0}^{p_{\rm T}=1}$ versus $\langle p_{\rm T}\rangle$. For all the particle species, a rise of $\langle p_{\rm T}\rangle$ towards the lower value of $S_{0}$ is observed; this can be attributed to larger jet contributions in such events. For the charged particles, $\langle p_{\rm T}\rangle$ decreases from lower to higher values of $S_{0}$, which becomes clear for the heavier particles. However, event selection based on $S_{0}$ is found to affect neutral particle $\langle p_{\rm T}\rangle$ differently. Here, $\langle p_{\rm T}\rangle$ of lambda hyperons ($\Lambda+\bar{\Lambda}$) and $K^{0}_{\rm S}$ first decreases and finally starts to increase as $S_{0}$ increases. This event selection bias based on $S_{0}$ for neutral and charged particles can be fixed when one considers $p_{\rm T}=1$ for the estimation of transverse spherocity. As can be seen in Fig.~\ref{fig:meanpTS0} right panel, all the particles, independent of the charge of the particle, retain a similar rising trend of $\langle p_{\rm T}\rangle$ throughout $S_{0}^{p_{\rm T}=1}$. The rise of $\langle p_{\rm T}\rangle$ becomes prominent towards $S_{0}^{p_{\rm T}=1}>0.85$ as one goes from lower to higher values of $S_{0}^{p_{\rm T}=1}$. This is because the low $S_{0}^{p_{\rm T}=1}$ region gets higher contribution from the jets while high values of $S_{0}^{p_{\rm T}=1}$ may show enhanced flow like effect leading to higher values of $\langle p_{\rm T}\rangle$. The effect becomes clearer as the particle mass increases.

Figure~\ref{fig:meanpTS0Nch} shows $\langle p_{\rm T}\rangle$ versus event multiplicity ($dN_{\rm ch}/d\eta$) for different classes of $S_{0}$ in $pp$ collisions at $\sqrt{s}=13$ TeV with ALICE. With the increase in event multiplicity, $\langle p_{\rm T}\rangle$ rises. In PYTHIA~8, this is explained by CR, also shown in Fig.~\ref{fig:meanpTNch}. When $S_{0}$-integrated data are compared with the isotropic case, a suppression in $\langle p_{\rm T}\rangle$ is observed within systematics. In contrast, the jetty events show an enhanced value of $\langle p_{\rm T}\rangle$ throughout the multiplicity region. This feature is captured by PYTHIA~8, as shown in Fig.~\ref{fig:meanpTS0} for all charged hadrons.

%\begin{figure*}[ht!]
%\begin{center}
%\includegraphics[width = 0.49\linewidth]{RTFigures/meanpT-RT.pdf}
%\includegraphics[width = 0.49\linewidth]{RTFigures/meanpT-RTmin.pdf}
%\includegraphics[width = 0.49\linewidth]{RTFigures/meanpT-RTmax.pdf}
%\end{center}
%\caption{Mean transverse momentum of all charged and identified hadrons as a function of $R_{\rm T}$~(top), $R_{\rm T}^{\rm min}$~(upper right), and $R_{\rm T}^{\rm max}$~(bottom) in $pp$ collisions at $\sqrt{s}=13$ TeV using PYTHIA8.}
%\label{fig:meanpTRT}
%\end{figure*}

\begin{figure*}[ht!]
\begin{center}
\includegraphics[scale=0.7]{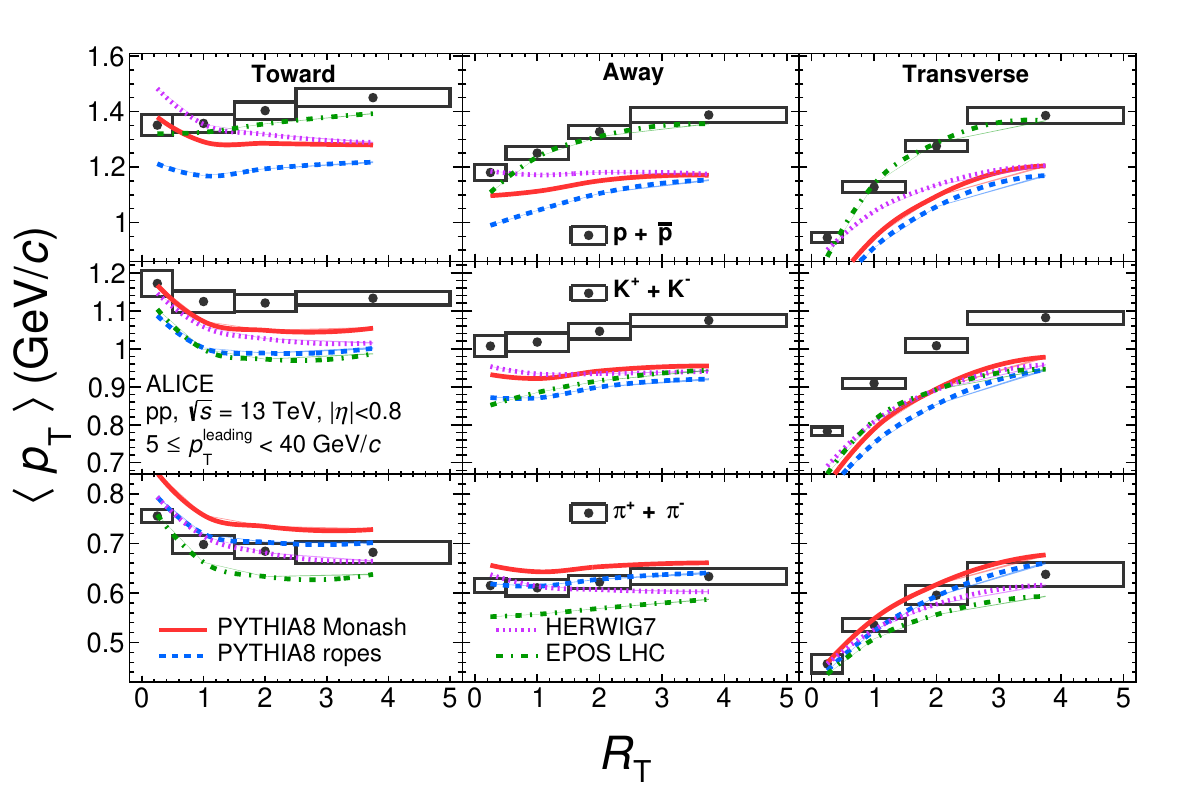}
\end{center}
\caption{Average transverse momentum ($\langle p_{\rm T}\rangle$) for charged protons, kaons and pions, shown from upper to lower panels, measured in different topological regions as a function of $R_{\rm T}$ in $pp$ collisions at $\sqrt{s}=13$ TeV with ALICE~\cite{ALICE:2023yuk}. Model comparisons for the corresponding measurements are shown using Ropes and Monash tunes of PYTHIA8, along with HERWIG7 and EPOS LHC~\cite{ALICE:2023yuk}.}
\label{fig:meanpTRTALICE}
\end{figure*}

Figure~\ref{fig:meanpTRTALICE} shows the average transverse momentum of charged protons, kaons, and pions measured in toward, away, and transverse regions as a function of $R_{\rm T}$ in $pp$ collisions at $\sqrt{s}=13$ TeV with ALICE~\cite{ALICE:2023yuk}. In the toward region, for the charged kaons and pions, as one moves from low $R_{\rm T}$ (low UE) to a higher $R_{\rm T}$ (large UE) region, $\langle p_{\rm T}\rangle$ decreases. This represents the presence of jet-fragmentations, which are low-mass particles (pions and kaons) having large transverse momentum in the $R_{\rm T}$ (low UE) region. Further, with an increase in $R_{\rm T}$, the UE contribution and due to the jet-dilution effect, a saturation of $\langle p_{\rm T}\rangle$ with an increase in $R_{\rm T}$ is observed. Conversely, $\langle p_{\rm T}\rangle$ for protons shows a rise with $R_{\rm T}$ which can be attributed to enhanced radial flow effects with increasing $R_{\rm T}$. The decrease of $\langle p_{\rm T}\rangle$ with $R_{\rm T}$ for pions and kaons is well explained by all the considered models. However, the $\langle p_{\rm T}\rangle$ for protons is correctly explained only with EPOS LHC.

In the away-side region, the particle production mechanism is similar to the toward regions and is dominated by the away-side jet. Here, for all the particle species, a rise in $\langle p_{\rm T}\rangle$ is inferred with an increase in $R_{\rm T}$. This is because, in the away-side region, the contribution from the particle fragmentation is smaller as compared to the toward region and with a rise in $R_{\rm T}$, the UE dominated the particle production leading to dominating-radial flow effects which causes a rise in $\langle p_{\rm T}\rangle$ with $R_{T}$. The rise of $\langle p_{\rm T}\rangle$ is smaller for pions than for kaons than for protons. Here, EPOS LHC explains the $\langle p_{\rm T}\rangle$ for protons but underestimates $\langle p_{\rm T}\rangle$ for pions and kaons. 

In the transverse region, we see a clear mass ordering of $\langle p_{\rm T}\rangle$ for different particle species. A strong rise in $\langle p_{\rm T}\rangle$ with an increase in $R_{\rm T}$ is observed for all the particle species. The rise of $\langle p_{\rm T}\rangle$ is stronger for protons and kaons than for pions. A similar observation is made as a function of charged-particle multiplicity~\cite{ALICE:2018pal}. This behavior is an indication of autocorrelation bias caused by measuring both particles and $R_{\rm T}$ in a similar region of $\Delta \phi$. The high multiplicity requirement in the transverse region increases the probability of having a jet in the same region. Here, all the considered models give a qualitative description of $\langle p_{\rm T}\rangle$ versus $R_{\rm T}$ for pions, kaons, and protons.

\begin{figure}[ht!]
\begin{center}
\includegraphics[scale=0.4]{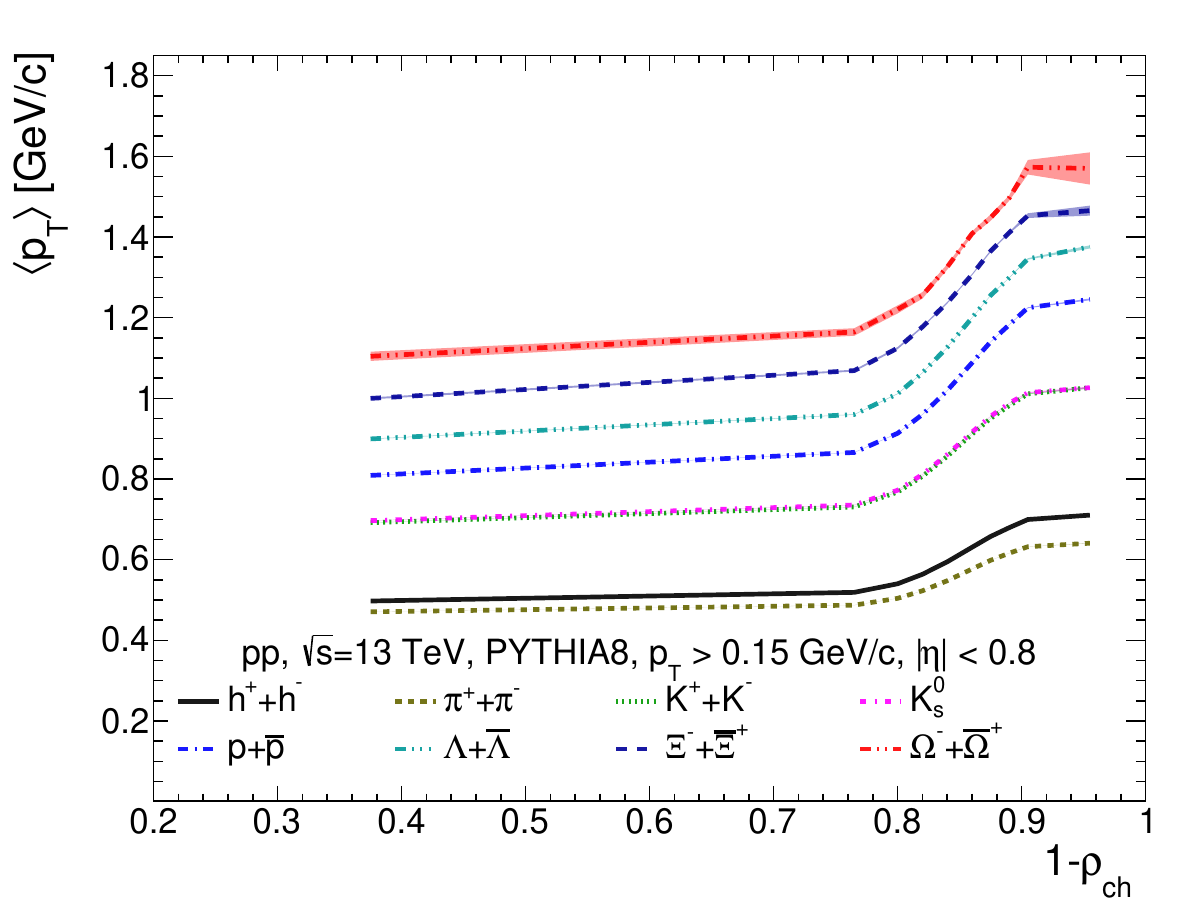}
\end{center}
\caption{Mean transverse momentum of all charged and identified hadrons as a function of charged-particle flattenicity ($\rho_{\rm ch}$) in $pp$ collisions at $\sqrt{s}=13$ TeV using PYTHIA8.}
\label{fig:meanpTflat}
\end{figure}

Figure~\ref{fig:meanpTflat} shows $\langle p_{\rm T}\rangle$ of particles measured in $|\eta|<0.8$ and $p_{\rm T}>0.15$ GeV/$c$ as a function of charged-particle flattenicity ($\rho_{\rm ch}$). In Fig.~\ref{fig:meanpTflat}, with an increase in the value of $\rho_{\rm ch}$ corresponding value of $\langle p_{\rm T}\rangle$ also increases for all the particle species shown in the figure. In addition, the value of $\langle p_{\rm T}\rangle$ follows a clear mass dependence where the heavier mass particles possess a larger increasing slope with an increase in the value of $\rho_{\rm ch}$ compared to the particles with a lower mass. Further, a saturation behavior in $\langle p_{\rm T}\rangle$ is observed towards $(1-\rho_{\rm ch})\gtrsim 0.9$. Flattenicity is one of the observables that is found to closely follow the trend of increasing $\langle p_{\rm T}\rangle$ for event selection based on $N_{\rm mpi}$ and the qualitative enhancement feature of $\langle p_{\rm T}\rangle$ is observed only in $0.75<(1-\rho_{\rm ch})<0.9$, where the most of the events lie. \\

\begin{itshape}
In summary,
\begin{itemize}
\item The increase of average transverse momentum with charged particle multiplicity is believed to be a hint for radial flow-like effects in small systems. In this section, such features are discussed with different event shapes in $pp$ collisions via experimental measurements and model predictions. Such a rise of $\langle p_{\rm T}\rangle$ is found to depend on particle mass.

\item It is observed that the increase of  $\langle p_{\rm T}\rangle$ with charged particle flattenicity is similar to that of MPI, indicating the most recent event classifier is the ideal tool to probe  QGP-like behavior in small systems. 

\end{itemize}
\end{itshape}

\subsection {Baryon-to-meson ratios and radial flow}
\noindent

%\subsubsection{Light flavor (LF) sector}
In this section, we present the event shape observable dependence of baryon to meson ratios as a function of $p_{\rm T}$ in $pp$ collisions at $\sqrt{s}$ = 13 TeV using PYTHIA8.  The particle ratios as a function of $p_{\rm T}$ are the consequence of the transverse momentum spectra of the corresponding particle. Any modification that occurs in the particle ratios as a function of the event shape observables is simply due to the spectral shape modification as a function of the corresponding event shape. It is observed that in heavy-ion collisions, the enhancement in baryon-to-meson ratios at intermediate $p_{\rm T}$ can be explained due to the interplay between the collective motion of the system and the hadronization mechanism. Furthermore, the amplitude and the position of the peak observed in baryon-to-meson ratios indicate the strength of the radial flow~\cite{ALICE:2019hno}. However, such enhancement is also observed in $pp$ collisions using PYTHIA8 due to the CR mechanism introduced in PYTHIA8 along with MPI. It is observed that the microscopic CR hadronization mechanism, along with MPI, suggests the flow-like features in $pp$ collisions~\cite{OrtizVelasquez:2013ofg}. \\

The charged-particle multiplicity dependence study of baryon-to-meson ratios provides information on radial flow as a function of charged-particle multiplicity. This motivates us to compare the baryon-to-meson ratios as a function of $p_{\rm T}$ with event selection based on several event shape observables such as the number of multi-partonic interactions, transverse sphericity, spherocity, relative activity classifier, and charged-particle flattenicity. Thus, this review provides an understanding of the radial flow as a function of these event activities.\\

In the present study, in baryon-to-meson ratios we mainly investigate ($ \rm p+\bar{p})/(\pi^{+}+\pi^{-}$), ($\Lambda+\bar{\Lambda})/2K^{0}_{\rm S}$, and ($ \rm p+\bar{p})/(\phi + \bar{\phi})$ ratios as a function of $p_{\rm T}$. For simplicity now onwards we refer these ratios as $\rm p/\pi$, $\Lambda/K^{0}_{\rm S}$, and $ \rm p/\phi $ respectively. \\
 
\begin{figure*}[ht!]
\begin{center}
\includegraphics[width = 0.49\linewidth]{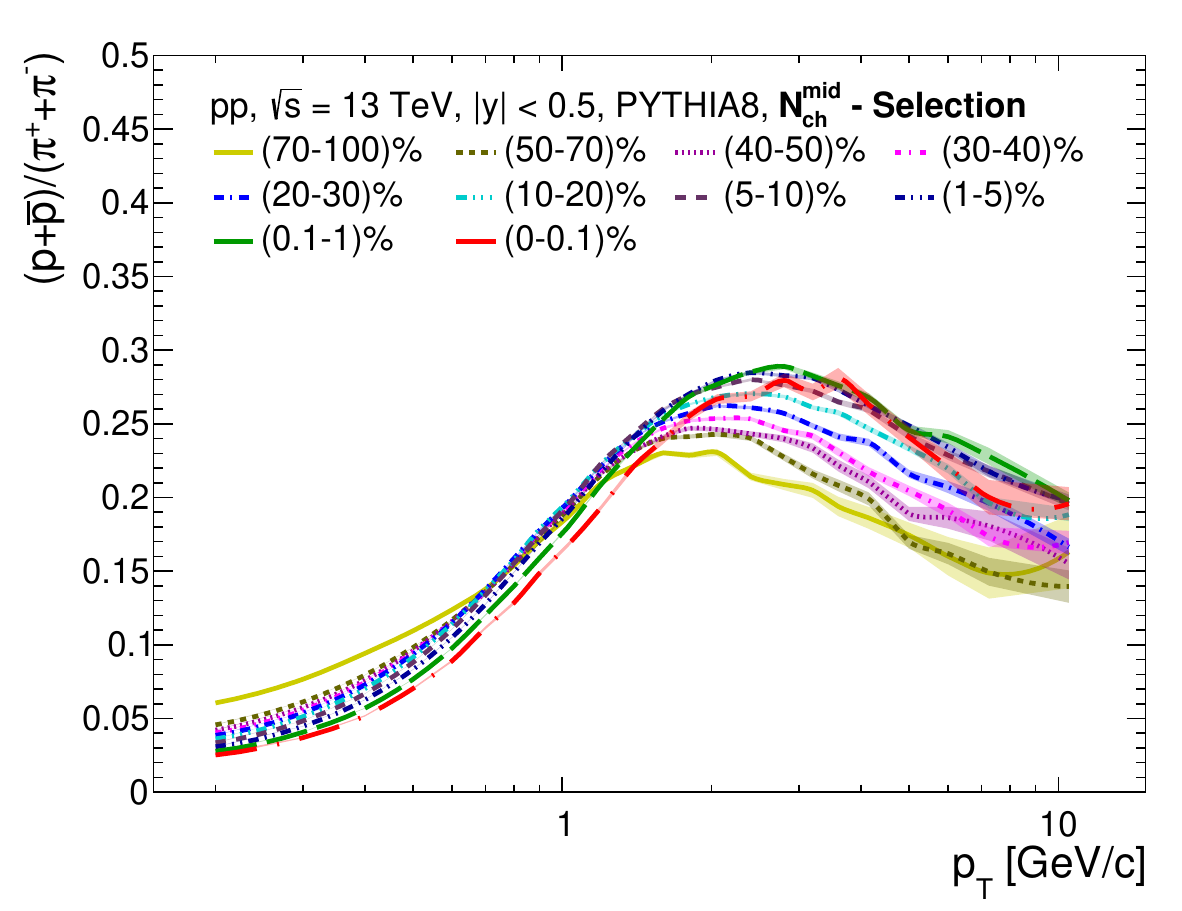}
\includegraphics[width = 0.49\linewidth]{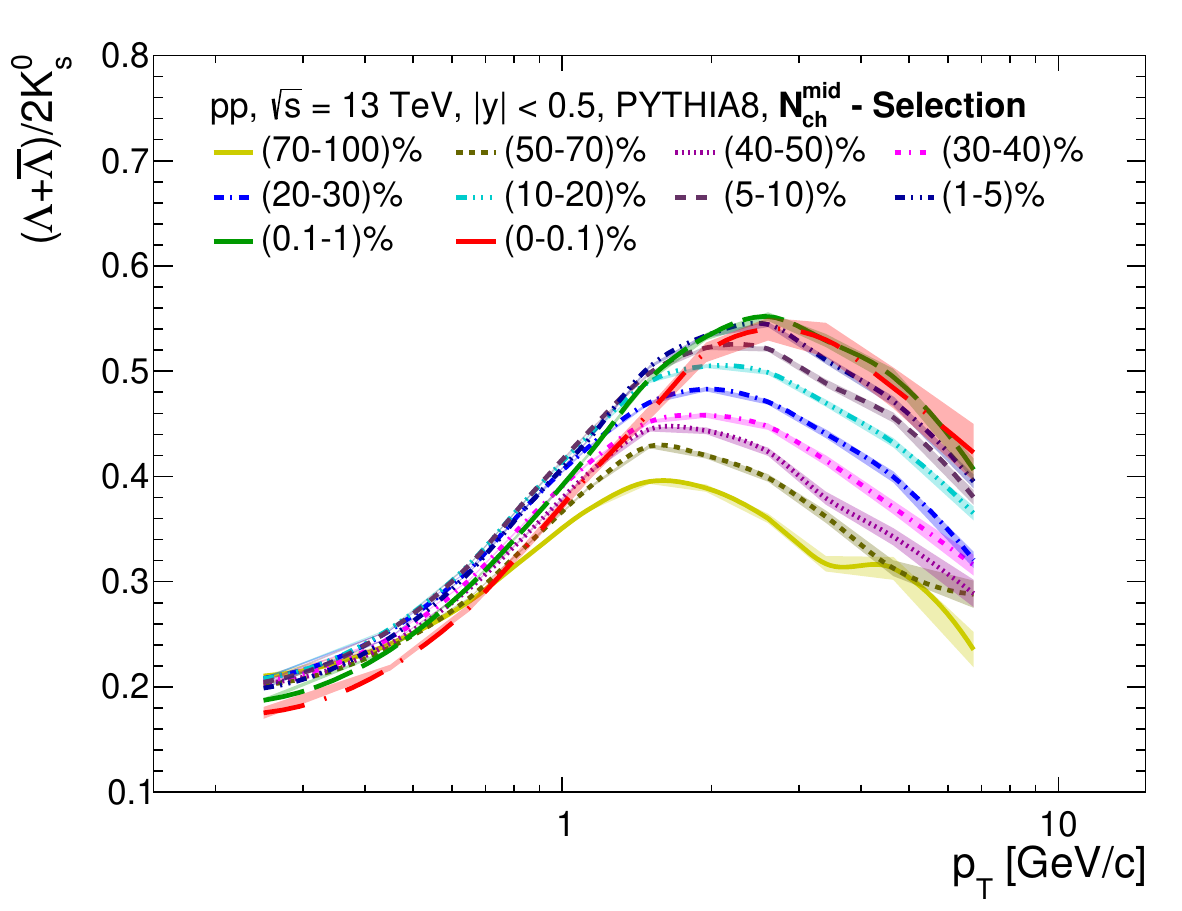}
\includegraphics[width = 0.49\linewidth]{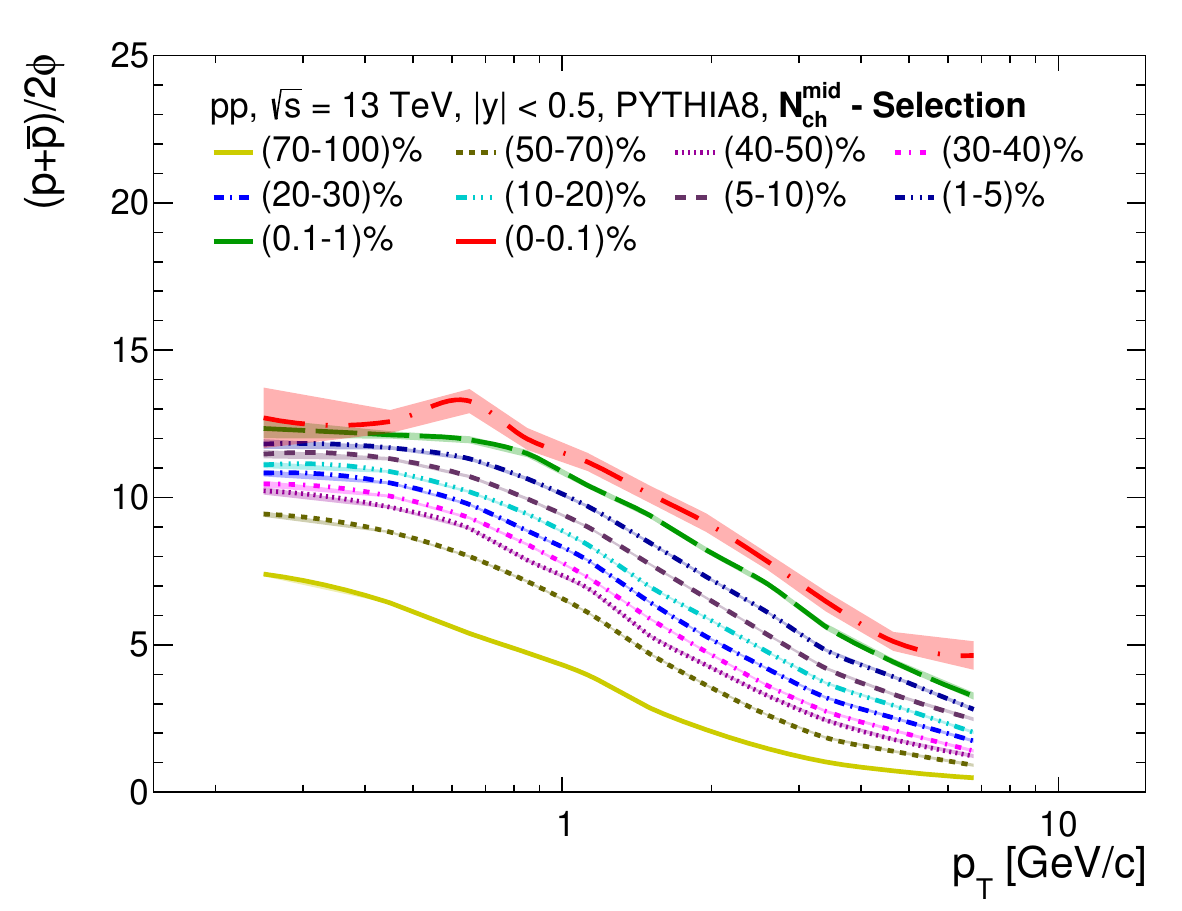}
\caption{$p_{\rm T}$-differential $\rm p/\pi$ (upper left) and $\Lambda/K^{0}_{\rm S}$ (upper right) and $\rm p/\phi$ (lower middle) ratios measured in $|y|<0.5$ for different mid-pseudorapidity charged-particle multiplicity ($\rm N_{\rm ch}^{mid}$) classes in $pp$ collisions at $\sqrt{s}$ = 13 TeV using PYTHIA8. The selection of events is based on $\rm N_{\rm ch}^{mid}$ classes and shown in the left panel of Fig.~\ref{fig:NchDist}.}
\label{fig:Nchratio}
\end{center}
\end{figure*}

Figure~\ref{fig:Nchratio} shows the $p_{\rm T}$-differential $\rm p/\pi$ (upper left), $\Lambda/K^{0}_{\rm S}$ (upper right), and $\rm p/\phi$ (lower middle) ratios in inelastic $pp$ collisions for different mid-pseudorapidity charged-particle multiplicity classes at $\sqrt{s}$ = 13 TeV using PYTHIA8. The mid-pseudorapidity charged-particle multiplicity class selections are shown in the left panel of Fig.~\ref{fig:NchDist}. The $\rm p/\pi$ and $\Lambda/K^{0}_{\rm S}$ ratios exhibits a bump structure in 1.5 $\lesssim p_{\rm T} \lesssim$ 8 GeV/c. This bump structure gradually increases with increasing charged-particle multiplicity. It is observed that the position of the peak shifts towards higher momentum when one goes from the lowest multiplicity to the highest multiplicity classes. The peak reaches a maximum of about 0.28 for  $\rm p/\pi$ and  0.4 for $\Lambda/K^{0}_{\rm S}$ at  $p_{\rm T} \approx 3.5 $ GeV/c for highest multiplicity class. Thus, Fig.~\ref{fig:Nchratio} shows a mass effect in the baryon to meson ratios since the $\Lambda/K^{0}_{\rm S}$  ratios exhibit a larger bump than $\rm p/\pi$ ratios for the highest multiplicity classes. Further, the distinction in the baryon-meson ratios for different multiplicity classes is visible towards high $p_{\rm T}$, i.e., $p_{\rm T} \gtrsim $ 3.0 GeV/c as shown in Fig.~\ref{fig:Nchratio}. \\

However, no such bump structure at intermediate $p_{\rm T}$ is observed in $\rm p/\phi$ ratios using PYTHIA8 with the same MPI and CR tune. Thus, we found that particles having similar masses but different quark content have different particle ratios. The lower middle panel of Fig.~\ref{fig:Nchratio} shows the $\rm p/\phi$ ratios decrease as a function of $p_{\rm T}$.  The $\rm p/\phi$ ratios greater than 1 indicate more protons are produced as compared to the $\phi$ mesons. It is observed that there is a clear charged-particle multiplicity dependence effect present in the $\rm p/\phi$ ratios as a function of $p_{\rm T}$. The $\rm p/\phi$ ratios increase with an increase in charged-particle multiplicity classes in $pp$ collisions. Moreover, we found that at lower values of $p_{\rm T}$ in the highest charged-particle multiplicity classes the $\rm p/\phi$ ratios are almost independent of $p_{\rm T}$. This indicates that spectral shapes only depend upon the particle mass, which is the main variable for determining the spectral shapes in most of the hydrodynamic models. This could be a possible signature of radial flow present in the system. It is important to note that although the spectral shape may change with the interactions present in the hadronic phase and it could modify the picture differently. However, it is believed that it will affect differently for protons and $\phi$ mesons as compared to the short-lived resonances.  \\

% It is observed that the particle spectra harden with increasing charged-particle multiplicity in $pp$ collisions. This implies the radial flow velocity increases with increasing the charged-particle multiplicity. Moreover, the baryon enhancement at intermediate $p_{\rm T}$ could be due to the redistribution of baryons and mesons over the momentum range. \\

It is worth mentioning that $\rm p/\phi$ ratios for other event shape classifiers behave similarly to charged particle multiplicity. Therefore, the $\rm p/\phi$ ratios for other event classifiers are not discussed in this review.
   
\begin{figure*}[ht!]
\begin{center}
\includegraphics[width = 0.49\linewidth]{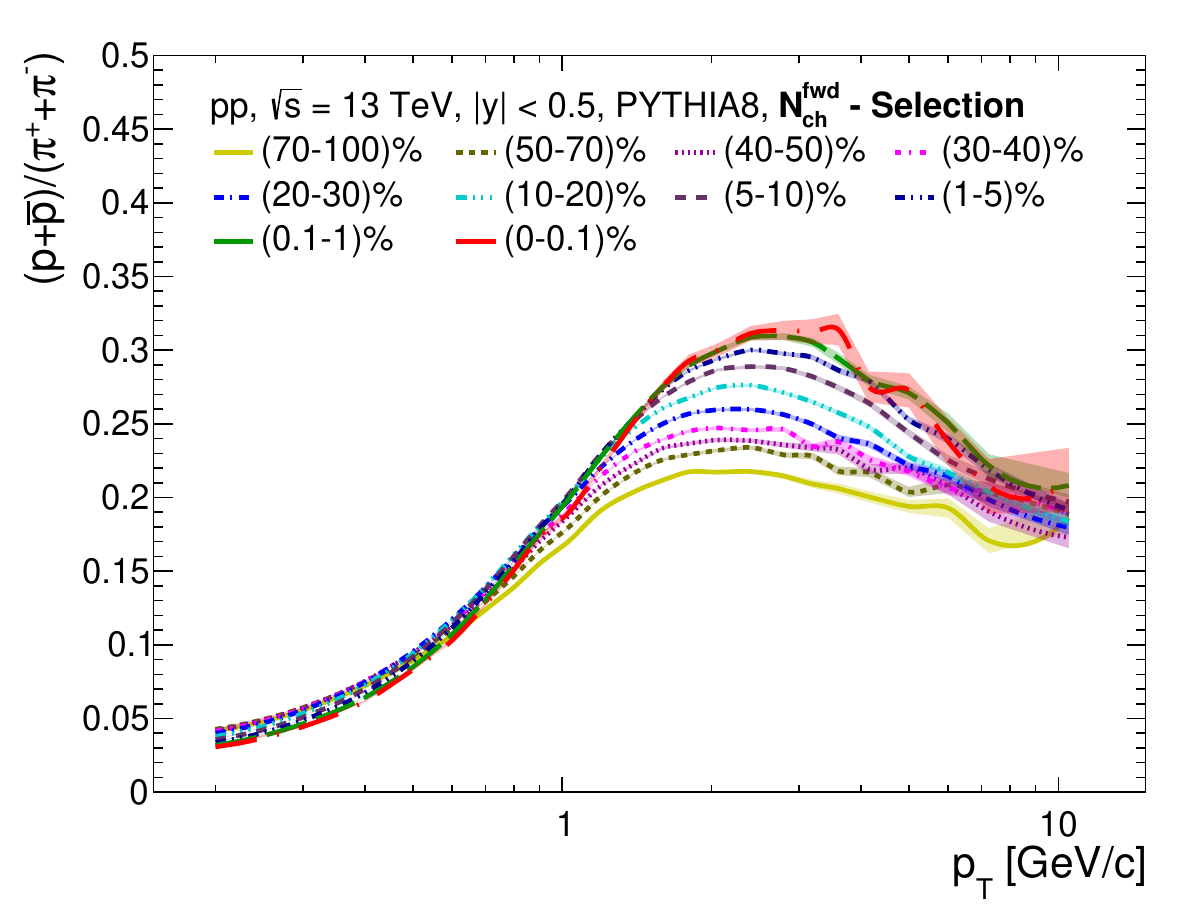}
\includegraphics[width = 0.49\linewidth]{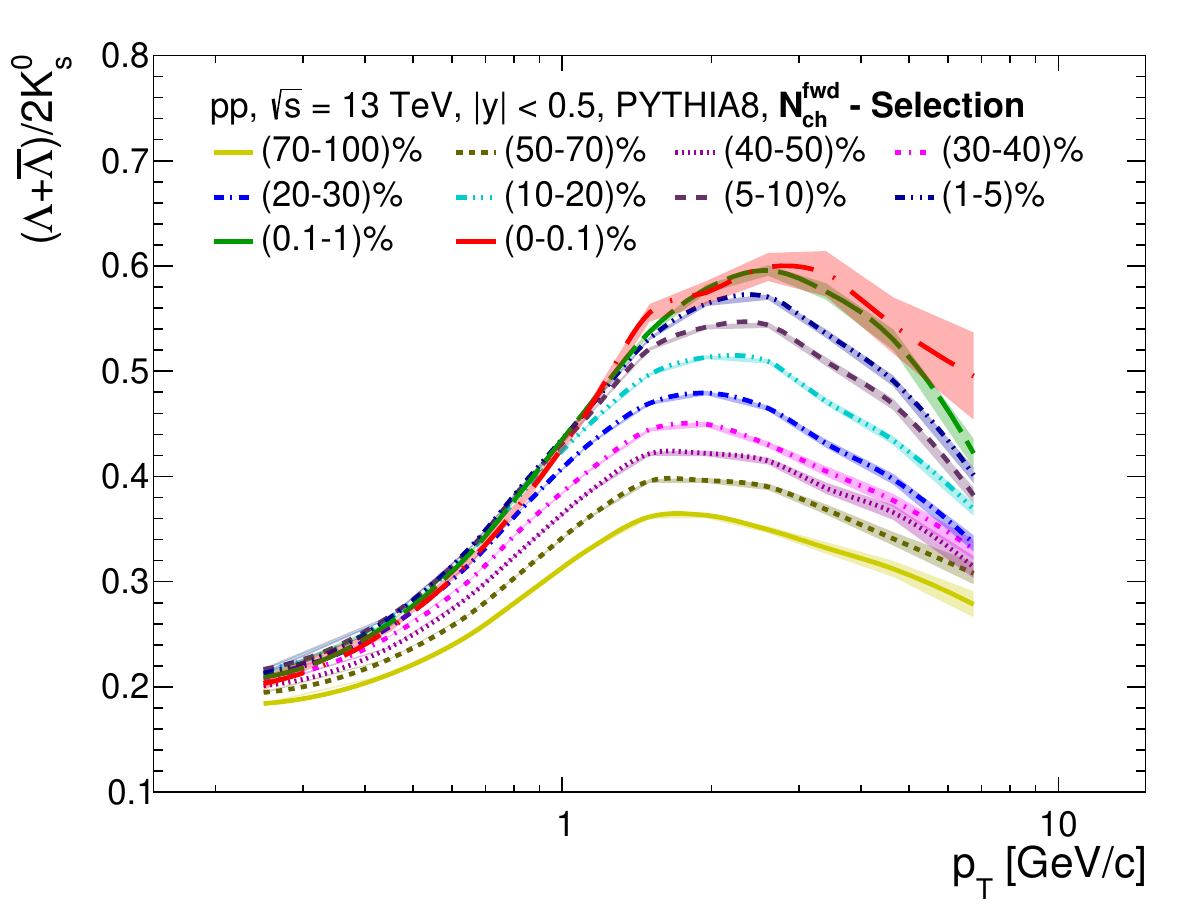}
\caption{$p_{\rm T}$-differential $\rm p/\pi$ (left) and $\Lambda/K^{0}_{\rm S}$ (right) ratios measured in $|y|<0.5$ for different forward pseudorapidity charged-particle multiplicity classes ($\rm N_{\rm ch}^{fwd}$) in $pp$ collisions at $\sqrt{s}$ = 13 TeV using PYTHIA8. The selection of events is based on $\rm N_{\rm ch}^{fwd}$ classes and shown in the right panel of Fig.~\ref{fig:NchDist}.}
\label{fig:V0Mratio}
\end{center}
\end{figure*}

\begin{figure*}[ht!]
\begin{center}
\includegraphics[width = 0.49\linewidth]{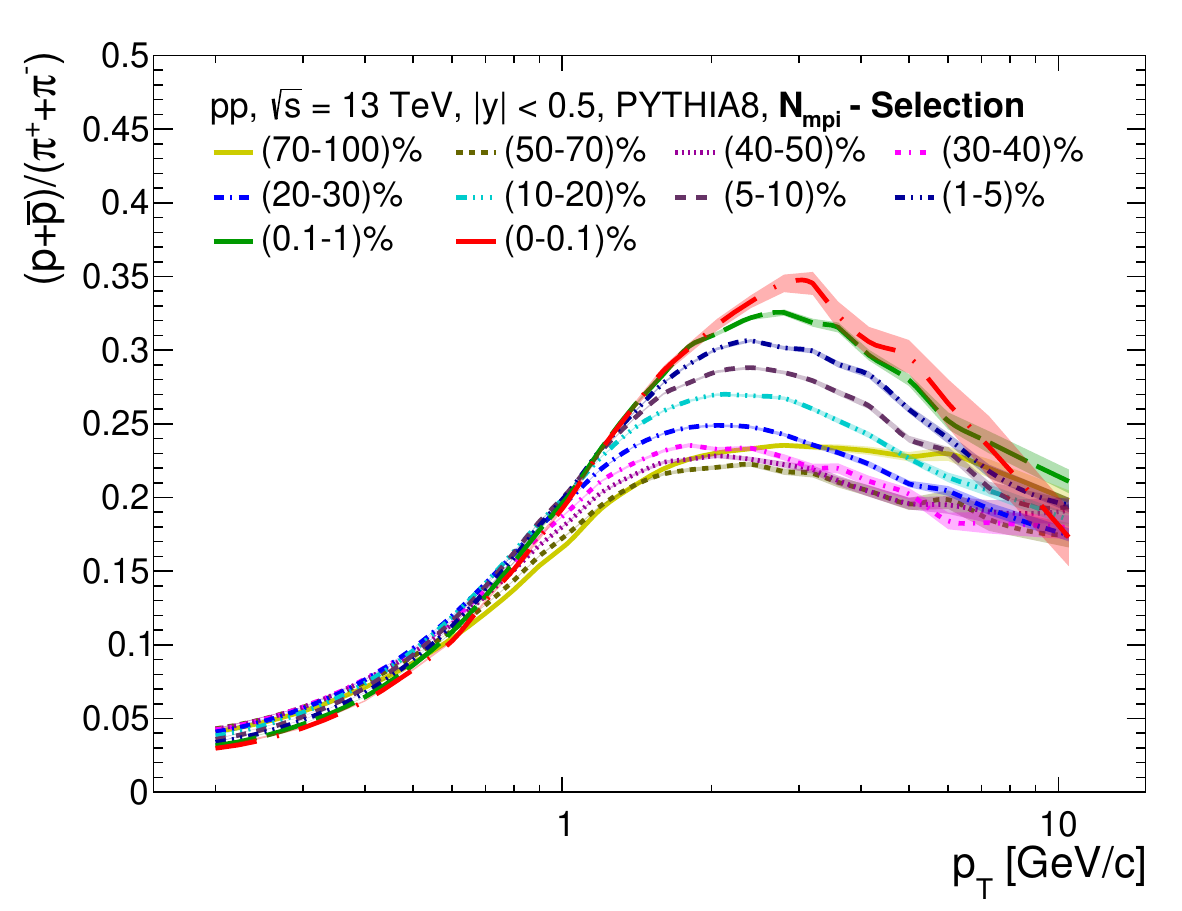}
\includegraphics[width = 0.49\linewidth]{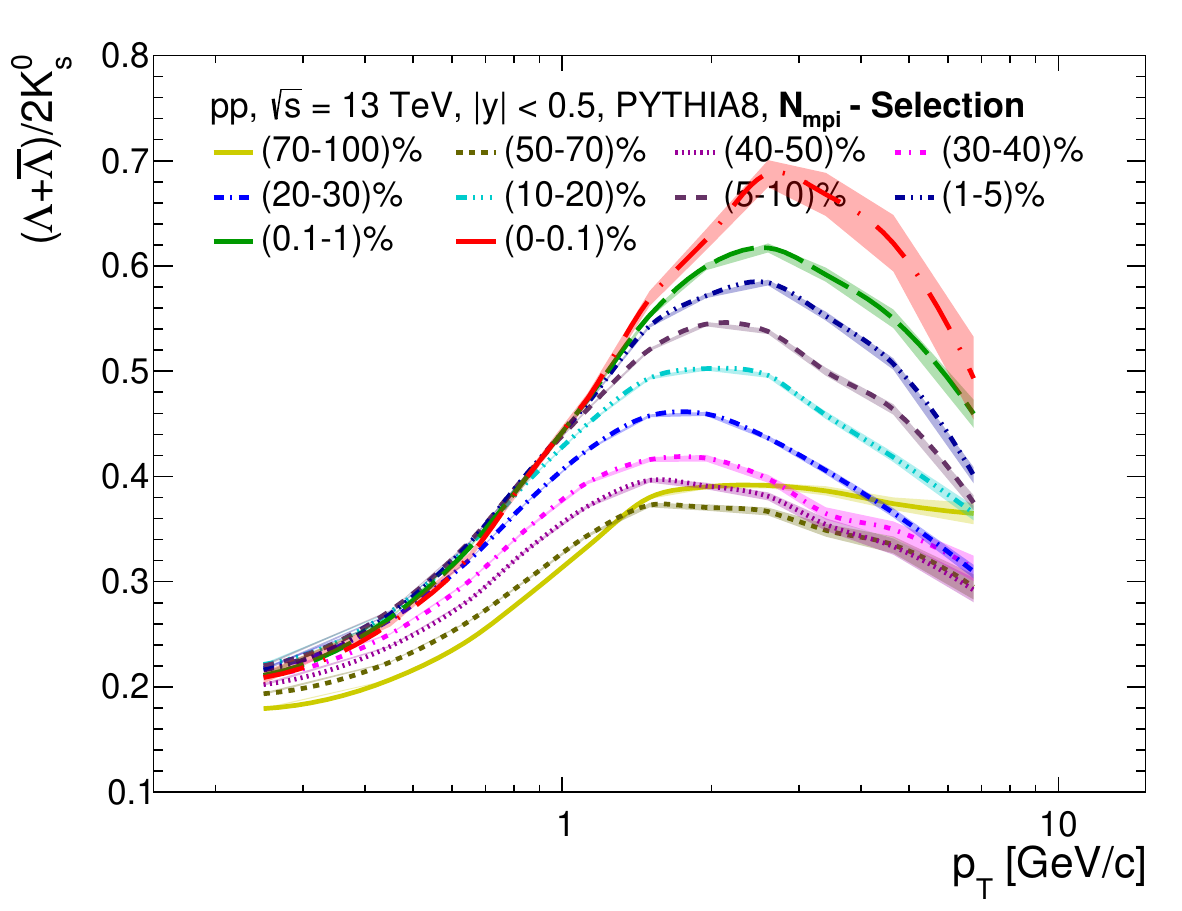}
\caption{$p_{\rm T}$-differential $\rm p/\pi$ (left) and $\Lambda/K^{0}_{\rm S}$ (right) ratios measured in $|y|<0.5$ for different $\rm N_{mpi}$ classes in $pp$ collisions at $\sqrt{s}$ = 13 TeV using PYTHIA8. The selection of events is based on $\rm N_{mpi}$ classes and shown in Fig.~\ref{fig:NmpiDist}.}
\label{fig:Nmpiratio}
\end{center}
\end{figure*}
Figure~\ref{fig:V0Mratio} shows the $\rm p/\pi$ (left) and $\Lambda/K^{0}_{\rm S}$ (right) ratios as a function of  $p_{\rm T}$ in $pp$ collisions for different forward pseudorapidity charged-particle multiplicity classes. In Fig.~\ref{fig:V0Mratio}, the forward pseudorapidity charged-particle multiplicity selections are according to the V0 detector acceptance, and class selections are displayed in the right panel of Fig.~\ref{fig:NchDist}. Since the charged-particle multiplicity at mid rapidity ($N_{\rm ch}^{\rm mid}$) follows a linear correlation with the charged-particle multiplicity at forward rapidity ($N_{\rm ch}^{\rm fwd}$), it is expected, the baryon to meson ratios in $N_{\rm ch}^{\rm fwd}$ follows the similar behavior with $N_{\rm ch}^{\rm mid}$. It is observed that although the $\rm p/\pi$ ratios at $N_{\rm ch}^{\rm fwd}$ has similar trend with $N_{\rm ch}^{\rm mid}$, a slight variation is observed towards the low $p_{\rm T}$ ($p_{\rm T} \lesssim $ 0.8 GeV/c). The $\rm p/\pi$ ratios for $N_{\rm ch}^{\rm mid}$ at low $p_{\rm T}$ are higher for the lowest multiplicity class, and they decrease with charged-particle multiplicity as shown in Fig.~\ref{fig:Nchratio}. While $\rm p/\pi$ ratios have nearly the same value for forward multiplicity classes at low $p_{\rm T}$. We observe that strange-hadron sector $\Lambda/K^{0}_{\rm S}$ ratios follow a similar trend as $\rm p/\pi$ ratios. However, the increasing trend of baryon-to-meson ratios with charged-particle multiplicity is clearly observed for all $p_{\rm T}$ bins. 

%Furthermore, the $\rm p/\phi$ ratios in forward pseudorapidity charged-particle multiplicity show a decreasing trend as a function of the  $p_{\rm T}$. However, the charged-particle multiplicity dependence in $\rm p/\phi$ ratios is a bit diluted in the forward pseudorapidity region, whereas a clear multiplicity dependence was observed for mid-pseudorapidity charged-particle multiplicity classes.  \\

Figure~\ref{fig:Nmpiratio} shows the $\rm p/\pi$ (left)  and $\Lambda/K^{0}_{\rm S}$ (right)  ratios as a function of  $p_{\rm T}$ for different $N_{\rm mpi}$ classes. The $N_{\rm mpi}$ class selections for the minimum bias sample are displayed in Fig.~\ref{fig:NmpiDist}. It is observed that events with the highest number of multi-partonic interactions show higher $\rm p/\pi$ ratios at intermediate $p_{\rm T}$ and the ratio decreases with a decrease in $N_{\rm mpi}$. Since $N_{\rm mpi}$ has a positive correlation with the $N_{\rm ch}^{\rm mid}$ and $N_{\rm ch}^{\rm fwd}$, it is expected the evolution of baryon-to-meson ratios with the number of MPI will be qualitatively similar to the charged-particle multiplicity. However, it is evident to note that the events with less number of multi-partonic interactions (i.e, (70-100)\% class) do not exhibit the bump structure at intermediate $p_{\rm T}$, it stays almost constant with $p_{\rm T}$. Thus, events with less MPI activity do not show flow-like patterns in $pp$ collisions. \

Apart from charged-particle multiplicity and MPI activity, we also investigate the $p_{\rm T}$-dependent baryon to meson ratios for the other event classifiers such as transverse sphericity ($S_{\rm T}$), transverse spherocity ($S_{0}$), relative activity classifier ($R_{\rm T}$) and charged-particle flattenicity ($\rho_{\rm ch}$).\\

\begin{figure*}[ht!]
\begin{center}
\includegraphics[width = 0.49\linewidth]{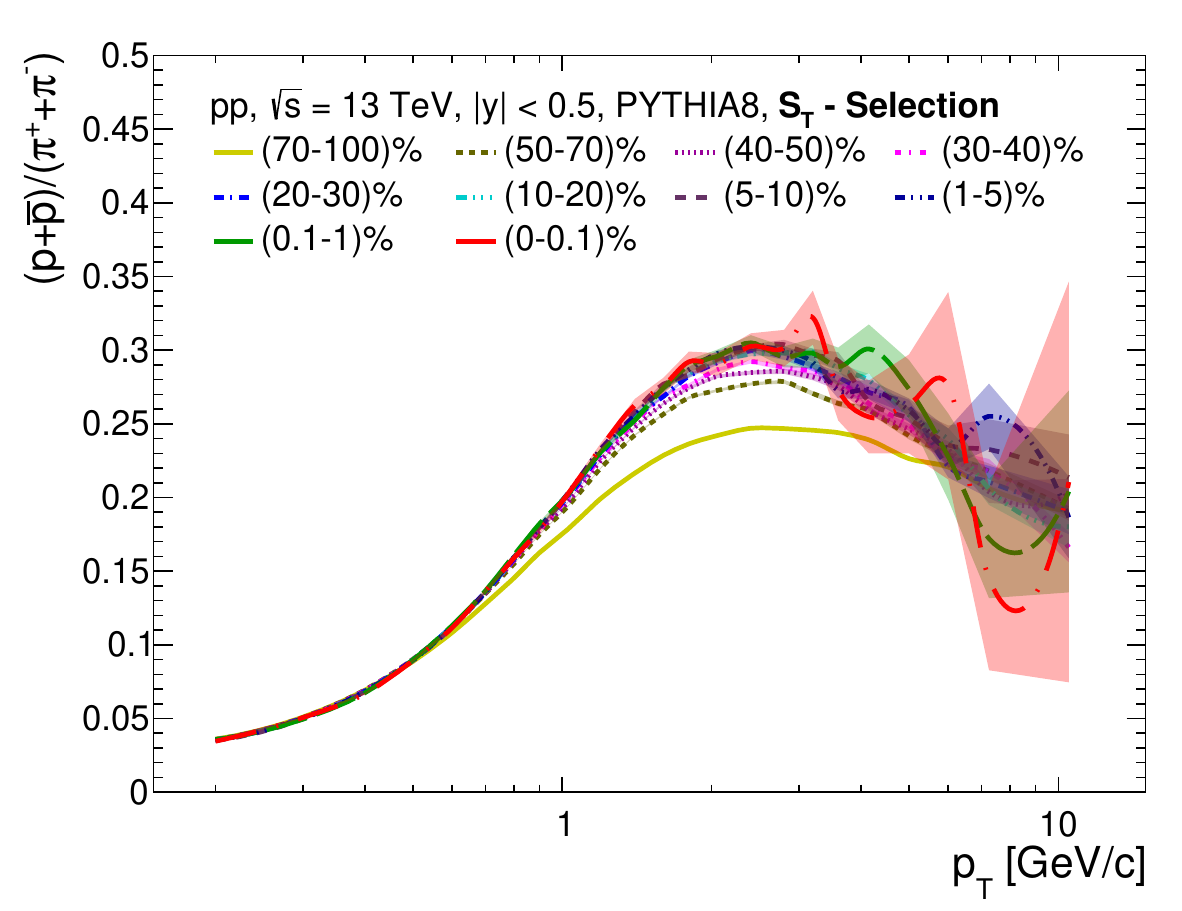}
\includegraphics[width = 0.49\linewidth]{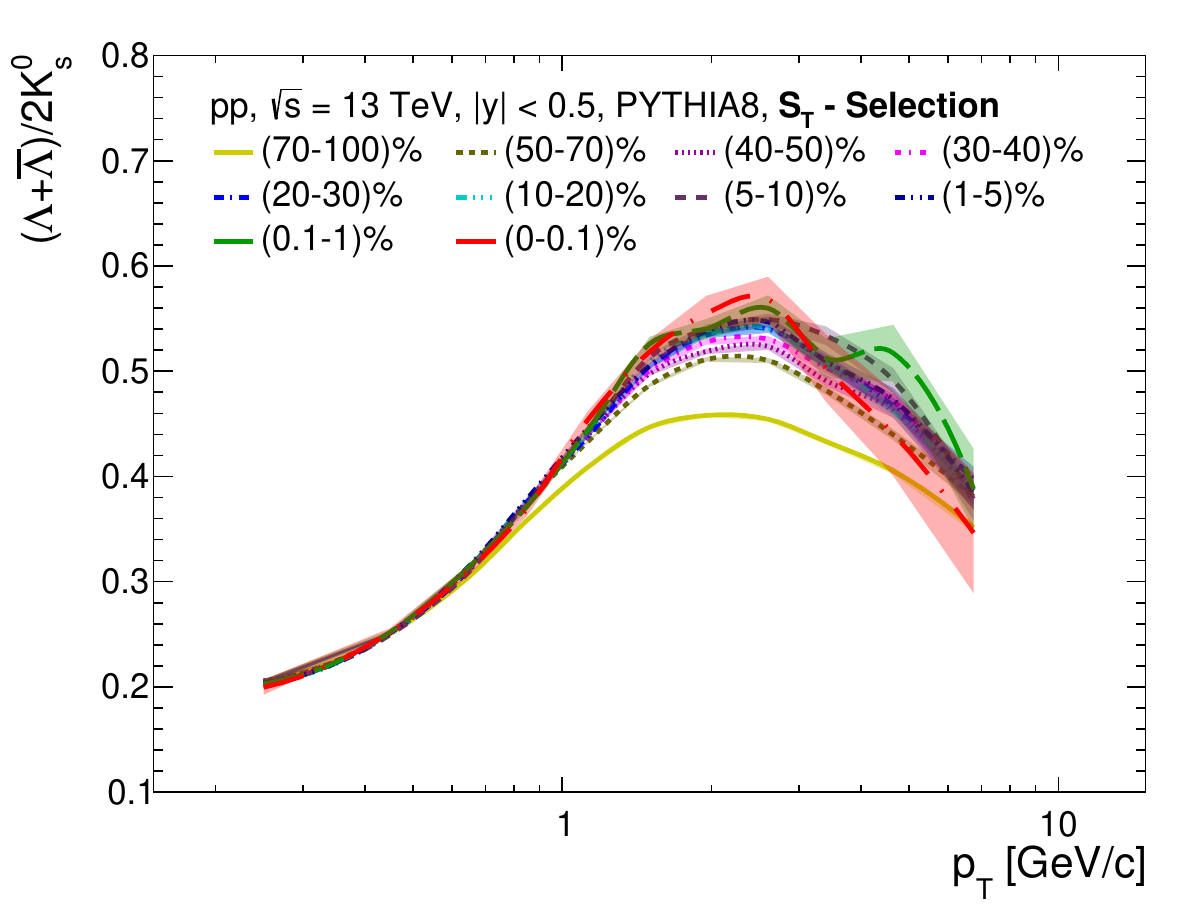}
\caption{$p_{\rm T}$-differential $\rm p/\pi$ (left) and $\Lambda/K^{0}_{\rm S}$ (right) ratios measured in $|y|<0.5$ for different unweighted transverse sphericity ($S_{\rm T}$) classes in $pp$ collisions at $\sqrt{s}$ = 13 TeV using PYTHIA8. The selection of events is based on $S_{\rm T}$ classes and shown in the right panel of  Fig.~\ref{fig:STdist}}
\label{fig:STratio}
\end{center}
\end{figure*}

Figure~\ref{fig:STratio} presents the $\rm p/\pi$ (left) and $\Lambda/K^{0}_{\rm S}$ (right) ratios as a function of  $p_{\rm T}$ for different $S_{\rm T}$ classes. The $S_{\rm T}$ class selections are shown in Fig.~\ref{fig:STdist}. From Fig.~\ref{fig:STratio} it is observed that the isotropic events  $\rm p/\pi$ ratios show an enhancement feature at the intermediate $p_{\rm T}$, while jetty events $\rm p/\pi$ ratios are flat as a function of $p_{\rm T}$. However, unlike the charged-particle multiplicity and/or the number of multi-partonic interactions, the dependence of transverse sphericity classes on $\rm p/\pi$ ratios is found to be weak. 
Similar observation is also seen for $\Lambda/K^{0}_{\rm S}$ with $p_{\rm T}$ for different $S_{\rm T}$ classes. 

%Moreover, the lower middle panel of Fig.~\ref{fig:STratio} shows $\rm p/\phi$ ratios decrease as a function of $p_{\rm T}$ for various $S_{\rm T}$ classes. Interestingly, it is observed that for (70-100)\% $S_{\rm T}$ class the $\rm p/\phi$ ratios have a minimum value with $p_{\rm T}$ $\lesssim$ 2.0 and it reaches a maximum value for $p_{\rm T}$ $\gtrsim$ 2.0 among all $S_{\rm T}$ classes.

\begin{figure*}[ht!]
\begin{center}
\includegraphics[width = 0.49\linewidth]{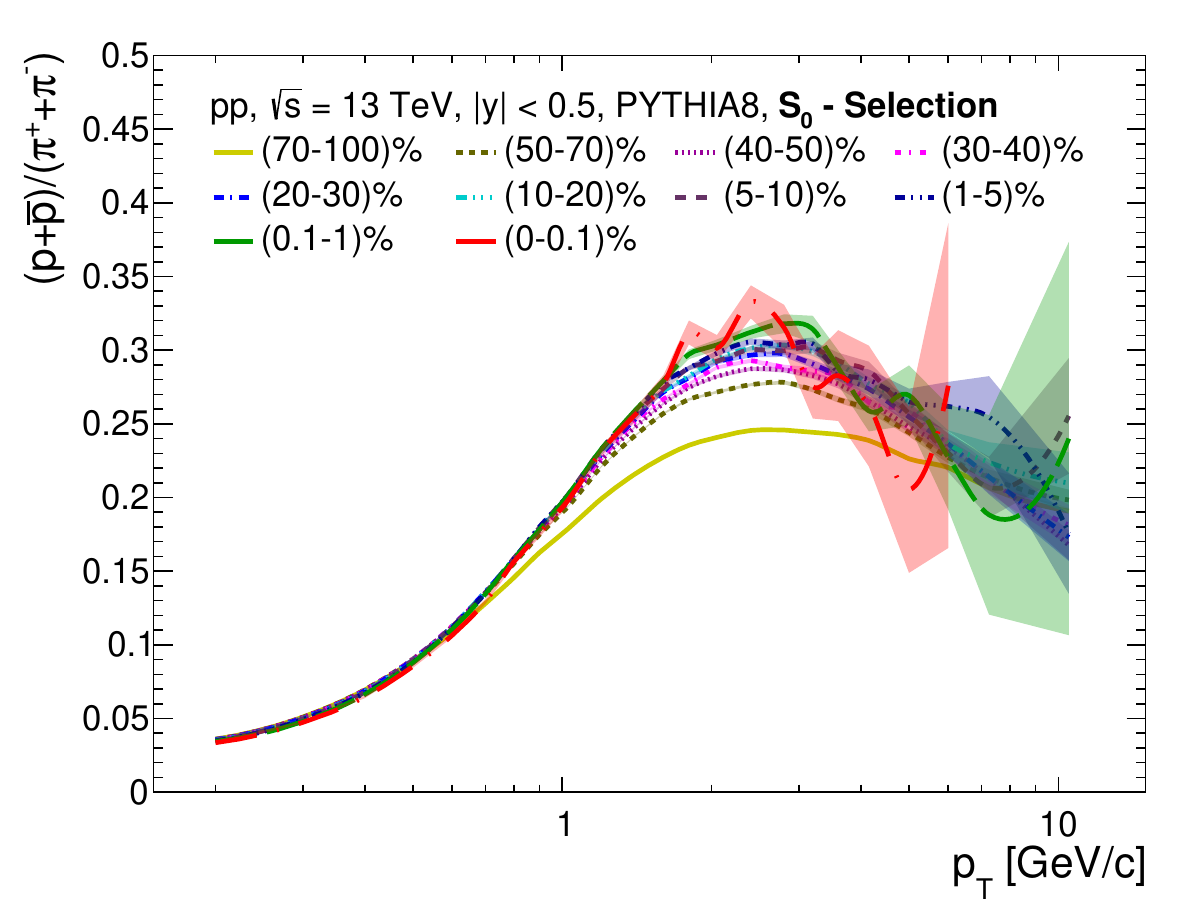}
\includegraphics[width = 0.49\linewidth]{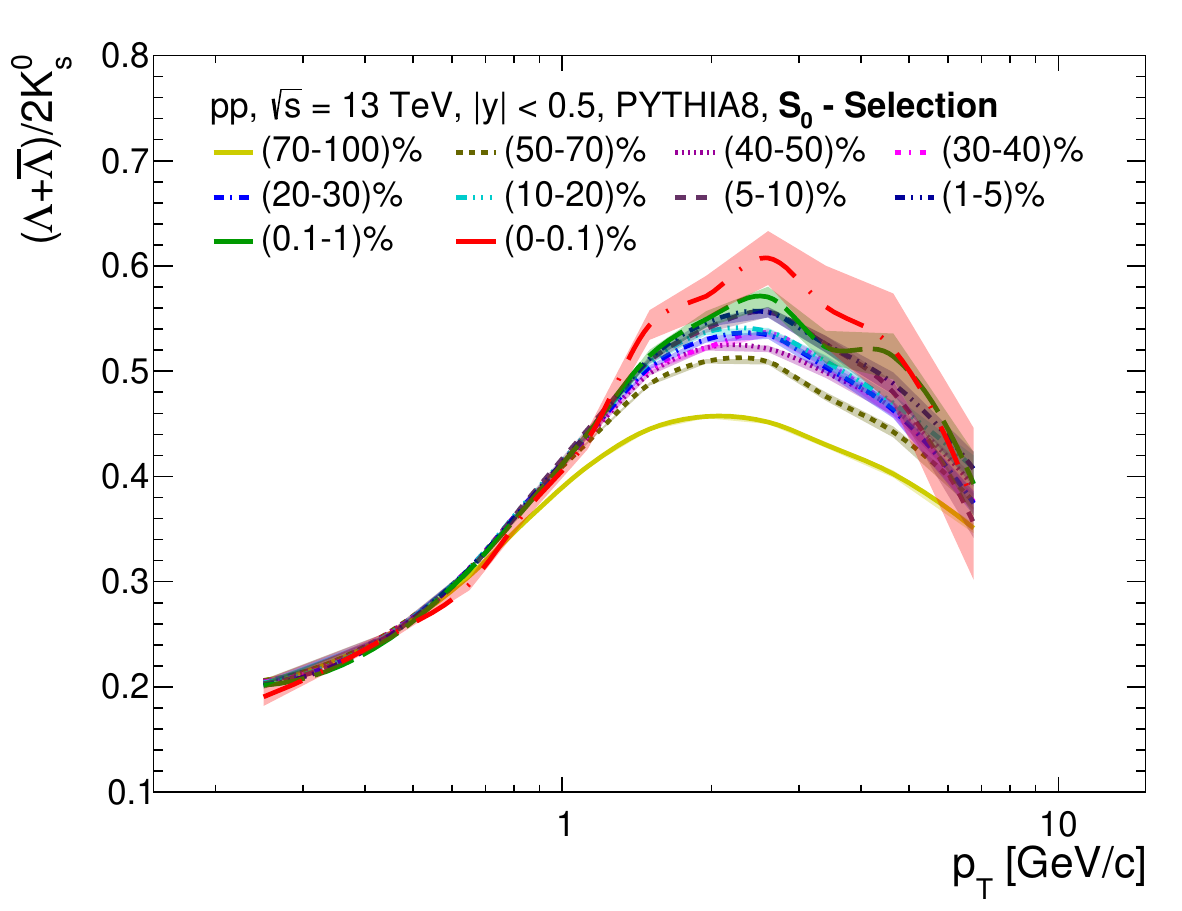}
\caption{$p_{\rm T}$-differential $\rm p/\pi$ (left) and $\Lambda/K^{0}_{\rm S}$ (right) ratios measured in $|y|<0.5$ for different weighted transverse spherocity ($S_{0}$) classes in $pp$ collisions at $\sqrt{s}$ = 13 TeV using PYTHIA8. The selection of events is based on $S_{0}$ classes and shown in the right panel of Fig.~\ref{fig:S0dist}.}
\label{fig:S0ratio}
\end{center}
\end{figure*}

%\begin{figure*}[ht!]
%\begin{center}
%\includegraphics[width = 0.49\linewidth]{SpherocityFigures/S0pt1protonpionratio.pdf}
%\includegraphics[width = 0.49\linewidth]{SpherocityFigures/S0pt1lambdaK0sratio.pdf}
%\includegraphics[width = 0.49\linewidth]{SpherocityFigures/S0pt1protonphiratio.pdf}
%\caption{$p_{\rm T}$-differential $\rm p/\pi$ (left) and $\Lambda/K^{0}_{\rm S}$ (right) ratios measured in $|y|<0.5$ for different unweighted transverse spherocity ($S_{0}^{p_{\rm T} = 1}$) classes in $pp$ collisions at $\sqrt{s}$ = 13 TeV using PYTHIA8. The selection of events is based on $S_{0}^{p_{\rm T} = 1}$ classes and shown in the right panel of Fig.~\ref{fig:S0dist}}
%\label{fig:S0pt1ratio}
%\end{center}
%\end{figure*}

Figure~\ref{fig:S0ratio} highlights $\rm p/\pi$ (left) and $\Lambda/K^{0}_{\rm S}$ (right) ratios as a function of  $p_{\rm T}$ for different $S_{0}$ classes. The $S_{0}$ class selections are shown in the left panel of Fig.~\ref{fig:S0dist}. The $p_{\rm T}$-differential baryon-to-meson ratios show higher enhancement for isotropic events compared to jetty events at intermediate $p_{\rm T}$ in the minimum bias sample. We found that the $\rm p/\pi$ and $\Lambda/K^{0}_{\rm S}$ ratios decrease rapidly for the highest (0-0.1)\% spherocity class towards high $\rm p_{\rm T}$. 

%The lower middle panel of Fig.~\ref{fig:S0ratio} shows $\rm p/\phi$ ratios decrease as a function of $p_{\rm T}$ for various $S_{0}$ classes. Interestingly, it is observed that for top central $S_{0}$ classes the $\rm p/\phi$ ratios are constant with $p_{\rm T}$ $\lesssim$ 1.0 within uncertainties and then it decreases with $p_{\rm T}$.In addition, it is observed that events with jetty type are found to have higher $\rm p/\phi$ value for $p_{\rm T}$ $\gtrsim$ 2.0, while it is minimum for $p_{\rm T}$ $\lesssim$ 2.0.Thus, with the transverse spherocity event classifiers, one can separate the events which have flow-like behavior. 

%Figure~\ref{fig:S0pt1ratio} shows $\rm p/\pi$ (left) and $\Lambda/K^{0}_{\rm S}$ (right) ratios as a function of $p_{\rm T}$ for different $S_{0}^{p_{\rm T} = 1}$ classes. The $S_{0}^{p_{\rm T} = 1}$ class selections are shown in the right panel of Fig.~\ref{fig:S0dist}. The $p_{\rm T}$-differential baryon to meson ratios show higher enhancement for isotropic events compared to jetty events at intermediate $p_{\rm T}$ in the minimum bias sample. However, it is important to note that unlike $S_{0}$, $S_{0}^{p_{\rm T} = 1}$ the rapid decrease of the $\rm p/\pi$ ratios at (0-0.1)\% spherocity class is towards high $p_{\rm T}$ is not observed.

%In addition, the $\rm p/\phi$ ratios decrease with  $p_{\rm T}$ and overall all spherocity event classes predict the same $\rm p/\phi$ ratios as a function of $p_{\rm T}$.\\

\begin{figure*}[ht!]
\begin{center}
\includegraphics[scale=0.7]{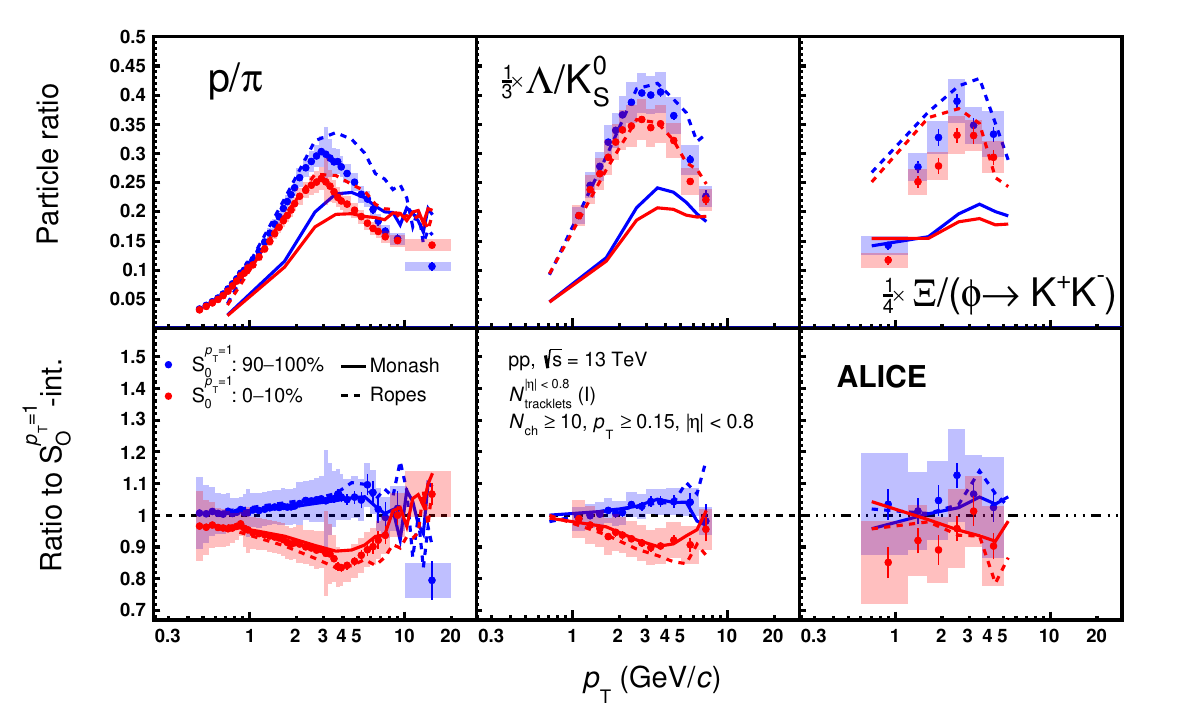}
\caption{$p_{\rm T}$-differential $p/\pi$ (upper left), $\Lambda/K^{0}_{\rm S}$ (upper middle), and $\Xi/\phi$ (upper right) ratios for different classes of $S_{0}^{p_{\rm T}=1}$ for the top 0-1\% multiplicity class measured by $N_{\rm tracklets}^{|\eta|<0.8}$ in $pp$ collisions at $\sqrt{s}=13$ TeV with ALICE and compared with PYTHIA~8 Color Ropes and Monash. The lower panel shows the ratio to $S_{0}^{p_{\rm T}=1}$-integrated case~\cite{ALICE:2023bga}.}
\label{fig:S0pt1ratiodata}
\end{center}
\end{figure*}

In Fig.~\ref{fig:S0pt1ratiodata}, we show that ALICE measurements for the $p_{\rm T}$-differential $p/\pi$, $\Lambda/K^{0}_{\rm S}$, and $\Xi/\phi$ ratios for the extreme classes of $S_{0}^{p_{\rm T}=1}$ of top 0-1\% multiplicity class measured by $N_{\rm tracklets}^{|\eta|<0.8}$ in $pp$ collisions at $\sqrt{s}=13$ TeV. A clear impact of $S_{0}^{p_{\rm T}=1}$ selection can be observed for the $p/\pi$ ratios. In contrast, for $\Lambda/K^{0}_{\rm S}$, and $\Xi/\phi$ ratios, the impact of $S_{0}^{p_{\rm T}=1}$ selection can only be observed within the systematic uncertainties. PYTHIA~8 Monash results fail to reproduce the data, while Ropes qualitatively reproduces the data (evident from the upper panel). The bottom panel shows the ratios of the extreme classes of $S_{0}^{p_{\rm T}=1}$ to $S_{0}^{p_{\rm T}=1}$-integrated events. Here, one finds that both models manage to reproduce the data.

\begin{figure*}[ht!]
\begin{center}
\includegraphics[scale=0.7]{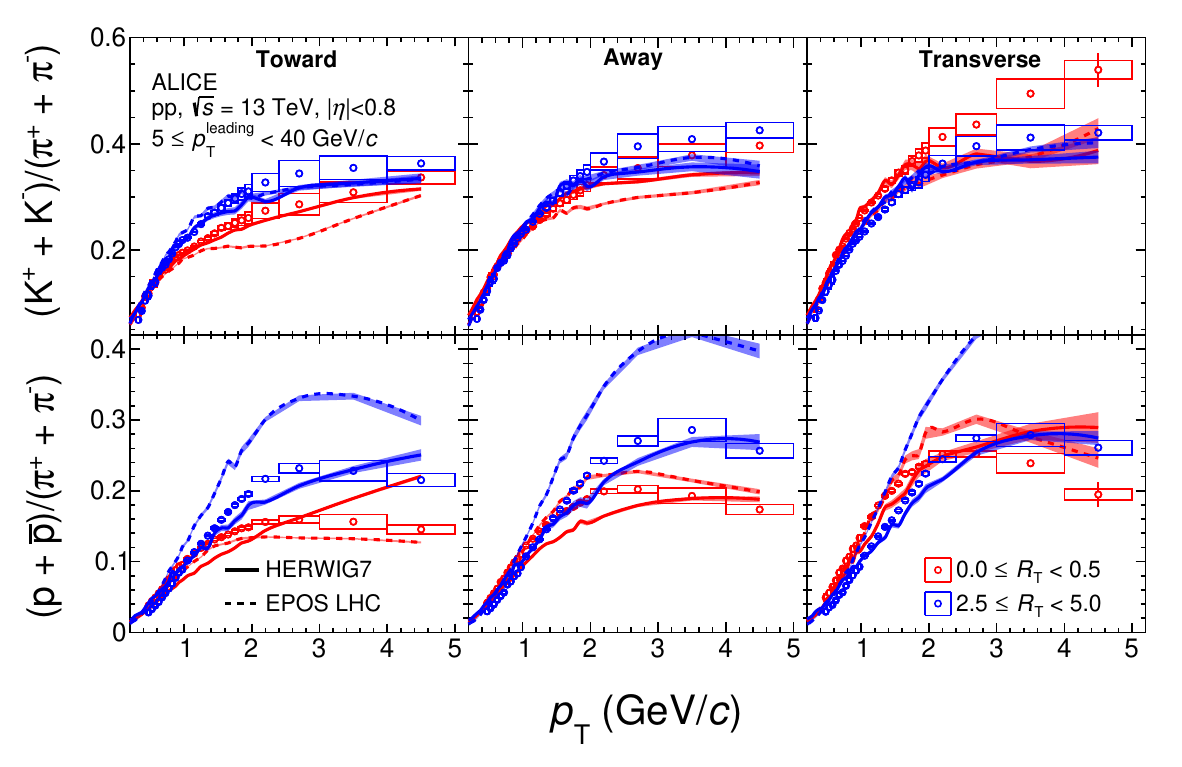}
\caption{$p_{\rm T}$-differential $K/\pi$ (upper) and $p/\pi$ (lower) ratios measured in different topological regions as a function of $R_{\rm T}$ in $pp$ collisions at $\sqrt{s}=13$ TeV with ALICE. Comparisons are made with similar measurements performed with EPOS LHC and HERWIG7~\cite{ALICE:2022fnb}.}
\label{fig:RTratiodata}
\end{center}
\end{figure*}

Now, using the charged-particle multiplicity information, we investigate the $p_{\rm T}$-dependent baryon to meson ratios for another underlying event activity called relative transverse activity classifier ($R_{\rm T}$) in the transverse region as depicted in Fig.~\ref{fig:myregions}. 
Figure~\ref{fig:RTratiodata} shows the $p_{\rm T}$ dependence of $K/\pi$ (upper panel) and $p/\pi$ (lower panel) ratios as a function of $R_{\rm T}$ in $pp$ collisions at $\sqrt{s}=13$ TeV with ALICE along with comparisons from EPOS LHC and HERWIG7. The $K/\pi$ ratios show similar behavior in both toward and away regions, which increases with an increase in $R_{\rm T}$. However, the $K/\pi$ ratios increase with a decrease in $R_{\rm T}$ or UE activity in the transverse region. In the lower panel, one finds that with an increase in $R_{\rm T}$, the contribution of UE activity in both toward and away regions for $p_{\rm T}>1$~GeV/c increases, leading to a rise in $p/\pi$ ratios. This increase in $p/\pi$ ratios with the increase in UE activity can be attributed to a growth in collective radial flow. Further, the $p/\pi$ ratios in the transverse regions have a mild dependence on $R_{\rm T}$. Here, towards the lower $p_{\rm T}$ region, an impression of proton production with respect to pions is observed for the events with larger UE activity. This can be attributed to the radial flow effects, which boost the protons towards higher $p_{\rm T}$ regions. Further, a jet hardening effect with an increase in transverse multiplicity is also visible for the high-$R_{\rm T}$ events which shifts the peak of $p/\pi$ ratios of high $R_{\rm T}$ events towards a higher $p_{\rm T}$ as compared to the events with low $R_{\rm T}$ events~\cite{Ortiz:2016kpz}.
%\begin{figure*}[ht!]
%\begin{center}
%\includegraphics[width = 0.49\linewidth]{RTFigures/RTmaxprotonpionratio.pdf}
%\includegraphics[width = 0.49\linewidth]{RTFigures/RTmaxlambdaK0sratio.pdf}
%\includegraphics[width = 0.49\linewidth]{RTFigures/RTmaxprotonphiratio.pdf}
%\caption{$p_{\rm T}$-differential $\rm p/\pi$ (left) and $\Lambda/K^{0}_{\rm S}$ (right) ratios measured in $|y|<0.5$ for different maxiumum relative transverse activity classifier ($R_{\rm T}^{\rm max}$) classes in $pp$ collisions at $\sqrt{s}$ = 13 TeV using PYTHIA8. The selection of events is based on  $R_{\rm T}^{\rm max}$ classes and shown in the middle panel of Fig.~\ref{fig:RTdist}}
%\label{fig:RTmaxratio}
%\end{center}
%\end{figure*}

%\begin{figure*}[ht!]
%\begin{center}
%\includegraphics[width = 0.49\linewidth]{RTFigures/RTminprotonpionratio.pdf}
%\includegraphics[width = 0.49\linewidth]{RTFigures/RTminlambdaK0sratio.pdf}
%\includegraphics[width = 0.49\linewidth]{RTFigures/RTminprotonphiratio.pdf}
%\caption{$p_{\rm T}$-differential $\rm p/\pi$ (left) and $\Lambda/K^{0}_{\rm S}$ (right) ratios  measured in $|y|<0.5$ for different minimum relative transverse activity classifier ($R_{\rm T}^{\rm min}$) classes in $pp$ collisions at $\sqrt{s}$ = 13 TeV using PYTHIA8. The selection of events is based on  $R_{\rm T}^{\rm min}$ classes and shown in the right panel of Fig.~\ref{fig:RTdist}}
%\label{fig:RTminratio}
%\end{center}
%\end{figure*}

The definition of flattenicity (Eq.~\eqref{eq:flatNch}) indicates that such an event classifier strongly depends upon the selection of charged-particle multiplicity in the forward pseudorapidity region. The limit ($1-\rho_{\rm ch}) \rightarrow$ 1 correspond to high multiplicity events and ($1-\rho_{\rm ch}) \rightarrow$ 0 correspond to low multiplicity events. The correlation of charged-particle flattenicity event shape observable with $N_{\rm mpi}$ is shown in Fig.~\ref{fig:FlatvsMPI}. One interesting point is to note that the correlation between ($1-\rho_{\rm ch}$) rises sharply with $N_{\rm mpi}$ after 0.7, indicating the flattenicity serves as a better tool among all event classifiers to separate the isotropic (soft) and jetty (hard) events. 
\begin{figure*}[ht!]
\begin{center}
\includegraphics[scale=0.7]{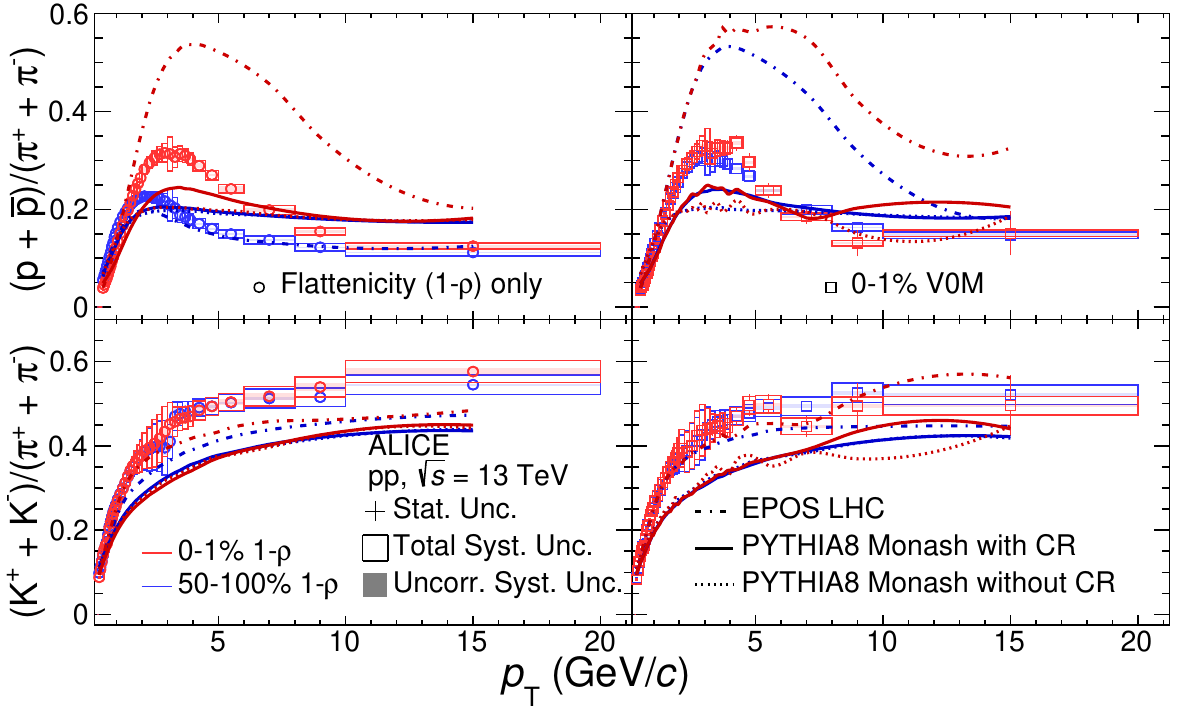}
\caption{$p_{\rm T}$-differential proton to pion (upper) and kaon to pion (lower) ratios for different classes of flattenicity in minimum bias (left) and (0-1)\% V0M class (right) in $pp$ collisions at $\sqrt{s}=13$ TeV with ALICE~\cite{ALICE:2024vaf}.}
\label{fig:flatratiodata}
\end{center}
\end{figure*}
Figure~\ref{fig:flatratiodata} shows the $p_{\rm T}$ dependence of proton to pion ($p/\pi$) (upper) and kaon to pion ($K/\pi$) (lower) ratios for different classes of flattenicity in minimum bias (left) and (0-1)\% V0M class (right) in $pp$ collisions at $\sqrt{s}=13$ TeV with ALICE~\cite{ALICE:2024vaf}. In the lower panel, $K/\pi$ increases with $p_{\rm T}$, and event selection based on charged-particle flattenicity does not seem to play any role for both the minimum bias and (0-1)\% V0M classes. However, for $p/\pi$, a clear distinction for the lowest and highest percentiles of flattenicity is observed for the minimum bias case, which is absent when the flattenicity-based event selection is performed for the (0-1)\% V0M class. PYTHIA 8 with Monash is found to give a good qualitative agreement with the experimental data. \\

\begin{itshape}
In summary,
\begin{itemize}
\item Baryon-to-meson ratios show a bump structure in the intermediate $p_{\rm T}$ region, hinting at radial flow-like features in small collision systems.  In this review, the $p_{\rm T}$ dependence of various measurements and simulation results of baryon to meson ratios is discussed as a function of different event shape classifiers in $pp$ collisions.

\item Event shape dependence of baryon to meson ratios is visible towards high $p_{\rm T}$, while such dependence is diluted towards low $p_{\rm T}$ for all considered classifiers. A bump structure in the intermediate $p_{\rm T}$ is observed for all event-shape classifiers, whose peak position and extent of enhancement depend on the particle mass and event classifier classes.
\end{itemize}

\end{itshape}

\subsection{Kinetic freeze-out properties}

The Tsallis Blast-Wave (TBW) analysis of transverse momentum spectra is another way to quantify the radial flow velocity along with the kinetic freeze-out temperature of the system formed in ultra-relativistic collisions. In the TBW model, the invariant differential yield of a hadron with mass $m$ is expressed as~\cite{Tang:2008ud}:
\begin{align}
    \frac{1}{2\pi m_{\rm T}}\frac{d^{2}N}{dm_{\rm T}dy}\Bigg|_{y=0}=&A\int_{-y_{b}}^{+y_{b}}m_{\rm T}\cosh(y_{s})dy_{s}\int_{-\pi}^{\pi}d\phi \\ \nonumber
    &\times \int_{0}^{R}rdr\Big[1+\frac{q-1}{T}\big\{m_{\rm T}\cosh(y_{s})\cosh(\rho)\\ \nonumber
    &-p_{\rm T}\sinh(\rho)\cos(\phi)\big\}\Big]^{-\frac{1}{(q-1)}}
\end{align}

Here, $T$ is the freeze-out temperature and $R$ is the boundary along the transverse radial direction. $A$ is the normalization constant, $y_{s}$ is the rapidity of the source, $y_{b}$ is the beam rapidity, and $\phi$ is the angle of particle emission relative to the fluid flow velocity. $m_{\rm T}=\sqrt{p_{\rm T}^{2}+m^2}$ is the transverse mass. $\rho=\tanh^{-1}\beta(r)$, where $\beta(r)=\beta_{S}\big(\frac{r}{R} \big)^{n}$ and $n$ is the flow profile index. Average transverse flow velocity, $\langle \beta\rangle =\frac{2}{2+n}\beta_{S}$. Here, $\beta_{S}$ is the surface velocity and $q$ is the non-extensive parameter that characterizes the deviations from the local thermal equilibrium. The TBW model can be extended to with the introduction of two separate parameters, namely, $q_{\rm M}$ for mesons and $q_{\rm B}$ for baryons, known as TBW4 framework, which provides a better fit to the identified hadron $p_{\rm T}$-spectra~\cite{He:2025bnp}.

The systematic study of freeze-out parameters using the above framework with different event shape variables would be helpful to understand their applicability to study the underlying collective dynamics in small collision systems. It is expected that an isotropic system with large $N_{\rm mpi}$ is expected to behave like a system close to thermal equilibrium, thus possessing small values of $q_{\rm M}$ or $q_{\rm B}$. In contrast, the non-equilibrium behavior is enhanced for the jetty events, which would imply large values of the non-extensive parameter.

\begin{figure}
    \centering
    \includegraphics[width=0.95\linewidth]{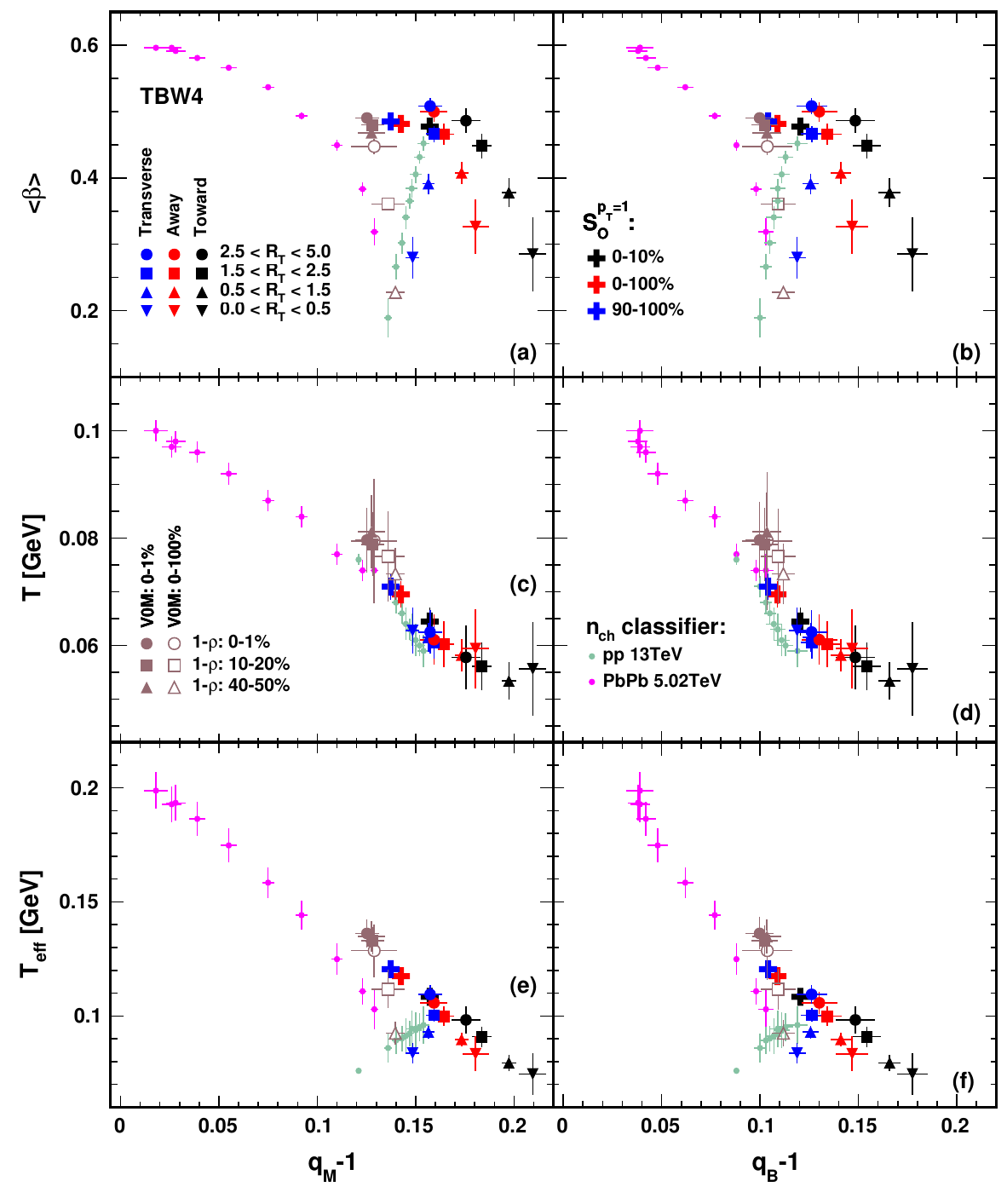}
    \caption{$\langle \beta\rangle$ (upper), $T$ (middle), and $T_{\text{eff}}$ (lower) as a function of $q_{\rm M}-1$ (left) and $q_{\rm B}-1$ (right) for different event classes based on $R_{\rm T}$, $S_{0}^{p_{\rm T}=1}$, $1-\rho$, $n_{\rm ch}$ (or $N_{\rm ch}^{\rm mid}$), and V0M (or $N_{\rm ch}^{\rm fwd}$) in $pp$ collisions~\cite{He:2025bnp}.}
    \label{fig:KFOvsESpp}
\end{figure}

Figure~\ref{fig:KFOvsESpp} shows the correlation of $\langle \beta\rangle$ (upper), $T$ (middle), and $T_{\text{eff}}$ ($T_{\rm eff}=T\sqrt{\frac{1+\langle\beta\rangle}{1-\langle\beta\rangle}}$) (lower) with $q_{\rm M}-1$ (left) and $q_{\rm B}-1$ (right), extracted using TBW4 fit to identified hadron $p_{\rm T}$-spectra in different event classes based on $R_{\rm T}$, $S_{0}^{p_{\rm T}=1}$, $1-\rho$, $n_{\rm ch}$ (or $N_{\rm ch}^{\rm mid}$), and V0M (or $N_{\rm ch}^{\rm fwd}$) in $pp$ collisions~\cite{He:2025bnp}. In the upper panel, where $\langle \beta\rangle$ is shown as a function of $q_{\rm M}-1$ and $q_{\rm B}-1$, the results from different topological regions tend to approach each other towards larger $R_{\rm T}$, which implies the dominance of a similar underlying effects in high-$R_{\rm T}$ events. However, for the transverse region, both $q_{\rm M}-1$ and $q_{\rm B}-1$ become larger with an increase in $R_{\rm T}$, contradictory to the observations in the toward and away regions. This implies that the transverse region is affected by strong fluctuations in the events with high activity. This behaviour is similar for events selected based on $n_{\rm ch}$ in $pp$ collisions, which implies the presence of biases from hard processes in high multiplicity events, which is applicable to events with large $n_{\rm ch}$ and $R_{\rm T}$. In contrast, the anti-correlation between $\langle \beta\rangle$ and $q$ in toward and away regions is understood as the interplay between the underlying event activity and hard processes with a change in $R_{\rm T}$. $S_{0}^{p_{\rm T}=1}$ dependence is shown only for the high-multiplicity events; therefore, $\langle\beta\rangle$ remains unchanged from isotropic to jetty events. In contrast, both $q_{\rm M}$ and $q_{\rm B}$ decreases from (0-10)\% to (90-100)\% $S_{0}^{p_{\rm T}=1}$ events approaching equilibrium. Event selection based on $1-\rho$ in $pp$ collisions for $\langle\beta\rangle$ versus $q$ qualitatively follows the trend of Pb--Pb collisions with a weak variation on $q$.

In the middle panel of Fig.~\ref{fig:KFOvsESpp}, which shows the correlation between $T$ and $q-1$, a universal scaling behavior among different collision systems with event selection based on different event classifiers can be observed. This scaling behaviour indicates the presence of a unified partonic evolution stage, where the partonic interactions in different systems are governed by similar QCD processes. Here, toward and away regions, with a decrease in $R_{\rm T}$, approach the largest $q$ and smallest $T$ values. This indicates large fluctuations in temperature in such events. In contrast, the events selected based on $1-\rho$, among all other event classifiers, show the largest $T$ values in $pp$ collisions with minute variations in $q-1$. In the lower panel of Fig.~\ref{fig:KFOvsESpp} a universal upper limit in $T_{\rm eff}$ versus $q$ as determined by the maximum $\langle \beta\rangle$ with variations in $T$ can be found. Here, central Pb--Pb collisions have the largest $T_{\rm eff}$ and the smallest $q$ while the toward region with the smallest $R_{\rm T}$ possesses the smallest $T_{\rm eff}$ and largest $q$.

The correlation study of different TBW4 parameters with different event shape classifiers shows that the parameter space can be largely expanded with the use of different event classifiers. Different event classifiers can tag different limits of the freeze-out parameters in $pp$ collisions, which can reach that of peripheral Pb--Pb collisions. \\

\begin{itshape}
In summary, 

\begin{itemize}
\item Along with baryon-to-meson ratios that are discussed in the previous section, the radial flow-like effects can be measured from the transverse momentum spectra itself with a statistical distribution such as the Tsallis Blast-Wave distribution.  In this section, the extracted kinetic freeze-out properties such as the kinetic freeze-out temperature, radial flow velocity from transverse momentum spectra are discussed as a function of different event shape variables such as $R_{\rm T}$, $S_{0}^{p_{\rm T}=1}$, $1-\rho$, $N_{\rm ch}^{\rm mid}$, and  $N_{\rm ch}^{\rm fwd}$.

\item It is evident from such a study that different event classifiers can tag different limits of the freeze-out parameters in pp collisions, which can reach that of peripheral Pb–Pb collisions.

\end{itemize}

\end{itshape}

\subsection {Strangeness enhancement}

\begin{figure}[ht!]
\begin{center}
\includegraphics[scale=0.4]{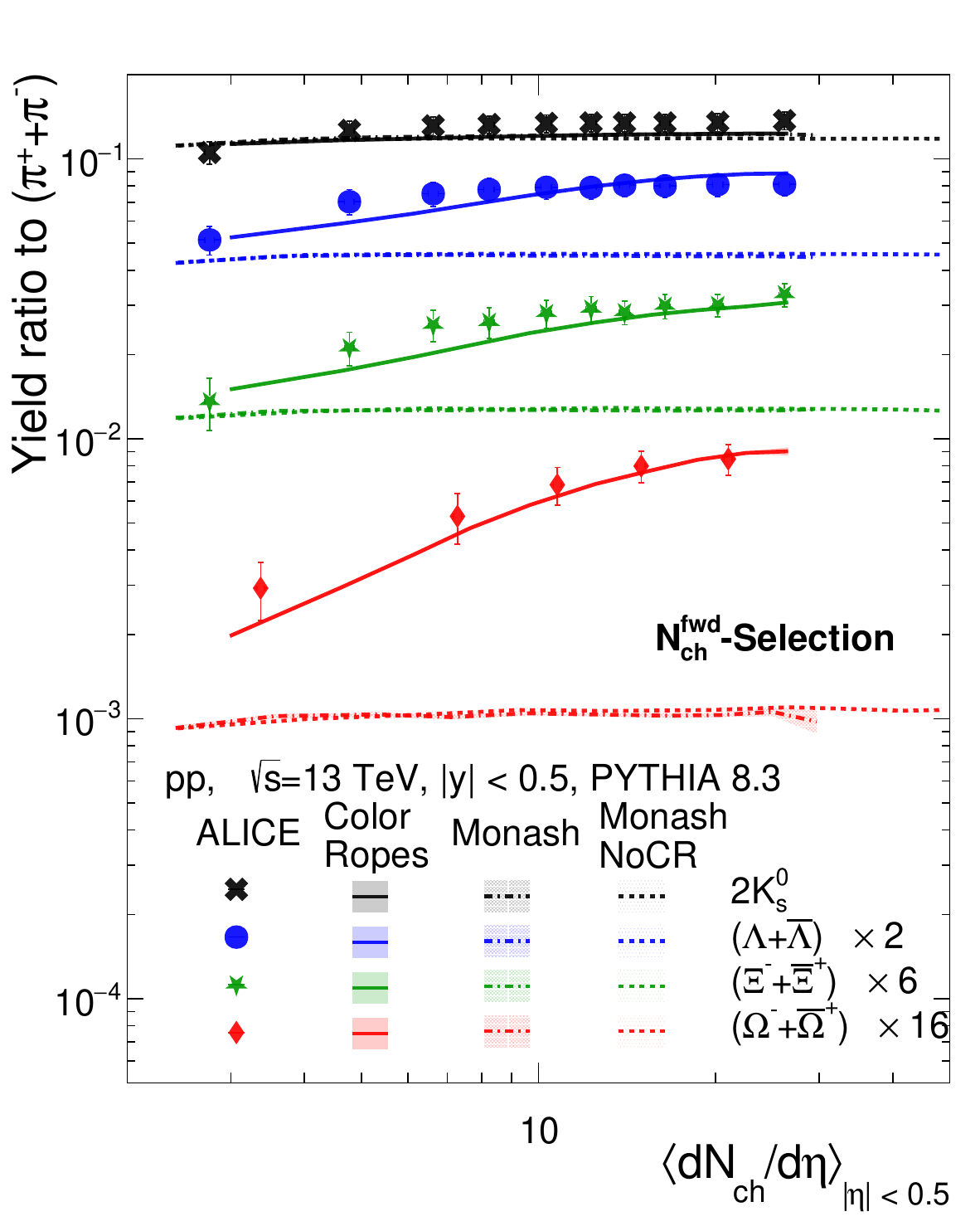}
\end{center}
\caption{$p_{\rm T}$-integrated yield ratios of strange hadrons to pions measured in $|y|<0.5$ versus average charged-particle multiplicity density ($\langle dN_{\rm ch}/d\eta\rangle$) measured at $|\eta|<0.5$ in $pp$ collisions at $\sqrt{s}=13$ TeV using Monash, Monash NoCR and Color Ropes tunes of PYTHIA~8~\cite{Prasad:2024gqq}. The results from PYTHIA~8 are compared with similar measurements at ALICE~\cite{ALICE:2020nkc}.}
\label{fig:strangedatacomp}
\end{figure}

In this section, we explore the event shape observable dependence of particle production of primary strange ($K^{0}_{\rm S}$, $\Lambda$, $\bar{\Lambda}$) and multi-strange ($\Xi^{+}$, $\Xi^{-}$,  $\Omega^{+}$, $\Omega^{-}$) hadrons in $pp$ collisions at $\sqrt{s}$ = 13 TeV using PYTHIA8. Recent experimental observation of $pp$ collisions indicates a significant enhancement in the ratio of strange to non-strange particles as a function of charged-particle multiplicity, similar to those observed in heavy-ion collisions~\cite{ALICE:2017jyt, ALICE:2020nkc}. This enhancement in $pp$ collisions creates a spark in the heavy-ion community and puts a question mark on the assumption of considering $pp$ as the reference system.\\

As discussed above, the charged-particle multiplicity serves as a probe to disentangle the hard and soft events. This motivates the study of the yield ratios of strange to non-strange hadrons as a function of other event shape classifiers such as the number of multi-partonic interactions, transverse sphericity, transverse spherocity, relative transverse activity classifier, charged-particle flattenicity, etc. Here, we mainly investigate $2K^{0}_{\rm S}/(\pi^{+}+\pi^{-})$, $(\Lambda+\bar{\Lambda})/ (\pi^{+}+\pi^{-})$, ($\Xi^{+} + \Xi^{-})/(\pi^{+}+\pi^{-})$, $2\phi/(\pi^{+}+\pi^{-})$, and $(\Omega^{+}, \Omega^{-})/(\pi^{+}+\pi^{-}))$ ratios as a function of different event shape observables. For simplicity now onwards we refer these ratios as $K^{0}_{\rm S}/\pi$, $\Lambda/\pi$, $ \Xi/\pi $, $\phi/\pi$, and $\Omega/\pi$ respectively. \\

Figure~\ref{fig:strangedatacomp} shows a comparison of results from Monash, Monash NoCR, and Color Ropes tunes of PYTHIA~8 with ALICE measurements of $p_{\rm T}$-integrated strange hadrons to pion ratios measured in $|y|<0.5$ as a function of charged-particle multiplicity density ($\langle dN_{\rm ch}/d\eta\rangle$) obtained in $|\eta|<0.5$ in $pp$ collisions at $\sqrt{s}=13~{\rm TeV}$. The ALICE results show a rise in the ratios of strange hadrons to pions with the increase in $\langle dN_{\rm ch}/d\eta\rangle$, which becomes significantly prominent for the hadrons with the higher number of valence strange quarks. Here, $K^{0}_{\rm S}/\pi$ does not show any rise with increase in $\langle dN_{\rm ch}/d\eta\rangle$. In contrast, $\Omega/\pi$, which has three valence strange quarks, shows the strongest rise with the increase in $\langle dN_{\rm ch}/d\eta\rangle$. PYTHIA~8   with Color Ropes is able to reproduce the ALICE measurements of strange hadrons to pion ratios quantitatively. On the contrary, PYTHIA~8 with Monash and Monash NoCR fail to explain the experimental measurements both qualitatively and quantitatively, except for $K^{0}_{\rm S}$. Henceforth, we shall limit our results for strange hadron to pion ratios only to PYTHIA~8 Color Ropes.

\begin{figure*}[ht!]
\begin{center}
\includegraphics[width = 0.49\linewidth]{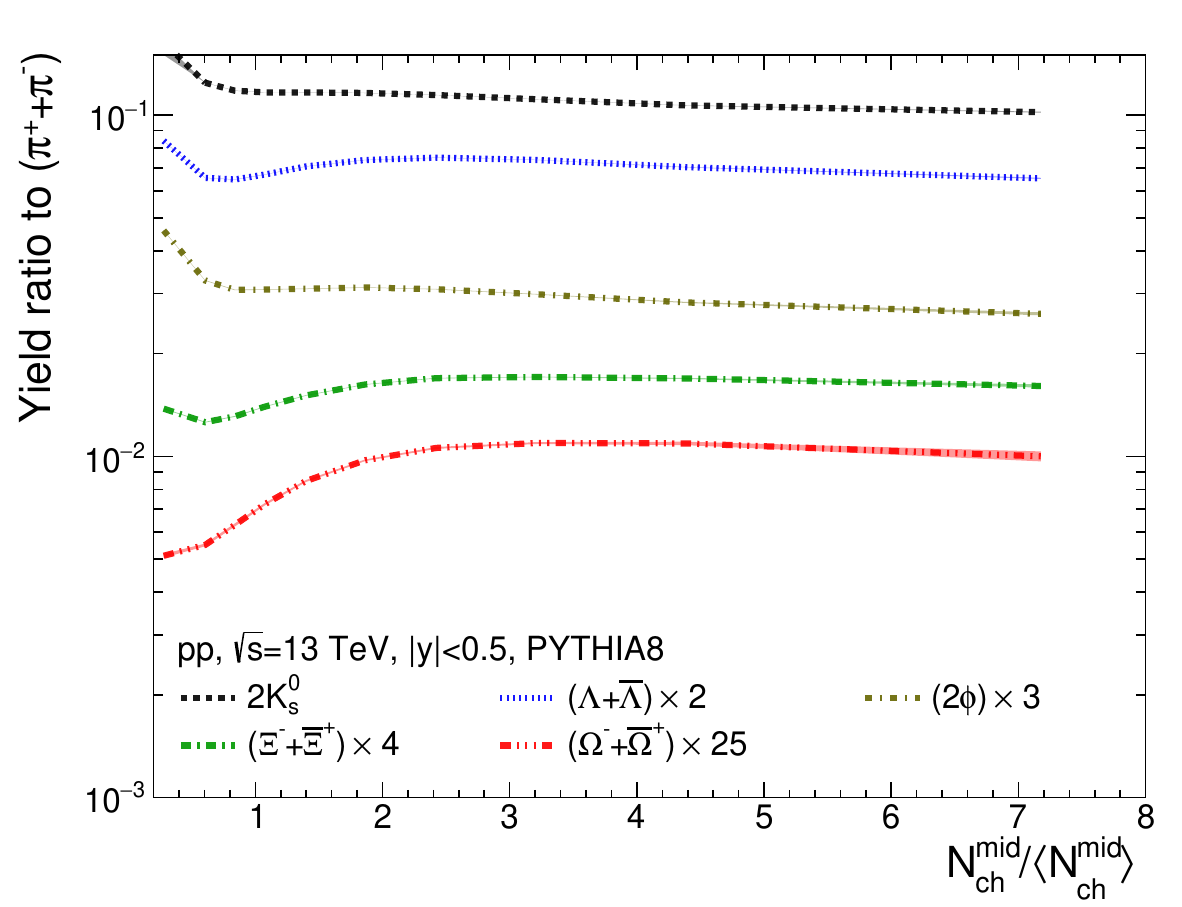}
\includegraphics[width = 0.49\linewidth]{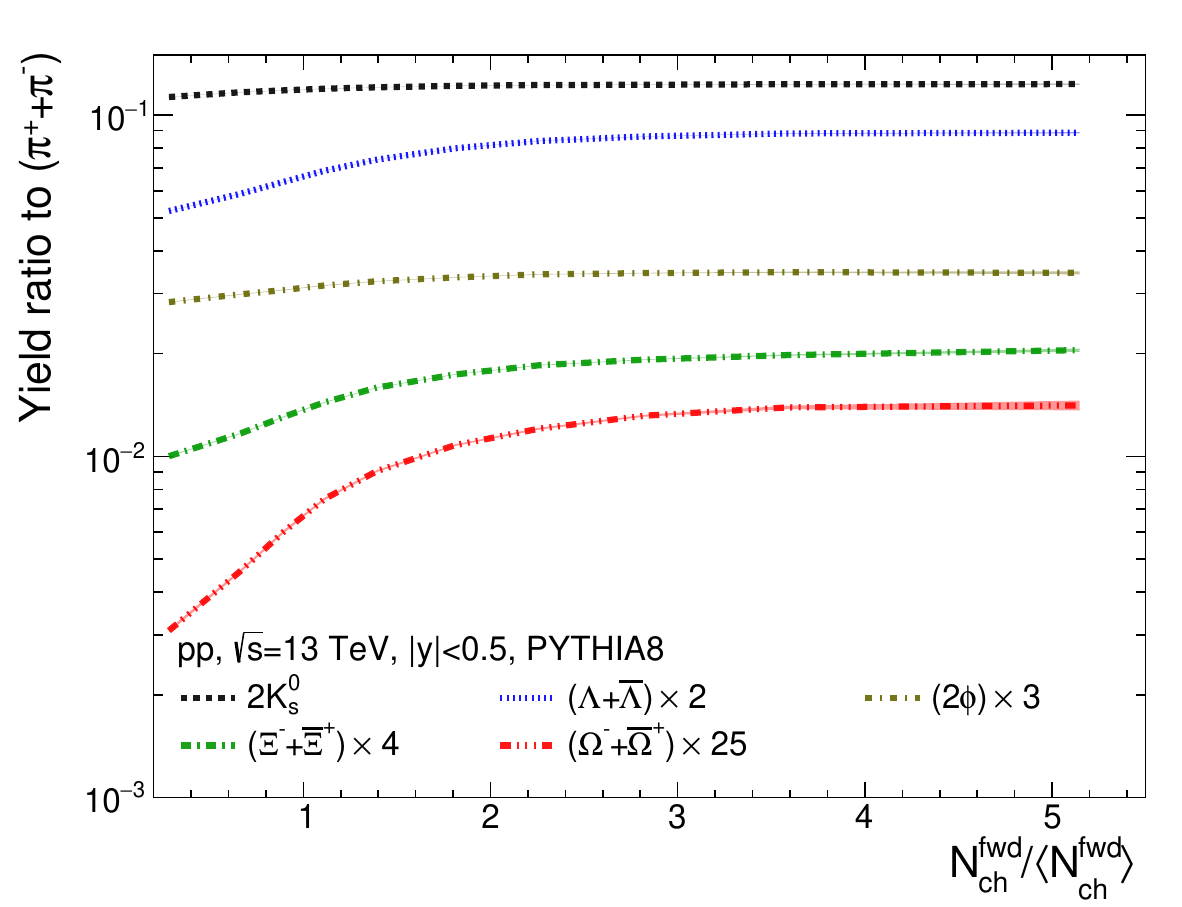}
\caption{$p_{\rm T}$-integrated yield ratios to pions ($\pi^{+}+\pi^{-}$) measured in $|y|<0.5$ as a function of charged-particle multiplicity measured in the mid-pseudorapidity ($N_{\rm ch}^{\rm mid}$) (left panel) and forward pseudorapidity ($N_{\rm ch}^{\rm fwd}$) (right panel) in $pp$ collisions at $\sqrt{s}=13$ TeV using PYTHIA8~\cite{Prasad:2024gqq}.}
\label{fig:strangeNch}
\end{center}
\end{figure*}

Figure~\ref{fig:strangeNch} shows the relative production rate of particle containing strange quarks measured using the $\Lambda/\pi$, $\Xi/\pi$, $\Omega/\pi$ ratios as a function of charged-particle multiplicity measured in the mid-pseudorapidity ($N_{\rm ch}^{\rm mid}$) (left panel) and forward pseudorapidity ($N_{\rm ch}^{\rm fwd}$) (right panel) region in $pp$ collisions at $\sqrt{s}=13$ TeV using PYTHIA8. From Fig.~\ref{fig:strangeNch}, it is observed that these integrated yield ratios increase with charged-particle multiplicity. This effect is commonly called \enquote{strangeness enhancement} in the heavy-ion community.

\begin{figure}[ht!]
\begin{center}
\includegraphics[scale=0.4]{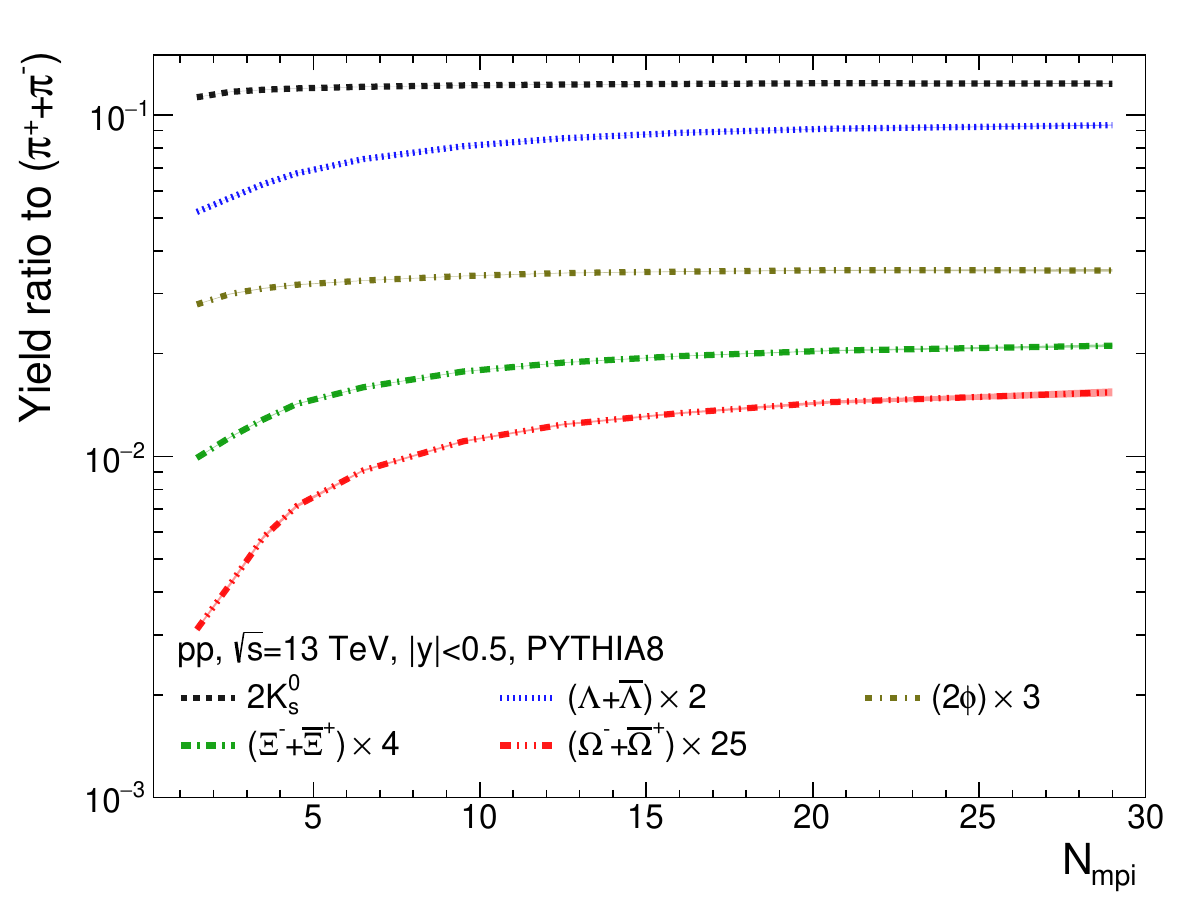}
\caption{$p_{\rm T}$-integrated yield ratios to pions ($\pi^{+}+\pi^{-}$) measured in $|y|<0.5$ as a function of  $N_{\rm mpi}$ in $pp$ collisions at $\sqrt{s}=13$ TeV using PYTHIA8~\cite{Prasad:2024gqq}.}
\label{fig:strangempi}
\end{center}
\end{figure}

 One more intriguing observation is that the strangeness enhancement effect is proportional to the strangeness content of the hadrons. From Fig.~\ref{fig:strangeNch} it is observed that the rate of enhancement of these ratios with multiplicity is higher for triply-strange $\Omega$ hyperons, and it gradually decreases for hadrons containing two and one strange quark quantum number, such as $\Xi$ and $\Lambda$, etc. Although $\phi$ has no net strangeness quantum number, a slight increasing behavior of $\phi$ to pion ratios is observed at lower forward multiplicity classes, $N_{\rm ch}^{\rm fwd}$. This implies that the $\phi$ particle production can not be described solely by considering net strangeness or the number of strange quark constituents. In addition, the left panel of Fig.~\ref{fig:strangeNch} indicates a decrease in the strange to non-strange ratios in the first charged-particle multiplicity bins. This could be attributed to the event selection bias developed while selecting charged-particle multiplicity in the mid-pseudorapidity region. Figure~\ref{fig:strangeNch} depicts the enhancement effect reaches a saturation behavior towards $N_{\rm ch}^{\rm mid}/\langle N_{\rm ch}^{\rm mid}\rangle \; \geq 2$ for all considered hadrons ratios, while it is found to be particle species dependent for $N_{\rm ch}^{\rm fwd}$. Therefore, it can be inferred that the event selection with charged-particle multiplicity measured in different windows of pseudorapidity regions plays an important role in describing the dynamics of the strange hadron production. Furthermore, Fig.~\ref{fig:strangempi} shows the  $K^{0}_{\rm S}/\pi$, $\Lambda/\pi$, $ \Xi/\pi $, $\phi/\pi$, and $\Omega/\pi$ ratios as a function of number of multi-partonic interactions in $pp$ collisions at $\sqrt{s}=13$ TeV using PYTHIA8. From Fig.~\ref{fig:strangempi}, it is observed that the strange production in PYTHIA8 using Color Ropes and CR is sensitive to MPI in $pp$ collisions. Higher MPI refers to those events in $pp$ collisions that have a larger number of final-state charged-particle multiplicities. Hence, the strange to non-strange ratios as a function of $N_{\rm mpi}$ show a similar trend as $N_{\rm ch}^{\rm fwd}$.   \\
 
\begin{figure*}[ht!]
\begin{center}
\includegraphics[width = 0.49\linewidth]{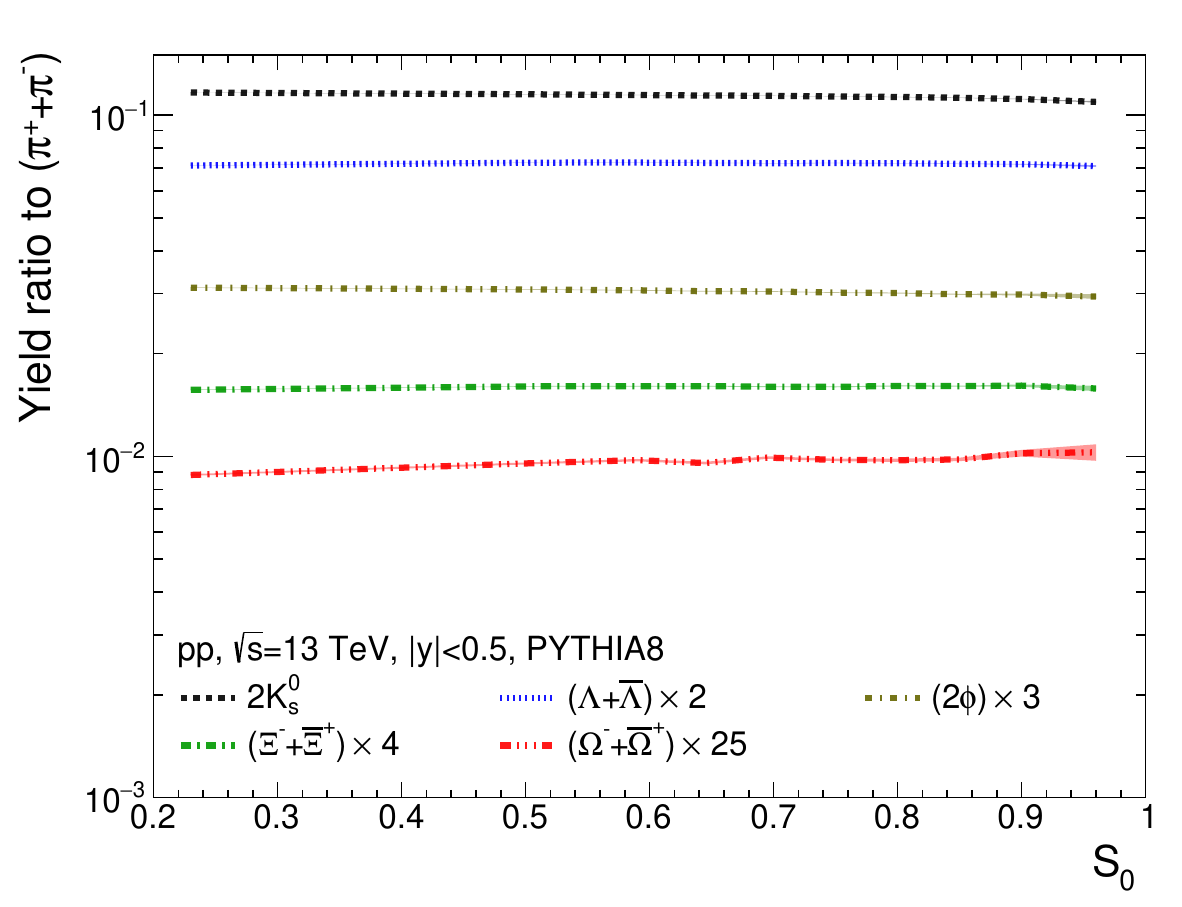}
\includegraphics[width = 0.49\linewidth]{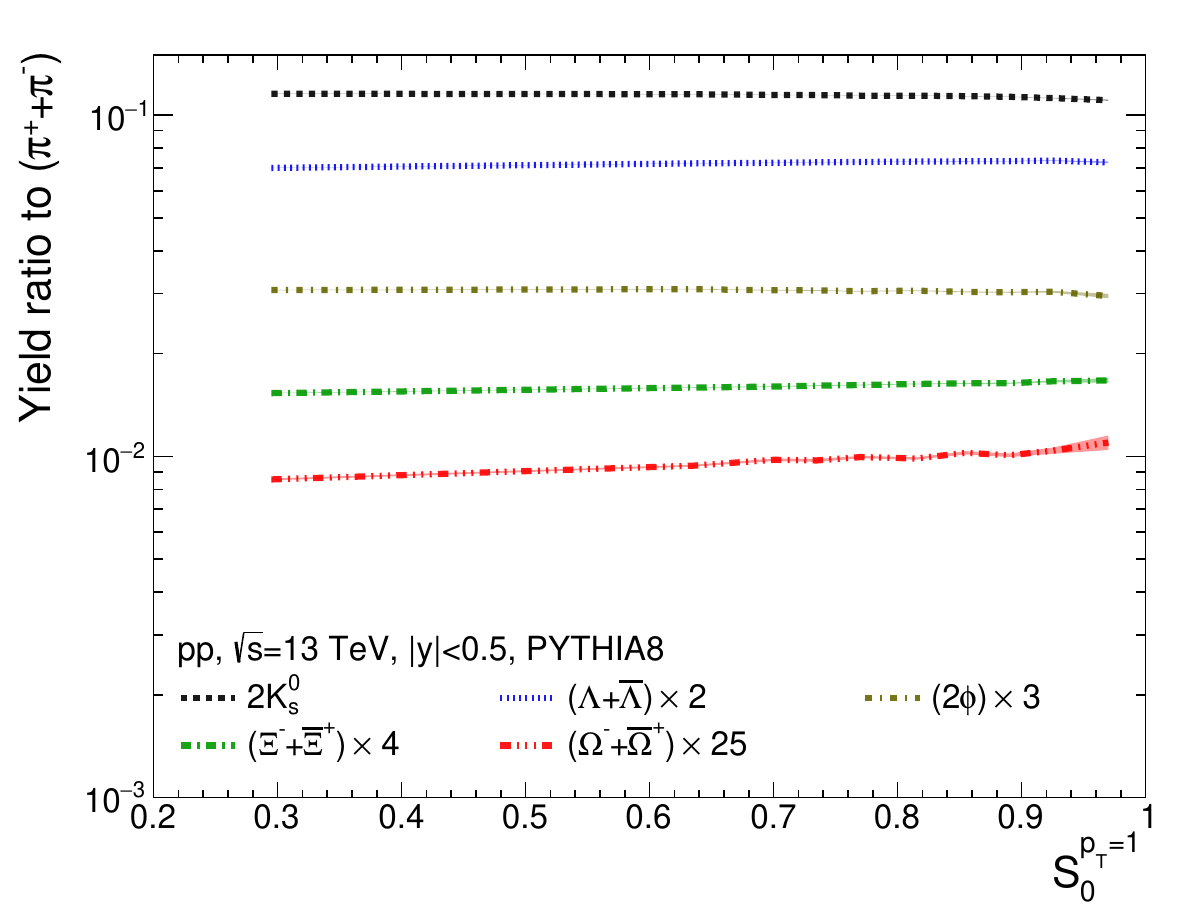}
\caption{$p_{\rm T}$-integrated yield ratios to pions ($\pi^{+}+\pi^{-}$) measured in $|y|<0.5$ as a function of $p_{\rm T}-$weighted transverse spherocity ($S_{0}$) (left panel) and unweighted transverse spherocity ($S_{0}^{p_{\rm T}=1}$) (right panel) in $pp$ collisions at $\sqrt{s}=13$ TeV using PYTHIA8~\cite{Prasad:2024gqq}.}
\label{fig:strangespherocity}
\end{center}
\end{figure*}

\begin{figure}[ht!]
\begin{center}
\includegraphics[scale=0.4]{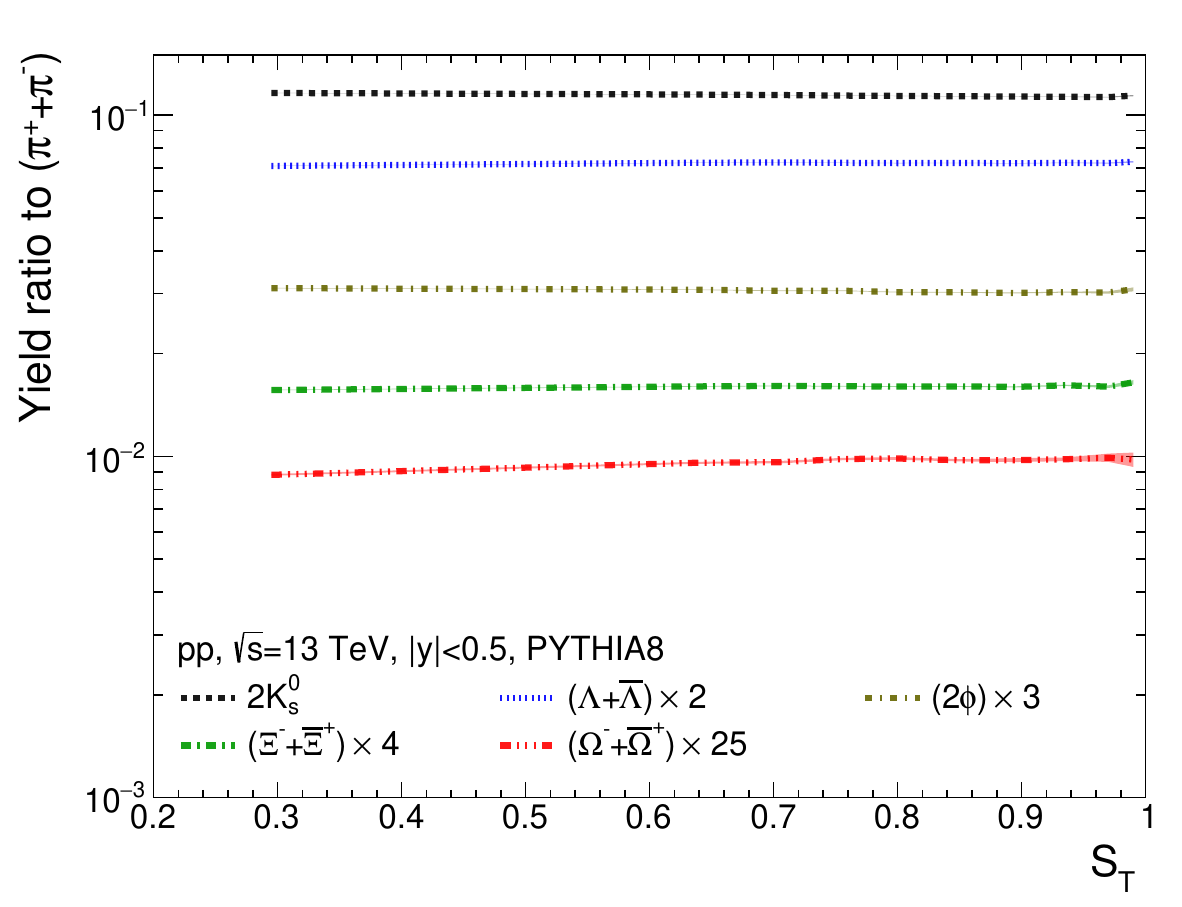}
\caption{$p_{\rm T}$-integrated yield ratio to pions ($\pi^{+}+\pi^{-}$) measured in $|y|<0.5$ as a function of transverse sphericity ($S_{\rm T}$) in $pp$ collisions at $\sqrt{s}=13$ TeV using PYTHIA8~\cite{Prasad:2024gqq}.}
\label{fig:strangesphericity}
\end{center}
\end{figure}

Figure~\ref{fig:strangespherocity} shows the strange to non-strange ratios as a function of $S_{0}$ (left panel) and $S_{0}^{p_{\rm T}=1}$ (right panel) in $pp$ collisions at $\sqrt{s}=13$ TeV using PYTHIA8.  From Fig.~\ref{fig:strangespherocity}, it is observed that the $K^{0}_{\rm S}/\pi$, $\Lambda/\pi$, $\Xi/\pi$, $\phi/\pi$, and $\Omega/\pi$ ratios remains almost constant with $S_{0}$ and $S_{0}^{p_{\rm T}=1}$. Furthermore, similar observation has also been observed for $K^{0}_{\rm S}/\pi$, $\Lambda/\pi$, $\Xi/\pi$, $\phi/\pi$, and $\Omega/\pi$ ratios as a function of $S_{\rm T}$ as shown in Fig.~\ref{fig:strangesphericity}. So, in conclusion, we found that events having isotropic emission of particles have almost similar behavior of strange to non-strange ratios, with the events dominated by jets. The absence of $S_{0}$, $S_{0}^{p_{\rm T}=1}$ and $S_{\rm T}$ dependence on strangeness enhancement can be attributed to the application of $N_{\rm ch}^{\rm mid}>10$ cut on the definitions of $S_{0}$, $S_{0}^{p_{\rm T}=1}$ and $S_{\rm T}$. Therefore, the transverse sphericity, weighted and unweighted transverse spherocity event classifiers are inadequate to study the strangeness enhancement characteristics in $pp$ collisions using the PYTHIA8 model. 

\begin{figure*}[ht!]
\begin{center}
\includegraphics[width = 0.49\linewidth]{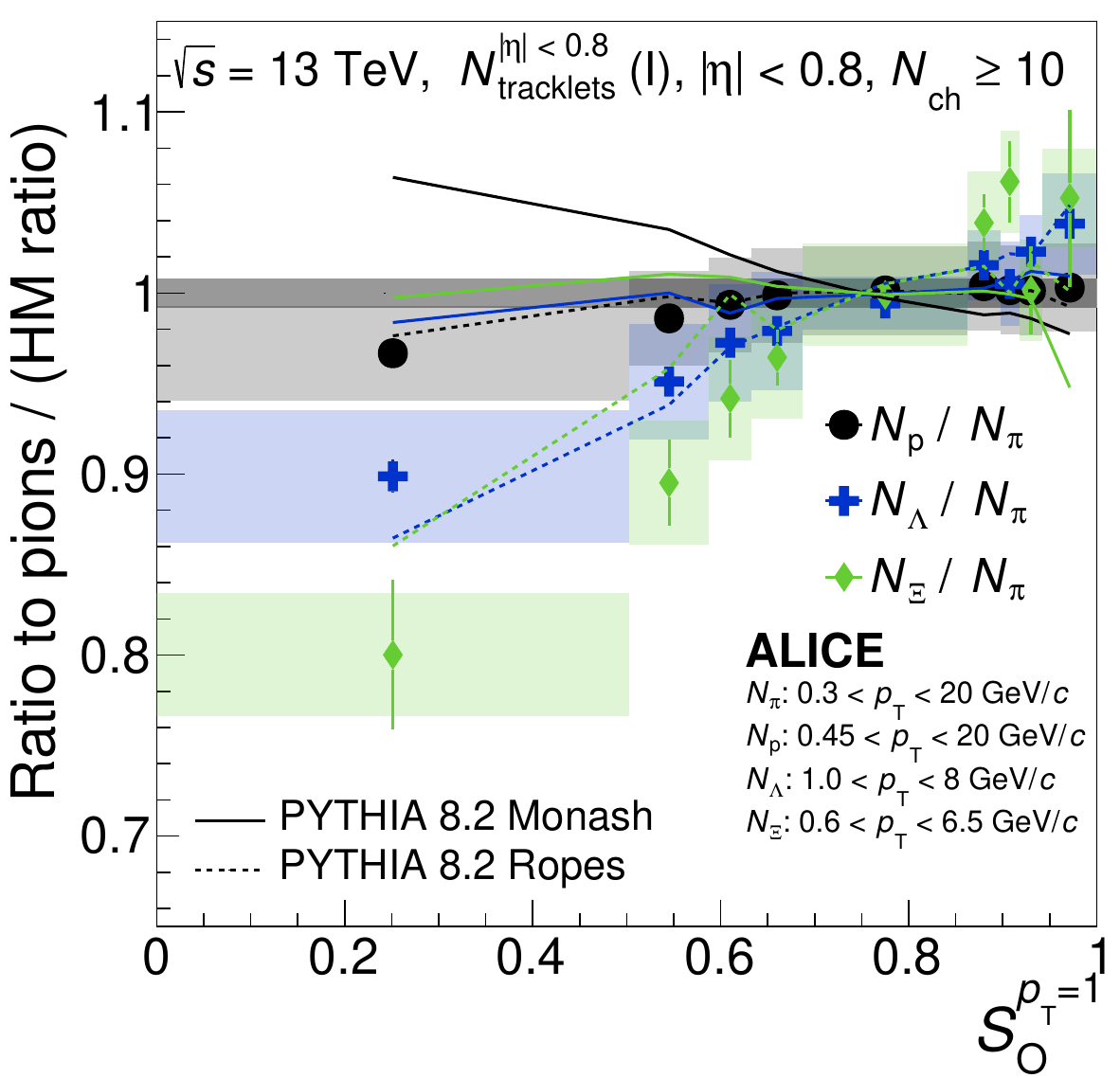}
\includegraphics[width = 0.49\linewidth]{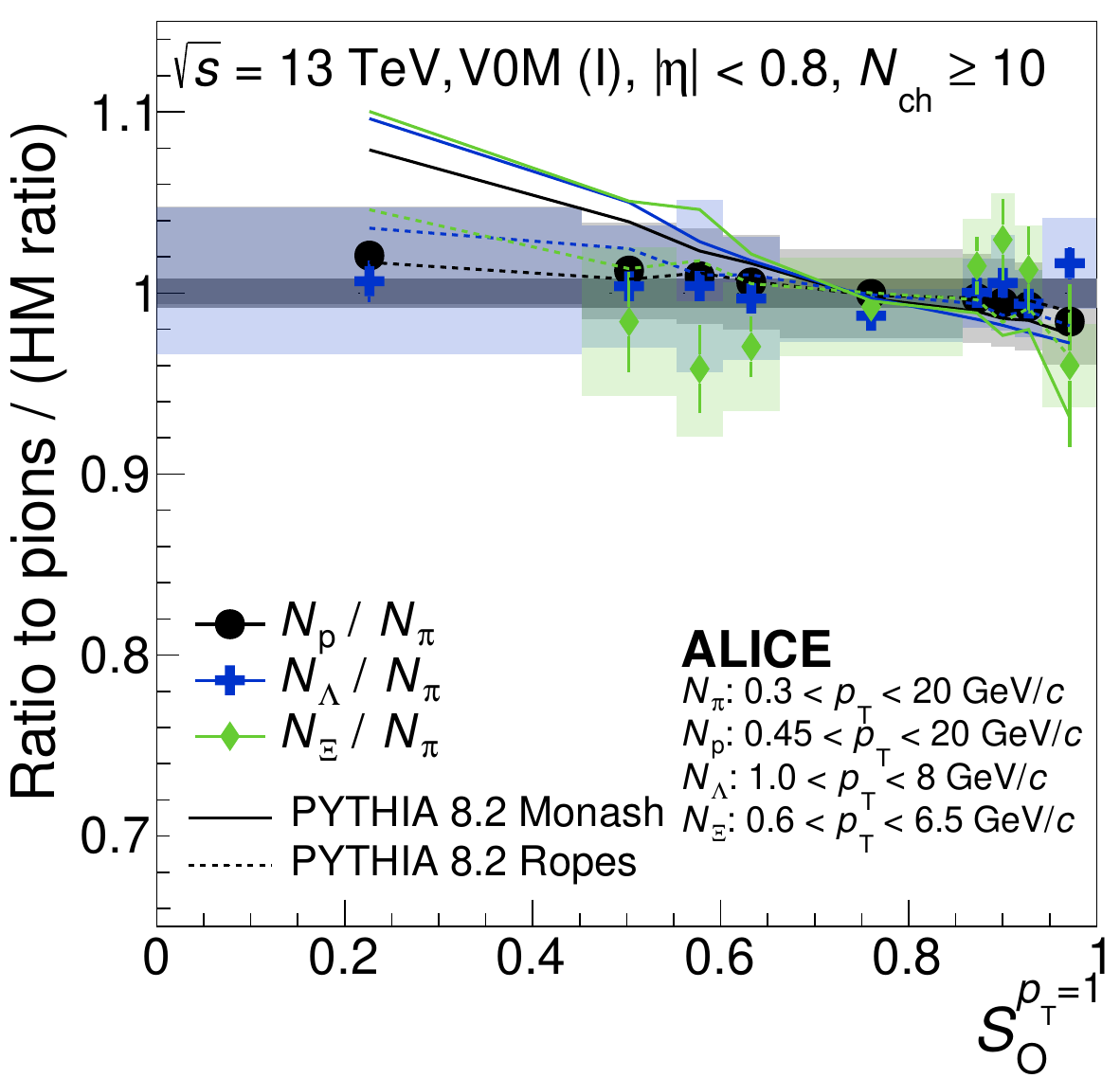}
\caption{Ratio of particle yield to pions as a function of unweighted transverse spherocity in high multiplicity events based $N_{\rm tracklets}^{|\eta|<0.8}$ (left panel) and V0M (right panel) in $pp$ collisions at $\sqrt{s}=13$ TeV with ALICE~\cite{ALICE:2023bga}.}
\label{fig:strangesphericityNch}
\end{center}
\end{figure*}

Figure~\ref{fig:strangesphericityNch} shows the $p_{\rm T}$-integrated particle yield ratios to pions normalized to corresponding value in high multiplicity (HM) events as a function of $S_{0}^{p_{\rm T}=1}$ in $pp$ collisions at $\sqrt{s}=13$ TeV with ALICE. The high multiplicity events are determined based on the number of SPD tracklets ($N_{\rm tracklets}^{|\eta|<0.8}$) at midrapidity (left panel) and multiplicity in the V0 region (V0M) (right panel). In the left panel with the high multiplicity events selected based on $N_{\rm tracklets}^{|\eta|<0.8}$, within the systematic uncertainties, we observe an enhancement of particle yield ratios to pions with the increase in $S_{0}^{p_{\rm T}=1}$ which is absent in the right panel where the high-multiplicity events are selected based on V0M. Further, in the left panel, this enhancement is stronger for $\Xi$ and weakest for protons. This indicates that, although the high multiplicity events have contributions from both isotropic events and multi-jet topologies, with $S_{0}^{p_{\rm T}=1}$, one can choose isotropic events having large strange hadron production. However, this works well with event selection based on charged particles at midrapidity, where $S_{0}^{p_{\rm T}=1}$ is estimated. This is because, $S_{0}^{p_{\rm T}=1}$ selection on top (0-1)\% $N_{\rm ch}^{\rm mid}$ events probes a higher value of $\langle N_{\rm mpi}\rangle$ as compared to similar $S_{0}^{p_{\rm T}=1}$ selection on top (0-1)\% $N_{\rm ch}^{\rm fwd}$ events, as shown in Fig.~\ref{fig:S0vsAvgMPI}.

\begin{figure*}[ht!]
\begin{center}
\includegraphics[width = 0.49\linewidth]{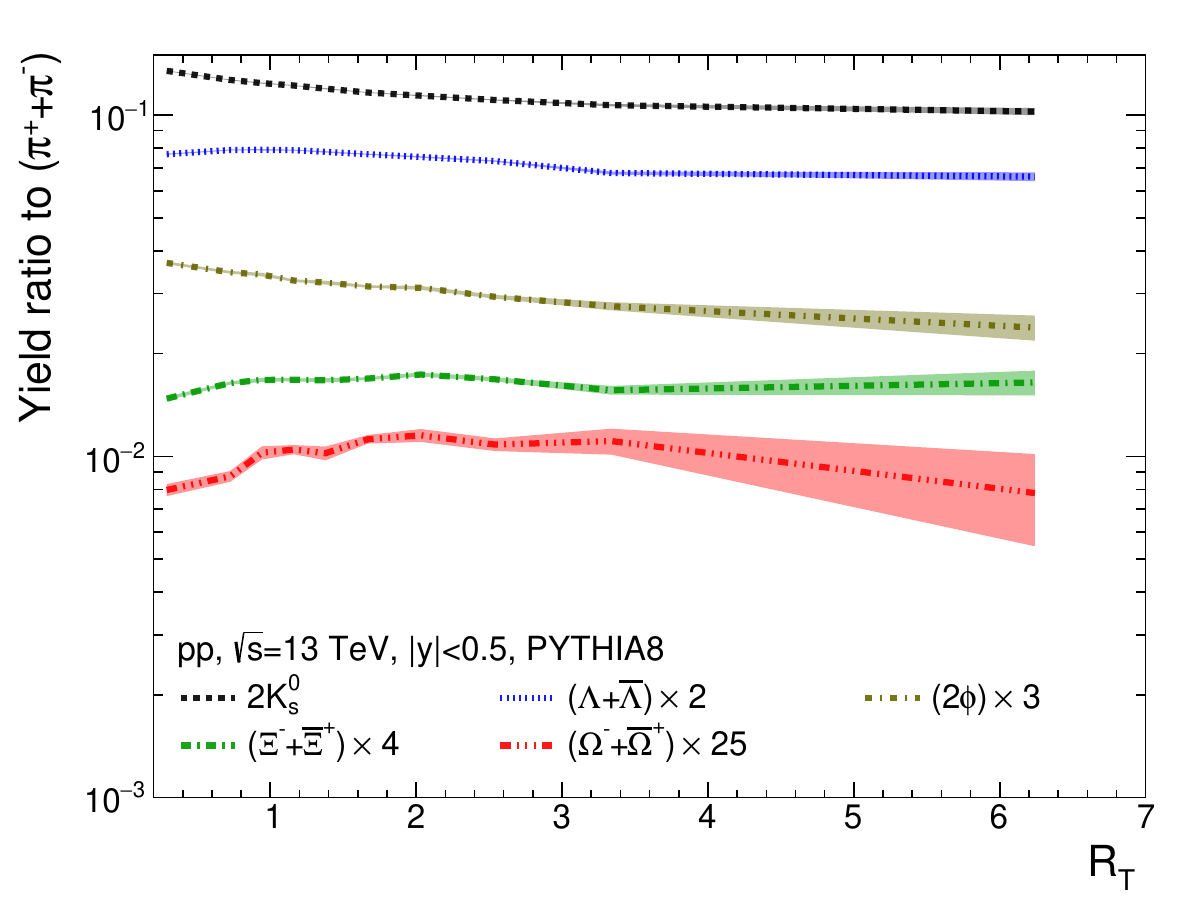}
\includegraphics[width = 0.49\linewidth]{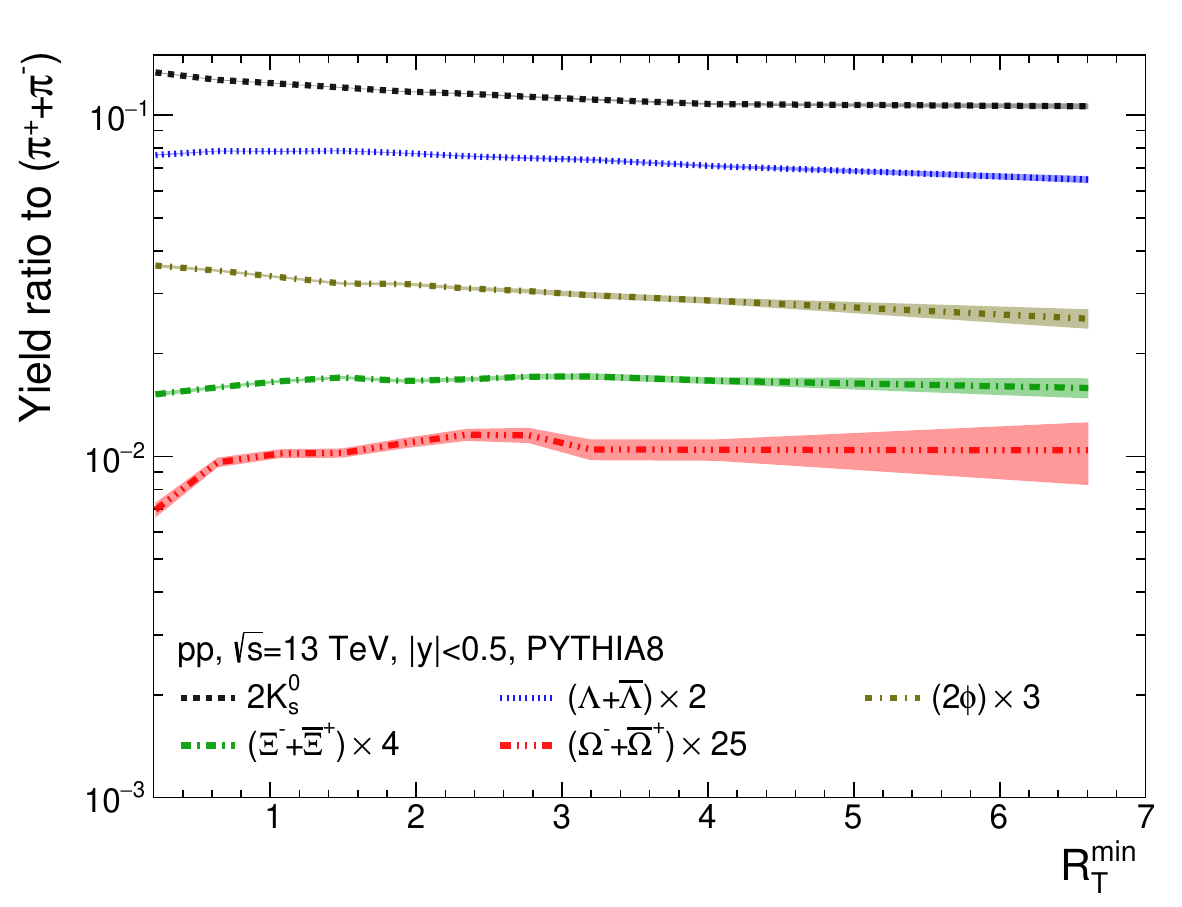}
\includegraphics[width = 0.49\linewidth]{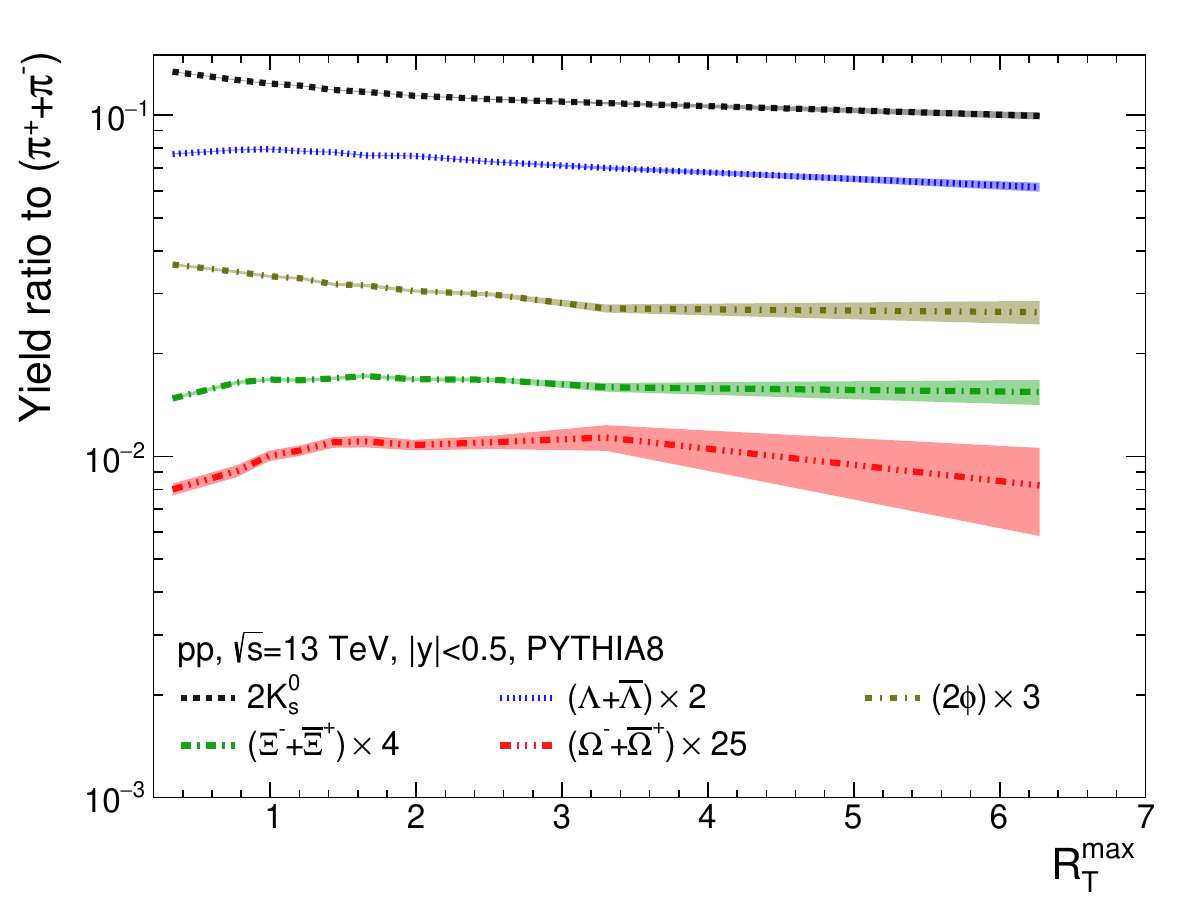}
\caption{$p_{\rm T}$-integrated yield ratio to pions ($\pi^{+}+\pi^{-}$) measured in $|y|<0.5$ as a function of $R_{\rm T}$~(upper left), $R_{\rm T}^{\rm min}$~(upper right), and $R_{\rm T}^{\rm max}$~(lower middle) in $pp$ collisions at $\sqrt{s}=13$ TeV using PYTHIA8~\cite{Prasad:2024gqq}.}
\label{fig:strangeRT}
\end{center}
\end{figure*}

Figure~\ref{fig:strangeRT} shows the $K^{0}_{\rm S}/\pi$, $\Lambda/\pi$, $\Xi/\pi$, $\phi/\pi$, and $\Omega/\pi$ ratios as a function of $R_{\rm T}$~(upper left), $R_{\rm T}^{\rm min}$~(upper right), and $R_{\rm T}^{\rm max}$~(lower middle) in $pp$ collisions at $\sqrt{s}=13$ TeV using PYTHIA8.  It is observed that these ratios remain almost constant with $R_{\rm T}$, $R_{\rm T}^{\rm min}$, $R_{\rm T}^{\rm max}$. However, a slightly increasing trend is observed in the $\Omega/\pi$ ratios for  $R_{\rm T}$ and $R_{\rm T}^{\rm max}$ $\gtrsim$ 3.2 within uncertainty. So, similar to sphericity and spherocity, the strangeness enhancement feature is also found to be marginal with the relative transverse activity classifiers.

\begin{figure}[ht!]
\begin{center}
\includegraphics[scale=0.4]{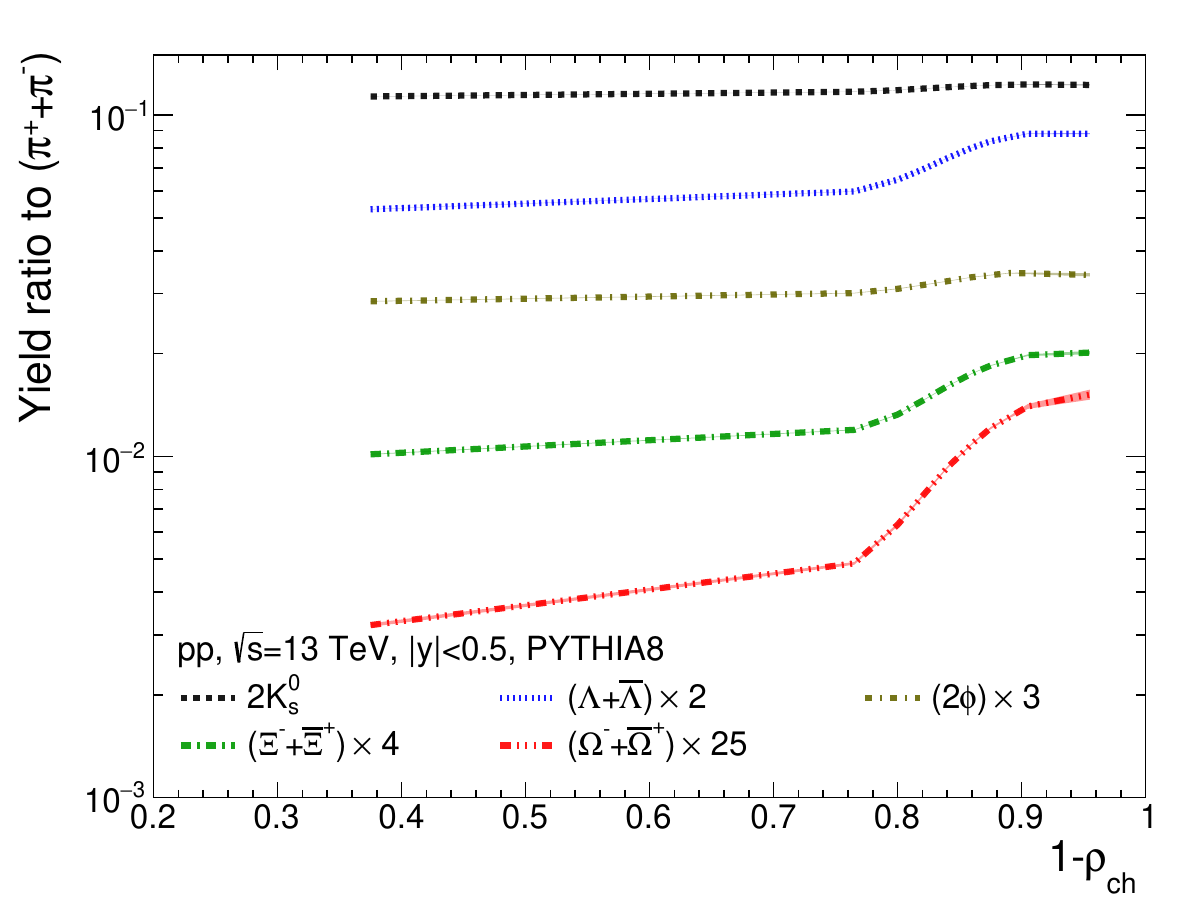}
\caption{$p_{\rm T}$-integrated yield ratio to pions ($\pi^{+}+\pi^{-}$) measured in $|y|<0.5$ as a function of charged-particle flattenicity ($\rho_{\rm ch}$) in $pp$ collisions at $\sqrt{s}=13$ TeV using PYTHIA8~\cite{Prasad:2024gqq}.}
\label{fig:strangeflat}
\end{center}
\end{figure}

The charged-particle flattenicity dependence of $K^{0}_{\rm S}/\pi$, $\Lambda/\pi$, $\Xi/\pi$, $\phi/\pi$, and $\Omega/\pi$ ratios is shown in Fig.~\ref{fig:strangeflat} for $pp$ collisions at $\sqrt{s}=13$ TeV using PYTHIA8. It is interesting to note that these ratios tend to increase as a function of (1-$\rho_{\rm ch}$) for all considered strange particles, except for $K_S^{0}$. A rapid increase in the slope of the ratios is observed in Fig.~\ref{fig:strangeflat} for the lower values of $\rho_{\rm ch}$ ($0.18 \lesssim \rho_{\rm ch} \lesssim 0.36 $) compared to the higher values of $\rho_{\rm ch}$ ($0.36 \lesssim \rho_{\rm ch} \lesssim 0.7 $). It is found that the rate of increase of the strange to non-strange particle ratios is strange quantum number dependent; the triple-strange baryons, such as $\Omega$, have more slope compared to the single and double strange baryons, such as  $\Xi$, and $\Lambda$. 

The enhanced production of strangeness with different event classifiers can be further probed with a comparison of the double ratio of strange-to-non-strange hadrons, where the ratio is scaled with the same observable in minimum bias collisions. This study is explicitly performed in Ref.~\cite{Prasad:2024gqq}.

\begin{itshape}
In summary,

\begin{itemize}
\item In $pp$ collisions, strangeness enhancement can be probed by measuring strange to non-strange particle yield ratios as a function of event classifiers. In this review, various measurements and simulation results of strange to non-strange particle ratios are discussed as a function of different event shape classifiers.

\item It is observed that strange to non-strange particle ratios show a flat trend as a function of sphericity, spherocity, relative transverse classifiers, while an enhancement feature is observed as a function of charged particle multiplicity measured at both mid and forward rapidity, MPI, and charged-particle flattenicity classes. This enhancement depends on the strangeness content of the hadrons.
\end{itemize}

\end{itshape}

\subsection{Particle correlations and anisotropic flow}
\label{sec:anisotropicflow}
So far, we have discussed the sources of collective radial expansion, also known as the radial flow, of the system formed in collider experiments. In this section, we discuss anisotropic flow, which is one of the major contributors to the collectivity of the system and is mostly driven by the geometry and fluctuations of the collision overlap region. The anisotropic flow of the system can be quantified by the coefficients of the Fourier expansion of the azimuthal distribution of particles in the final state, given as follows~\cite{Ollitrault:1992bk}.
\begin{equation}
    \frac{dN}{d\phi}\propto 1 + 2\sum_{i=1}^{\infty}v_{n}\cos[n(\phi-\psi_{n})]
    \label{eq:flowfourier}
\end{equation}
Here, $v_{n}$ and $\psi_{n}$ are the $n$th order anisotropic flow coefficients and symmetry plane angles, respectively. $v_{1}$, $v_{2}$, and $v_{3}$, etc., are called directed, elliptic, and triangular flow, respectively. As the name suggests, the directed flow ($v_{1}$) quantifies the collective sideward deflection of particles with respect to the reaction plane~\cite{Jiang:2021foj, Ollitrault:1992bk}. The contribution of $v_{1}$ in symmetric systems at the LHC energies is small~\cite{Ollitrault:1992bk}. In contrast, in non-central collisions, the collision overlap geometry is elliptic, leading to large values of $v_{2}$ in heavy-ion collisions. However, in small systems such as $pp$ and p-Pb collisions, the density fluctuations dominate over the contribution of geometry in the final state flow. Further, the collectivity itself in small systems bears a big question mark, and thus, the study of anisotropic flow in $pp$ and p-Pb collision systems is interesting. 
\begin{figure*}[ht!]
\begin{center}
\includegraphics[scale=0.7]{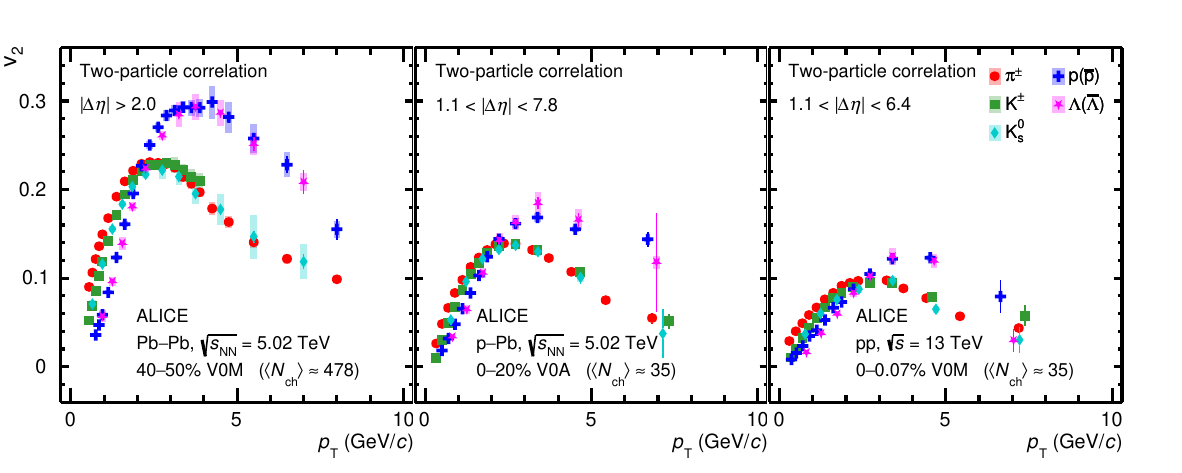}
\caption{Elliptic flow ($v_2$) of mesons ($\pi^{\pm}$, $K^{\pm}$, $K^{0}_{\rm S}$) and baryons ($p+\bar{p}$, $\Lambda+\bar{\Lambda}$) with $|\eta|<0.8$ in Pb--Pb collisions at $\sqrt{s_{\rm NN}}=5.02$ TeV (left), p-Pb collisions at $\sqrt{s_{\rm NN}}=5.02$ TeV (upper right) and $pp$ collisions at $\sqrt{s}=13$ TeV (right) using two-particle correlation method. Centrality and multiplicity classes are chosen based on the multiplicity deposited in the V0 detector of ALICE~\cite{ALICE:2024vzv}.}
\label{fig:vnsystems}
\end{center}
\end{figure*}
Figure~\ref{fig:vnsystems} shows elliptic flow of mesons ($\pi^{\pm}$, $K^{\pm}$, $K^{0}_{\rm S}$) and baryons ($p+\bar{p}$, $\Lambda+\bar{\Lambda}$) with $|\eta|<0.8$ in Pb--Pb, p--Pb and $pp$ collisions~\cite{ALICE:2024vzv}. The measurement is performed using a two-particle correlation method in different collision energies and multiplicity regions for the collision systems.
%, discussed in Sec.~\ref{sec:flowandtwoparticlecorrelation} in different collision energies and multiplicity regions for the collision systems.
The two-particle correlation method requires the estimation of the two-particle correlation function, which provides an estimation of the probability of finding a pair of particles separated by a given pseudorapidity difference ($\Delta\eta$) and azimuthal angle difference ($\Delta\phi$).
The one-dimensional two-particle correlation function can be defined as:

\begin{equation}
     C(\Delta\phi)=\frac{dN_{\rm pairs}}{d\Delta\phi} \propto \Bigg[1+2\sum_{n=1}^{\infty}v_{n,n}(p_{\rm T}^{a},p_{\rm T}^{b})\cos{n\Delta\phi}\Bigg].
     \label{eqn-cdelfourier}
\end{equation}

Here, $v_{n,n}$ is the two-particle flow coefficient. $v_{n,n}$ are symmetric with respect to $p_{\rm T}^{a}$ (transverse momentum of trigger particle) and $p_{\rm T}^{b}$ (transverse momentum of associated particle). $v_{n,n}$ can be calculated by the discrete Fourier transformation using the following expression.
\begin{equation}
   v_{n,n}(p_{\rm T}^{a},p_{\rm T}^{b})=\langle\cos{n\Delta\phi}\rangle= \frac{\sum_{m=1}^{N}\cos{(n\Delta\phi_{m})}C(\Delta\phi_{m})}{\sum_{m=1}^{N}C(\Delta\phi_{m})}.
\end{equation}
Here, $N$ is the number of bins in the $\Delta\phi$ distribution. Consequently, one can determine $v_{n}$ from $v_{n,n}$ from the following expression
\begin{equation}
    v_{n}(p_{\rm T}^{a})=\frac{v_{n,n}(p_{\rm T}^{a},p_{\rm T}^{b})}{\sqrt{v_{n,n}(p_{\rm T}^{b},p_{\rm T}^{b})}}
    \label{eqn-vnntovn}
\end{equation}

Due to the collective behavior of the system formed in heavy-ion collisions, in the left panel for Pb--Pb, we observe a clear mass ordering of particles below $p_{\rm T}<2$~GeV/c. This provides significant evidence of hydrodynamic flow in heavy-ion collisions. Interestingly, in small systems such as $pp$ and p-Pb, for high-multiplicity events, similar observations are made, which signifies the presence of radial flow in small systems. Further, for $p_{\rm T}>2$ GeV/c, the $v_{2}$ of different particles cross each other, and for $p_{\rm T}>2.5$ GeV/c, a clear baryon-meson grouping is observed in all the collision systems shown in Fig.~\ref{fig:vnsystems}. This baryon-meson grouping in heavy-ion collisions provides evidence of particle production through the quark-coalescence mechanism~\cite{STAR:2003wqp}. However, elliptic flow measurements in small systems at other multiplicity classes do not exhibit similar behavior (mass-ordering at low-$p_{\rm T}$ and baryon-meson separation at intermediate $p_{\rm T}$) as observed in heavy-ion collisions~\cite{CMS:2018loe}. The absence of such behaviour in elliptic flow measurements in these collisions at lower multiplicity regions could be influenced by the domination of jetty events in the measurements. Thus, an event shape-based study of anisotropic flow coefficients in small systems is timely to properly understand the microscopic origin of flow signatures in small collision systems.

In a master's thesis at Lund University~\cite{9160009}, a detailed investigation of two-particle correlations using event-shape classifiers for proton-proton collisions was performed. In these preliminary studies, while comparing spherocity, $R_{\rm T}$, and flattenicity, it is concluded that flattenicity selects flatter events than spherocity, both when looking at the low and high flattenicity bins. Also, it was found that flattenicity appears to be able to select the most isotropic events out of the three investigated classifiers. 

\begin{figure}
    \centering
\includegraphics[width=0.95\linewidth]{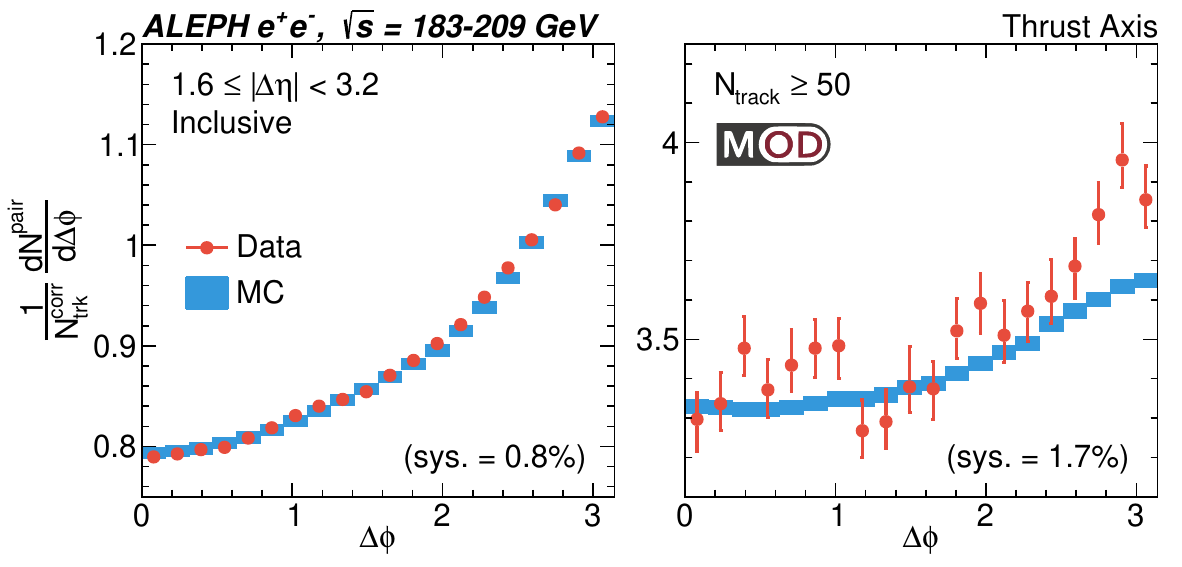}
    \caption{The azimuthal associated yield for the long-range region 1.6 $ < |\Delta \eta| < 3.2$ is shown for $N_{trk} \geq$ 5 (left) and  $N_{trk} \geq$ 50 (right). The red dots with statistical error bars are from experimental data, while the blue line with statistical error bars is obtained from the PYTHIA 6 model~\cite{Chen:2023njr}}
    \label{fig:assoyield}
\end{figure}

While understanding the collective dynamics in small collision systems with event-shape, a study is performed in $e^{+}e^{-}$ collisions by measuring the two-particle azimuthal correlation of charged particles using the event-shape observable thrust~\cite{Chen:2023njr}. The long-range near-side correlation is measured using archived hadronic $e^{+}e^{-}$ data collected by ALEPH LEP-II at center-of-mass energies  $\sqrt{s}$ = 183 - 209 GeV. As discussed in Sec.~\ref{sec:definitions}, the event-shape observable thrust measures the out-of-event-plane energy flow. The thrust-axis is used to define the coordinate system. The determination of thrust axis and the exact procedure used to find the two-particle correlation functions can be found in Ref.~\cite{Chen:2023njr}. The left and right panel of Fig.~\ref{fig:assoyield} show the associated azimuthal yield for $N_{\rm trk} \geq$ 5 (inclusive) and  $N_{\rm trk} \geq$ 50 (high-multiplicity), respectively. For inclusive measurement, the MC simulation based on the PYTHIA 6 model qualitatively describes the experimental data. However, high multiplicity class data exhibits a long-range near-side structure, while the MC simulation does not reproduce it well. In particular, the data have a significant slope compared to the PYTHIA 6 simulation at higher azimuthal angle $\Delta \phi$.

\begin{figure}
    \centering
\includegraphics[width=0.95\linewidth]{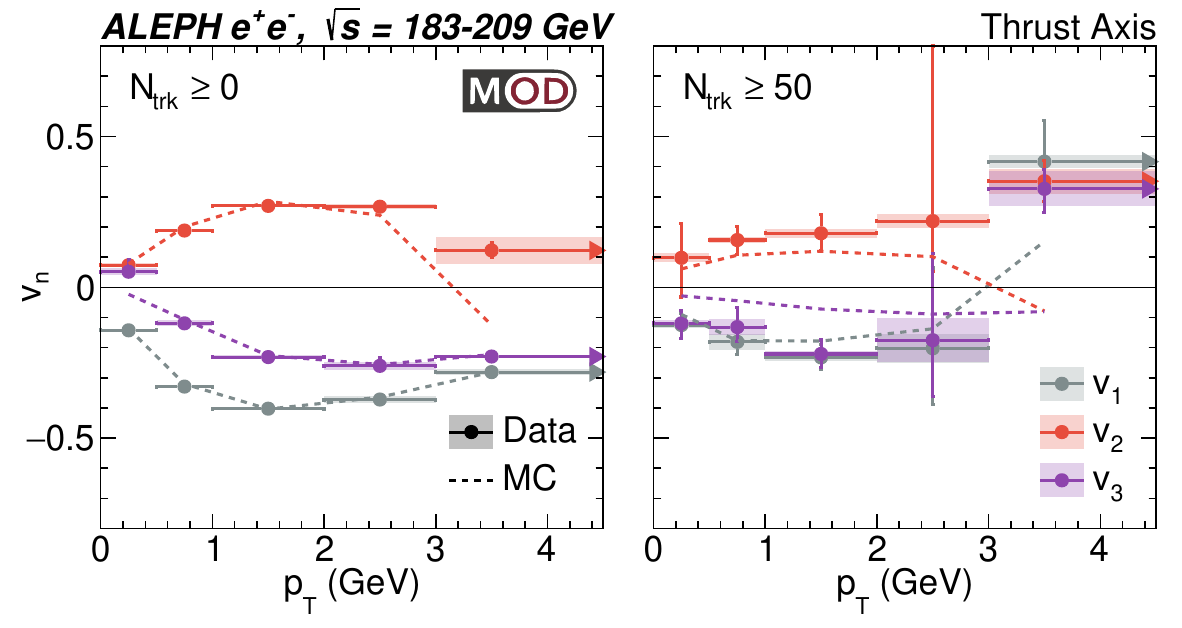}
    \caption{$v_{n}$ as a function of $p_{\rm T}$ for $N_{\rm trk} \geq$ 0 (left) and  $N_{\rm trk} \geq$ 50 (right) for the thrust axis analysis using LEP-II data. The marker with black, red, and purple error bars represents $v_1$, $v_2$, and $v_3$ measured from the data. The dashed line with the corresponding color represents the results obtained from the MC simulation~\cite{Chen:2023njr}}
    \label{fig:vn}
\end{figure}

Furthermore, for the first time, the two-particle correlation function is decomposed in a Fourier series in $e^{+}e^{-}$ collisions. The resulting anisotropic flow coefficients $v_1$, $v_2$, and $v_3$ are obtained as a function of $p_{\rm T}$ and shown in Fig.~\ref{fig:vn}. The flow coefficients $v_1$, $v_2$, and $v_3$ measured from the data are compared with the PYTHIA 6 simulation results. It is observed that PYTHIA 6 has qualitative agreement with data for multiplicity integrated events, while for  $N_{\rm trk} \geq$ 50 the PYTHIA 6 predicts a smaller magnitude for $|v_n|$. The interpretation of the positive $v_2$ and the negative $v_3$ in $e^{+}e^{-}$ collisions cannot be explained by the geometry-driven interpretation. Hence, this motivates further theoretical and experimental eﬀorts to identify the physical mechanisms that could be responsible for the emergence of collectivity in such systems. Future experiments such as the Electron-Ion Collider at BNL, the Future Circular
Collider at CERN may provide new and more differential measurements to clarify the origin of long-range near-side correlations in small systems.

\begin{itshape}
The key takeaway messages of this section are summarized below:

\begin{itemize}
\item The measurement of anisotropic flow coefficients in small collision systems (i.e, $pp$, and $e^{+}e^{-}$) is interesting because one of the dominant sources of collectivity in these collisions is assumed to be the density fluctuations rather than the collision geometry, which is traditionally believed for heavy-ion collisions. 
\item Recent measurement of $v_{2}$ in high-multiplicity $pp$ collisions by ALICE and $v_{1}$, $v_{2}$, and $v_{3}$ in $e^{+}e^{-}$ collisions by ALEPH collaboration trigger emmense interest in the community to understand the origin of collectivity in such collision system. Here, the event shape classifiers play a role in understanding the collective phenomena by separating the events based on their energy flow.
\end{itemize}
\end{itshape}

{\it In summary, this section shows that event shape classifiers have recently become central to soft QCD studies at the LHC, although their origins are from hard QCD phenomena. With clear region definitions and $\eta$ gaps, these classifiers suppress autocorrelations and reveal how global geometry correlates with soft activity and multiparton interactions. Isotropic selections identified by event shape observables pick out collisions with larger mean transverse momentum, enhanced strangeness, and increased baryon-to-meson ratios. Correlation and flow measurements made with the same selections indicate that long-range structure persists, within current precision, once biases are controlled. The key message is that heavy-ion-like and QGP-like behavior grows with enhanced soft and underlying event activity, which motivates the use of event shape classifiers alongside bare multiplicity. More broadly, event shape classifiers provide precise tools to test alternative explanations for QGP-like signatures in small collision systems.}

Let us now move to the hard QCD probes at the LHC and their interplay with event shape classifiers.

\section{Hard QCD Probes with event shape classifiers}
\label{hardprobes}
Hard QCD probes are usually referred to as the observables that involve large momentum transfers, where perturbative calculations are possible as the coupling constant is significantly small. In the hard regime of QCD, jets and high-momentum particles are produced from parton scatterings. Here, one can take advantage of the infrared and collinear safety features of event shape classifiers for the pQCD predictions. A brief review of the dependence of hard probes on the event shapes at the LHC is presented in this section. 

\subsection{Heavy flavor production}
\begin{figure*}[ht!]
    \centering
    \includegraphics[width=0.49\linewidth]{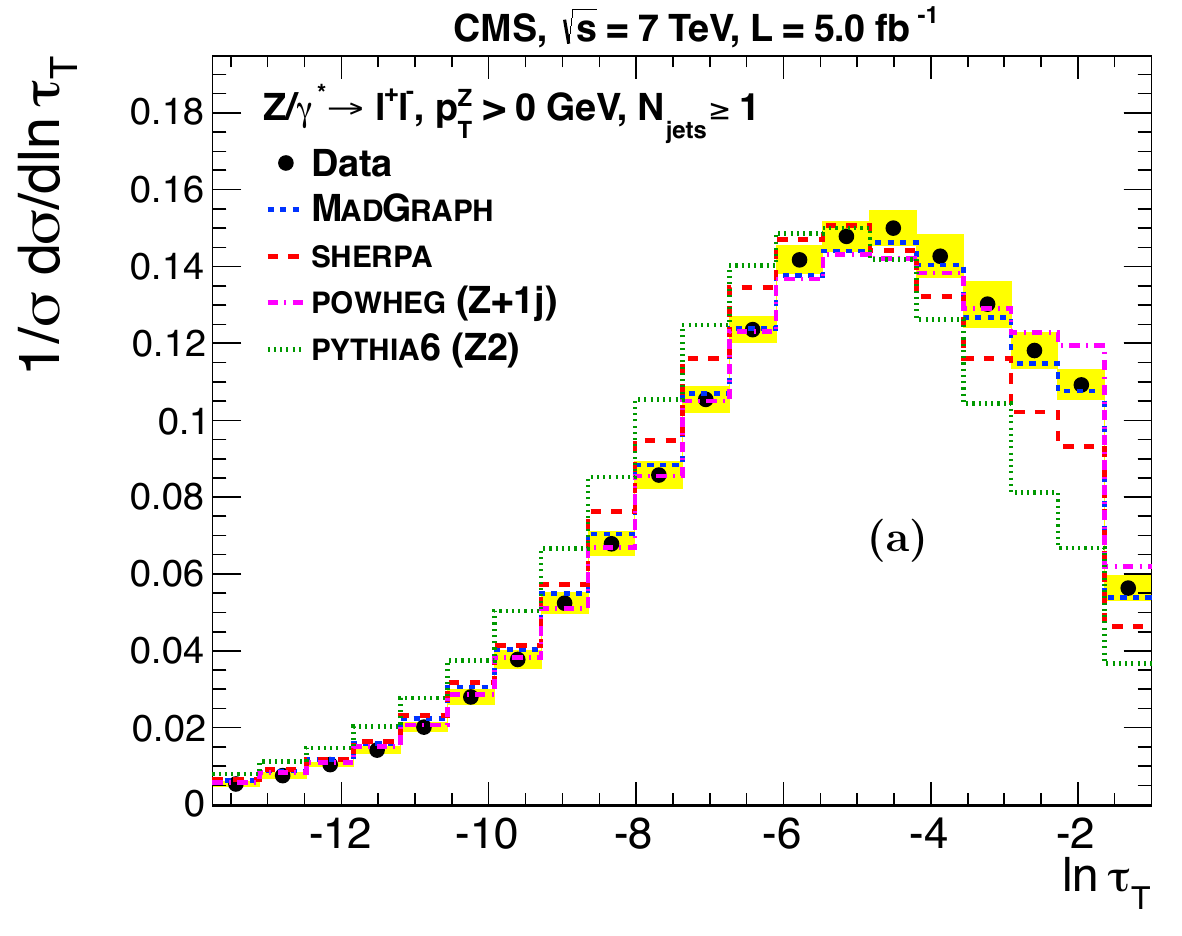}
    \includegraphics[width=0.49\linewidth]{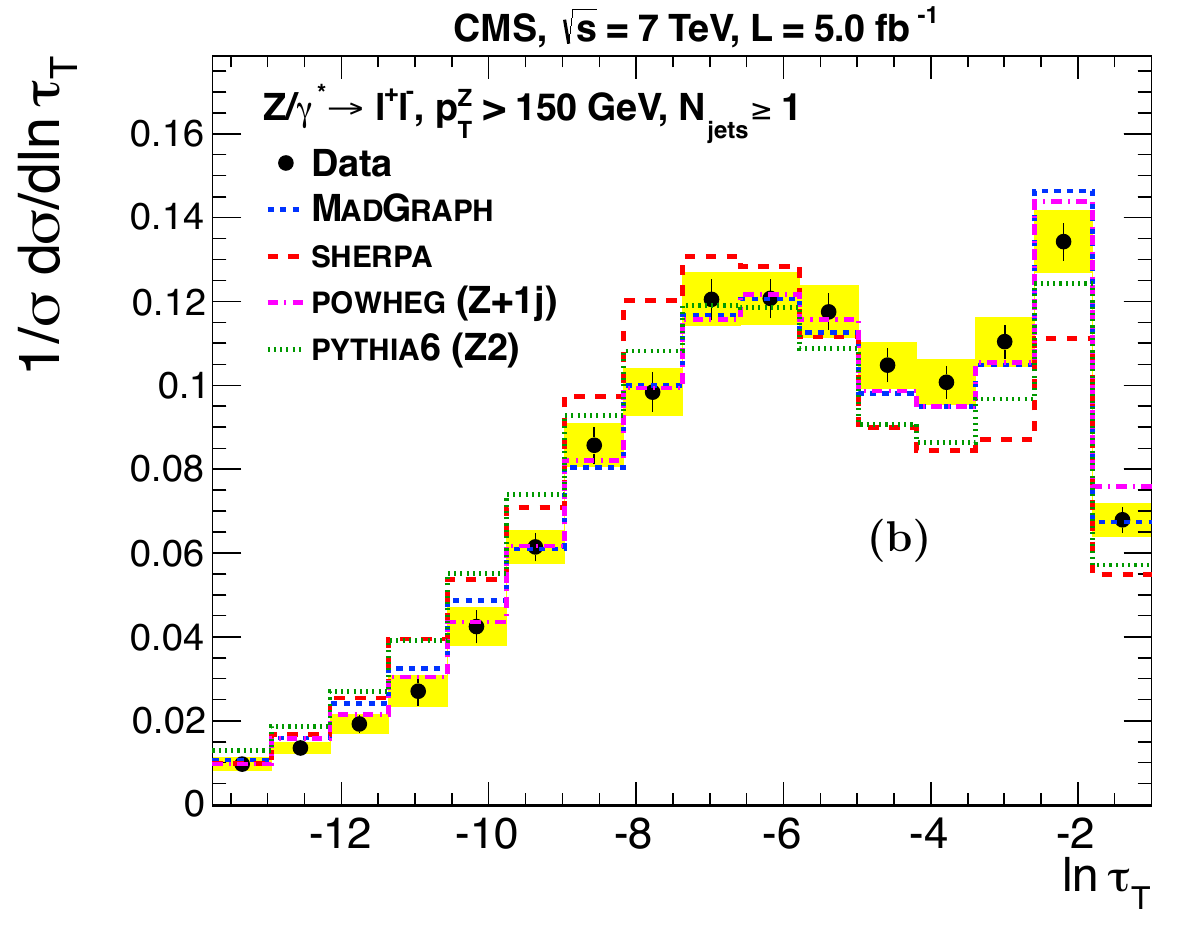}
    \caption{Normalized distribution of $\ln\tau_{\rm T}$ for all $p_{\rm T}^{\rm Z}$ (left) and $p_{\rm T}^{\rm Z}>150$ GeV/$c$ in $pp$ collisions at $\sqrt{s}=7$ TeV with CMS experiment~\cite{CMS:2013lua}.}
    \label{fig:ZvsES}
\end{figure*}

The production of heavier particles, such as the charm and bottom hadrons or the $\rm W^{\pm}$ or $\rm Z$ bosons, etc., which involve large momentum transfers, affects the distribution of the event shape in the final state. To prove this, Ref.~\cite{CMS:2013lua} shows that $\rm Z$-boson is strongly correlated with the production of jets in the opposite azimuthal direction. This correlation gets weaker, and the distribution of relative azimuthal angle between the jets and $\rm Z$ boson becomes isotropic when the number of jets increases. A similar observation is also made when the event with a large transverse momentum ($p_{\rm T}^{\rm Z}>150$ GeV/$c$) of $\rm Z$-boson is selected. These effects are well reflected in the distribution of transverse thrust ($\tau_{\rm T}$, see Section~\ref{sec:thrust} for definition). This is evident from the left and right panels of Fig.~\ref{fig:ZvsES}, which shows the distribution of $\ln\tau_{\rm T}$ calculated using the jets and the $\rm Z$ boson, for (a) all $p_{\rm T}^{\rm Z}$ and (b) $p_{\rm T}^{\rm Z}>150$ GeV/$c$, respectively. Here, $p_{\rm T}^{\rm Z}$ is the transverse momentum of $\rm Z$-boson, which is reconstructed using its decay to di-lepton pairs.
One can clearly see an additional peak structure for the $p_{\rm T}^{\rm Z}>150$ GeV/$c$ case in the right panel near $\ln \tau_{\rm T}\sim -2$, which is indicative of accumulation of large number of events with a large spherical component, corresponding to production of two or more jets.

\begin{figure}
    \centering
    \includegraphics[width=0.49\linewidth]{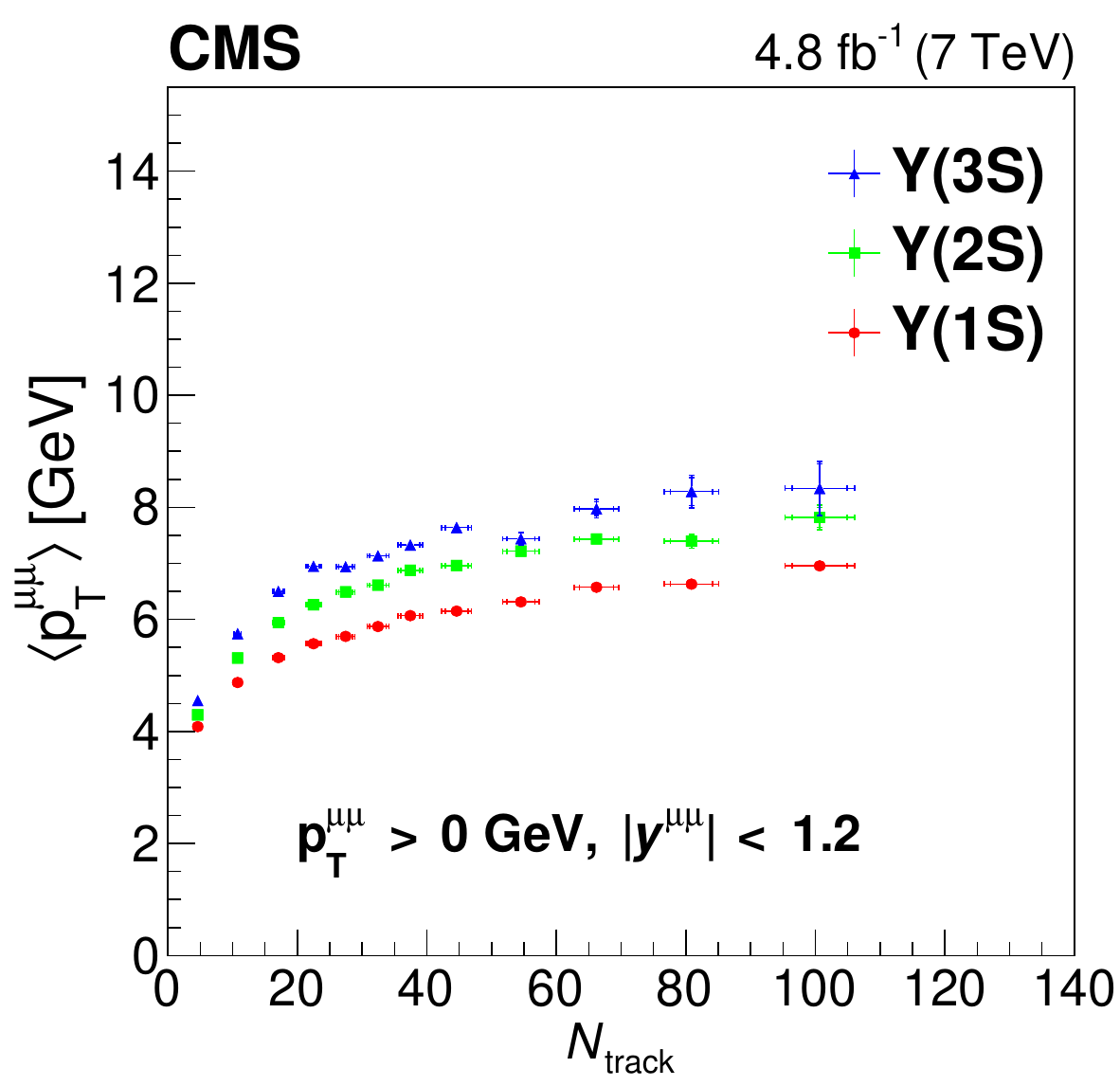}
    \caption{$\langle p_{\rm T}^{\mu\mu}\rangle$ as a function of $N_{\rm track}$ for $\Upsilon(\rm 1S)$, $\Upsilon(\rm 2S)$ and $\Upsilon(\rm 3S)$ in $pp$ collisions at $\sqrt{s}=7$ TeV with CMS experiment~\cite{CMS:2020fae}.}
    \label{fig:UpsilonnSmeanpTvsNtracks}
\end{figure}

Let us now move to the discussion of event shape dependence of another heavy flavour, $\Upsilon(\rm nS)$. Figure~\ref{fig:UpsilonnSmeanpTvsNtracks} shows average transverse momentum ($\langle p_{\rm T}^{\mu\mu}\rangle$) of $\Upsilon$(nS) candidates as a function of number of charged tracks ($N_{\rm track}$) in $pp$ collisions at $\sqrt{s}=7$ TeV with the CMS experiment~\cite{CMS:2020fae}. A hierarchal structure of  $\langle p_{\rm T}^{\mu\mu}\rangle$ is observed where $\langle p_{\rm T}^{\mu\mu}\rangle$ increases more rapidly with $N_{\rm track}$ with an increase in the mass of corresponding $\Upsilon(\rm nS)$. This feature is similar to that observed in Section~\ref{sec:avgpT}.

\begin{figure*}
    \centering
    \includegraphics[width=0.49\linewidth]{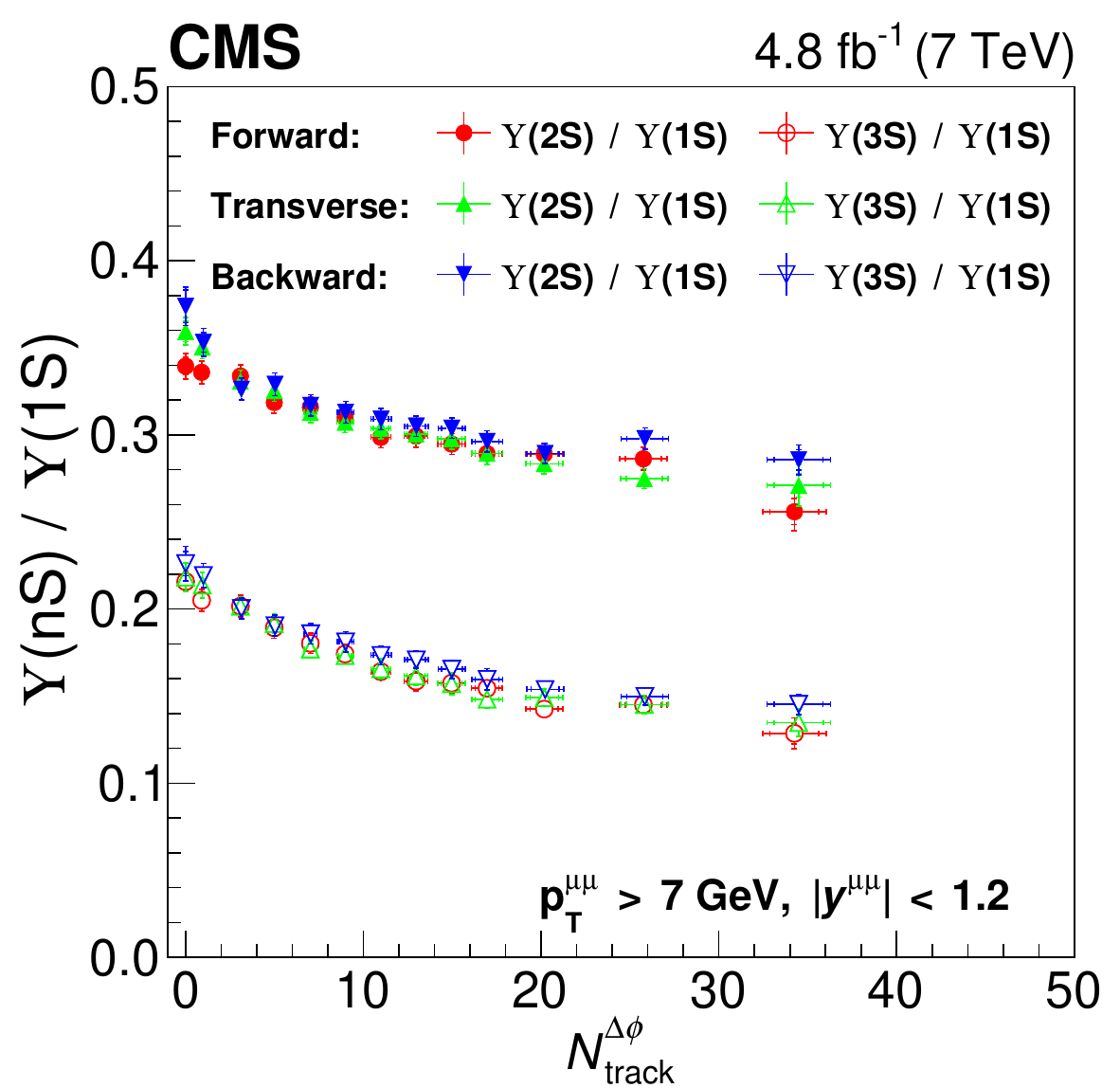}
    \includegraphics[width=0.49\linewidth]{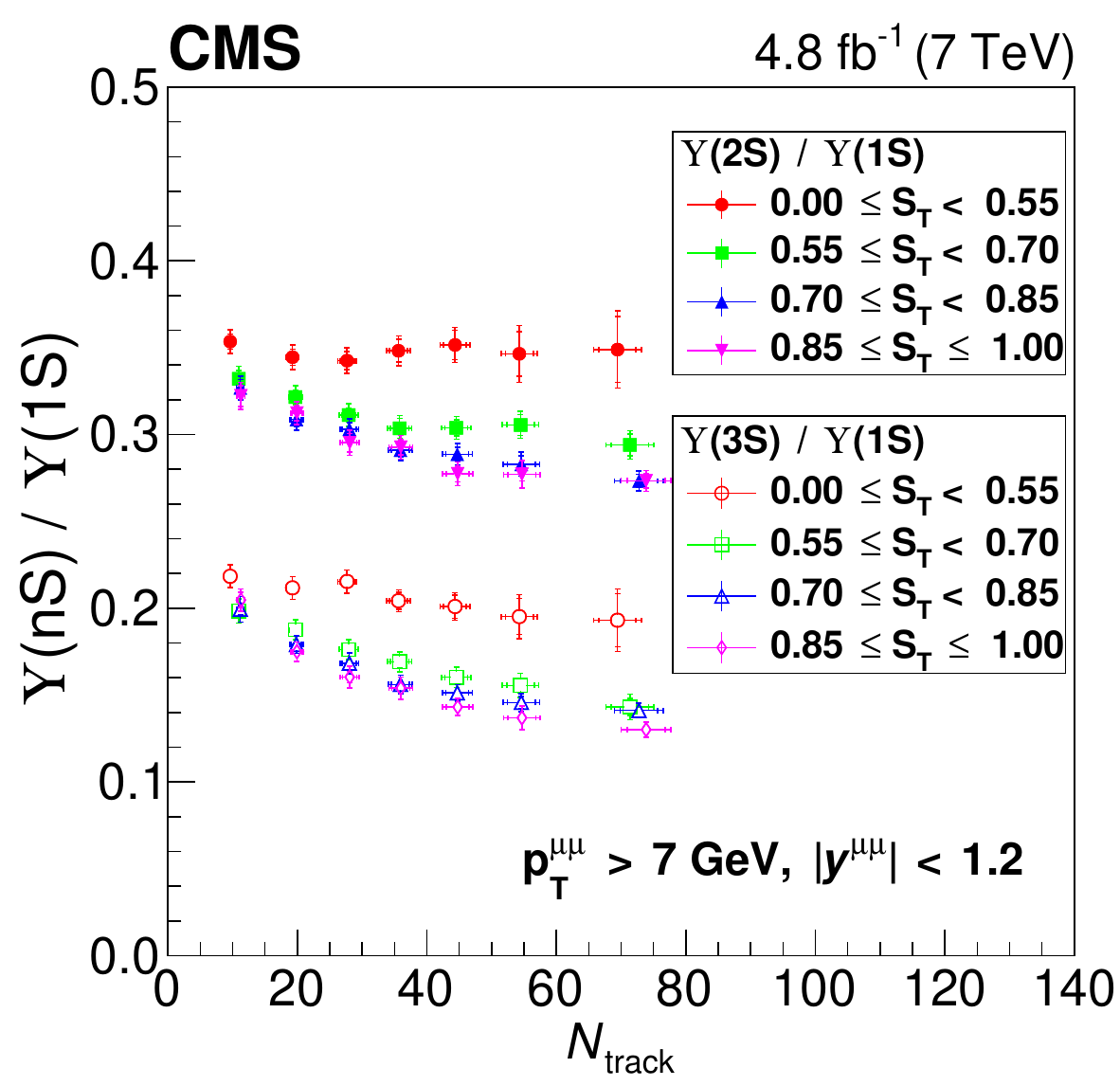}
    \caption{(Left) Yield ratios, $\Upsilon(\rm nS)/\Upsilon(1S)$ as a function of the number of tracks in forward, transverse, and backward regions in $pp$ collisions at $\sqrt{s}=7$ TeV with CMS experiment~\cite{CMS:2020fae}. (Right) Yield ratios, $\Upsilon(\rm nS)/\Upsilon(1S)$ as a function of $N_{\rm track}$ for different values of $S_{\rm T}$ in $pp$ collisions at $\sqrt{s}=7$ TeV with CMS experiment~\cite{CMS:2020fae}.}
    \label{fig:UpsilonvsNT}
\end{figure*}

To understand the relation between the underlying event properties and production of $\Upsilon(\rm nS)$, the relative yield of $\Upsilon(\rm nS)$ is studied as a function of particle multiplicity in different topological regions with respect to $\Upsilon(\rm nS)$, as shown in the left panel of Fig.~\ref{fig:UpsilonvsNT}. Here, the regions `forward' and `backward' are similar to `toward' and `away' regions (see Section~\ref{sec:RTdefn}), respectively, the only difference is that instead of a trigger charged particle, the regions are defined based on the relative azimuthal angle ($\Delta\phi$) of charged particles with respect to the direction of $\Upsilon(\rm nS)$. Here the relative yield of $\Upsilon(\rm 2S)$ and  $\Upsilon(\rm 3S)$ with respect to $\Upsilon(\rm 1S)$ decrease as a function of $N_{\rm track}^{\Delta\phi}$ in all three regions. The difference between the relative yields, $\Upsilon(\rm nS)/\Upsilon(1S)$ appears towards lower $N_{\rm track}^{\Delta\phi}$ events, where the yield is higher for the backward azimuthal region. A similar behavior in the transverse regions highlights the connection of production of $\Upsilon(\rm nS)$ on underlying event activity rather than dependence on the particle activity along the direction of $\Upsilon(\rm nS)$, which would affect only the forward region.

The above discussion on the connection between underlying event activity and the production of $\Upsilon(\rm nS)$ is confirmed when studied using transverse sphericity, as shown in the right panel of Fig.~\ref{fig:UpsilonvsNT}. Here the relative yield, $\Upsilon(\rm nS)/\Upsilon(1S)$, is studied as a function of $N_{\rm tracks}$ for different classes of $S_{\rm T}$. As can be seen from Fig.~\ref{fig:UpsilonvsNT} right, for the events with low-$S_{\rm T}$ (jetty events with low underlying event activity), $\Upsilon(\rm nS)/\Upsilon(1S)$ varies weakly with $N_{\rm tracks}$. In contrast, for events with larger $S_{\rm T}$ (isotropic events with significant underlying event activity), $\Upsilon(\rm nS)/\Upsilon(1S)$ decreases with an increase in $N_{\rm tracks}$. This suggests that the decreasing trend is an effect of underlying event activity.

{\it In summary, the studies of hard probes such as $\rm Z$-boson and the bottomonia with event shape classifiers by CMS experiment are instrumental in highlighting the fact that the event shape classifiers cleanly separate the hard recoil from the soft environment and provide a precision tool for testing and tuning models in topologies with a well-defined pQCD hard scale.}

\subsection{Measurements of jet production}

Jets, defined as the collimated emission of particles, are one of the important sources to understand the hard QCD processes occurring at the spatial scales smaller than the radius of a proton. The jets originate from quarks or gluons produced in ultra-relativistic collisions and subsequently hadronize through fragmentation to produce a collimated spray of hadrons. The production of jets involves large momentum transfers, and they can influence the particle multiplicity measured in the final state as well as the shape of the events.

\begin{figure*}
    \centering
    \includegraphics[width=0.49\linewidth]{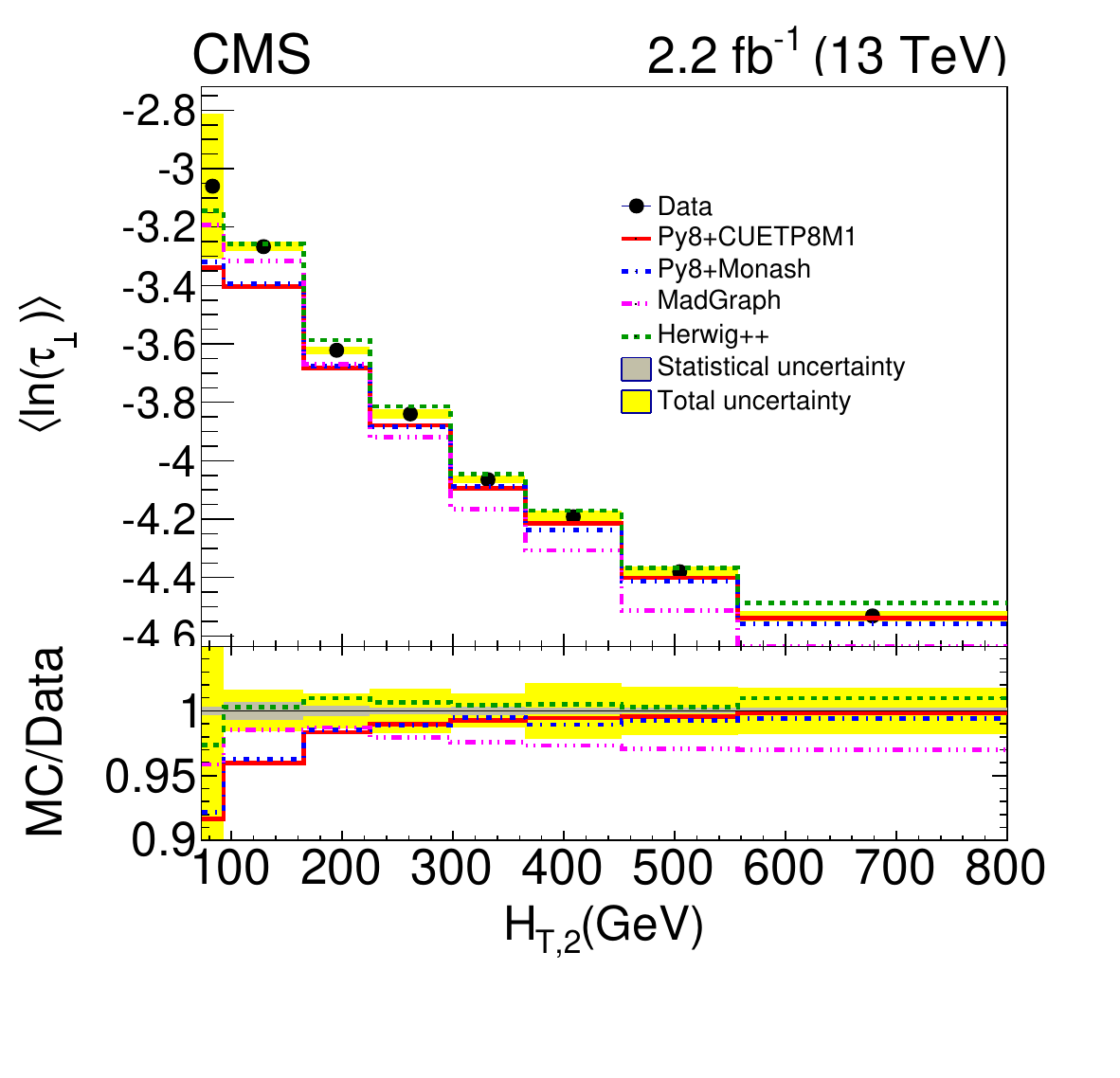}
    \includegraphics[width=0.49\linewidth]{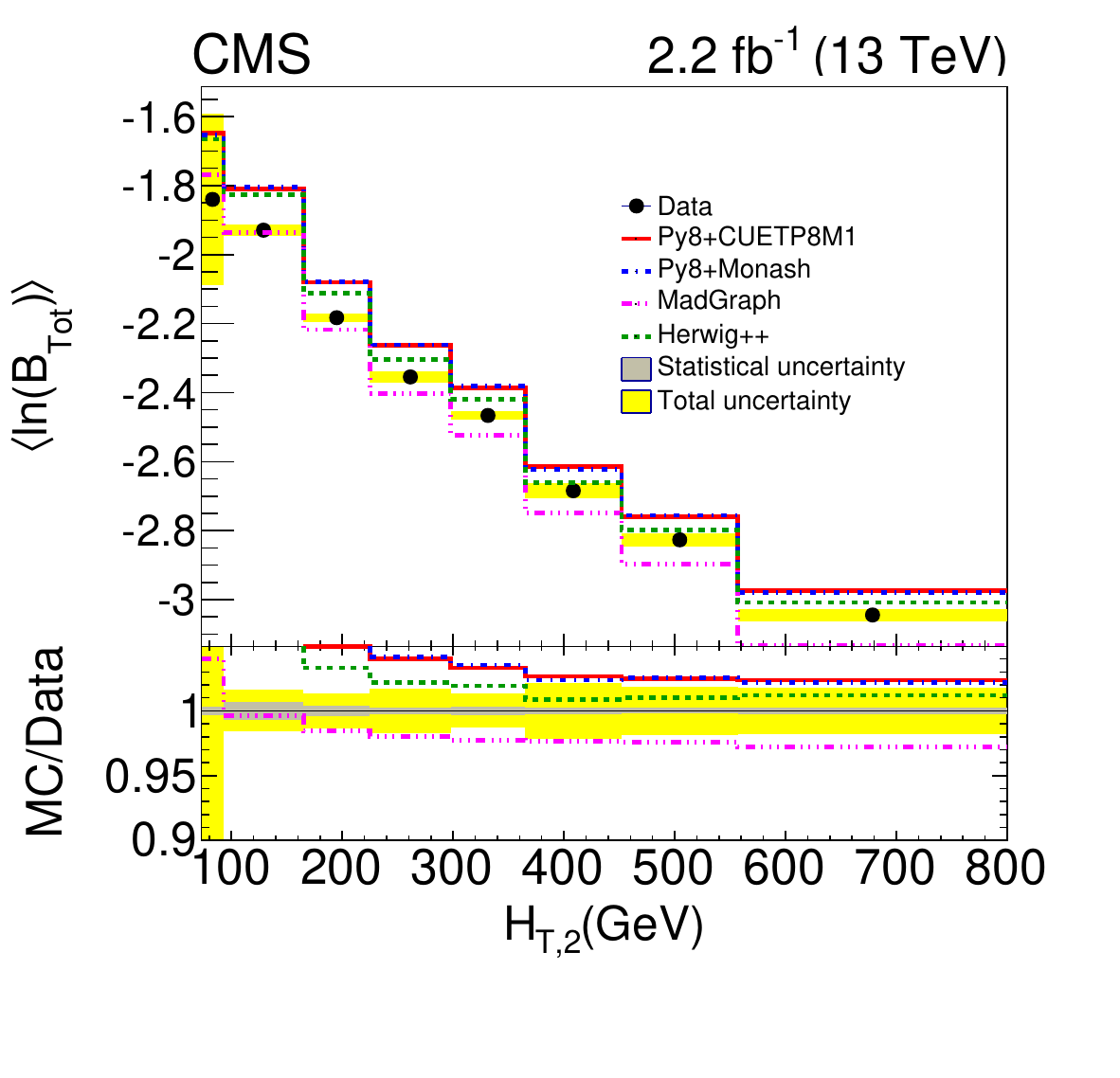}
    \includegraphics[width=0.49\linewidth]{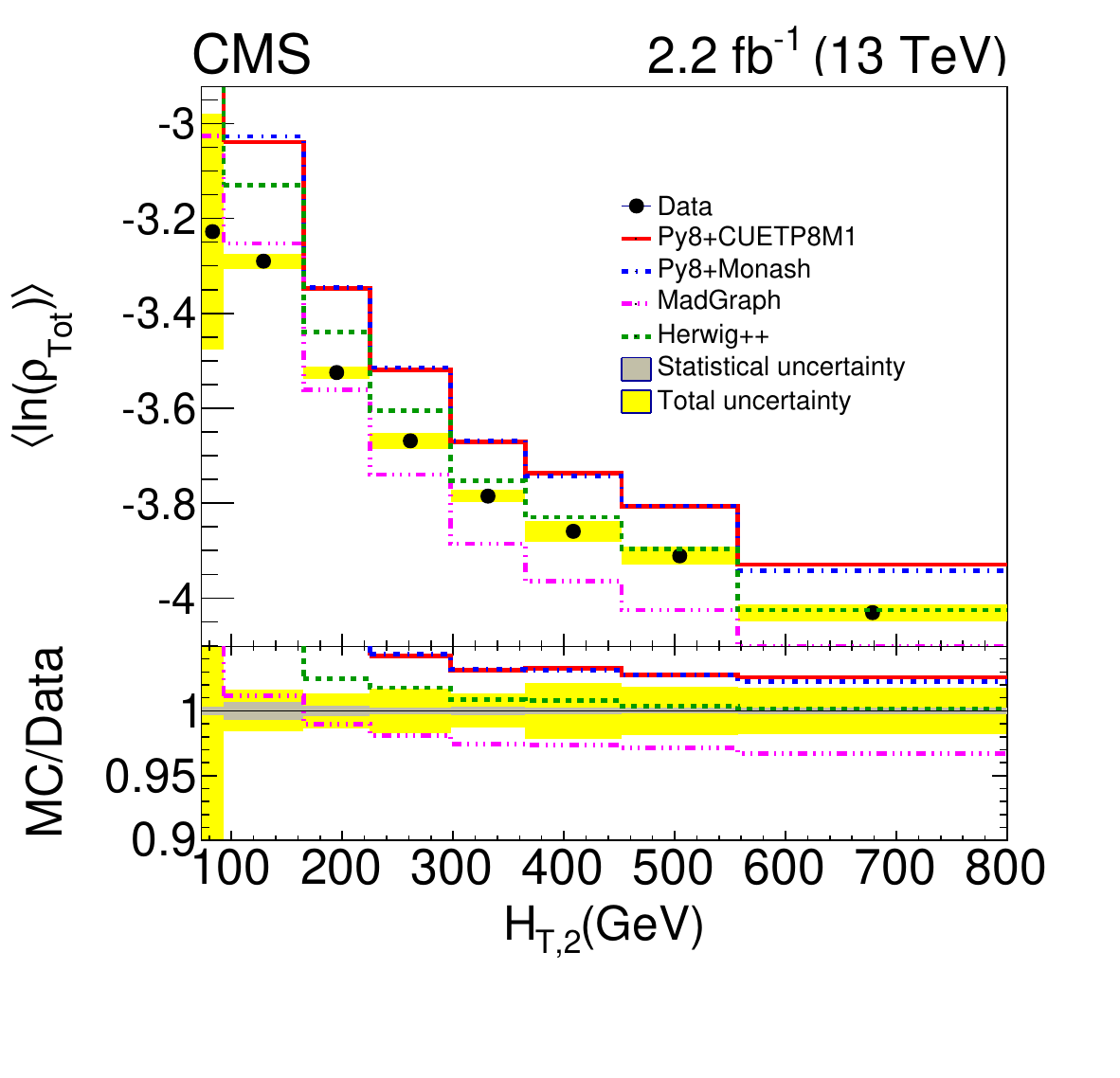}
    \includegraphics[width=0.49\linewidth]{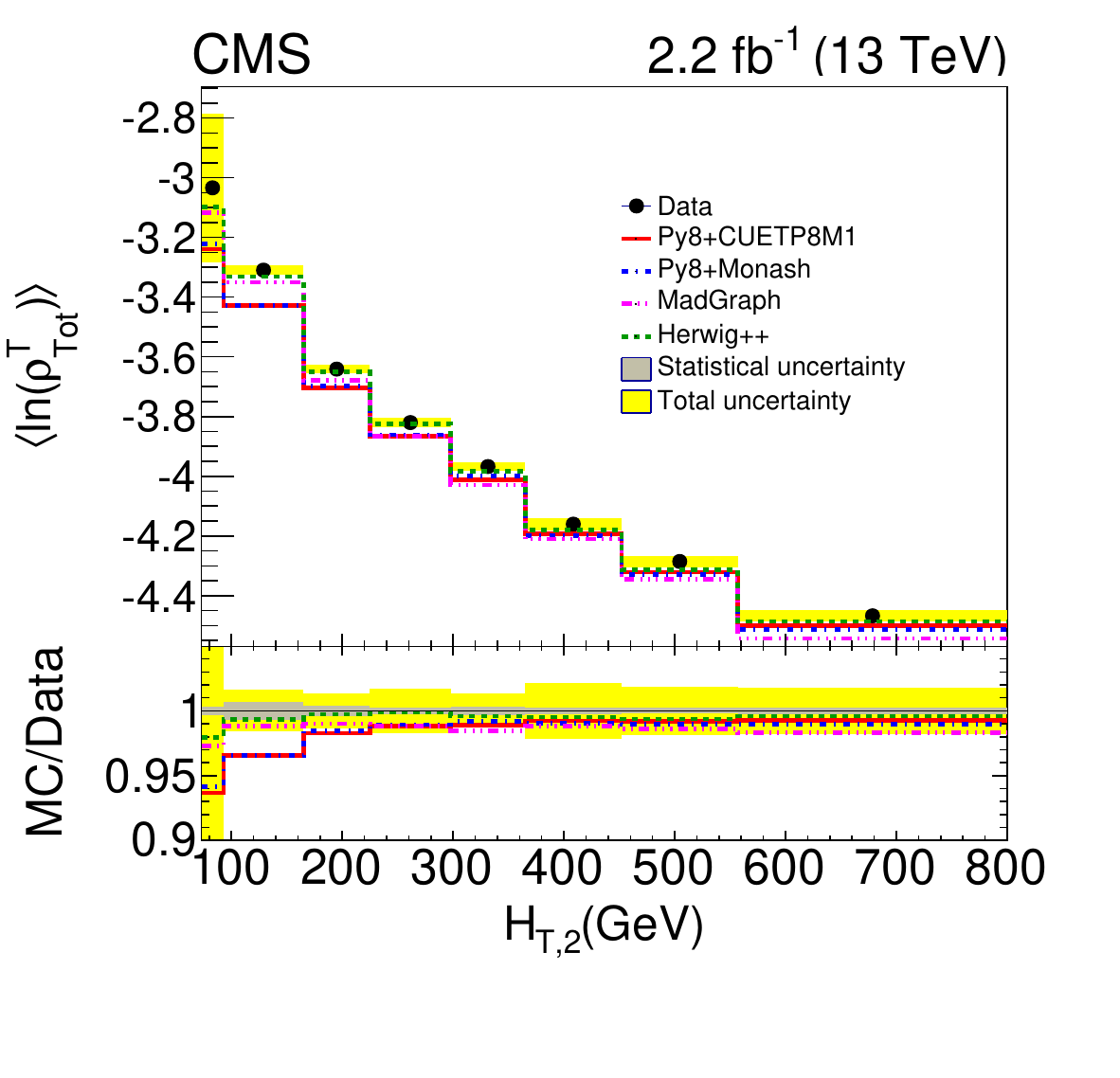}
    \caption{$\langle \ln(\tau_{\perp})\rangle$ (upper left), $\langle\ln(B_{\rm Tot})\rangle$ (upper right), $\langle \ln(\rho_{\rm Tot})\rangle$ (lower left), and $\langle \ln(\rho_{\rm Tot}^{\rm  T})\rangle$ (lower right) as a function of $H_{\rm T, 2}$ in $pp$ collisions at $\sqrt{s}=13$ TeV with CMS Experiment and compared with calculations from different MC event generators~\cite{CMS:2018svp}. $H_{\rm T, 2}=(p_{\rm T,jet1}+p_{\rm T,jet2})/2$, where $p_{\rm T,jet1}$ and $p_{\rm T,jet2}$, respectively, are the transverse momentum of hightest and second highest $p_{\rm T}$ jets.}
    \label{fig:jetHTvsES}
\end{figure*}

\begin{figure*}
\centering
\includegraphics[width=0.49\linewidth]{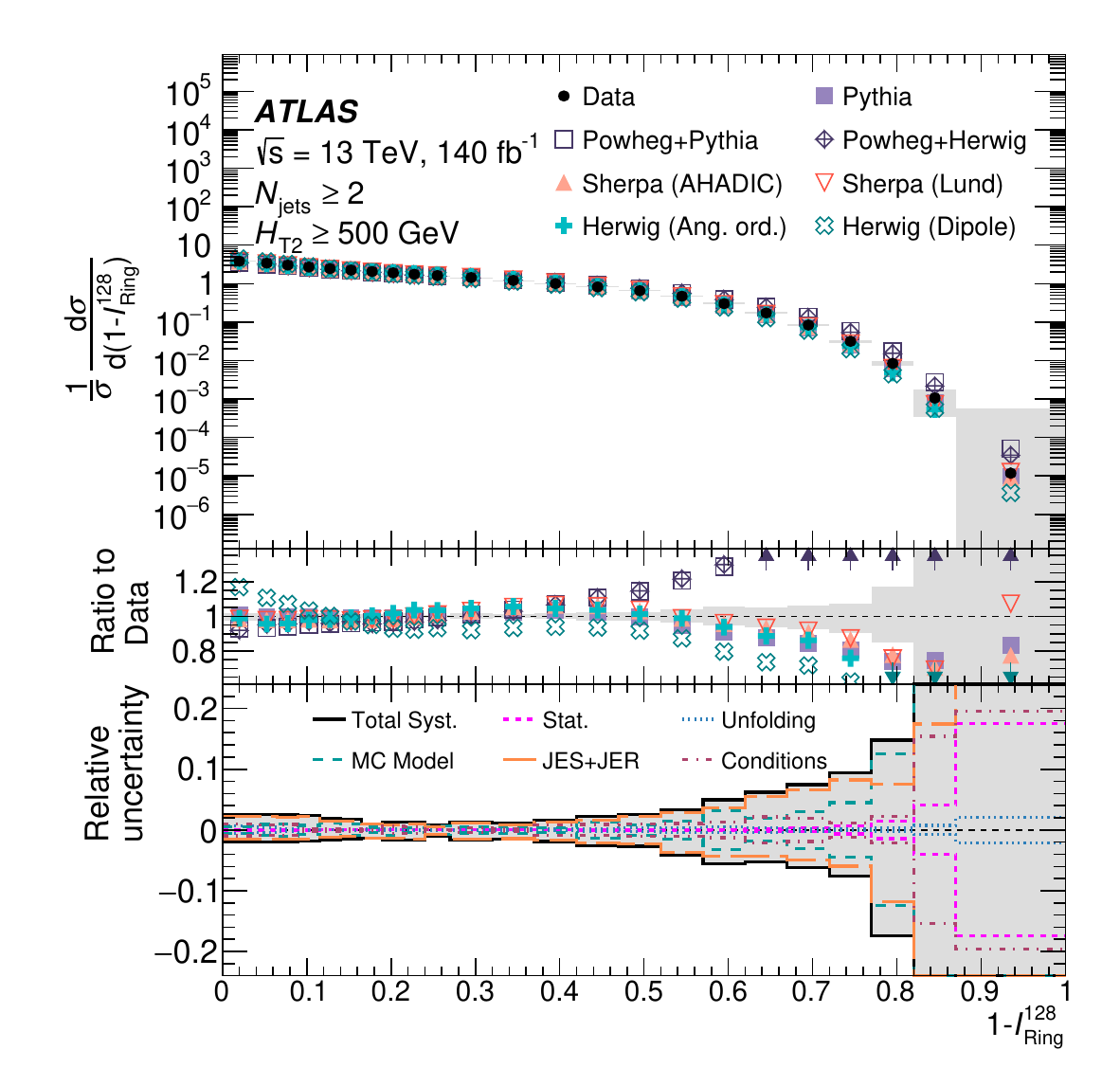}
\includegraphics[width=0.49\linewidth]{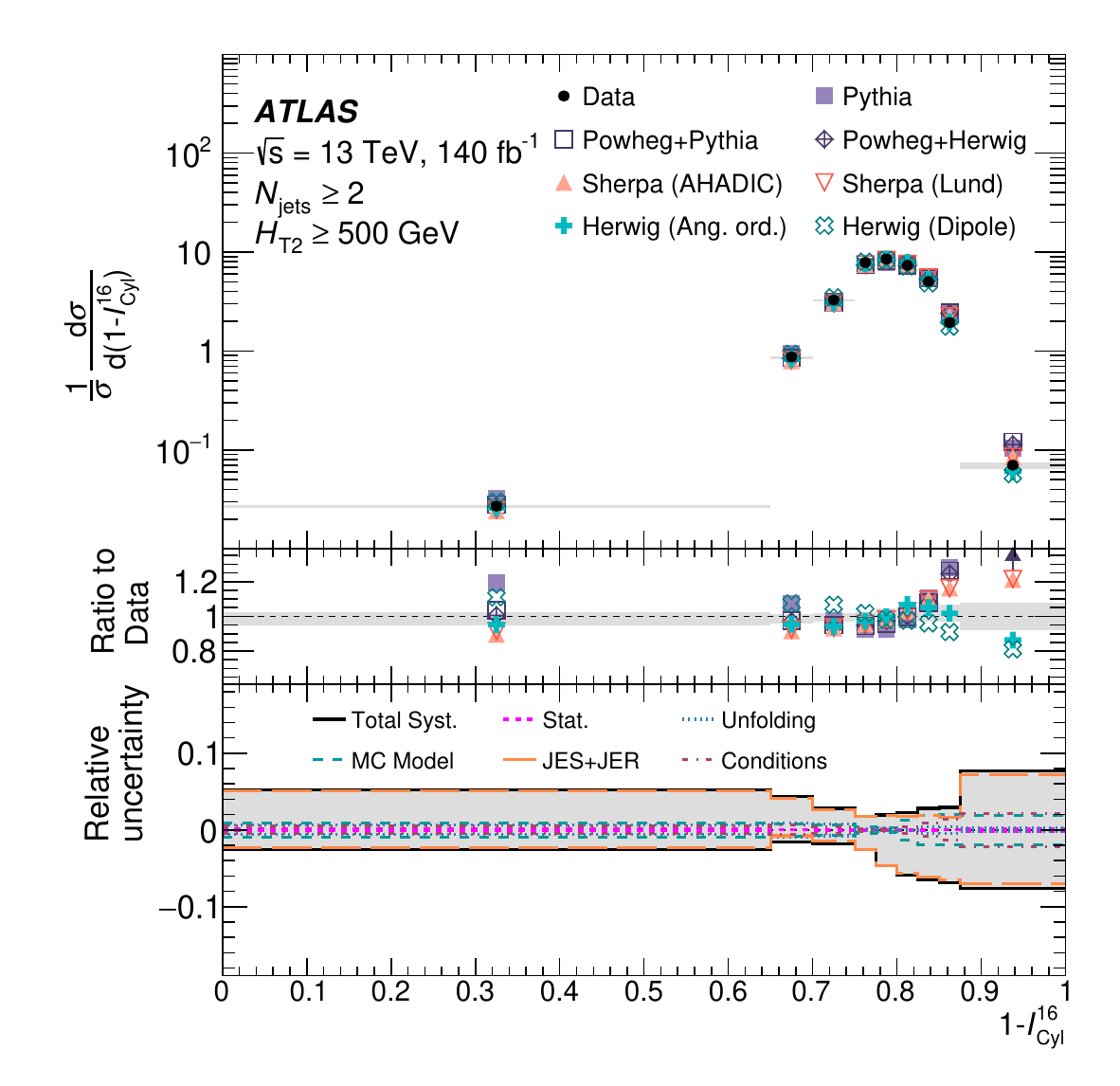}

\caption{The shape-normalized $\mathcal{I^{ \rm N = 128}_{ \rm Ring}}$ (left panel),  $\mathcal{I^{\rm N = 16}_{\rm Cyl}}$ (right panel) cross section in data (closed circles), compared with predictions from several Monte Carlo generators. Events with $H_{\rm T, 2} \geq$ 500 GeV and $N_{\rm jet} \geq$ 2 are included. The middle panels display the ratios of different event generator predictions to the unfolded data. The lower panels summarise the various
sources of systematic uncertainty in the measurement~\cite{ATLAS:2023mny}}.
\label{fig:ATLASEventisotropy}
\end{figure*}

The influence of jet production to the measured event shapes is also testified in Fig.~\ref{fig:jetHTvsES}, which shows the evolution of $\langle \ln(\tau_{\perp})\rangle$ (upper left), $\langle\ln(B_{\rm Tot})\rangle$ (upper right), $\langle \ln(\rho_{\rm Tot})\rangle$ (lower left), and $\langle \ln(\rho_{\rm Tot}^{\rm  T})\rangle$ (lower right) with $H_{\rm T, 2}$ in $pp$ collisions at $\sqrt{s}=13$ TeV with CMS Experiment. Here, $H_{\rm T, 2}$ is the average of the transverse momentum of the highest and second highest $p_{\rm T}$ jets. The event shape observables shown in Fig.~\ref{fig:jetHTvsES} are calculated using the jets present in the event. Higher values of these event shape observables indicate the isotropic or spherical distribution of jets, which occurs mostly in the case of multi-jet topology in the event. In contrast, a lower value corresponds to di-jet structures. A higher value of $H_{\rm T, 2}$ indicates that the initial partons are more boosted, which makes the events less spherical, thereby reducing the values of event shape observables. Moreover, the value of the strong coupling constant ($\alpha_{s}$) decreases with an increase in $H_{\rm T, 2}$, which hampers the emission of hard gluons and spoils the multijet production and the spherical nature of the jet emission in the event.

The MPI parameters in PYTHIA8 CUETP8M1~\cite{CMS:2015wcf} and Monash are comparable, the agreement of which with CMS data improves with increasing $H_{\rm T, 2}$. The details about this specific tunes of PYTHIA can be found in appendix~\ref{appendixpythia}. Moreover, these two models show good agreement with the CMS data except for the lowest values of $H_{\rm T, 2}$ for $\langle \ln(\tau_{\perp})\rangle$ and $\langle \ln(\rho_{\rm Tot}^{\rm  T})\rangle$ which are estimated in the transverse plane. On the other hand, the disagreement of PYTHIA8 CUETP8M1 and Monash with CMS data is large for   $\langle\ln(B_{\rm Tot})\rangle$ and $\langle \ln(\rho_{\rm Tot})\rangle$, which incorporate both transverse and longitudinal components. This indicates that the energy flow in the transverse plane is well incorporated for PYTHIA8 CUETP8M1 and Monash which is not well for the longitudinal plane. On the other hand, HERWIG++ explains all the event classifiers better than PYTHIA8 suggesting a overall description of energy flow in both transverse and longitudinal directions. The predictions from MadGraph for the event shape is better at lower values of $H_{\rm T, 2}$ which indicates a good description of energy flow in both longitudinal and transverse directions for the multi-parton hard scattering processes.

Furthermore, ATLAS has recently performed jet-related event shape studies by measuring the differential cross section of event isotropy observables $\mathcal{I}$ (defined in Sec.~\ref{sec:eventisotropy}) using different reference geometries. The left panel of Fig.~\ref{fig:ATLASEventisotropy} shows the shape-normalized
differential cross section for a ‘ring-like’ geometry with $N$ = 128 points ($\mathcal{I}^{\rm N = 128}_{\rm Ring}$) and events with  $H_{\rm T, 2} \geq$ 500 GeV and $N_{\rm jet} \geq$ 2. While the right panel of Fig.~\ref{fig:ATLASEventisotropy} shows the shape-normalized
differential cross section for a ‘cylindrical’  geometry with $N$ = 16 points ($\mathcal{I}^{\rm N = 16}_{\rm Cyl}$). Here, $N$ represents a finite number of particles used in any reference geometry. It is evident that the cylindrical event isotropy distribution exhibits different characteristics than the ring-like geometries. Events with dijet systems produce the smallest values of this observable, while multijet events with isotropic energy arrangements produce the largest values. The Powheg+Pythia and Powheg+Herwig
predictions overestimate the measured cross section for isotropic events, while other MC generators underestimate the 1 - $\mathcal{I}^{\rm N = 128}_{\rm Ring}$ distribution. The Sherpa hadronisation models (AHADIC vs. Lund) do not exhibit any notable differences and are found to be in qualitative agreement with experimental data.  However, the Herwig dipole model predicts relatively more dijet-like events than the Herwig angle-ordered model. For the cylindrical $1-\mathcal{I}^{\rm N = 16}_{\rm Cyl}$ distribution none of the MC predictions accurately describe this observable. Although the best descriptions occur near the peak of the distribution around $(1-\mathcal{I}^{\rm N = 16}_{\rm Cyl}) \sim$ 0.8, the model’s prediction for the region away from the peak shows deviations from the experimental data. The predictions from the Pythia, Powheg+Pythia, and Powheg+Herwig samples are consistent except at low values, where the Pythia sample overestimates the observed cross section. Once again, no sensitivity to the hadronisation models implemented in Sherpa is observed. 

{\it This subsection highlights how the jet-based event shapes provide a clean and sensitive handle to separate hard recoil from soft activity and to constrain shower, hadronisation, and multiparton interaction modeling in a way that simple jet counts cannot.}  

\subsection{Ridge-like structure for jets in high-multiplicity $pp$ collisions}

\begin{figure*}
    \centering
    \includegraphics[width=0.60\linewidth]{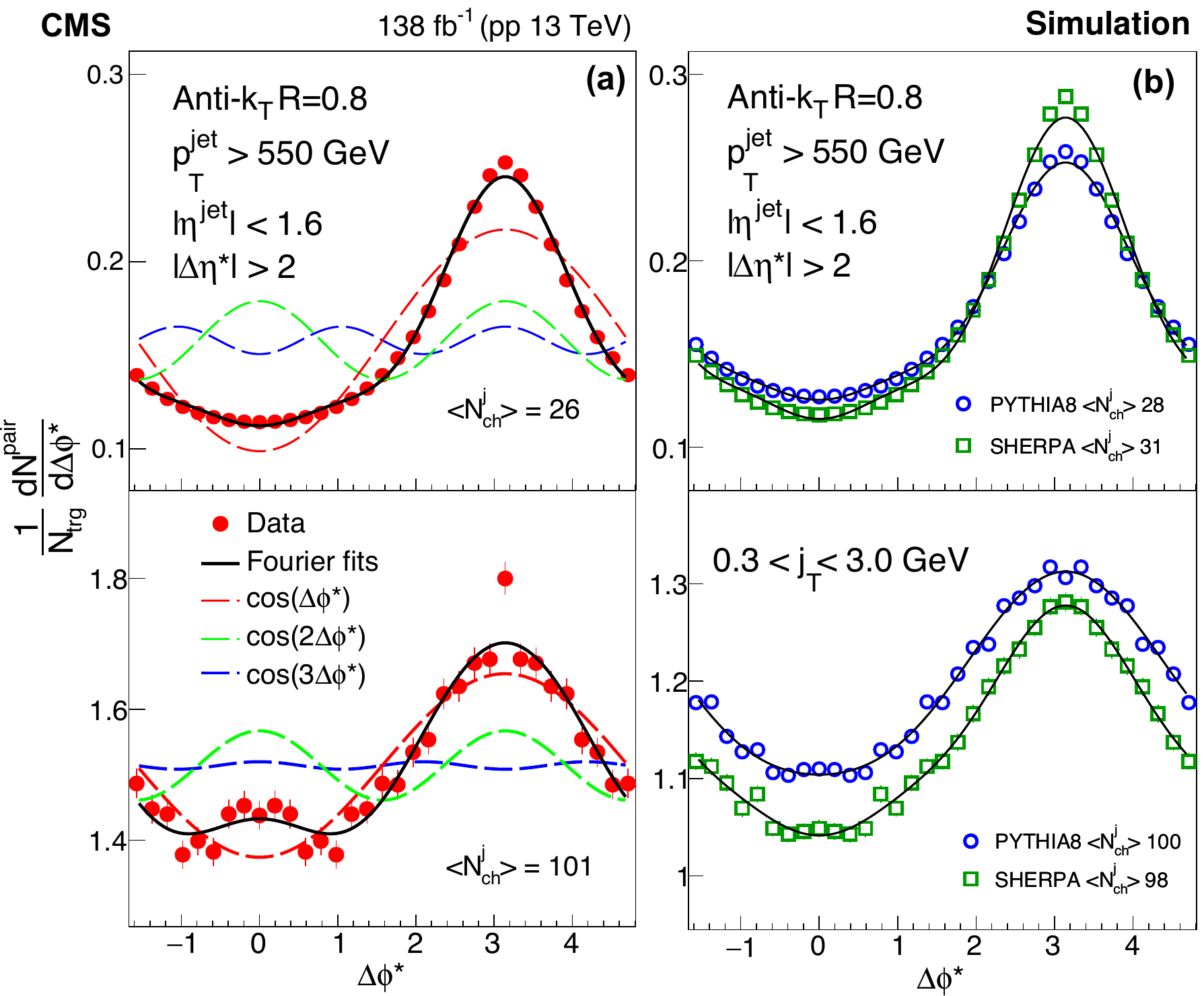}
    \includegraphics[width=0.345\linewidth]{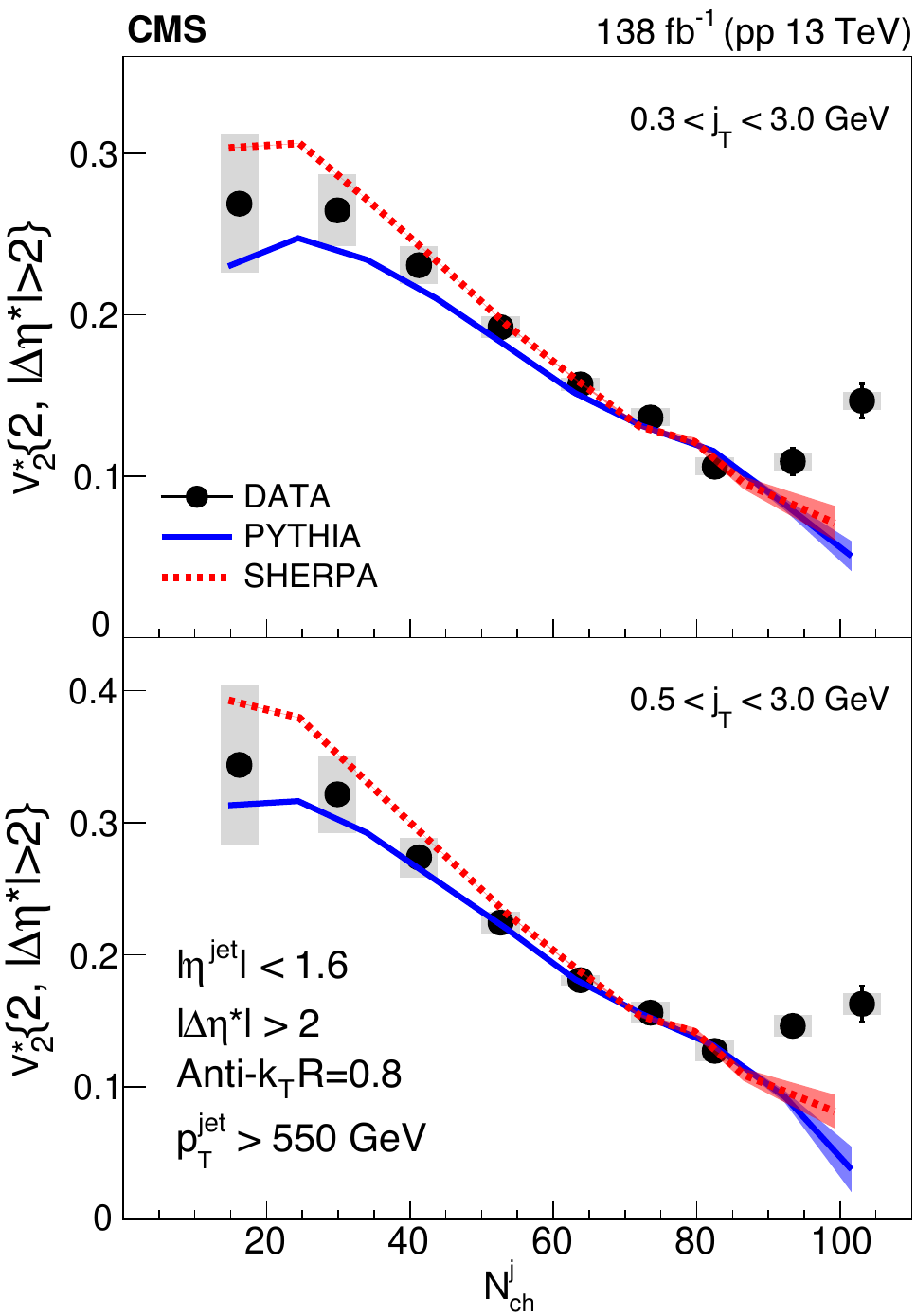}
    \caption{1D two particle azimuthal correlations in $\Delta\phi^{*}$ for $|\Delta\eta^{*}|>2$ calculated using CMS data (left panel), PYTHIA8 and SHERPA (middle panel) in $pp$ collisions at $\sqrt{s}=13$ TeV~\cite{CMS:2023iam}. The right panel shows the elliptic flow coefficient ($v_{2}^{*}$) obtained using the two-particle azimuthal correlation method as a function of $N_{\rm ch}^{\rm j}$~\cite{CMS:2023iam}.}
    \label{fig:jetv2}
\end{figure*}

Motivated by the observation of collective dynamics in small systems as discussed in the previous section, a natural question arises about the minimum system size required for QCD collective effects to develop. To answer this question, Ref.~\cite{Baty:2021ugw} states that the collective effects can emerge from an initial system as small as an energetic parton that can fragment and hadronize in the vacuum. Thus, it becomes crucial to investigate whether a possible collective correlation can build up within the partonic constituents of a jet. Figure~\ref{fig:jetv2} shows the results from the CMS experiment to search for collective effects inside individual jets produced in $pp$ collisions at $\sqrt{s}=13$ TeV~\cite{CMS:2023iam}. For this study, the coordinate system is defined with respect to the jet axis as the $z$-axis. The direction of the jet acts as a proxy for the direction of the initial parton that initializes the jet. Consequently, the momentum vector of the charged particles is defined in a new basis, $\vec{p}^{*}=(j_{\rm T},\;\eta^{*},\;\phi^{*})$. Here, $j_{\rm T}$, $\eta^{*}$, and $\phi^{*}$ are the transverse momentum, pseudorapidity, and azimuthal angles, respectively, measured with respect to the jet axis. This implies that a particle emitted along the direction of the jet would have $\eta^{*}=\infty$ and perpendicular to the jet axis would have $\eta^{*}=0$. Only events having a jet with $p_{\rm T}^{\rm jet}>550$ GeV/$c$ and $|\eta|<1.6$ are considered. Moreover, the results are presented as a function of in-jet charged particle multiplicity, $N_{\rm ch}^{\rm j}$~\cite{CMS:2023iam}. For the estimation of particle correlations and elliptic flow, charged particles of the jets with $p_{\rm T}>0.3$ GeV/$c$ in $|\eta|<2.4$ are considered.

The left and middle panels of Fig.~\ref{fig:jetv2} shows the $\Delta\phi^{*}$ distribution of charged particles with $0.3<j_{\rm T}<3.0$ GeV/$c$ and $|\Delta\eta^{*}|>2$. The CMS data is presented in the left panel, whereas the results of MC simulation with PYTHIA8 and SHERPA are shown in the middle panel of Fig.~\ref{fig:jetv2}. In the CMS data, an indication of near-side enhancement is seen in the lower panel with $\langle N_{\rm ch}^{\rm j}\rangle=101$, which is absent in the upper panel with smaller $\langle N_{\rm ch}^{\rm j}\rangle=26$. This enhancement is a testimony of a possible `ridge-like' structure and indicates a possible collective behavior. Moreover, the enhancement becomes less obvious for the results with PYTHIA8 and SHERPA simulations.

The right panel of Fig.~\ref{fig:jetv2} shows the elliptic anisotropy ($v_{2}^{*}$) in the jet basis as a function of $N_{\rm ch}^{\rm j}$. Possible $j_{\rm T}$ dependence is also investigated with a comparison of upper and lower panels with $0.3<j_{\rm T}<3.0$ GeV/$c$ and $0.5<j_{\rm T}<3.0$ GeV/$c$, respectively. In both the upper and lower panels, the MC event generators qualitatively explains the experimental CMS measurements upto $N_{\rm ch}^{\rm j}\sim 80$. The PYTHIA8 and SHERPA predict a monotonic decrease throughout $N_{\rm ch}^{\rm j}$, which persists beyond $N_{\rm ch}^{\rm j}\sim 80$, which does not explain the CMS measurements which shows a rise beyond $N_{\rm ch}^{\rm j}$. This non-monotonic increase is unexpected when few-body processes are the dominating sources of particle correlations, as in PYTHIA8 and SHERPA. Thus, this may indicate the onset of non-perturbative QCD process of particle fragmentation in vacuum and emergence of collective effects driven by final state re-scatterings, as stated in Refs.~\cite{Baty:2021ugw, CMS:2023iam}. The understanding of this phenomenon requires in-depth theoretical calculations and experimental measurements, which includes but not limited to the event shape analysis separating the pQCD dominated processes from non-pQCD dominated processes and separation of dijet events and events with multi-jet topology.

{\it This subsection highlights that for high multiplicity jets in pp collisions, CMS data show a near-side enhancement in the azimuthal correlation, which is a feature usually referred to as collective-like. These measurements point to additional collective-like correlations or final state rescatterings in high multiplicity jet environments and motivate event shape-based selections to separate perturbative multi-jet topologies from soft contributions in a controlled way.}

\subsection{Search for jet-quenching in small systems}

\begin{figure*}[ht!]
\begin{center}
\includegraphics[width = 0.98\linewidth]{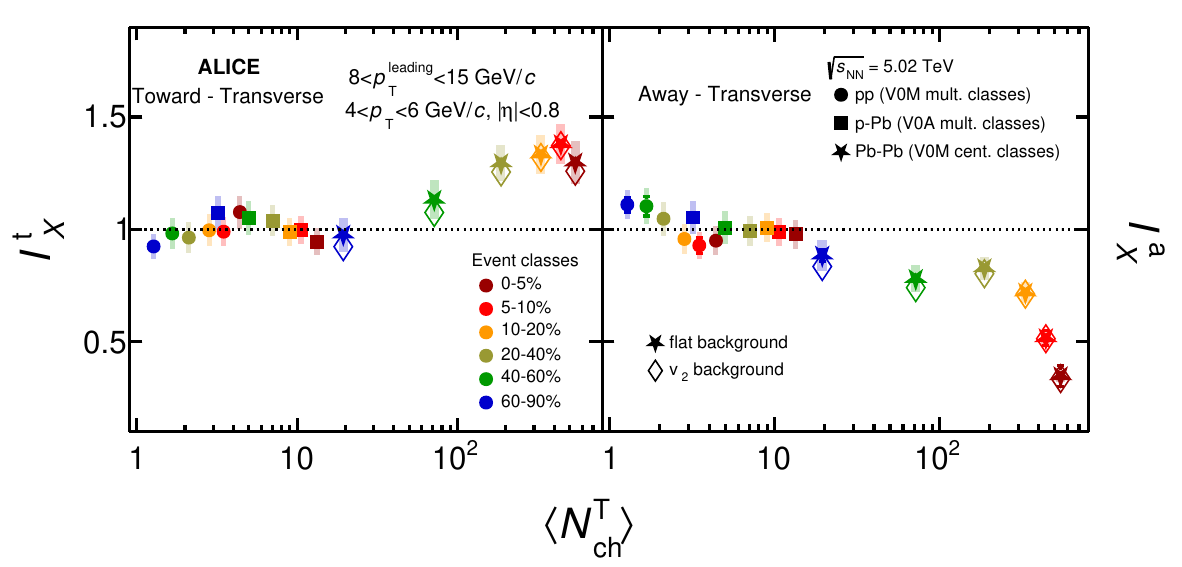}
\caption{$I^{t}_{X}$ (left panel) and $I^{a}_{X}$ (right panel) as a function of average charged-particle in transverse region in $4<p_{\rm T}<6$~GeV/c for different multiplicity classes in $pp$, p--Pb and Pb--Pb collisions at $\sqrt{s_{\rm NN}}$ = 5.02 TeV~\cite{ALICE:2022qxg}.}
\label{fig:IAANTALICE}
\end{center}
\end{figure*}

Let us now move to one of the missing signals of QGP behavior in small systems, $i.e.,$ jet quenching or jet-like region modification. So far, jet quenching effects have not been observed in small collision systems~\cite{ALICE:2018vuu,ALICE:2017svf, ALICE:2023plt, ALICE:2023jye}. 

The possible modification of the charged particles produced in the hard scattering in $pp$, p--Pb, and Pb--Pb collisions, the $p_{\rm T}$ distributions in the toward and away regions can be obtained after the subtraction of the $p_{\rm T}$ spectra in the transverse region. The subtracted yields (${\rm{d}}N_{\rm{ch}}^{\rm{st,sa}}/{\rm{d}}p_{\rm T}$) can be further normalised to those measured in minimum-bias (MB) $pp$ collisions, which can be sensitive to jet quenching effects. The ratio can be given as,
\begin{equation}
I_{X}^{\rm{t,a}} =
\frac{({\rm{d}}N_{\rm{ch}}^{\rm{st,sa}}/{\rm{d}}p_{\rm T})|_{X}}{ ({\rm{d}}N_{\rm{ch}}^{\rm{st,sa}}/{\rm{d}}p_{\rm T})|_{\rm {pp,MB}}},
\label{eq0}
\end{equation}
where $X$ is the collision system and the event multiplicity class. In Eq.~\eqref{eq0}, the superscripts `t' and `a' stand for the $p_{\rm T}$-differential yields in the toward and away regions, respectively. Further, the superscripts `st' and `sa' denote the $p_{\rm T}$-differential yields in the toward and away regions, respectively, obtained after the subtraction of $p_{\rm T}$ spectra in the transverse region. 
With this ratio, the hard process $p_{\rm T}$ spectra in the toward and away regions are isolated, and thus allowing one to study possible jet-like region modifications to the produced particles due to medium effects in  $pp$, p--Pb, and Pb--Pb collisions. In heavy-ion collisions, this ratio is similar to the studies using the quantity $I_{\rm AA}$~\cite{Adams:2006yt,ALICE:2011gpa,ALICE:2016gso}, where jets produced in the early stages of the collision propagate through the hot and dense QGP. Their interaction with the medium leads to parton-energy loss, which is often referred to as jet quenching~\cite{Qin:2015srf}. This effect manifests in the suppression of the high-\pt~charged-particle yield in the away region~\cite{Adams:2006yt,ALICE:2011gpa}. ALICE experiment measured $I^{t}_{X}$ and $I^{a}_{X}$ using an event classifier, $i.e.,$ average charged-particle multiplicity in transverse region in 4 $<p_{\rm T}<$ 6 GeV/c for different multiplicity classes in $pp$, p--Pb and Pb--Pb collisions at $\sqrt{s_{\rm NN}}$ = 5.02 TeV~\cite{ALICE:2022qxg}, as shown in Fig.~\ref{fig:IAANTALICE}. One observes that the Pb--Pb collisions clearly show jet-like region modifications, while for both $pp$ and p--Pb collisions, the $I^{t}_{X}$ and $I^{a}_{X}$ values are found to be unity, indicating no jet-like region modification within the precision of Run 2 statistics. 

\begin{figure}
    \centering
    \includegraphics[width=0.75\linewidth]{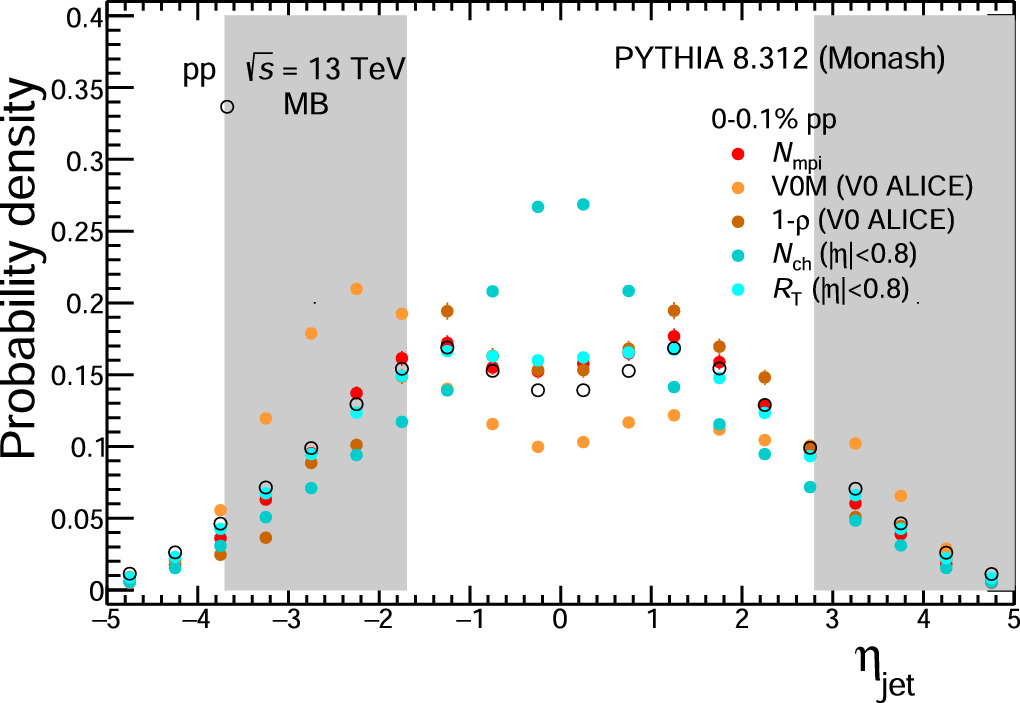}
    \caption{Probability distribution of jets recoiling from a high-$p_{\rm T}$ hadron for various event-activity estimators in the 0–0.1\% $pp$ collisions~\cite{Mendez:2025dqz}}
    \label{fig:recoiljet}
\end{figure}

In Run 3 and Run 4 of the LHC, thanks to huge statistics, one can focus on the sample with very high-multiplicity $pp$ collisions (0-0.1\% event class). As highlighted before, the search for jet quenching in small systems at high-multiplicity would require the event estimators to bias the jet fragmentation as little as possible. To study the correlation between event activity estimators and jet fragment multiplicity, following a similar strategy that was adopted in ALICE measurement of di-jet acoplanarity~\cite{ALICE:2023plt}, the authors in Ref.~\cite{Mendez:2025dqz} have employed FastJet~\cite{Cacciari:2011ma} in the PYTHIA8 simulations. For this study, the leading particle is chosen within $|\eta| <$  0.8 and the transverse momentum within 10–30 GeV/c, and the recoil jet is reconstructed using the anti-$k_{\rm T}$ algorithm implemented in FastJet. The transverse momentum of the recoil jet is required to be greater than 25 GeV/c, and this should be found in the opposite hemisphere relative to the leading particle. Figure~\ref{fig:recoiljet} shows the probability distribution of jets recoiling from a high-$p_{\rm T}$ hadron for various event-activity estimators in the 0–0.1\% $pp$ collisions. Here, the V0M distribution (ALICE estimator with both V0 detectors) shows an asymmetry, indicating a bias toward events with a recoil jet in either of the V0 detector regions. Since the acceptance of one of the V0 detectors of ALICE is narrower than that of others, there is a significant enhancement in one of the regions. The simulation results with charged-particle multiplicity at mid-rapidity and $R_{\rm T}$ show an enhancement in the region where these estimators are calculated, which indicates an autocorrelation or a bias towards coplanar jets. Charged particle flattenicity event estimator shows the absence of any of these biases, and it has a resemblance to results from minimum bias collisions and has a similar behavior to that of $N_{\rm MPI}$ estimator. This demonstrates its potential to improve future jet acoplanarity measurements in experiments~\cite{ALICE:2023plt}.

{\it This subsection highlights the ALICE measurements of event shape and region-based selections that are used to search for jet quenching in small collision systems by forming subtracted spectra in the toward and away regions and normalising to minimum bias pp through the ratios $I_X^{t}$ and $I_X^{a}$. Pb--Pb collisions show clear suppression consistent with parton energy loss, whereas p--Pb and pp collisions remain consistent with unity within Run 2 precision, indicating no observable jet quenching signal so far. The subsection also highlights that future studies should target very high multiplicity pp collisions while minimising estimator-induced biases in jet fragmentation, for example, by preferring charged-particle flattenicity-like measures when studying acoplanarity and recoil distributions.}

\subsection{Particle production in jet-like region}

As discussed in the previous sections, the absence of conclusive evidence of jet quenching in pp and p--Pb collisions questions the nature of the system created in these collision systems and the physical processes that govern the energy loss in these environments. Often, it is assumed that the energy loss effects in small collision systems are small enough to be measured in current experimental precision. Additionally, it is also argued that the significant selection biases may hinder these measurements. Therefore, to isolate the elusive jet modifications in small systems from the fluctuating UE background, different experimental techniques have been proposed. One of these methods incorporates the definition of in-jet particle production as the difference in the particle yields between the toward and transverse regions~\cite{ALICE:2023yuk, Peng:2025mpf, Sahu:2025mla}, i.e.,
\begin{equation}
    N^{\text{In-Jet}}=N^{\text{Toward}}-N^{\text{Transverse}}.
\end{equation}
This difference in the yield between the toward region, possessing particle production through fragmentation of jets and UE activity, and the transverse region, having dominating contributions from UE activity, would statistically reduce the influence of background UE activity.
\begin{figure}[ht!]
    \centering
    \includegraphics[width=0.9\linewidth]{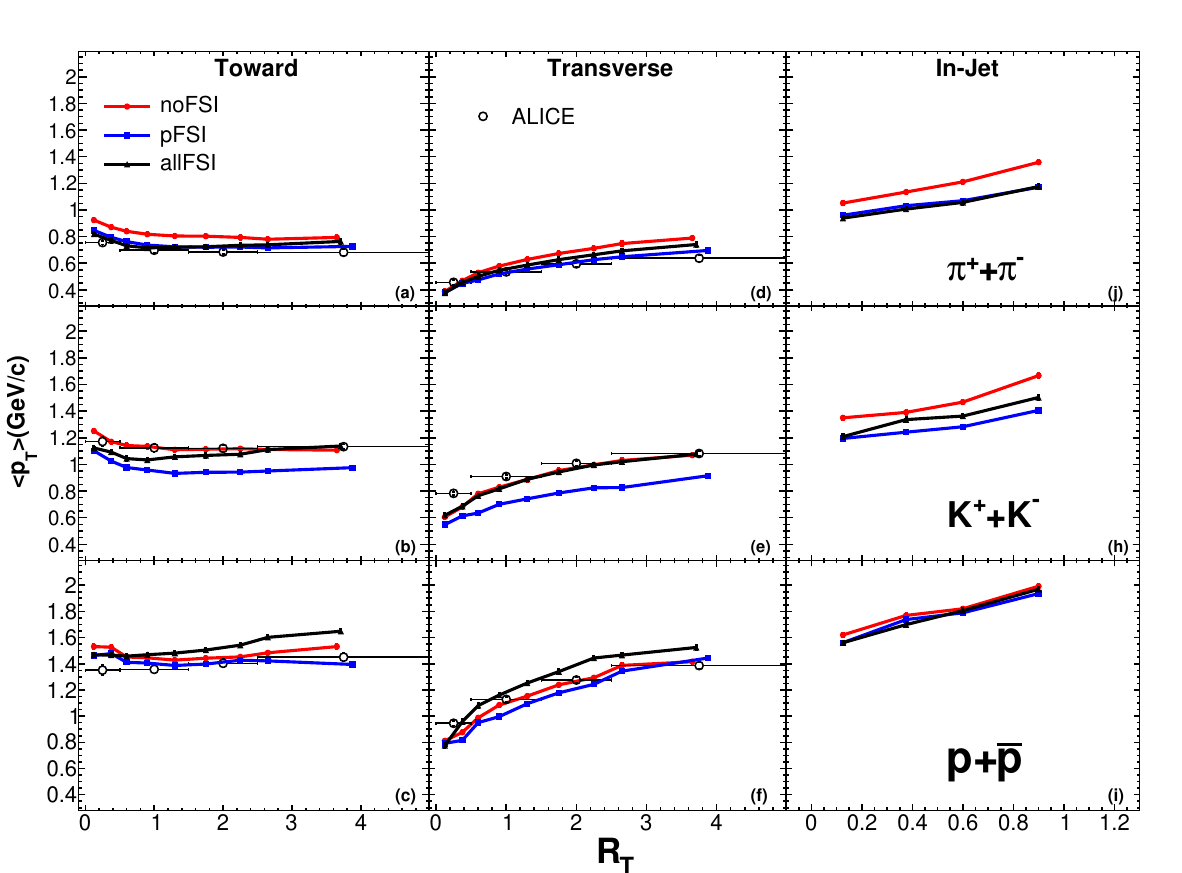}
    \caption{Average transverse momentum of $\pi^{+}+\pi^{-}$, $K^{+}+K^{-}$ and $p+\bar{p}$ as a function of $R_{\rm T}$ in toward, transverse and in-jet regions in $pp$ collisions at $\sqrt{s}=13$ TeV using PYTHIA8+AMPT and compared with corresponding measurements from ALICE~\cite{ALICE:2023yuk, Peng:2025mpf}.}
    \label{fig:injetavgpTvsRT}
\end{figure}

Figure~\ref{fig:injetavgpTvsRT} shows toward, transverse and in-jet $\langle p_{\rm T}\rangle$ as a function of $R_{\rm T}$ for charged pions, kaons and protons in $pp$ collisions at $\sqrt{s}=13$ TeV using AMPT based on PYTHIA8 initial conditions (PYTHIA8+AMPT)~\cite{Peng:2025mpf}. Here the results are shown considering the following scenarios and compared with experimental measurements with ALICE~\cite{ALICE:2023yuk}.
Here, noFSI refers to both partonic and hadronic rescatterings being disabled (no final state interactions), pFSI refers to only partonic rescattering being enabled, and the hadronic rescattering is disabled (partonic final state interaction), and allFSI refers to both partonic and hadronic scatterings being enabled (all final state interactions).

From Fig.~\ref{fig:injetavgpTvsRT}, it can be observed that $\langle p_{\rm T}\rangle$ in the toward and transverse regions is affected by the final state interactions (FSI). With the inclusion of partonic FSI, $\langle p_{\rm T}\rangle$ becomes smaller, indicating the parton-level energy loss. In contrast, the effect of hadron-level FSI is found to be substantial only for charged kaons and protons, while the effect is small for the charged pions. The impact of hadron-level FSI is strongest for the protons, where a rise is observed with $R_{\rm T}$ in the toward region. In-jet $\langle p_{\rm T}\rangle$ is higher as compared to toward and transverse regions, which indicates a strong jet-fragmentation effect. Additionally, in-jet $\langle p_{\rm T}\rangle$ for protons is less sensitive to FSI, while in comparison, charged pions and kaons show a FSI dependence similar to that in toward and transverse regions. This difference in the in-jet $\langle p_{\rm T}\rangle$ dependence on FSI for different particles may arise due to the fact that the energy loss modification is mainly restricted to the low-$p_{\rm T}$ regions and $\langle p_{\rm T}\rangle$ for protons are large enough to be affected by the variations in FSI.

\begin{figure}
    \centering
    \includegraphics[width=0.9\linewidth]{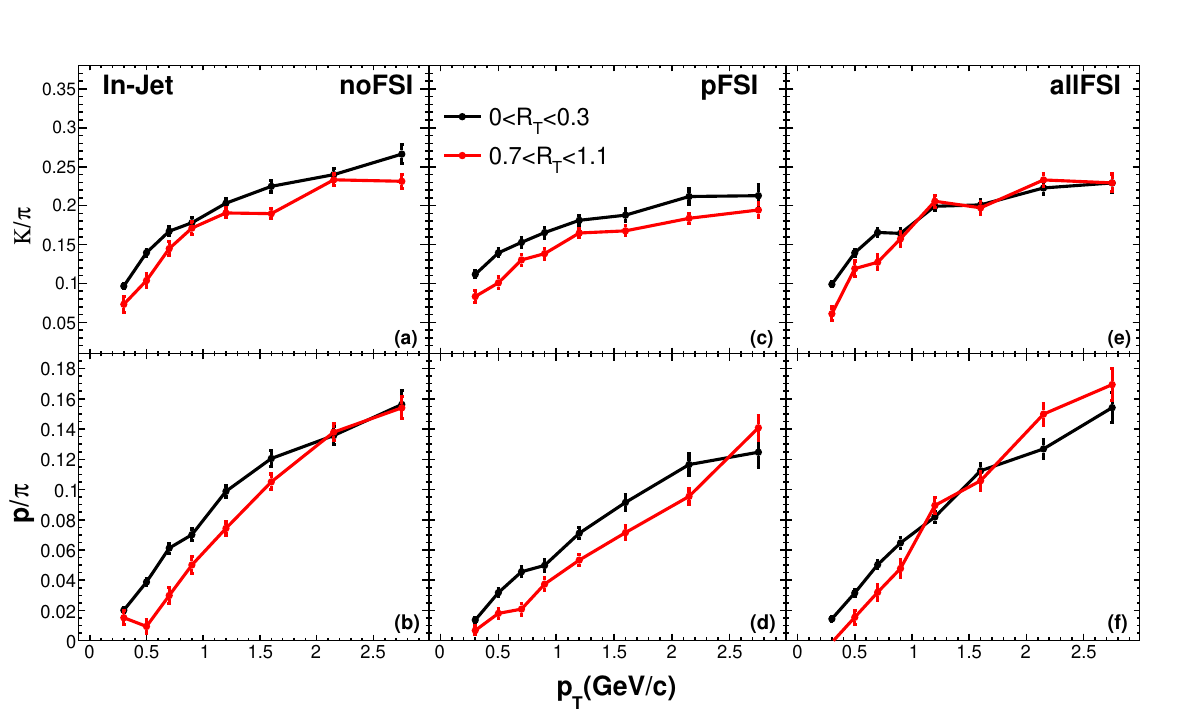}
    \caption{In-jet $p_{\rm T}$-differential yield ratios, $K/\pi$ (upper) and $p/\pi$ (lower) for different event classes selected based on $R_{\rm T}$ in $pp$ collisions at $\sqrt{s}=13$ TeV using PYTHIA8+AMPT~\cite{Peng:2025mpf}.}
    \label{fig:injetPRvsRT}
\end{figure}

To obtain insight into the production of baryon versus meson and strange to non-strange production, Fig.~\ref{fig:injetPRvsRT} shows in-jet $p_{\rm T}$-differential yield ratios of charged kaons to pions (upper) and protons to pions (lower) for different FSI in different regions of $R_{\rm T}$ in $pp$ collisions at $\sqrt{s}=13$ TeV using PYTHIA8+AMPT. As imparted in Ref.~\cite{Peng:2025mpf}, the in-jet particle yield ratios are usually smaller than those of toward regions, which indicates that the particle production dominated by pure jet-fragmentation results in a smaller baryon-to-meson and strange-to-nonstrange production ratio. For the case with noFSI, both $p/\pi$ and $K/\pi$ become smaller with an increase in $R_{\rm T}$ throughout the considered $p_{\rm T}$ regions, which can be interpreted as the result of depleted gluon radiations in the leading jet with an increase in UE activity. The limited phase space for gluon radiation of the leading jet reduces the string tension during string fragmentation, thereby limiting the production of strange quarks and diquarks. The inclusion of partonic FSI diminishes the particle ratios even further, as shown in the middle panel. Conversely, the inclusion of hadronic FSI enhances the particle ratios at higher $p_{\rm T}$ in the large $R_{\rm T}$ events, where a crossing behaviour is observed between low and high $R_{\rm T}$ events. This $R_{\rm T}$ dependent crossing at intermediate $p_{\rm T}$ can be considered as an important signature of jet-medium interaction in high multiplicity small system collisions, which can potentially reveal effects of energy loss in small systems. This study shows the importance of both partonic and hadronic interactions that shape the jet modification feature in small systems~\cite{Peng:2025mpf}.

{\it This subsection defines an in-jet observable as the difference between yields in the toward and transverse regions, so that the underlying event largely cancels and the jet-related component is isolated. The results highlight the need for transverse region subtraction combined with less biased activity estimators such as charged-particle flattenicity as a practical strategy to search for or constrain quenching-like effects in pp and p--Pb collisions within current precision.}

{\it The hard QCD section highlights how event shape selections are useful when a hard scale is present. In heavy flavor and jet-based studies, they provide precision tools to separate the hard recoil from the soft environment and are useful in testing and tuning models in topologies with a well-defined pQCD scale. The correlation measurements in the jet frame for high multiplicity jets display near-side structure that standard simulations like PYTHIA do not reproduce, which motivates topology-based selections in future studies. This section also highlights that future searches for jet quenching in small collision systems should target very high multiplicity pp collisions while minimising estimator-induced biases in jet fragmentation.}

\section{Studies in heavy-ion collisions}
\label{sec:HIC}

While discussing the studies related to small systems, it is now equally important to understand the dynamics of event shape observables in heavy-ion collisions, where the formation of QGP is most likely to occur. To bring all collision systems on an equal footing, we will now move our discussion towards extending the event classifiers to heavy-ion collisions. So far, there is no experimental study in such directions, except using flow vectors~\cite{ATLAS:2015qwl}. However, there are quite a few theoretical studies involving transverse spherocity in heavy-ion collisions. We provide the following compilation of some of the important studies at both experimental and theoretical frontiers on heavy-ion collisions in the following sections.
%As a starting point, we discuss the implementation of reduced flow vectors followed by some explorations in phenomenological studies of transverse spherocity in heavy-ion collisions.  

\subsection{Two-particle correlations and anisotropic flow}
\label{sec:flowandtwoparticlecorrelation}

As stated earlier, the reduced flow vectors are one of the observables often used in heavy-ion collisions to select events based upon different magnitudes of anisotropic flow coefficients. For example, $q_2$, the second-order reduced flow vector, is often used to select events with smaller or larger values of elliptic flow. Similarly, the third-order reduced flow vector, $q_3$, is useful to identify the triangular events. In this section, we review some of the usages of reduced flow vectors to study particle correlations and anisotropic flow, and compare with those of event selections based on transverse spherocity.

\begin{figure*}
    \centering
    \includegraphics[width = 0.98\linewidth]{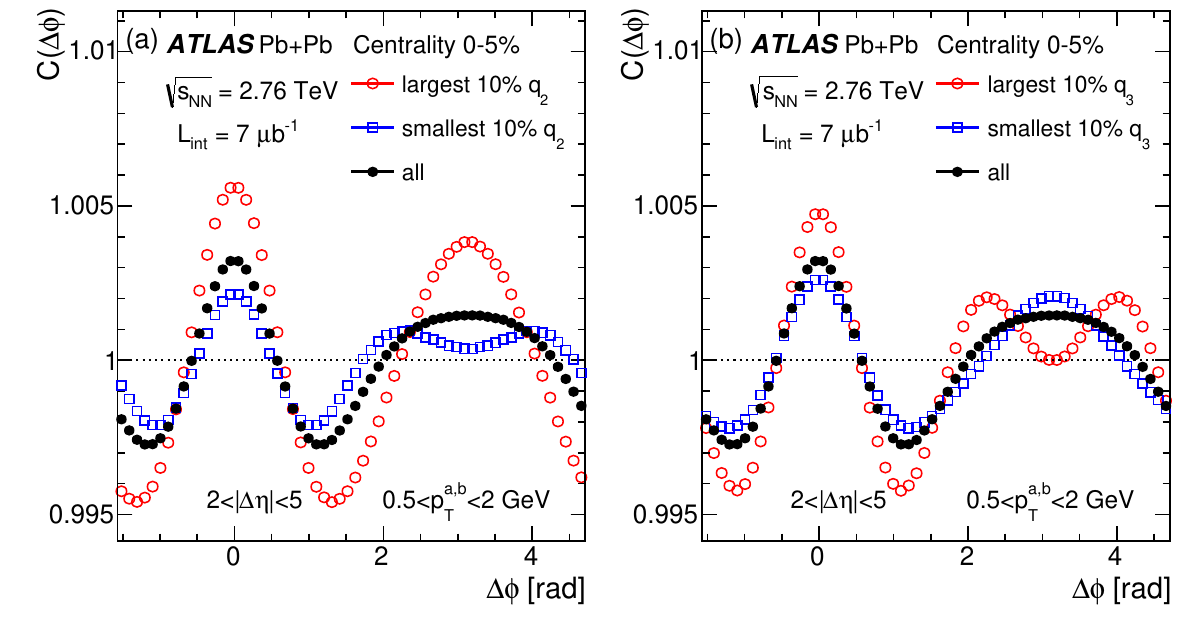}
    \caption{Azimuthal dependence of two-particle correlation function ($C(\Delta\phi)$) versus $q_{2}$ (left) and $q_3$ (right) for 0-5\% centrality class in Pb-Pb collisions at $\sqrt{s_{\rm NN}}=2.76$ TeV with ATLAS~\cite{ATLAS:2015qwl}.}
    \label{fig:qnvscorrelation}
\end{figure*}

Figure~\ref{fig:qnvscorrelation} shows the two-particle correlation function as a function of relative azimuthal angle for different percentiles of events selected with $q_2$ (left) and $q_3$ (right) in Pb-Pb collisions at $\sqrt{s_{\rm NN}}=2.76$ TeV with ATLAS~\cite{ATLAS:2015qwl}. In the left panel, one selects the events having the largest (red open circles) and smallest (blue open squares) elliptic events with $q_2$. One finds that the events with the largest values of $q_2$ show an amplified two-distinct peak structure as compared to when no $q_2$ based event selection was applied. Further, a three-peak structure in $C(\Delta\phi)$ is observed when the smallest $q_2$ events are selected, which corresponds to events dominated by triangular flow. Similarly, in the right panel, the two-particle correlation function as a function of relative azimuthal angle is shown for different event percentiles of $q_3$. Here, the behavior is opposite to what one observes in the left panel, where events are selected based on $q_2$. Two-peak structure in the away-side region of the two-particle correlation function for the largest $q_2$ events hints at a large contribution of triangular flow, $v_3$.

\begin{figure*}
    \centering
    \includegraphics[width = 0.43\linewidth]{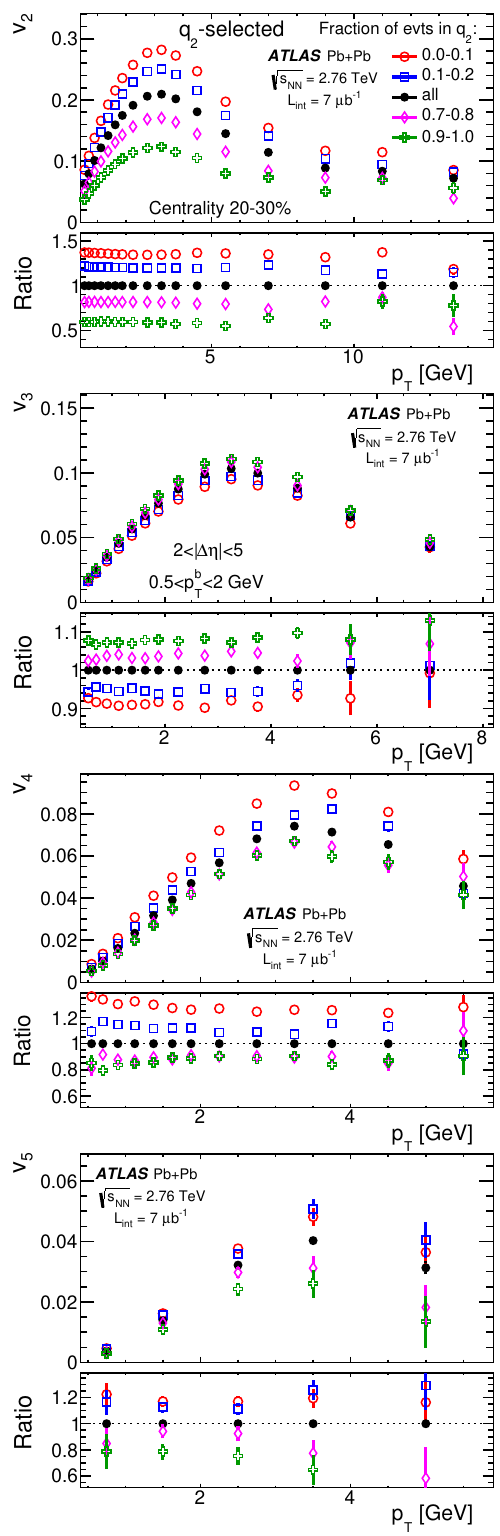}
    \includegraphics[width = 0.43\linewidth]{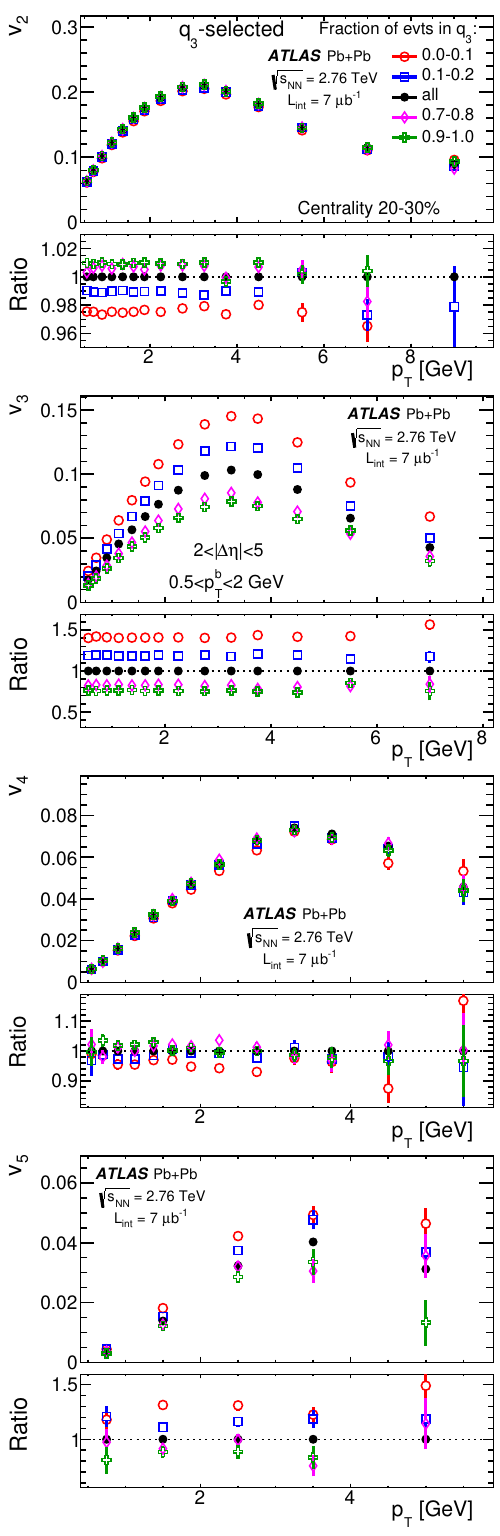}
    \caption{$p_{\rm T}$-differential anisotropic flow coefficients versus $q_{2}$ (left) and $q_3$ (right) in Pb-Pb collisions at $\sqrt{s_{\rm NN}}=2.76$ TeV with ATLAS~\cite{ATLAS:2015qwl}.}
    \label{fig:qnvsvn}
\end{figure*}

The observations made in Fig.~\ref{fig:qnvscorrelation} are reflected in the extracted flow coefficients. Fig.~\ref{fig:qnvsvn} shows $p_{\rm T}$-differential $v_2$, $v_3$, $v_4$ and $v_5$, from top to bottom, respectively, as a function of events selected with $q_2$ (left) and $q_3$ (right) in Pb-Pb collisions at $\sqrt{s_{\rm NN}}=2.76$ TeV with ATLAS~\cite{ATLAS:2015qwl}. As inferred from Fig.~\ref{fig:qnvscorrelation}, in Fig.~\ref{fig:qnvsvn}, one observes large $v_2$ and smaller $v_3$ for the largest $q_2$ events. Similarly, a higher $v_3$ and smaller $v_2$ are observed for events selected by $q_3$. These observations hint at an anti-correlation between $v_2$ and $v_3$. Further, the event selection with $q_2$ affects both $v_4$ and $v_5$, which shows higher values for the event with large $q_2$. This indicates a non-linear contribution of $v_2$ on $v_4$ and $v_5$. In contrast, the event selection with $q_3$ has a small impact on $v_4$, where large $v_3$ events possess small values of $v_4$. This can be attributed to the positive correlation between $v_2$ and $v_4$, and the anti-correlation between $v_2$ and $v_3$. Further, a positive correlation between $q_3$ and $v_5$ signifies a non-linear contribution of $v_3$ on $v_5$ measurements. Thus, the studies of reduced flow vectors in heavy-ion collisions are useful to quantify the contributions of $v_2$ and $v_3$ on higher-order anisotropic flow coefficients, which can also be removed following the method discussed in Ref.~\cite{ATLAS:2015qwl}. A similar measurement is also performed with ALICE~\cite{ALICE:2015lib}, where $v_{2}$ with event selection with $q_2$ from two different detectors. Here, $q_2$ is measured with the charged particles in both midrapidity using the TPC and forward rapidity using the ALICE V0C detectors. One finds that $q_2$ estimated with TPC selects events with stronger elliptic flow in the midrapidity region as compared to $q_2$ estimated with the ALICE V0C detector. The main contributor to this difference is attributed to the selectivity of $q_2$ from ALICE TPC, which is different from that of ALICE V0C. This is explicitly checked in Ref.~\cite{ALICE:2015lib} by relaxing the selection itself or rejecting a random fraction of tracks for the computation of $q_{2}$ from ALICE TPC. It is observed that the selection of (65-100)\% for the largest $q_{2}$ sample ((0-55)\% of the smallest $q_2$ sample) measured with ALICE TPC, or by randomly rejecting 70\% of the ALICE TPC tracks leads to average variation of $v_2$ in $0.2<p_{\rm T}<4$ GeV/c comparable to the one obtained by selecting standard largest (smallest) 10\% $q_2$ events measured with ALICE V0C~\cite{ALICE:2015lib}. 

Before moving ahead with the studies of particle correlations and anisotropic flow coefficients in heavy-ion collisions using transverse spherocity, let us study the variations in the distributions of transverse spherocity when migrating from pp to Pb--Pb collisions.

\begin{figure}[ht!]
\begin{center}
\includegraphics[scale=0.4]{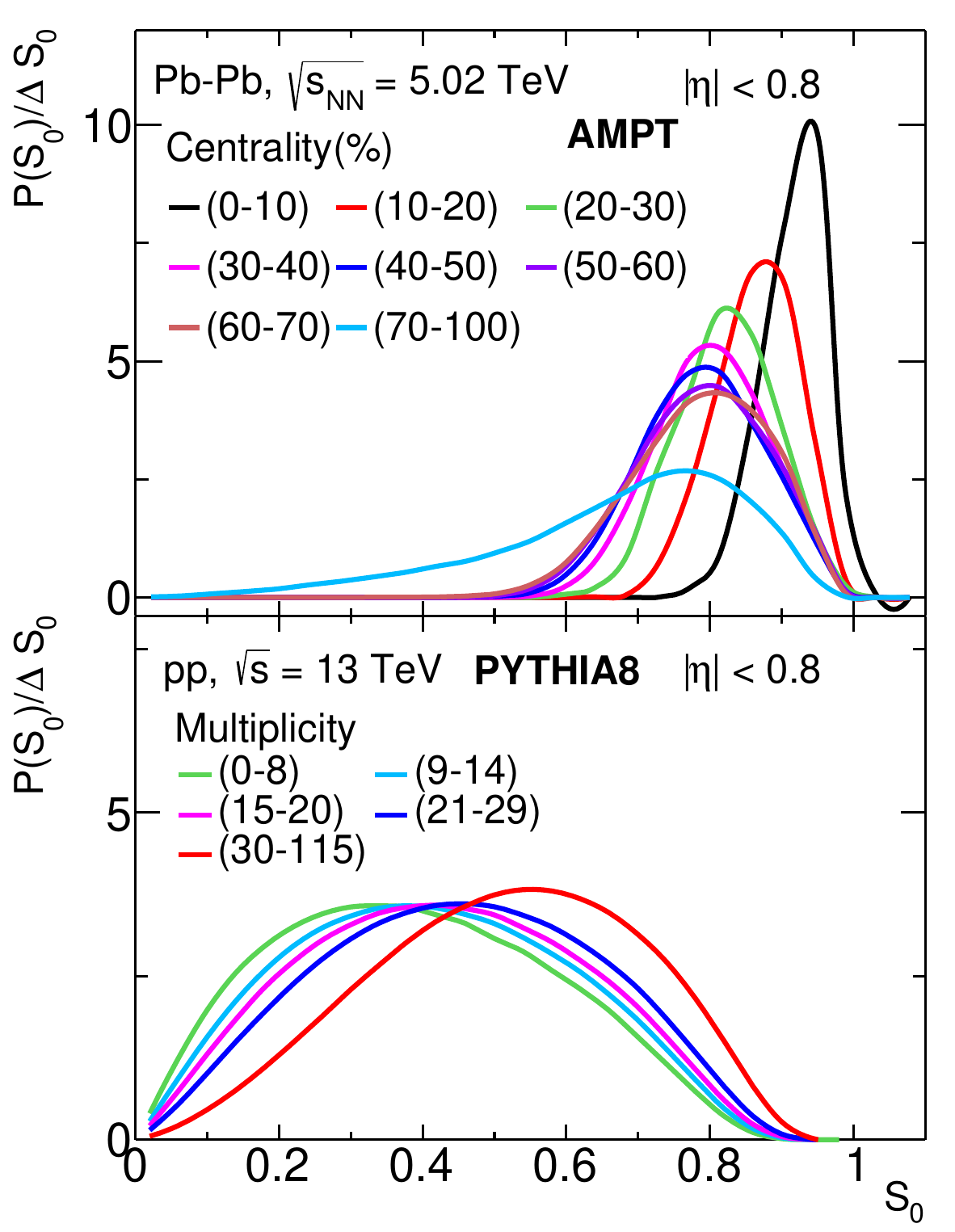}
\caption{Top panel: $p_{\rm T}$-weighted transverse spherocity ($S_{0}$) distribution for different classes of collision centrality in Pb--Pb collisions at $\sqrt{s_{\rm NN}}=5.02$ TeV using AMPT. Bottom panel: $p_{\rm T}$-weighted transverse spherocity ($S_{0}$) distribution for different multiplicity classes in V0 acceptance region of ALICE detector in $pp$ collisions at $\sqrt{s}=13$ TeV using PYTHIA8~\cite{Prasad:2021bdq}.}
\label{fig:ppvsPbPbsphero}
\end{center}
\end{figure}

Figure~\ref{fig:ppvsPbPbsphero} top (bottom) panel shows $S_{0}$ distribution for different centrality (multiplicity) classes in Pb--Pb ($pp$) collisions at $\sqrt{s_{\rm NN}}=5.02$ TeV ($\sqrt{s}=13$ TeV) obtained using AMPT (PYTHIA8). It is to be noted that the multiplicity in $pp$ collisions is estimated in the V0 acceptance of the ALICE detector during Run 1 and Run 2, and the centrality selection in AMPT for Pb--Pb collisions is performed using geometrical impact parameter slicing. In $pp$ collisions, $S_{0}$ distribution shifts towards the isotropic limit with increasing multiplicity selection; however, the peak of the distributions still lies towards the jetty event, except for the class having the highest multiplicity. This is due to the fact that a higher value of $N_{\rm ch}^{\rm fwd}$ corresponds to events having higher $N_{\rm mpi}$. A similar centrality dependence is observed in Pb--Pb collisions, where the distribution of transverse spherocity shifts towards the isotropic limits as one moves towards the central collisions. The peaks of the transverse spherocity distribution in Pb--Pb collisions always lie towards the extreme isotropic limit, where the formation of a QCD medium is viable. While both AMPT and PYTHIA8 have differences in the underlying processes, using transverse spherocity, one can probe for similar event topology, whether soft or hard interactions dominate underlying physical processes.

\begin{figure*}[ht!]
\begin{center}
\includegraphics[width = 0.98\linewidth]{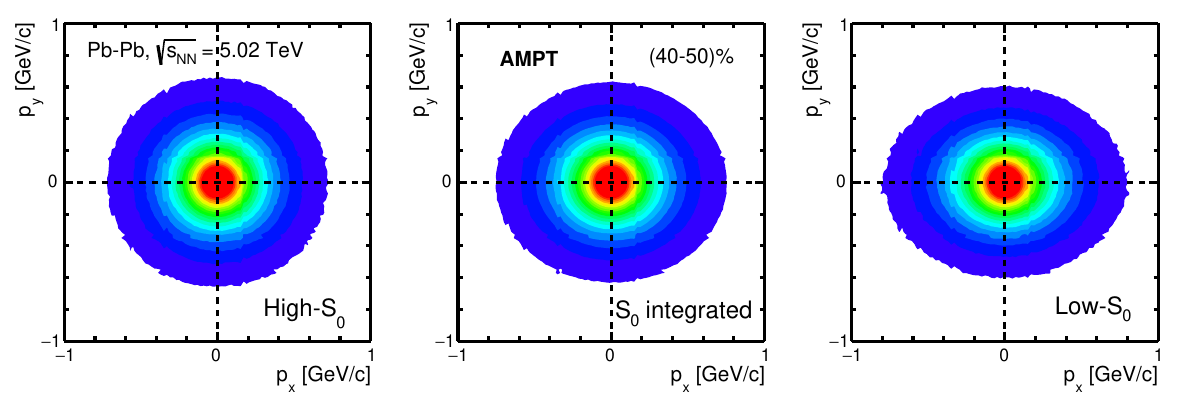}
\caption{Transverse momentum space correlation ($p_{x}$ vs $p_{y}$) for different classes of transverse spherocity and (40-50)\% centrality class in Pb--Pb collisions at $\sqrt{s_{\rm NN}}=5.02$ TeV using AMPT. High-$S_{0}$ and low-$S_{0}$ events have the highest 20\% and lowest 20\% value in the transverse spherocity distribution, respectively~\cite{Mallick:2020ium}.}
\label{fig:pxvspysphero}
\end{center}
\end{figure*}

Figure~\ref{fig:pxvspysphero} shows the transverse momentum space correlation ($p_{x}$ vs $p_{y}$) for different classes of $S_{0}$ in (40-50)\% Pb--Pb collisions at $\sqrt{s_{\rm NN}}=5.02$ TeV using AMPT. In the upper right figure, the $S_{0}$-integrated events, a clear momentum-space anisotropy can be seen, which indicates the presence of initial-stage spatial anisotropy. The finite initial state spatial anisotropy can be understood as the presence of an almond-shaped overlap region in the mid-central collisions, where the presence of a larger pressure gradient along the $x-$axis than the $y-$axis. This consequently leads to the emission of particles with larger $p_{x}$, i.e., $p_{x}>p_{y}$. As one moves towards the high-$S_0$ events, the correlation almost vanishes, showing a circular geometry. This indicates an absence of momentum-space anisotropy in the final state and almost zero elliptic flow for the high-$S_0$ events. However, for the low-$S_0$ events, a higher momentum space anisotropy is observed, where the transverse momentum correlation becomes more elliptical compared to the $S_0$-integrated events. This feature of the transverse spherocity testifies to its applicability from small systems to heavy-ion collisions and as a probe to understand the formation of a QCD medium.

\begin{figure}[ht!]
\begin{center}
\includegraphics[scale=0.4]{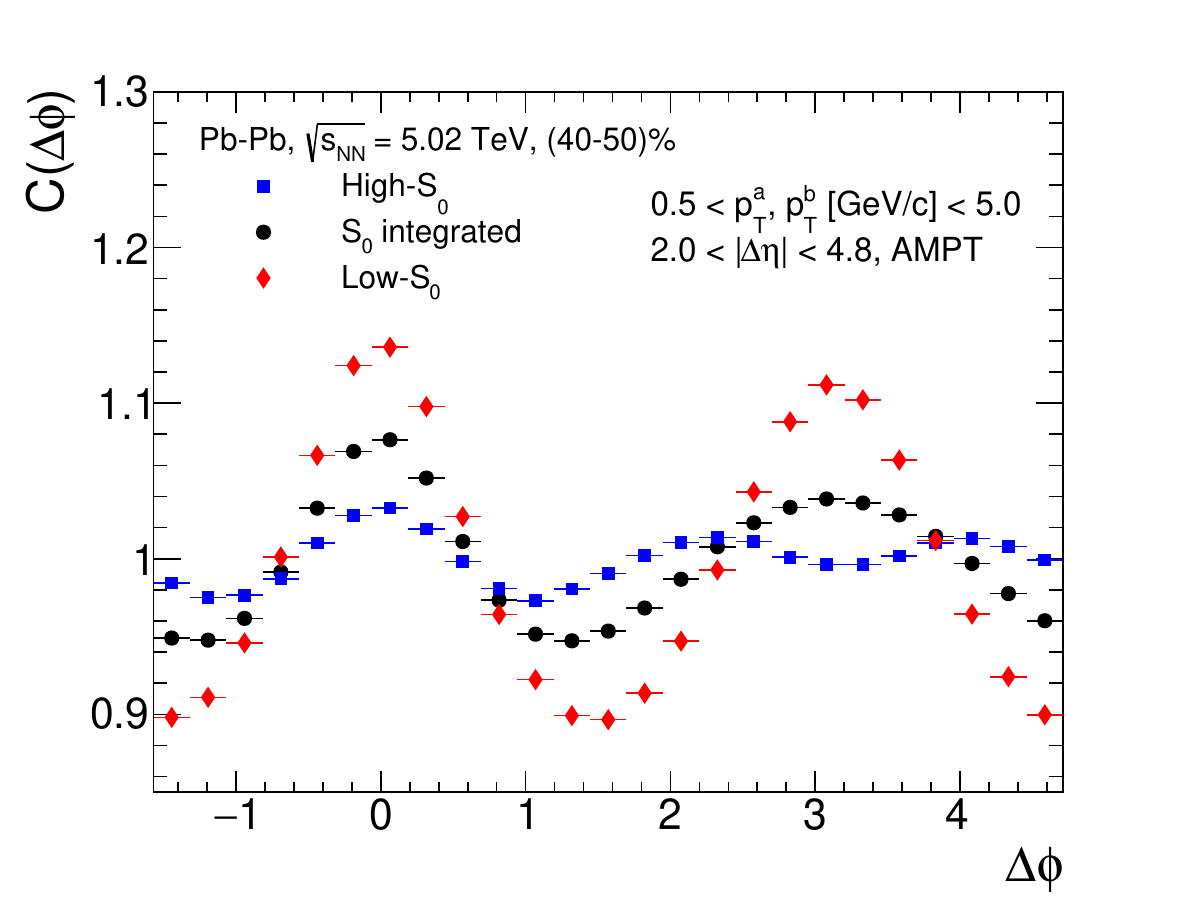}
\caption{Two-particle azimuthal correlation function for different transverse spherocity classes in (40-50)\% centrality Pb--Pb collisions at $\sqrt{s_{\rm NN}}=5.02$ TeV using AMPT~\cite{Mallick:2020ium, Prasad:2022zbr}.}
\label{fig:CDeltaPhisphero}
\end{center}
\end{figure}

One of the key signatures for collectivity is the anisotropic flow, which can be obtained using two-particle azimuthal correlations. As already highlighted in Fig.~\ref{fig:pxvspysphero}, the initial state spatial anisotropy can be probed using event shape observables, which translates to final state momentum anisotropy. So, the study of anisotropic flow with event shape would give an experimental handle to probe the collectivity.
Figure~\ref{fig:CDeltaPhisphero} shows the two-particle azimuthal correlation function ($C(\Delta\phi)$) as a function of relative azimuthal angle for different transverse spherocity classes in (40-50)\% centrality Pb--Pb collisions using AMPT. In Fig.~\ref{fig:CDeltaPhisphero}, the peak and width of the correlation function vary strongly with respect to the choice of transverse spherocity selection. This indicates that using transverse spherocity, one can distinguish events based on geometrical shapes. In addition, one can observe that for low-$S_0$ events, the two-particle azimuthal correlation function has a larger peak compared to $S_0$-integrated and high-$S_0$ events have the smallest peak structure. This indicates the presence of a larger elliptic flow for low-$S_0$ events. Furthermore, one observes two peaks in the away-side region for high-$S_0$ events, indicating the presence of a larger triangular flow in such events.

\begin{figure*}[ht!]
\begin{center}
\includegraphics[width = 0.49\linewidth]{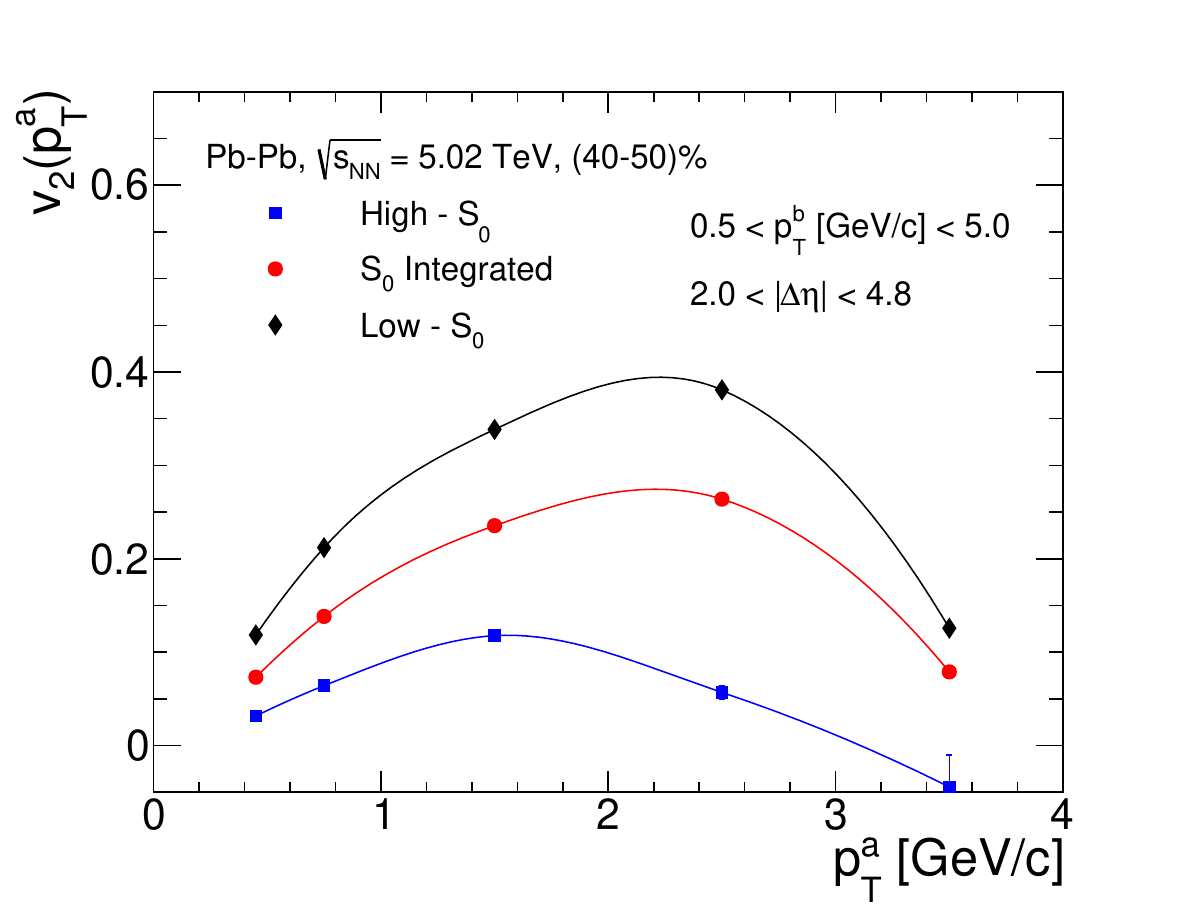}
\includegraphics[width = 0.49\linewidth]{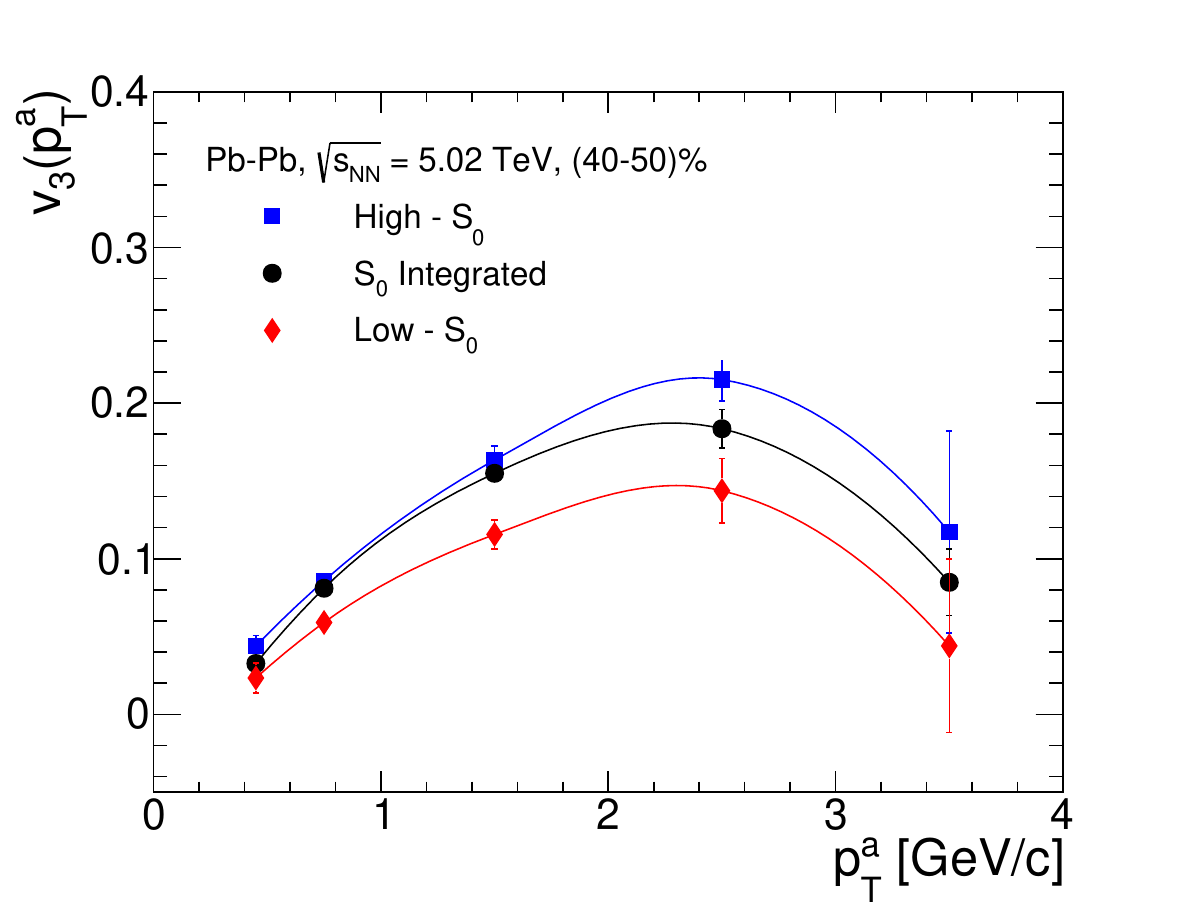}
\caption{Elliptic (top) and triangular (bottom) flow as a function of transverse momentum in (40-50)\% centrality for different spherocity selections in Pb--Pb collisions at $\sqrt{s_{\rm NN}}=5.02$ TeV using AMPT~\cite{Prasad:2022zbr, Mallick:2020ium}.}
\label{fig:PbPbvnptsphero}
\end{center}
\end{figure*}

\begin{figure*}[ht!]
\begin{center}
\includegraphics[width = 0.49\linewidth]{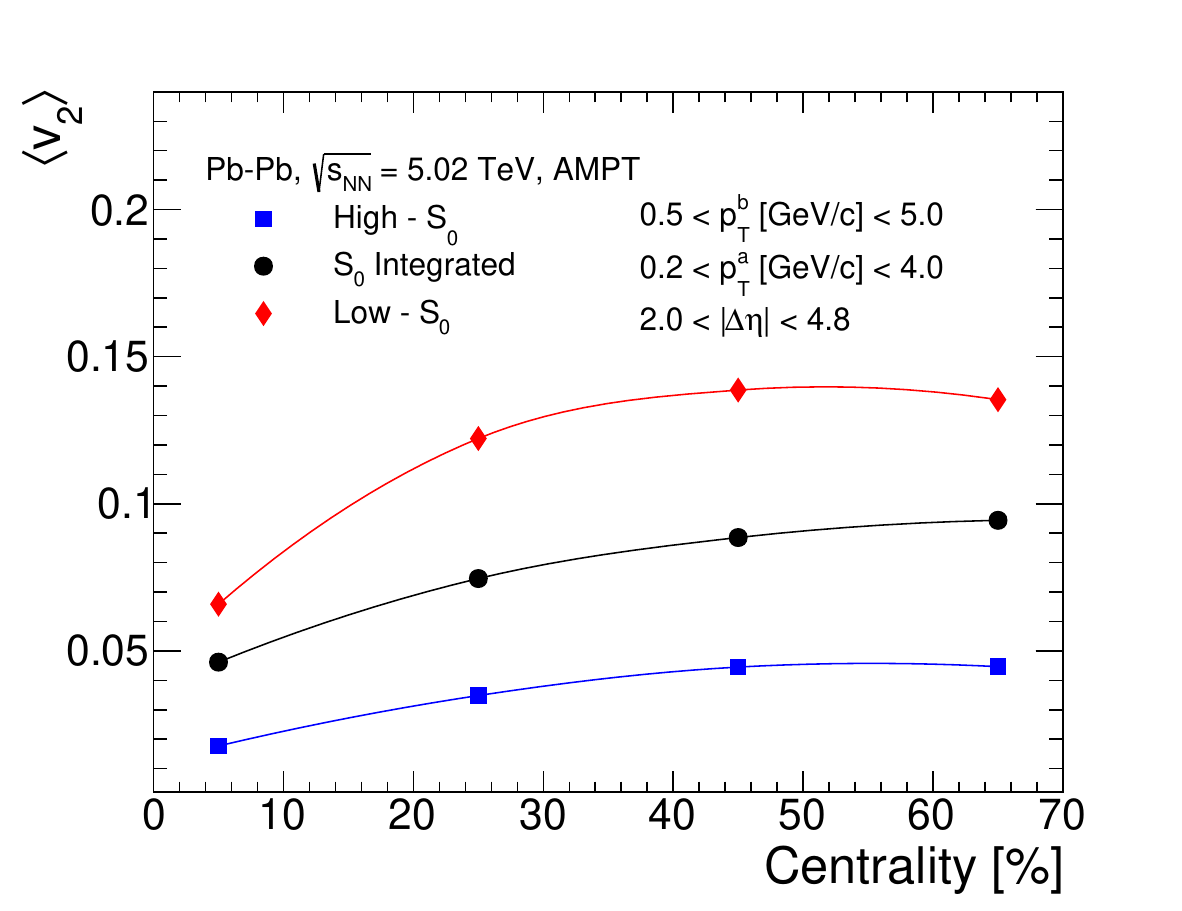}
\includegraphics[width = 0.49\linewidth]{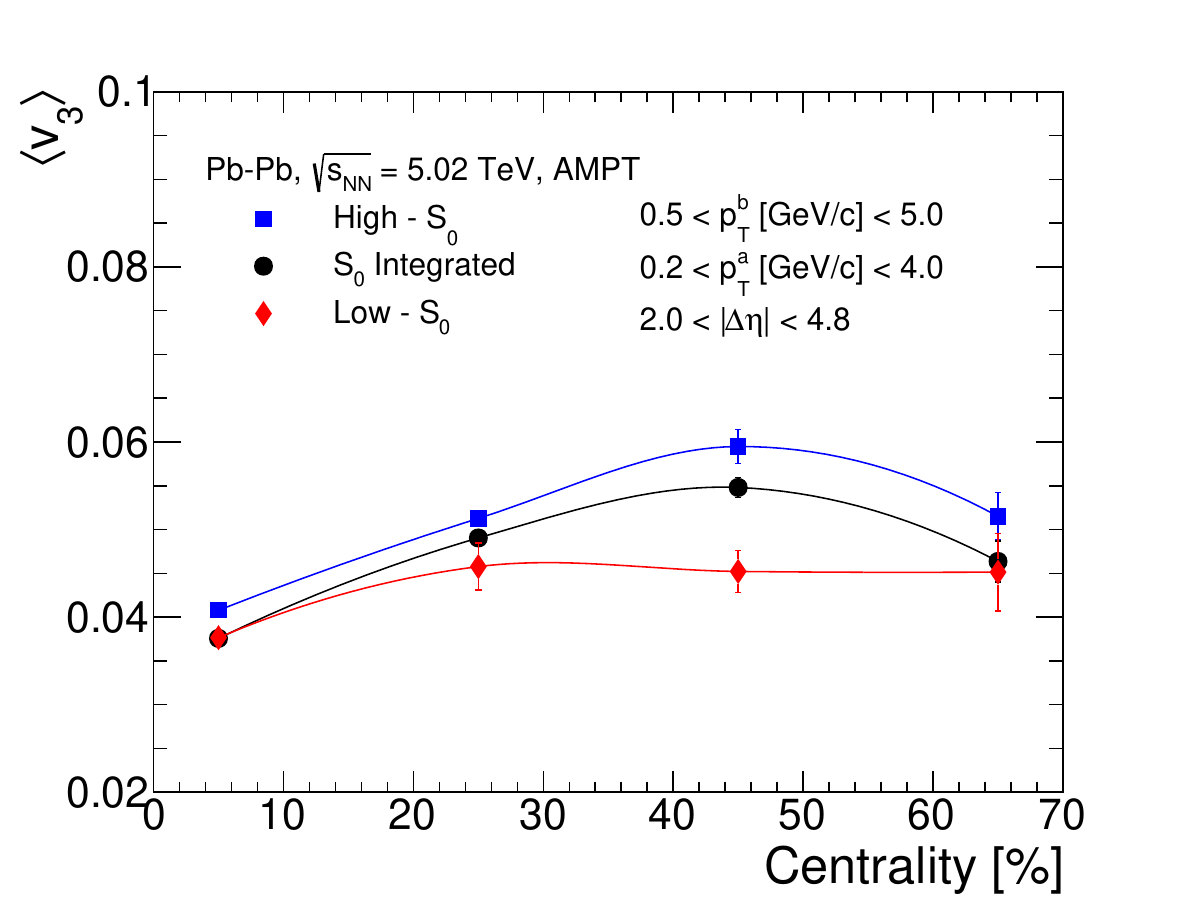}
\caption{Elliptic (top) and triangular (bottom) flow as a function of centrality for different spherocity selections in Pb--Pb collisions at $\sqrt{s_{\rm NN}}=5.02$ TeV using AMPT~\cite{Prasad:2022zbr}.}
\label{fig:PbPbvncentsphero}
\end{center}
\end{figure*}

Figure~\ref{fig:PbPbvnptsphero} shows the elliptic flow ($v_{2}$) (top) and triangular flow ($v_{3}$) (bottom) as a function of transverse momentum for different classes of transverse spherocity in Pb--Pb collisions at $\sqrt{s_{\rm NN}}=5.02$ TeV using AMPT. The results in Fig.~\ref{fig:PbPbvnptsphero} are limited to (40-50)\% centrality class only as the mid-central events are expected to have the highest contribution from both elliptic flow and triangular flow, which could help us to identify the transverse spherocity dependence clearly. As inferred from Fig.~\ref{fig:CDeltaPhisphero}, where the signals for the two-particle correlation function are highest for the low-$S_0$ events, indicating a larger value of elliptic flow. A similar direct conclusion can be found in Fig.~\ref{fig:PbPbvnptsphero}. One finds that, as events go from high-$S_0$ events to low-$S_0$ events, the value of elliptic flow increases. This shows that elliptic flow has a direct anti-correlation with transverse spherocity selection. However, in the bottom panel of Fig.~\ref{fig:PbPbvnptsphero}, the triangular flow is found to have the highest value for the high-$S_0$ class of events and possesses a minimum value for the low-$S_0$ events, as inferred from Fig.~\ref{fig:CDeltaPhisphero}. Triangular flow is found to have a positive correlation with the transverse spherocity selection.

Figure~\ref{fig:PbPbvncentsphero} shows the elliptic flow ($v_{2}$) (top) and triangular flow ($v_{3}$) (bottom) as a function of centrality for different classes of transverse spherocity in Pb--Pb collisions at $\sqrt{s_{\rm NN}}=5.02$ TeV using AMPT. In heavy-ion collisions, the pressure gradient formed during the collision is responsible for transforming the initial spatial anisotropy to the final momentum space anisotropy. In most central collisions, the collision geometry is almost spherical, which is reflected in the low value of anisotropic flow coefficients. As one moves towards the mid-central collisions, the value of both elliptic and triangular flow starts to rise because the collision overlap region retains an almond-shaped structure, as seen in the figure. However, for the peripheral collisions, due to the low system size, the lifetime of the QCD medium is very short, which leads to a lower value of anisotropic flow. The important thing to notice in the figure is that transverse spherocity plays a crucial role in the event selections for the anisotropic flow coefficients. This is because the anisotropic flow coefficients are found to be very sensitive to transverse spherocity selection.

In Ref.~\cite{Sambataro:2022sns}, the authors have presented a study of event shape engineering of D meson azimuthal anisotropy with reduced flow vectors. This study highlights a correlation between the heavy flavor azimuthal anisotropy and the collective expansion of bulk matter. In addition, they also highlight the increasing sensitivity of the event shape with the collision centrality. 

{\it Reduced flow vectors provide a useful way to separate events by geometry in heavy-ion collisions, and the selections visibly highlight the variation in two-particle correlations and elliptic flow. ATLAS measurements highlight that the large reduced flow vector $q_2$  enhances the elliptic flow. Model studies with transverse spherocity in Pb--Pb collisions reproduce the same patterns, supporting topology-based selection as a robust handle on final state anisotropy.}

\subsection{Eccentricity and elliptic flow fluctuations}
\begin{figure*}
    \centering
    \includegraphics[width=0.49\linewidth]{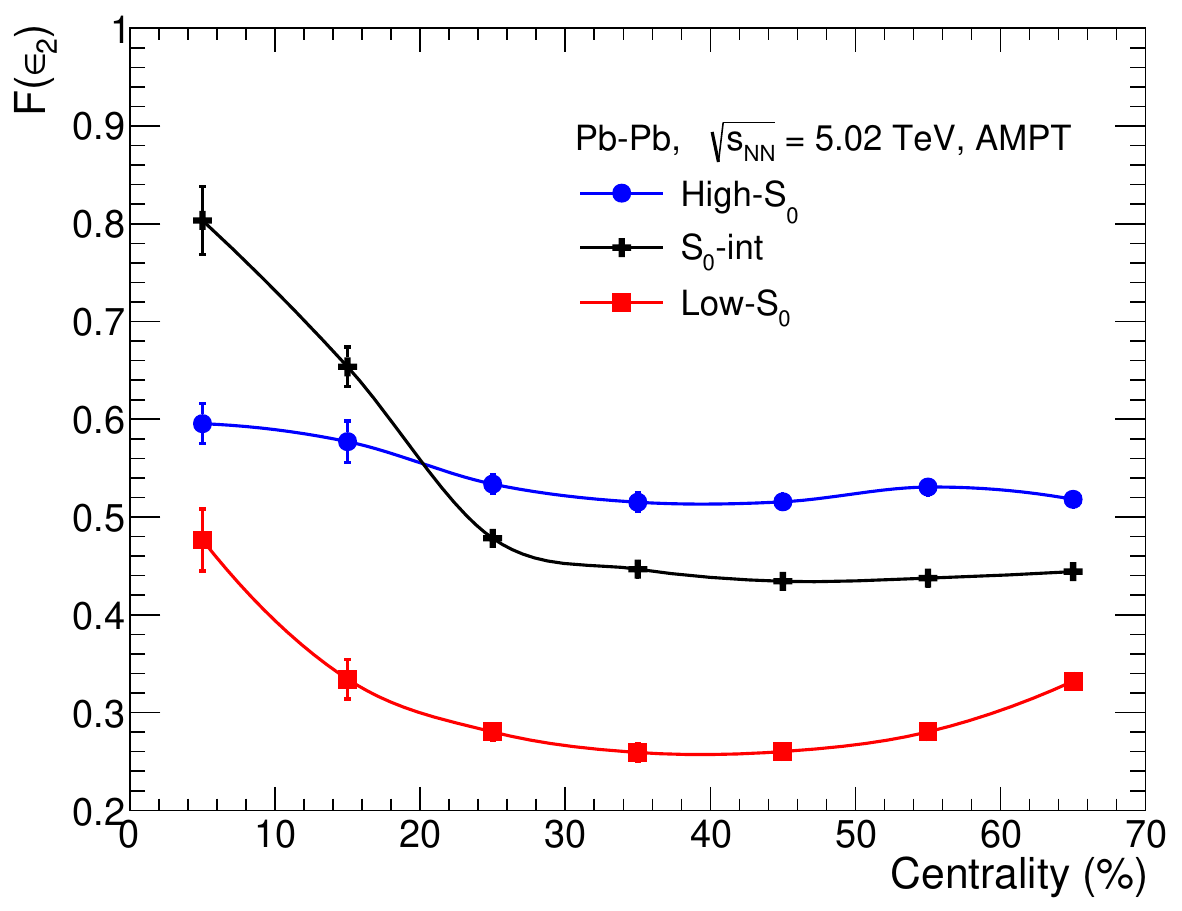}
    \includegraphics[width=0.49\linewidth]{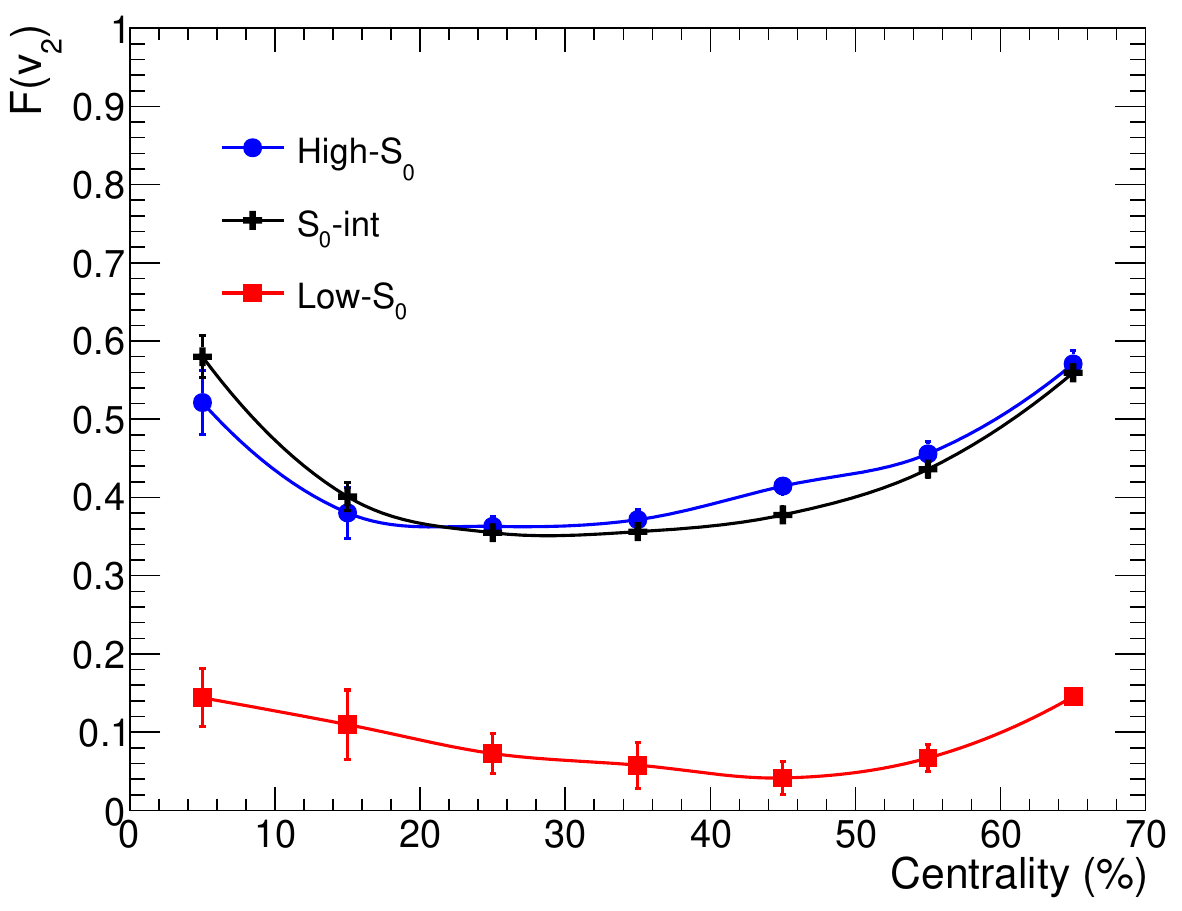}
    \caption{$F(\epsilon_2)$ (left) and $F(v_2)$ (right) as a function of collision centrality for different classes of transverse spherocity in Pb--Pb collisions at $\sqrt{s_{\rm NN}}=5.02\;\text{TeV}$ using AMPT~\cite{Prasad:2025ezg}.}
    \label{fig:fe2fv2}
\end{figure*}

The strong correlation of $S_{0}$ with $v_{2}$ shown in the previous section can be exploited to constrain the probability distributions of $v_{2}$, thereby reducing their event-by-event fluctuations. The left panel of Fig.~\ref{fig:fe2fv2} shows the relative eccentricity fluctuations, $F(\epsilon_2)=\sigma_{\epsilon_2}/\langle\epsilon_2\rangle$, as a function of centrality for different classes of transverse spherocity in Pb--Pb collisions at $\sqrt{s_{\rm NN}}=5.02$ TeV using AMPT. Here, $\sigma_{\epsilon_2}=\sqrt{\langle\epsilon_{2}^{2}\rangle-\langle\epsilon_{2}\rangle^{2}}$ is the eccentricity fluctuation, $\langle\epsilon_2\rangle$ is the event-average eccentricity and $\sqrt{\langle\epsilon_{2}^{2}\rangle}$ is the RMS value. Towards the most central collisions, the collision overlap region is close to a circle in the transverse plane, where the eccentricity of the collision geometry is small. In this region, the measured finite values of initial eccentricity are dominated by event-by-event eccentricity fluctuations, driven by the density fluctuations. Thus, the relative eccentricity fluctuations are the highest in the central collisions. However, as one moves towards the mid-central collisions, the collision geometry becomes elliptic, and the corresponding contribution of eccentricity fluctuation reduces in the measured value of eccentricity. It is interesting to see that for the $S_0$ integrated case $F(\epsilon_2)$ is highest compared to other spherocity classes in (0-20)\% centrality. Towards the mid-central and peripheral collisions, high-$S_0$ events are found to have the largest values of $F(\epsilon_2)$, where the collision overlap geometry is expected to be isotropic. $F(\epsilon_2)$ is found to be smallest for the low-$S_0$ classes where the collision geometry is elliptic~\cite{Prasad:2022zbr} compared to other $S_0$ classes.

Relative elliptic flow fluctuations, $F(v_{2})=\sigma_{v_2}/\langle v_{2}\rangle$, is shown in the right panel of Fig.~\ref{fig:fe2fv2}. Here, $\sigma_{v_2}=\frac{\sqrt{v_{2}^{2}\{2, |\Delta\eta|>1.0\}-v_{2}^{2}\{4\}}}{2}$ is the elliptic flow fluctuations and $\langle v_{2}\rangle=\frac{\sqrt{v_{2}^{2}\{2, |\Delta\eta|>1.0\}+v_{2}^{2}\{4\}}}{2}$ is the RMS value of elliptic flow estimated from two- and four- particle Q-cumulant method~\cite{Prasad:2025ezg}. For a particular spherocity class, $F(v_{2})$ is higher for the central and peripheral collisions and shows a minimum in the mid-central collisions. This is expected as the collision geometry is isotropic, having a smaller value of $v_2$, in the central collisions where the contribution is dominated by event-by-event fluctuations rather than elliptic geometry. On the other hand, in the peripheral collisions, although the collision overlap region is driven by an initial large elliptic geometry, a smaller size of the system results in large fluctuations of elliptic flow. Similar to $F(\epsilon_2)$, low-$S_0$ events possess smaller $F(v_2)$ and within uncertainties, one can observe a larger $F(v_2)$ in the (0-20)\% centrality class for the $S_0$-integrated case as compared to other classes of transverse spherocity.

{\it This subsection highlights how the event shape selections control geometric fluctuations in heavy-ion collisions and provide cleaner constraints on initial conditions and transport than centrality alone.}

\subsection{Symmetry plane correlations}
After the discussion on anisotropic flow coefficients, their fluctuations, and the impact of event shape selection on these coefficients, it is equally important to understand the significance of interplay between the event shape and the symmetry plane angles, $\psi_n$. However, unlike $v_n$, the estimation of $\psi_n$ is not trivial in experiments due to uncertainty in the estimation of the reaction plane. Therefore, it is pointless to estimate $\psi_n$ in heavy-ion collisions, which would always be `0', due to their random orientation for each event. However, the studies of symmetry plane correlation bear a great amount of significance. Similar to anisotropic flow coefficients, which depend upon the initial spatial anisotropy and are sensitive to the transport properties of the medium, the symmetry plane correlations (SPCs) can provide an independent measure of the initial correlations among the participant planes and are sensitive to the transport properties of the medium formed. The calculation of anisotropic flow coefficients ($v_{n}$) along with the symmetry plane angles ($\psi_n$) can be performed using the following expression~\cite{ALICE:2023wdn, Bhalerao:2011yg,ALICE:2024vzv}.
\begin{equation}
v_{n_1}^{a_1}v_{n_2}^{a_2}...v_{n_k}^{a_k}e^{i(a_1n_1\psi_{n_1}+a_2n_2\psi_{n_2}+...+a_kn_k\psi_{n_k})}= \langle e^{i(n_1\phi_1+n_2\phi_2+...+n_l\phi_l)}\rangle
\label{eq:vnpsin}
\end{equation}
Here, $a_i$ denotes the number of times the harmonics $n_i$ or equivalently $-n_i$ occurs\footnote{Due to the azimuthal symmetry of particle production, i.e., invariance of Eq.~\eqref{eq:flowfourier} for $\phi\rightarrow-\phi$, allows $v_{n}=v_{-n}$ and $\psi_n=\psi_{-n}$. This is useful to construct the symmetry plane correlations among specific harmonics.}. The choice of $n_{i}$ and $a_{i}$ are made such that $\sum_{i=1}^{k}a_{i}n_{i}=0$.

The estimation of SPCs is performed using the Gaussian Estimator (GE) method~\cite{ALICE:2023wdn, Bhalerao:2011yg,ALICE:2024vzv,ALICE:2024fus}, defined as follows.

\begin{align}
\label{main_eqn}
&\langle \cos(a_1n_1\psi_1+a_2n_2\psi_2+...+a_kn_k\psi_k)\rangle_{\rm GE}\nonumber\\
&=\sqrt{\frac{\pi}{4}}\frac{\langle{v_{n_1}^{a_1}v_{n_2}^{a_2}...v_{n_k}^{a_k}\cos(a_1n_1\psi_1+a_2n_2\psi_2+...+a_kn_k\psi_k)}\rangle}{\sqrt{\left\langle{v_{n_1}^{2a_1}v_{n_2}^{2a_2}...v_{n_k}^{2a_k}}\right\rangle}}
\end{align}
\begin{figure*}
    \centering
    \includegraphics[width=0.32\linewidth]{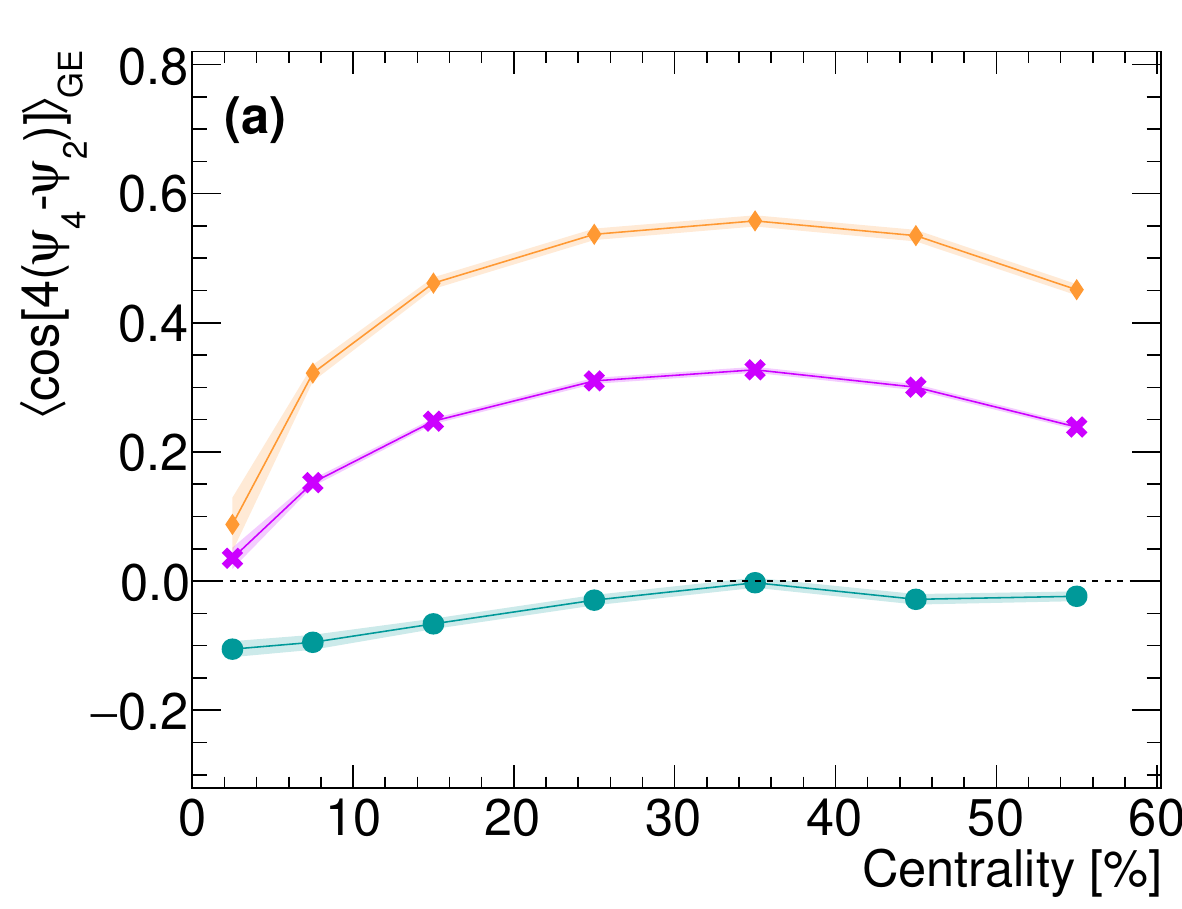}
    \includegraphics[width=0.32\linewidth]{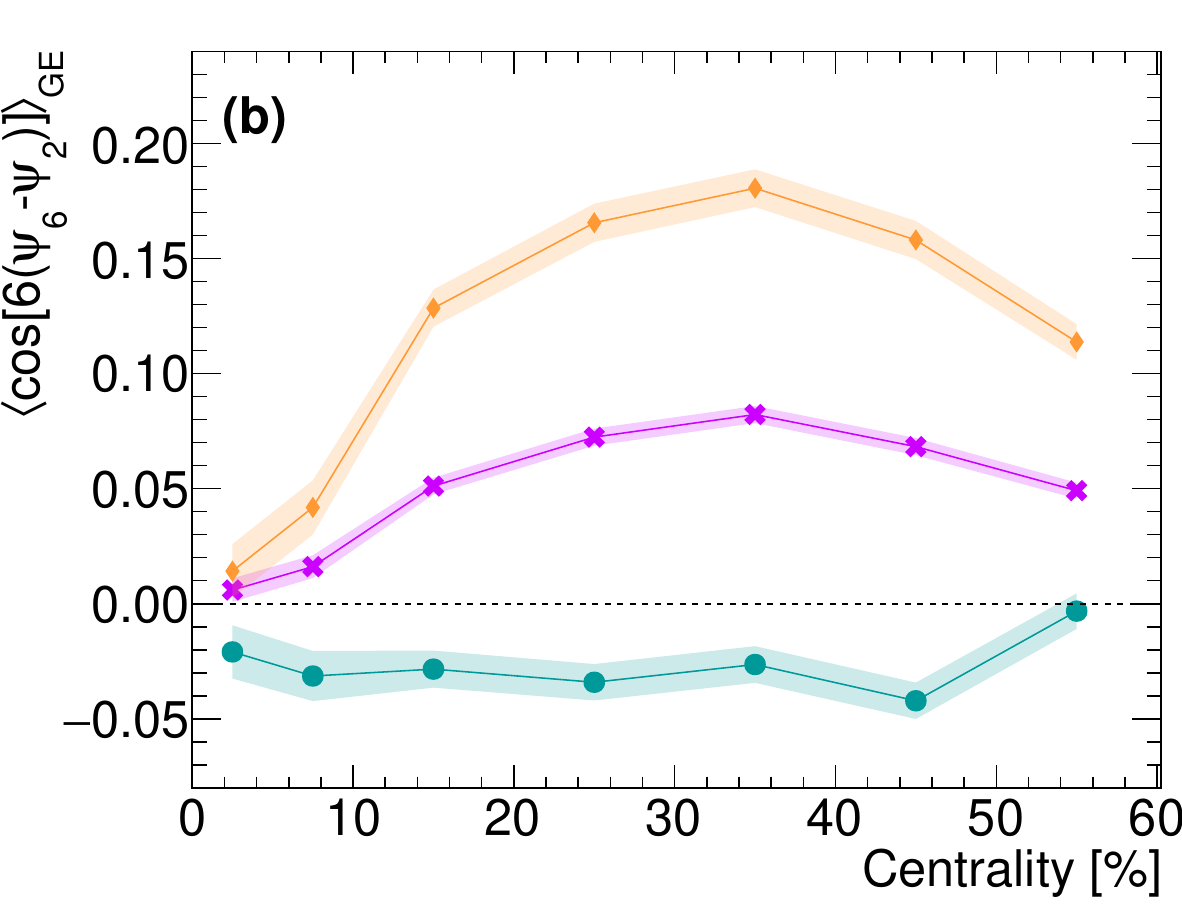}
    \includegraphics[width=0.32\linewidth]{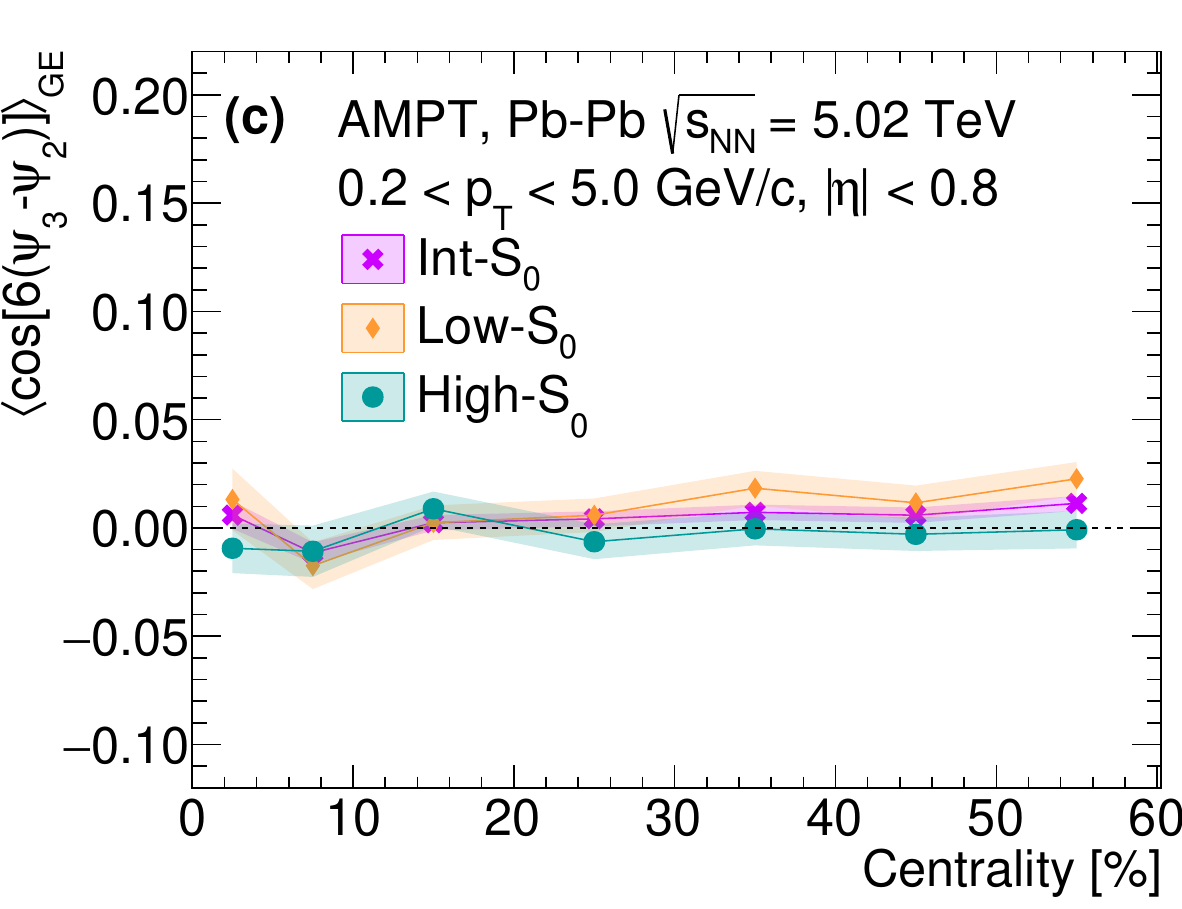}
    \caption{Transverse spherocity and collision centrality dependence of symmetry plane correlations using Gaussian Estimator in Pb-Pb collisions at $\sqrt{s_{\rm NN}}=5.02$ TeV using AMPT~\cite{Tripathy:2025npe}.}
    \label{fig:SPCsvsS0}
\end{figure*}

Here, $\langle\dots\rangle$ denotes the event average over a particular set of events. Figure~\ref{fig:SPCsvsS0} shows the centrality and spherocity dependence of SPCs in Pb-Pb collisions at $\sqrt{s_{\rm NN}}=5.02$ TeV using AMPT. Figures ~\ref{fig:SPCsvsS0} (a), (b) and (c) show $\langle\cos[4(\psi_4-\psi_2)]\rangle_{\rm GE}$, $\langle\cos[6(\psi_6-\psi_2)]\rangle_{\rm GE}$, and $\langle\cos[6(\psi_3-\psi_2)]\rangle_{\rm GE}$, respectively. The strength of the correlations for the spherocity-integrated events decrease from $\langle\cos[4(\psi_4-\psi_2)]\rangle_{\rm GE}$ to $\langle\cos[6(\psi_6-\psi_2)]\rangle_{\rm GE}$, and $\langle\cos[6(\psi_3-\psi_2)]\rangle_{\rm GE}$, which require 3, 4 and 5 particle correlations, respectively, for their estimation. These initial state correlations and their event-by-event fluctuations can be traced back to the correlation among the participant planes and their event-by-event fluctuations. The initial state fluctuations are random in nature and arise due to the finite size of the participants in the collision. According to the central limit theorem, the average of random fluctuations converges to a Gaussian distribution when the number of sampling is large. For a distribution close to Gaussian, the second-order cumulant has the largest values, the values of other moments are smaller, and vanish for higher-order moments. Here, the number of particle correlations is related to the order of particle cumulants, which explains the quantitative behavior of different SPCs. Looking at the spherocity dependence, one finds that for a non-vanishing choice of SPCs, low-$S_0$ events have larger correlations as compared to high-$S_0$ events, which are sometimes negative. This can be intuitively understood since the low-$S_0$ events have an elliptic geometry, the symmetry planes align themselves with $\psi_2$, leading to a rise in SPCs. On the other hand, since high-$S_0$ events have isotropic emission of particles, the symmetry planes do not have any specific direction of alignment, which leads to smaller values. Comparing the SPCs in Fig.~\ref{fig:SPCsvsS0}, to the corresponding participant plane correlations (PPCs) in Ref.~\cite{Tripathy:2025npe}, it can be noted that the system response to the evolutions of SPCs from PPCs is nonlinear with respect to the choice of transverse spherocity event in a similar centrality bin. 
%This behaviour is similar to the observations made with respect to the evolution of anisotropic flow coefficients in different $S_0$ and centrality bins in Ref.~\cite{}. 
This indicates that the system response to the evolution of SPCs is affected by the $S_0$ selection or the event selection via $S_0$ inherently biases the sample towards a specific symmetry, causing higher or lower values of SPCs for low or high $S_0$ events. A more differential event shape analysis of SPCs in events generated with different models can be helpful to make stronger conclusions. It is interesting to note that, although $v_2$ and $v_3$ are anticorrelated, one does not observe any correlation between $\psi_2$ and $\psi_3$, for all classes of $S_0$~\cite{ATLAS:2015qwl, Jia:2014jca, Niemi:2015qia, Qian:2016pau, ALICE:2017kwu, ALICE:2016kpq, Niemi:2012aj}. 

{\it This subsection shows that the event selection via $S_0$ affects the measurement of SPCs, while PPCs are unaffected. 
The results highlight the applicability of transverse spherocity to study system response to the evolution of different SPCs from corresponding PPCs in heavy-ion collisions. In a similar manner, the studies of SPCs with different event classifiers can be used in future explorations to bias the selection to specific symmetry planes, which can improve our understanding of SPCs.}

\subsection{Chiral magnetic effect (CME)}

Recently, another interesting physics phenomenon, called chiral magnetic effect (CME), has been probed using an event-shape engineering (ESE) technique at LHC energy from both experimental and phenomenological fronts~\cite{ALICE:2017sss, CMS:2017lrw, Milton:2021wku, Xu:2023elq, Li:2024gdz, Schukraft:2012ah}. The CME is a phenomenon in which electric charge is separated by a strong magnetic field from local domains of chirality imbalance and parity violation in quantum chromodynamics. On average, the local imbalance of chirality is perpendicular to the reaction plane, and the sign of topological charge is equally probable to be positive or negative. Thus, the charge separation averaged over many events is zero, which makes the CME measurement in the experiment challenging and may be possible only via the correlation techniques. Various CME observables, such as the two-particle correlator ($\delta_{\alpha \beta}$), three-particle correlator ($\gamma_{\alpha \beta}$), have been proposed to observe the CME signal. However, such measurements are dominated by background contributions. Examples of such background sources are the local charge
conservation (LCC) coupled with elliptic flow, momentum conservation, and directed-flow fluctuations, etc. Therefore, the event shape engineering technique is proposed to disentangle background contributions from the potential CME signal. With the ESE method, one can select events with eccentricity values larger or smaller than the average in a given centrality class. The event shape variable known as the second-order reduced flow vector $q_{2}$ (defined in earlier sections) is used for event shape selections. The details about the event and track selection for CME measurement in ALICE can be found in Ref.~\cite{ALICE:2017sss}, and CMS can be found in Ref.~\cite{CMS:2017lrw}. \\

\begin{figure*}[ht!]
\centering
\includegraphics[width = 0.6\linewidth]{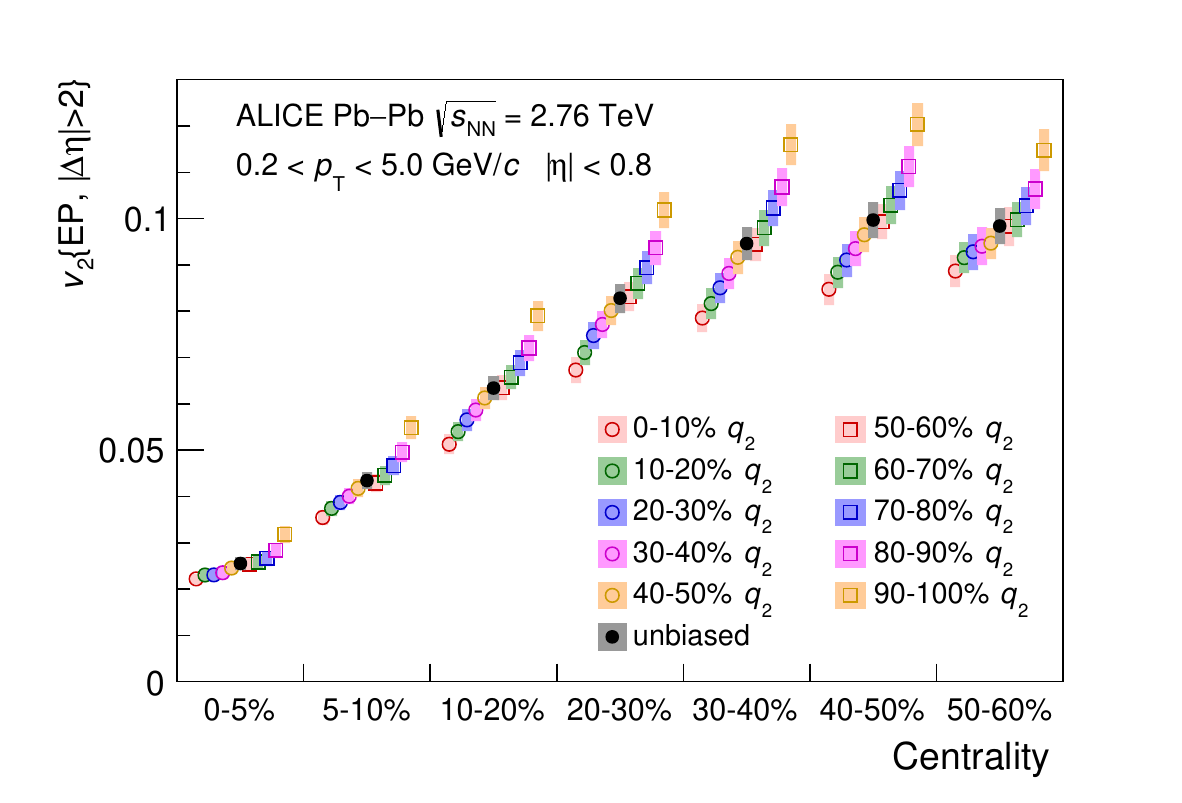}
\caption{The centrality dependence of
elliptic flow ($v_{2}$) for unidentified charged particles with shape-selected and unbiased events with ALICE~\cite{ALICE:2017sss}. The event selection is based on $q_{2}$ determined in the V0C, with the lowest (highest) value corresponding to 0–10\% (90–100)\% $q_{2}$.}
\label{fig:CME1}
\end{figure*}

\begin{figure*}[ht!]
\centering
\includegraphics[width = 0.6\linewidth]{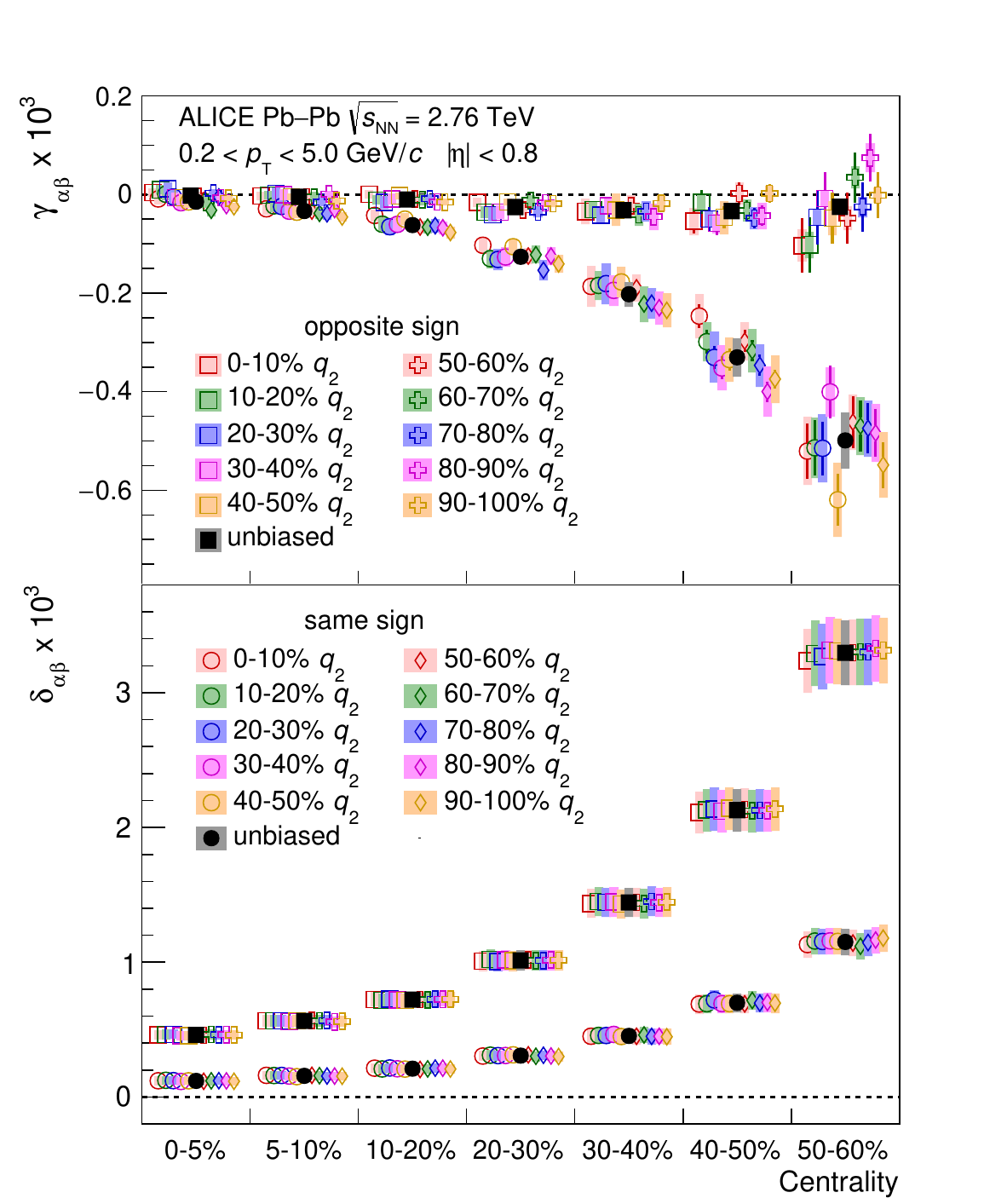}
\caption{ The centrality dependence of $\delta_{\alpha \beta}$ (top panel) and $\gamma_{\alpha \beta}$ (bottom panel) with same and opposite charge for shape-selected and unbiased events with ALICE~\cite{ALICE:2017sss}. The event selection is based on $q_{2}$ determined in the V0C, with the lowest (highest) value corresponding to 0–10\% (90–100)\% $q_{2}$.}

\label{fig:CME2}
\end{figure*}

 Figure~\ref{fig:CME1} shows the measured elliptic flow $v_{2}$ for unidentified charged particles averaged over 0.2 $< p_{\rm T} < $ 5.0 GeV/c for the second-order reduced flow vector $q_{2}$ based selected events differ from the average by up to 25\%. This indicates that with the ESE method experimentally, one can select events with desired initial spatial anisotropy. The effect of $q_{2}$ selection is clearly visible in both the central 0-5\% (lower $q_{2}$) and peripheral 50-60\% (higher $q_{2}$) collisions. Furthermore, the top and bottom panel of  Fig.~\ref{fig:CME2} shows the centrality dependence of two-particle correlator ($\delta_{\alpha \beta}$) and three-particle correlator ($\gamma_{\alpha \beta}$), respectively, with the same and opposite charge for shape-selected and unbiased events.
 The top panel of Fig.~\ref{fig:CME2} shows that 
 the correlation of pairs with the same charge is stronger than the correlation for pairs
of opposite charge, while the bottom panel shows that the correlation for the same charge pairs is smaller than for the
opposite charge combinations. The pattern of the correlations of pairs with the same and opposite charge reveals a charge separation with respect to the event plane. Figure~\ref{fig:CME2} depicts the magnitude of the same and opposite charge pair correlations for $\delta_{\alpha \beta}$ and $\gamma_{\alpha \beta}$ are insensitive to the event-shape selection in a given centrality bin. In order to investigate the charge separation effect, the difference of same and opposite sign charge pair correlations is measured as a function of $v_{2}$ in Fig.3 of Ref.~\cite{ALICE:2017sss} and found to be linearly scaled with $v_2$. This dependence on $v_2$ points to a large background contribution to $\gamma_{\alpha \beta}$.

\begin{figure*}[ht!]
\centering
\includegraphics[width = 0.6\linewidth]{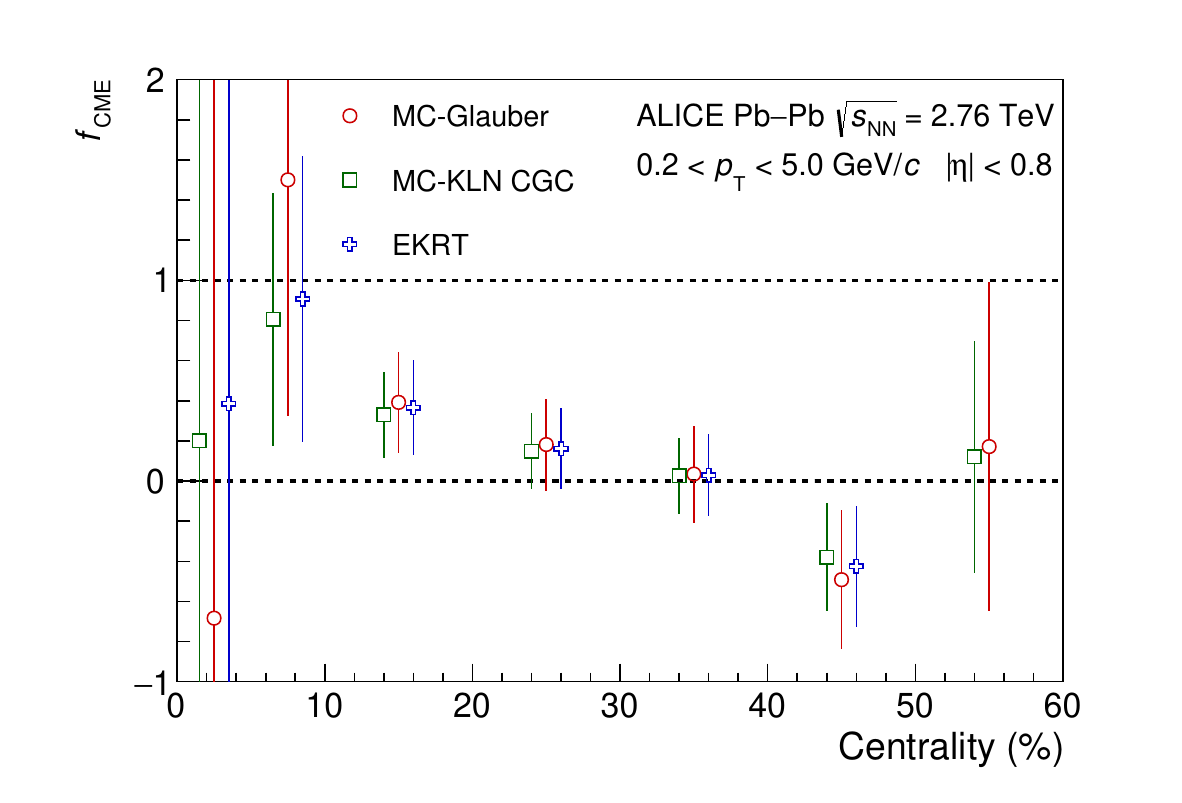}
\caption{Centrality dependence of the CME fraction ($f_{\rm CME}$) extracted from
the slope parameter of fits to data and MC-Glauber, MC-KLN CGC, and EKRT models, respectively. The dashed lines indicate the physical parameter space of the CME fraction. Points are slightly shifted along the horizontal axis for better visibility~\cite{ALICE:2017sss}.}

\label{fig:CME3}
\end{figure*}

To disentangle the potential CME signal from the background, various Monte Carlo simulations are adopted with the inclusion of a magnetic field. The dependence on $v_2$ of the difference between opposite and same charge pair correlations for $\gamma_{\alpha \beta}$ is fitted linear function to both the ALICE data and MC models. Finally, the CME fraction ($f_{\rm CME}$) is estimated for 10–50\% collision centrality in ALICE and shown in Fig.~\ref{fig:CME3}. The technical details of the fitting procedure and CME fraction estimation can be found in Ref.~\cite{ALICE:2017sss}. In conclusion, the event shape engineering technique has been applied to measure the dependence on $v_2$ of the charge-dependent
two- and three-particle correlators $\delta_{\alpha \beta}$ and $\gamma_{\alpha \beta}$ in heavy-ion collisions.

{\it This subsection highlights that, within current precision, there is no compelling evidence for a CME contribution at the LHC, and event shape classifiers can provide a precision tool to set bounds on any residual signal.}

\subsection{Particle ratios, Kinetic freeze-out temperature and radial flow}

\begin{figure*}
    \centering
    \includegraphics[width = 0.9\linewidth]{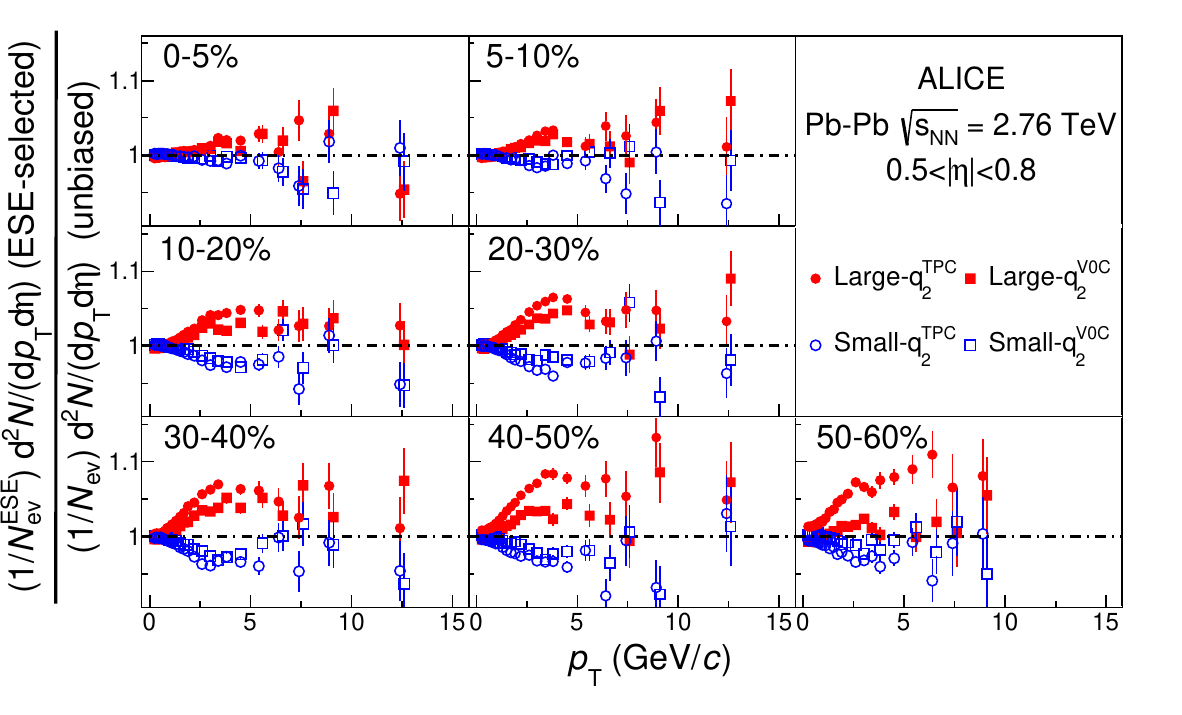}
    \caption{Ratio of $p_{\rm T}$-spectra of different event-shape-engineering (ESE) selected events using $q_{2}$ estimated with ALICE TPC and V0C detectors to $p_{\rm T}$-spectra of unbiased events for different centrality classes in Pb--Pb collisions at $\sqrt{s_{\rm NN}}=2.76$ TeV with ALICE~\cite{ALICE:2015lib}.}
    \label{fig:qnvspTspectra}
\end{figure*}

Figure~\ref{fig:qnvspTspectra} shows the ratio of $p_{\rm T}$-differential yield from $q_2$-selected events to that of unbiased events for different centrality classes in Pb--Pb collisions at $\sqrt{s_{\rm NN}}=2.76$ TeV with ALICE~\cite{ALICE:2015lib}. $q_2$ estimated from ALICE TPC is also compared to that measured with the V0C detector. To reduce the auto-correlation bias of measuring both $q_2$ from TPC and particle $p_{\rm T}$-spectra, $q_2$ measurement is performed with charged particles in $|\eta|<0.4$ while the $p_{\rm T}$-spectra is measured with particles in $0.5<|\eta|<0.8$. It is observed that large-$q_{2}$ events possess a harder $p_{\rm T}$-spectra as compared to small-$q_{2}$ events. Further, the transverse momentum spectra of the large (small) $q_2^{\rm TPC}$ are harder (softer) than that of events selected with $q_{2}^{\rm V0C}$.

\begin{figure}[ht!]
\begin{center}
\includegraphics[scale=0.4]{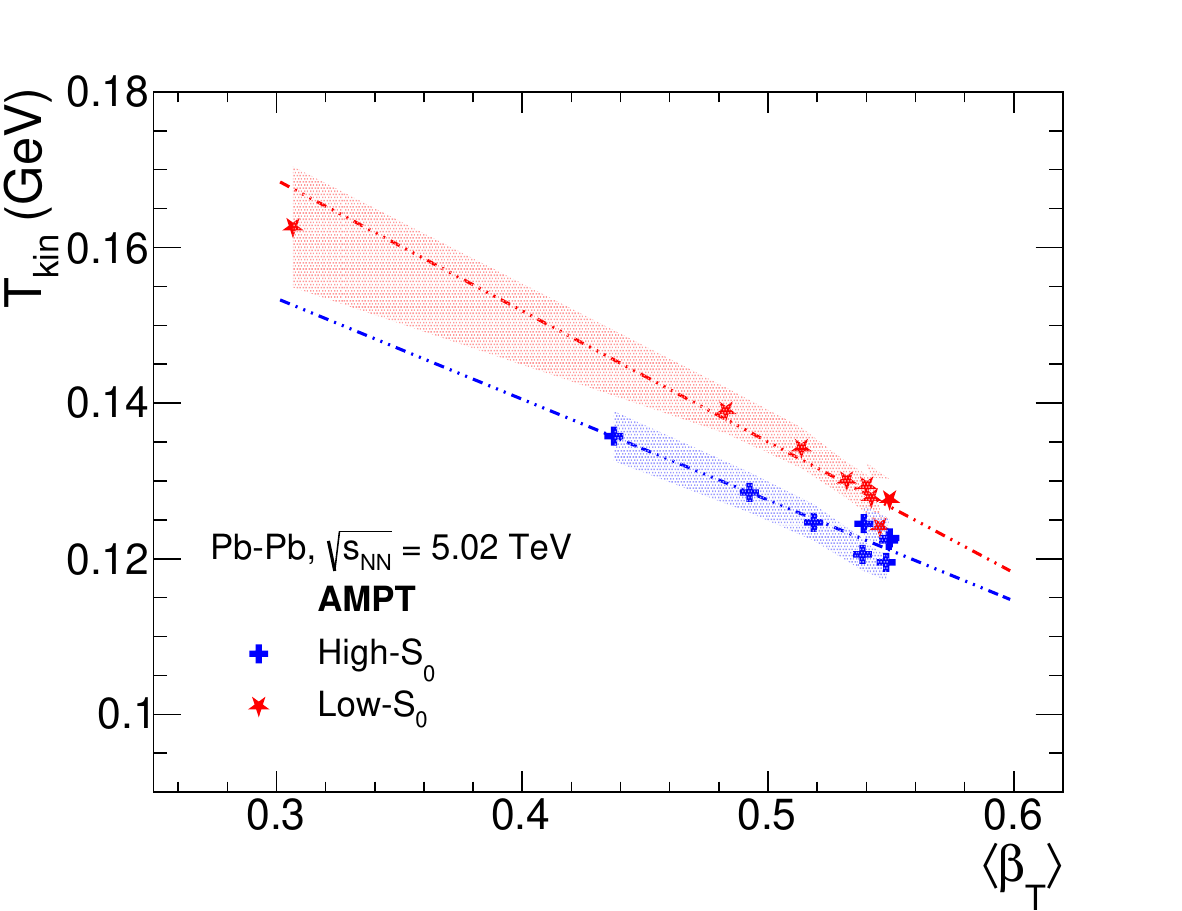}
\caption{Transverse spherocity dependence of kinetic freeze-out temperature ($T_{\rm kin}$) versus mean transverse radial flow velocity ($\beta_{\rm T}$) extracted from a simultaneous fit of Boltzmann-Gibbs Blastwave function to identified particle transverse momentum spectra in Pb--Pb collisions at $\sqrt{s_{\rm NN}}=5.02$ TeV using AMPT~\cite{Prasad:2021bdq}.}
\label{fig:PbPbbetavsTkinSphero}
\end{center}
\end{figure}

 One of the crucial observables used in the high-energy collisions is the study of kinetic freeze-out temperature ($T_{\rm kin}$) versus mean transverse radial flow velocity ($\langle\beta_{\rm T}\rangle$), which gives much more information about the system formed in such collisions. Both these observables are anti-correlated and are proportional to the system formed in heavy-ion collisions~\cite{ALICE:2019hno}. Figure~\ref{fig:PbPbbetavsTkinSphero} shows $T_{\rm kin}$ versus $\langle\beta_{\rm T}\rangle$ for different classes of transverse spherocity and centrality extracted from a simultaneous fit of Boltzmann-Gibbs Blastwave function to identified particle transverse momentum spectra in Pb--Pb collisions at $\sqrt{s_{\rm NN}}=5.02$ TeV using AMPT. For a particular class of transverse spherocity, as one moves from central to peripheral collisions, $T_{\rm kin}$ increases and the value of $\langle\beta_{\rm T}\rangle$ decreases. This is because, due to the large number of particles in the most central collisions, the outward pressure is large, which contributes to a larger value of $\langle\beta_{\rm T}\rangle$. On the other hand, due to the large multiplicity in the most central collisions, the system would take a longer time to reach kinetic freeze-out. However, as one moves towards the peripheral collisions, due to less multiplicity, one may find a lower value for $\langle\beta_{\rm T}\rangle$ and a higher value for $T_{\rm kin}$. As seen previously, transverse spherocity is proportional to the charged-particle multiplicity; thus, for the high-$S_0$ events, one may find a larger multiplicity than for low-$S_0$ events. Consequently, one may observe a higher value for $\langle\beta_{\rm T}\rangle$ and a lower value for $T_{\rm kin}$ for high-$S_0$ events in contrast to low-$S_0$ events, as shown in Fig.~\ref{fig:PbPbbetavsTkinSphero}.

 {\it This section highlights that extending event classifiers from small systems to heavy-ion collisions shows how event shape observables can enhance/suppress geometry-driven collectivity and soft production. Selections based on reduced flow vectors and on transverse spherocity modify the two-particle correlations, flow harmonics, symmetry plane correlations, and spectra in ways that are consistent with expectations from initial geometry and collective expansion. Event shape classifiers provide an important tool even in the largest systems at the LHC that complements centrality selection. In addition, they provide useful tools to compare across collision systems to bring the events on equal footing.}

\section{Machine learning and event shape classifiers}
\label{machinelearning}
 In recent years, machine learning (ML) has emerged as a powerful tool for classifying and characterising events that complement the traditional event-shape classifiers. While this review focuses on event-shape observables, it is important to shed some thoughts on the usage of event shape classifiers in the ML/AI era. 
 
 LHC analyses increasingly deploy data-trained event classifiers that learn event topology from low-level inputs (tracks, calorimeter cells, jet constituents). Theory-driven ML frameworks such as energy flow polynomials~\cite{Komiske:2017aww} and energy flow networks~\cite{Komiske:2018cqr}, explicitly encode infrared and collinear safety, providing a bridge between traditional event shapes and learned representations. At the experiment level, CMS demonstrated end-to-end event classification from detector images~\cite{Andrews:2018nwy}, and ATLAS has used sequence/graph models for event-topology identification~\cite{Jiang:2024vvn} and evaluated ML for real-time trigger decisions, typically benchmarking against global shape selections. In parallel, unsupervised/weakly-supervised anomaly detection (online and offline) learns deviations in event topology directly from data. Also, such ML studies are being attempted in preparatory studies for the forthcoming Electron–Ion Collider (EIC) experiment in the USA~\cite{Lee:2022kdn}. Attempts are also made to probe MPI using ML-based classifiers at the LHC~\cite{Ortiz:2020rwg,Ortiz:2021peu}.

In addition, the event shape studies can also be extended to heavy-ion collisions with the help of ML-based algorithms. Machine learning tools such as boosted decision trees and deep neural network models are quite popular in the high-energy physics community due to their accuracy and robustness in regression and classification tasks. These models are usually trained with a simulated dataset obtained from phenomenological models that best describe the real-world data. The training dataset contains a set of input variables and the corresponding output variable, or the target. The machine is trained to capture the input-output correlations; hence, a careful consideration of the input variables can be made, which can help in training the model. A few phenomenological studies were attempted to 
estimate transverse spherocity in heavy-ion collisions using ML~\cite{Mallick:2021wop}.  Charged-particle multiplicity and mean transverse momentum ($\langle p_{\rm T}\rangle$) are chosen as input observables. A follow-up study has been performed using different ML algorithms in Ref.~\cite{Basak:2025idd}, where the authors test different algorithms, models, and input parameters.  Such studies can serve as a baseline for future developments aimed at full-scale applications in collider experiments focused on event shape observables. Advanced ML models could also be explored for further testing. 

{\it While ML techniques are still being benchmarked against established event-shape variables, they represent an emerging trend in which data-trained event classifiers are beginning to complement and in some cases, replace traditional physics-inspired observables in high-energy physics analyses. However, we believe, with the advent of advanced ML models in the era of big data, the applicability of ML in various directions of the event-shape studies to understand multi-hadron production dynamics will remain a promising research domain.}

\section{Summary and Outlook}
\label{sec:summary}
\noindent

Collisions at GeV and TeV energies produce a wide variety of final states, ranging from jet-dominated configurations arising from hard partonic scatterings to high-partonic activity environments governed by soft QCD dynamics. A central challenge in collider experiments is to organize this diversity in a manner that is both experimentally driven and theoretically meaningful. Event topology classifiers provide a systematic framework that can characterize events based on the global structure of particle production by capturing the geometrical distribution of momentum and energy flow in the final state. Therefore, the event topology classifiers offer a model-independent method to study multi-hadron production dynamics and the correlations linking soft and hard QCD regimes.

Historically, the event shape studies in the pre-LHC era have provided very important physics insights in the understanding of multi-hadron production dynamics. Electron-positron colliders established thrust, jet broadenings, heavy jet mass, Fox–Wolfram moments, and related variables as infrared and collinear safe probes which connect directly to perturbative QCD calculations. Deep inelastic scattering at HERA showed how frame choice and hemisphere definitions highlight the scale dependence of power corrections. Hadron colliders before the LHC introduced the toward, transverse, and away methodology that separates the hard recoil from the soft environment. These lessons motivate a few sets of precise tools for the LHC era where geometry is reported with clear region definitions, rapidity separation, and explicit control over selection bias. 

As discussed in this review, event shape observables have proven to be valuable tools in understanding the underlying dynamics of $pp$ collisions. They offer key insights into QCD dynamics and serve as a bridge between theory and experimental measurements.  Event shape observables are expected to be instrumental in testing the QCD, as, by construction, they are collinear and infrared safe observables. Thus, they do not change their value if a parton is split into two collinear partons or an extra soft gluon is added. This is an important condition for the cancellation of divergences associated with such gluon emissions, which makes them ideal tools for making finite perturbative QCD predictions. In addition, these event-shape classifiers have shown a significant correlation with MPI, which makes them the ideal tool for the understanding of QGP-like effects. Thus, these are nearly perfect observables to be used as event estimators as they can probe both perturbative and non-perturbative QCD sectors. Thus, over the last decade, extensive theoretical and phenomenological research has been carried out in the domain of event shape observables in small systems; however, limited statistics have prevented experimentalists from using these observables to perform precision studies.

The review also highlights how event shape selections are useful when a hard scale is present. In heavy flavor and jet-based studies, the event shape classifiers provide precision tools to separate the hard recoil from the soft environment, and
they are useful in testing and tuning models in topologies with a well-defined
pQCD scale. The correlation measurements in the jet frame for high multiplicity jets display near-side structure that standard simulations like PYTHIA
do not reproduce, which motivates topology-based selections in future studies. For future searches for jet quenching in small collision systems, one should target very high multiplicity pp collisions while minimising estimator-induced biases in jet fragmentation.

Given that significantly higher statistics would be available in Runs 3 and 4 of the LHC with respect to Runs 1 and 2, all the above-discussed event-classifiers can be experimentally used to probe the discussed observables with a high level of precision. As shown in the manuscript, the usage of event shape observables can be even extended to heavy-ion collisions, which will also bring different collision systems on equal footing. In addition, it is also essential to have a common benchmark set for future measurements in LHC experiments, which would make results easier to compare across experiments and models. 

%{\it In summary, this review article will provide a baseline for future experimental studies, which will give a clear insight into the microscopic origin of QGP-like behavior in small collision systems at the LHC and hopefully, in the next decade, one can get a clear picture if small systems collectivity originates from macroscopic features like the formation of QGP or from microscopic mechanisms like string interactions.}

{\it The event topology classifiers highlight the role of global event geometry in shaping the multi-hadron production dynamics, rather than attributing observed features to a single mechanism. By sorting the events from jetty to isotropic configurations, event topology classifiers enable a unified way to study the particle production across both soft and hard QCD regimes. This event topology-based approach to understand multi-hadron production dynamics is therefore central to precision measurements at the LHC and beyond.}

\section{Appendix}
\label{appendix}
Event generators are the most crucial computational tools used in contemporary high-energy physics phenomenology, bridging the gap between theoretical models and experimental data. 
These act as the most effective substitutes for realistic experiments involving hadronic and nucleus-nucleus collisions (AA). These generators are based on the Monte Carlo techniques for simulating the collisions using physics processes to better understand the nature of the events. Based on the kind of colliding species involved and underlying known physics processes, there are several event generators like AMPT, PYTHIA, EPOS LHC, EPOS4, and HERWIG, etc, modeled for hadronic and nuclear collisions.  We employ PYTHIA8 and AMPT models as event generators to explore the possible correlation between all event shape classifiers discussed in Sec.~\ref{sec:definitions}. The detailed mechanism and event generation using these models are described below.

%\subsection{Experimental methodology}

\label{sec:method}
\noindent
\subsection{PYTHIA 8}
\label{appendixpythia}

In the present study, we use PYTHIA 8.308, an improved version of PYTHIA6, which includes the multi-partonic interactions scenario as one of the important improvements. Using the standard Monash 2013 Tune (Tune:pp = 14)~\cite{Skands:2014pea}, we generate 60 million events in $pp$ collisions at  $\sqrt{s}$ = 13 TeV for this analysis.  The inelastic and soft QCD events (SoftQCD:inelastic=on) are simulated; as a result, the total scattering cross section includes the contribution from all the single, double, and central diffractive components. We have considered the MPI (PartonLevel:MPI = on) along with mode 1 of color reconnection (ColourReconnection:mode = 1) and Beam Remnants (BeamRemnants:remnantMode = 1). The mode 1 of CR is a newer QCD-based scheme which builds on the dipole formation via the QCD color rules as well as the minimization of the string length. This scheme introduces the junction structure, as a result of which junctions are produced between three or four dipoles. A detailed mechanism can be found in Ref.~\cite{manual}. Additionally, along with the CR mechanism, we employ another hadronization mechanism for color strings known as rope hadronization (RH). The physical description of CR and RH can be found below.

\subsubsection{Color Reconnection}
The hadronisation through the fragmentation-based models, such as Lund's string fragmentation model~\cite{Andersson:1983jt}, can be envisaged as a stretched color flux tube between two partons originating from individual partonic scatterings leading to linear confinements through a massless relativistic string. With the increase in the potential energy of the strings, the quark-antiquark pair moves apart till a new quark-antiquark pair is formed. The hadrons can be formed by combining these quarks and antiquarks. The hadronisation through fragmentation in this scenario is independent of each individual scattering, also shown in Fig.~\ref{fig:CRCartoon} (a). However, in the presence of color reconnection, the strings connecting the partonic endpoints from each individual scattering can color reconnect~\cite{Argyropoulos:2014zoa}. This is clearly depicted in Fig.~\ref{fig:CRCartoon} (b). Therefore, the fragmentation of two individual scatterings is dependent on each other when the color reconnection is considered, which induces a rise in average transverse momentum with an increase in the number of individual scatterings or the number of multi-partonic interactions.

\subsubsection{Rope Hadronisation}
In relativistic collisions at LHC energies, a large number of multi-partonic interactions can lead to the production of several strings connecting the partonic endpoints in a smaller transverse region. Within the framework of rope hadronisation, these overlapping strings can act coherently to form a color rope having a large effective string tension. These string-overlap regions usually have a higher energy density as compared to the nearby regions, which can result in the production of a pressure gradient. In the overlap regions, this pressure gradient can push the strings in the outward direction. The pushing of strings outward is called string shoving, which mimics the flow-like pattern observed in heavy-ion collisions. Further, the breaking of strings with a higher effective tension can lead to the production of strange quarks and diquarks, leading to the enhanced production of baryons and strange hadrons in events having a large number of multi-partonic interactions.

The RH mechanism of PYTHIA8 has various tuning parameters.  We use similar settings to the string-shoving mechanism introduced in PYTHIA8. The values of the parameters used for rope hadronization are shown in Table~\ref{table:rope}. Furthermore, the flag partonvertex (PartonVertex:setVertex = on) is used to set impact-parameter plane vertices for partonic production by ISR, FSR, MPI, and beam remnants. In the simulated events, the hadronic level decay mode (HadronLevel:Decay = on) is enabled for all the resonances except the ones used in our study. 

\begin{figure*}
    \centering
    \includegraphics[width=0.49\linewidth]{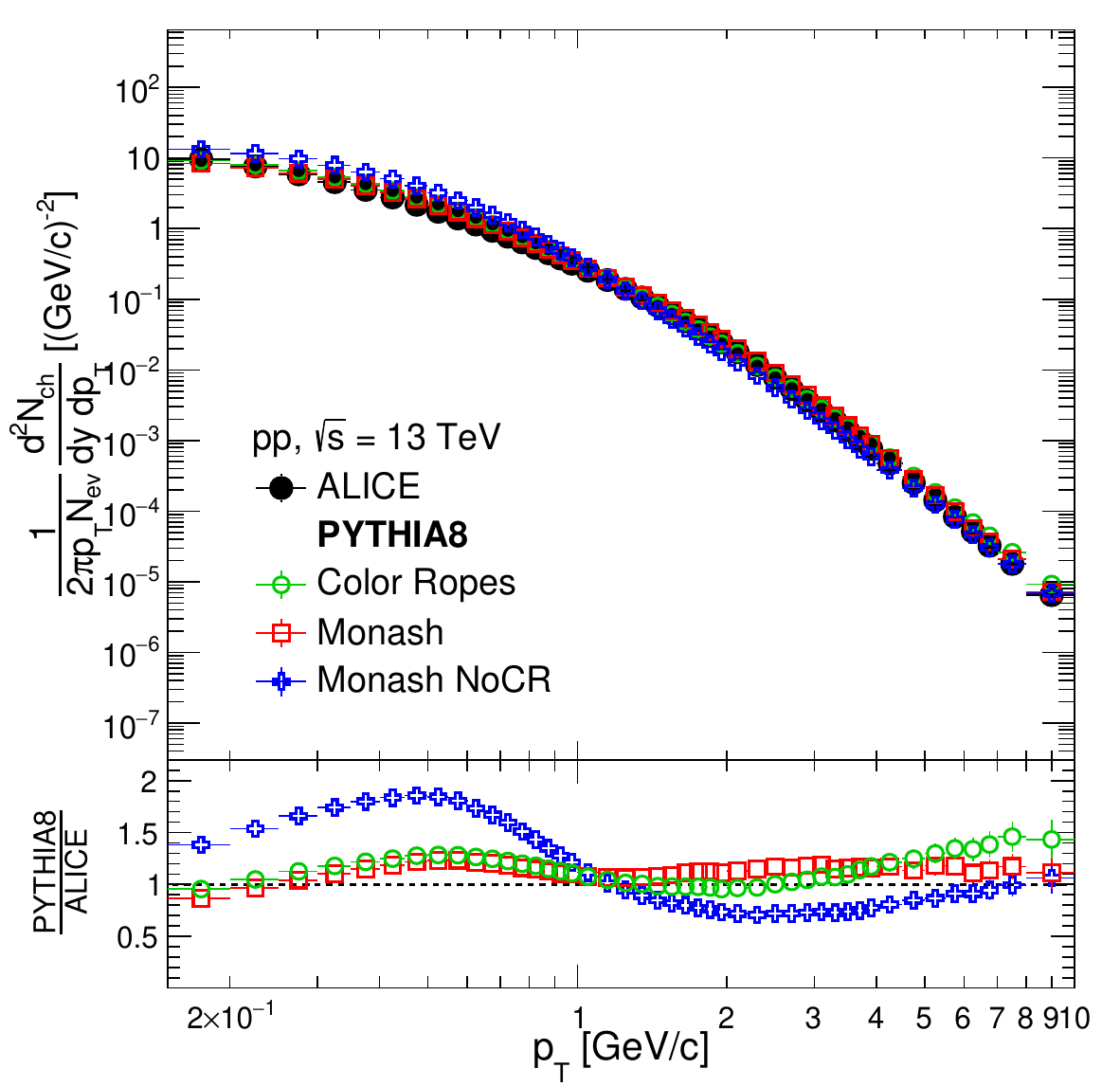}
    \includegraphics[width=0.49\linewidth]{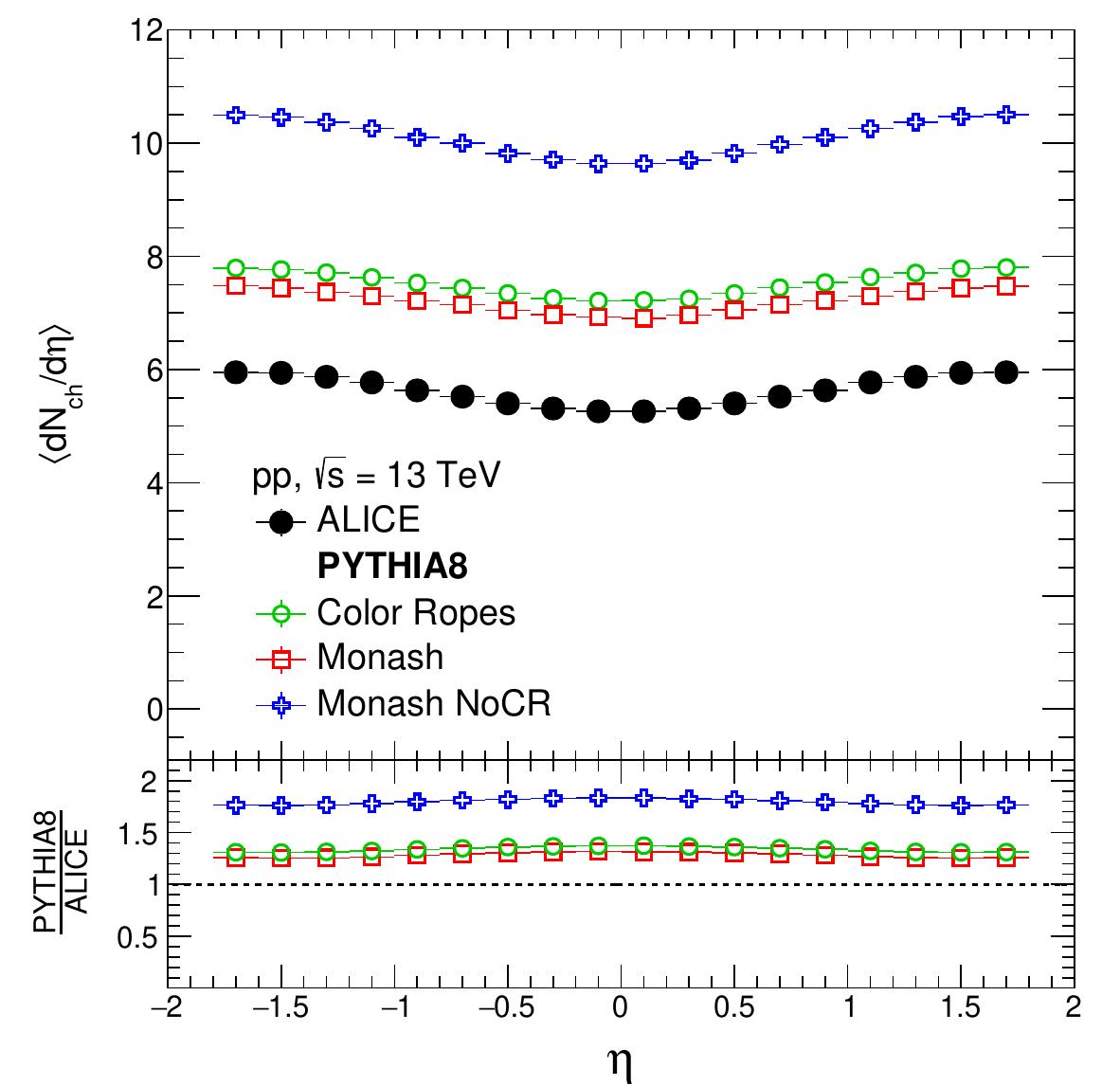}
    \caption{Transverse momentum (left) and pseudorapidity (right) differential yield distribution in minimum bias pp collisions at $\sqrt{s}=13$ TeV using PYTHIA8 Color Ropes, Monash and Monash NoCR compared with corresponding measurements with ALICE~\cite{ALICE:2015qqj}.}
    \label{fig:PythiaSpectCompALICE}
\end{figure*}

To check the compatibility of PYTHIA8 with experimental data, we have used the same tuning as used in one of our previous works described in Ref~\cite{Prasad:2024gqq}, where we have compared the transverse momentum and pseudorapidity spectra obtained from different tunes of PYTHIA8 with the ALICE experimental data~\cite{ALICE:2015qqj} for all charged particles in $pp$ collisions at $\sqrt{s}$ = 13 TeV, as shown in Fig.~\ref{fig:PythiaSpectCompALICE}. It can be observed that PYTHIA8 with Color Ropes and Monash show a better quantitative and qualitative agreement to experimental data than PYTHIA8 with Monash NoCR.

\begin{table}
\centering
\begin{tabular}{c|c|c|c|c|c|c|c|c|c|c|c|}
\hline
\multicolumn{10}{|c|}{Rope Hadronization}& Values\\
\hline
\multicolumn{10}{|c|}{ Ropewalk:RopeHadronization} &on\\
\multicolumn{10}{|c|}{ Ropewalk:doShoving} &on \\
\multicolumn{10}{|c|}{ Ropewalk:doFlavour} &on\\
\multicolumn{10}{|c|}{ Ropewalk:r0} &0.5\\
\multicolumn{10}{|c|}{ Ropewalk:m0} &0.2\\
\multicolumn{10}{|c|}{ Ropewalk:beta} &1.0 \\
\multicolumn{10}{|c|}{ Ropewalk:tInit} &1.0\\
\multicolumn{10}{|c|}{ Ropewalk:deltat} &0.05\\
\multicolumn{10}{|c|}{ Ropewalk:tShove} &10.0\\
\hline
\end{tabular}
   \caption[p]{The parameter values of the rope hadronization
model used with the color reconnection mechanism.}
\label{table:rope}
 \end{table}

Apart from the above-mentioned PYTHIA tunes, there are a few additional MC tunes, such as PYTHIA CP1, CP5, A3, A14, etc. are used in CMS and ATLAS experiments. These new tunes differ according to the order of the parton distribution functions (PDF) set used at leading (LO), next-to-leading (NLO), or next-to-next-to-leading (NNLO) orders in perturbative quantum chromodynamics, and the strong coupling evolution at LO or NLO. The details about these specific MC tunes are described below:

\begin{itemize}
    \item PYTHIA CP1: Tune CP1 uses the NNPDF3.1 PDF set at LO, with $\alpha_{s}$ values used for the simulation of MPI, hard scattering, FSR, and ISR equal to, respectively, 0.13, 0.13, 0.1365, and 0.1365, and running coupling according to an LO evolution~\cite{CMS:2019csb}.  

    \item PYTHIA CP5: Tune CP5 uses the NNPDF3.1 PDF set at NNLO, with $\alpha_{s}$ values used for the simulation of MPI, hard scattering, FSR, and ISR contributions equal to 0.118, and running coupling according to an NLO evolution and the ISR emissions ordered according to rapidity~\cite{CMS:2019csb}.

\item PYTHIA A3: Tune A3 uses the NNPDF2.3 PDF set at LO and running coupling according to an LO evolution QCD + QED processes. A3 tune is aimed at modelling of low-$p_{\rm T}$ QCD processes~\cite{ATLAS:2016puo}.

\item PYTHIA A14: Tune A14 (ATLAS 2014) uses the NNPDF3.1 PDF set at LO and running coupling according to an LO evolution of QCD  processes. This tune consists of four LO PDFs, such as CTEQ6L1, MSTW2008LO, NNPDF23LO, and HERAPDF15LO~\cite{TheATLAScollaboration:2014rfk}.

\item PYTHIA CUETP8M1: A new PYTHIA 8 tune read as \enquote{CMS UE Tune PYTHIA 8 Monash set 1}. This tune uses several PDF sets and UE tunes, a more detailed discussion can be found in Ref.~\cite{CMS:2015wcf}

These tunes are extracted by varying the various
parameters and fitting UE observables
at various collision energies~\cite{CMS:2019csb, ATLAS:2016puo, TheATLAScollaboration:2014rfk}.

\end{itemize}

%-- MPI on/off, RH on/off
%\subsection{Herwig}
\subsection{A multi-phase transport model (AMPT)}
\label{appendixampt}
In the present study, we use the string melting (SM) mode of AMPT (version 2.26t9b) as the particle anisotropic flow coefficients and spectra are well described at the intermediate-$p_{\rm T}$ region by the quark coalescence mechanism for hadronization \cite{Fries:2003vb,Fries:2003kq}. 

\begin{enumerate}
\item Initial conditions: The AMPT model employs the heavy-ion jet interaction generator (HIJING)\cite{Wang:1991hta} for initial conditions. In the HIJING model, the differential scattering cross section of the energetic minijet partons and soft strings with excited strings, which are further converted to partons, is calculated in $pp$ collisions and then parametrized into the heavy-ion collisions. These are incorporated via the nuclear overlap and shadowing function using an in-built Glauber model.
    
\item Parton transport: The produced partons are then transported into the parton transport part called Zhang's Parton Cascade (ZPC) model \cite{Zhang:1997ej}. In the String Melting version of AMPT (AMPT-SM), colored strings melt into low-momentum partons.
    
\item Hadronization: A spatial coalescence mechanism \cite{Lin:2001zk,He:2017tla} is used for the hadronization of the transported partons in AMPT-SM, whereas in the default AMPT version, a Lund string fragmentation mechanism is used for hadronization of the transported partons.

\item Hadron transport: The final evolution of produced hadrons is done using a relativistic transport mechanism through meson-meson, meson-baryon, and baryon-baryon interactions \cite{Li:2001xh,Greco:2003mm}.
\end{enumerate}

The AMPT settings in the current study are the same as reported in Ref. \cite{Tripathy:2018bib}. The choice of centrality selection has been done using the geometrical slicing of the impact parameter distribution. 

%\subsection{Other possible models} 

% Figure~\ref{fig:atlasfig} shows the average track density ($\langle d^2 N_{\rm ch}/d\eta;d|\Delta\phi|\rangle$) and average scalar sum $p_{\rm T}$ density $\langle d^2 \sum{p_{\rm T}}/d\eta;d|\Delta\phi|\rangle$ of tracks as a function of absolute relative azimuthal angle ($|\Delta\phi|$) of all charged hadrons having $p_{\rm T}>0.5$ GeV/$c$ and $|\eta|<2.5$ with respect to the leading particle jet from various models, which includes PYTHIA8 A14, PYTHIA8 A2, EPOS, PYTHIA8 Monash, and HERWIG++ EE5 compared with the ATLAS data (uncorrected) in \textit{pp} collisions at $\sqrt{s}=13$ TeV \cite{ATLASRTUE}. Two sets of events are selected based on the transverse momentum of the leading particles, \textit{i.e.,}, for $p_{\rm T}^{\rm lead}>5$ GeV/$c$ and $p_{\rm T}^{\rm lead}>1$ GeV/$c$.

\subsection{Other MC models}
\label{sec:otherMCmodels}
This subsection highlights several additional MC models that are relevant to this review.
\begin{enumerate}
\item HERWIG: It is a Monte Carlo event generator widely used to simulate parton showers, hadronization via the cluster model, and underlying event dynamics based on  perturbative QCD studies. The used HERWIG CH3 (CMS HERWIG) tune in Sec.~\ref{sec:Spherodefn} based on the NNPDF3.1 PDF set in all aspects of simulation. The NNLO PDF is used at the parton showering with $\alpha_{s}$ =  0.118, while a LO PDF is used at the MPI, and
beam remnant handling with $\alpha_{s}$ =  0.13~\cite{CMS:2020dqt}

\item EPOS LHC: It is an MC event generator used to simulate hadronic, heavy ion collisions as well as cosmic ray air shower interactions~\cite{Pierog:2013ria}. It is based on EPOS 1.99~\cite{Pierog:2009zt} which is tuned to reproduce experimental data at the LHC energies. EPOS is a hadronic model based on Gribov-Regge theory~\cite{Drescher:2000ha} which provides a consistent framework for calculating cross sections and simulating particle production, incorporating energy conservation in both aspects.

\item JETSET: It is an MC event generator widely used to generate $e^{+}+e^{-}\rightarrow {\rm hadrons}$, which employs the Lund string fragmentation model~\cite{Sjostrand:1993yb}. The matrix element (ME) component of JETSET models the hard scattering process, determining the initial parton configuration and kinematics according to perturbative QCD. In contrast, the parton shower (PS) component of JETSET simulates successive QCD radiation, producing cascades of softer quarks and gluons.

\item ARIADNE: It is an MC program that simulates QCD parton cascades using the color dipole model~\cite{Lonnblad:1992tz}. It provides an alternative to traditional parton‑shower algorithms for the evolution of partons before hadronization

\item COJETS: It is an MC generator used to simulate proton-proton and proton-antiproton interactions~\cite{Odorico:1989bi}. The interactions include parton-level pQCD calculations followed by electroweak interactions. Partons are then fragmented independently into jets of hadrons following the Field-Feynman model. The contributions from beam jets are added following a phase-space model. The model also includes QCD radiations from initial and final state partons.

\item PHOJET: It is an MC event generator based on the dual parton model with a hard and a soft component, which simulates hadron-hadron, photon-hadron, and photon-photon collisions~\cite{Engel:1994vs, Engel:1995yda}. The model can also be applied to photon-photon interactions in electron-hadron and electron-electron or electron-positron colliders.

\item MadGraph: MadGraph is a matrix-element event generator which automates the calculations of tree-level scattering amplitudes~\cite{Stelzer:1994ta}. Using user-defined initial and final states, it can generate all relevant Feynman diagrams and translate them into numerical code. This removes the need for tedious manual calculations, even for complex multi-particle processes. MadGraph is widely used to compute parton-level cross sections and to provide hard-process inputs for full event simulations when interfaced with parton shower and hadronization programs.

\item POWHEG: It is an MC event generator which combines NLO QCD calculations with parton-shower simulations to avoid the problem of negative event weights~\cite{Frixione:2007vw, Nason:2004rx}. This MC generator first generates the hardest radiation using NLO QCD calculations, followed by parton showers, which take care of the soft emissions. It can be interfaced with PYTHIA OR HERWIG to simulate close to realistic events in collider experiments~\cite{Frixione:2007vw}.

\item SHERPA: It is a general-purpose MC event generator that can simulate lepton-lepton, lepton-photon, photon-photon, lepton-hadron, and hadron-hadron collisions~\cite{SHERPA:gitlab}. It is equipped with a full simulation, which is split into well-defined event phases, based on QCD factorization theorems. This simulation of physics processes with this model covers all reactions in the Standard Model.

\item EKRT: It is a theoretical framework which is primarily used to simulate heavy-ion collisions~\cite{Eskola:1999fc, Niemi:2015qia}. It combines pQCD minijet production with gluon saturation and viscous hydrodynamics to model initial conditions and bulk evolution of the quark–gluon plasma.

\item MC-KLN CGC: It is an MC event generator based on the Color Glass Condensate (CGC) framework, which incorporates gluon saturation effects through local saturation scales determined by nuclear thickness functions~\cite{Kharzeev:2004if}. Here, the particle production is computed using $k_{\rm T}$-factorization, leading to event-by-event fluctuating initial gluon density profiles.

\item MC-Glauber: It is an MC implementation of the Glauber model, which describes the initial geometry of nuclear collisions. The description is given by sampling nucleon positions inside the colliding nuclei and determining interactions based on nucleon–nucleon collision probabilities~\cite{Loizides:2017ack}. The MC-Glauber model is widely used to define collision centrality and initial spatial eccentricities in heavy-ion collisions.

\item FASTJET: It is a computational tool that is mostly used in jet-based studies. It includes efficient native implementations of all widely used 2 $\rightarrow$ 1 sequential recombination jet algorithms for $pp$ and $e^{+}e^{-}$ collisions, as well as access to 3$^{\rm rd}$ party jet algorithms through a plugin mechanism, including all currently used jet cone algorithms~\cite{Cacciari:2011ma}.

\end{enumerate}

\section*{Acknowledgments}
 S.P. and B.S. acknowledge the doctoral fellowships from the UGC and CSIR, Government of India. S.T. acknowledges the CERN Research Fellowship and the funding received from the European Union’s Horizon Europe research and innovation programme under the Marie Skłodowska-Curie grant agreement No. 101149298. The authors acknowledge the DAE-DST, Government of India, funding under the Mega-Science Project—“Indian participation in the ALICE experiment at CERN” bearing Project No. SR/MF/PS-02/2021-
IITI (E-37123). The authors acknowledge Neelkamal Mallick for the discussion on the heavy-ion-related studies. S.T. acknowledges discussion with Alexander Philipp Kalweit and Alice Ohlson during the preparation of this manuscript.


\begin{thebibliography}{999}

\bibitem{Diehl:2011yj}
M.~Diehl, D.~Ostermeier and A.~Schafer, Elements of a theory for multiparton interactions in QCD,
JHEP \textbf{03}, 089 (2012)
[erratum: JHEP \textbf{03}, 001 (2016)].


%\cite{Field:2012kd}
\bibitem{Field:2012kd}
R.~Field, The underlying event in hadronic collisions,
Ann. Rev. Nucl. Part. Sci. \textbf{62}, 453 (2012).

%\cite{Christiansen:2024bhe}
\bibitem{Christiansen:2024bhe}
P.~Christiansen and P.~Van Mechelen,
Soft QCD Physics at the LHC: highlights and opportunities,
Annu. Rev. Nucl. Part. Sci. 75, 327 (2025).
%[arXiv:2412.02672 [hep-ex]].

\bibitem{Sjostrand:1987su}
T.~Sjostrand and M.~van Zijl, A Multiple Interaction Model for the Event Structure in Hadron Collisions,
Phys. Rev. D \textbf{36}, 2019 (1987).

%\cite{ZEUS:2007xwd}
\bibitem{ZEUS:2007xwd}
S.~Chekanov \textit{et al.} [ZEUS], Three- and four-jet final states in photoproduction at HERA,
Nucl. Phys. B \textbf{792}, 1 (2008).

%\cite{D0:2009apj}
\bibitem{D0:2009apj}
V.~M.~Abazov \textit{et al.} [D0], Double parton interactions in $\gamma$+3 jet events in $p \bar{p}$ collisions $\sqrt{s}=1.96$ TeV.,
Phys. Rev. D \textbf{81}, 052012 (2010).

\bibitem{Sjostrand:2014zea}
T.~Sj\"ostrand {\it et. al.}, An introduction to PYTHIA 8.2,
Comput. Phys. Commun. \textbf{191}, 159 (2015).

%\cite{CMS:2016fnw}
\bibitem{CMS:2016fnw}
V.~Khachatryan \textit{et al.} [CMS], Evidence for collectivity in $pp$ collisions at the LHC,
Phys. Lett. B \textbf{765}, 193 (2017).

%\cite{Grosse-Oetringhaus:2024bwr}
\bibitem{Grosse-Oetringhaus:2024bwr}
J.~F.~Grosse-Oetringhaus and U.~A.~Wiedemann,
``A Decade of Collectivity in Small Systems,''
[arXiv:2407.07484 [hep-ex]].
%27 citations counted in INSPIRE as of 03 Jun 2025

%\cite{ALICE:2019zfl}
\bibitem{ALICE:2019zfl}
S.~Acharya \textit{et al.} [ALICE], Investigations of Anisotropic Flow Using Multiparticle Azimuthal Correlations in $pp$, p-Pb, Xe-Xe, and Pb-Pb Collisions at the LHC,
Phys. Rev. Lett. \textbf{123}, 142301 (2019).

%\cite{CMS:2010ifv}
\bibitem{CMS:2010ifv}
V.~Khachatryan \textit{et al.} [CMS], Observation of Long-Range Near-Side Angular Correlations in Proton-Proton Collisions at the LHC,
JHEP \textbf{09}, 091 (2010).

%\cite{ATLAS:2015hzw}
\bibitem{ATLAS:2015hzw}
G.~Aad \textit{et al.} [ATLAS], Observation of Long-Range Elliptic Azimuthal Anisotropies in $\sqrt{s}=$13 and 2.76 TeV $pp$ Collisions with the ATLAS Detector,
Phys. Rev. Lett. \textbf{116}, 172301 (2016).

%\cite{CMS:2015fgy}
\bibitem{CMS:2015fgy}
V.~Khachatryan \textit{et al.} [CMS], Measurement of long-range near-side two-particle angular correlations in $pp$ collisions at $\sqrt s =$13 TeV,
Phys. Rev. Lett. \textbf{116} , 172302 (2016).

%\cite{ATLAS:2016yzd}
\bibitem{ATLAS:2016yzd}
M.~Aaboud \textit{et al.} [ATLAS], Measurements of long-range azimuthal anisotropies and associated Fourier coefficients for $pp$ collisions at $\sqrt{s}=5.02$ and $13$ TeV and $p$+Pb collisions at $\sqrt{s_{\mathrm{NN}}}=5.02$ TeV with the ATLAS detector,
Phys. Rev. C \textbf{96}, 024908  (2017).

%\cite{Busza:2018rrf}
\bibitem{Busza:2018rrf}
W.~Busza, K.~Rajagopal and W.~van der Schee, Heavy Ion Collisions: The Big Picture, and the Big Questions,
Ann. Rev. Nucl. Part. Sci. \textbf{68}, 339 (2018).

%\cite{ALICE:2025bwp}
\bibitem{ALICE:2025bwp}
S.~Acharya \textit{et al.} [ALICE], First observation of ultra-long-range azimuthal correlations in low multiplicity $pp$ and p-Pb collisions at the LHC,
[arXiv:2504.02359].


\bibitem{ALICE:2017jyt} 
J.~Adam {\it et al.} [ALICE], Enhanced production of multi-strange hadrons in high-multiplicity proton-proton collisions,
Nature Phys.\  {\bf 13}, 535 (2017).

%\cite{ALICE:2018pal}
\bibitem{ALICE:2018pal}
S.~Acharya \textit{et al.} [ALICE], Multiplicity dependence of light-flavor hadron production in $pp$ collisions at $\sqrt{s}$ = 7 TeV,
Phys. Rev. C \textbf{99}, 024906 (2019).

%\cite{ALICE:2019avo}
\bibitem{ALICE:2019avo}
S.~Acharya \textit{et al.} [ALICE], Multiplicity dependence of (multi-)strange hadron production in proton-proton collisions at $\sqrt{s}$ = 13 TeV,
Eur. Phys. J. C \textbf{80}, 167 (2020).



\bibitem{Nagle:2018nvi}
J.~L.~Nagle and W.~A.~Zajc, Small System Collectivity in Relativistic Hadronic and Nuclear Collisions,
Ann. Rev. Nucl. Part. Sci. \textbf{68}, 211 (2018).

%\cite{ALICE:2023plt}
\bibitem{ALICE:2023plt}
S.~Acharya \textit{et al.} [ALICE], Search for jet quenching effects in high-multiplicity $pp$ collisions at $ \sqrt{s} $ = 13 TeV via di-jet acoplanarity,
JHEP \textbf{05}, 229 (2024).


%\cite{Bierlich:2015rha}
\bibitem{Bierlich:2015rha}
C.~Bierlich and J.~R.~Christiansen, Effects of color reconnection on hadron flavor observables,
Phys. Rev. D \textbf{92}, 094010 (2015).

\bibitem{Bierlich:2014xba}
C.~Bierlich, G.~Gustafson, L.~L\"onnblad and A.~Tarasov, Effects of Overlapping Strings in $pp$ collisions,
JHEP \textbf{03}, 148 (2015).

\bibitem{Bierlich:2016vgw}
C.~Bierlich, G.~Gustafson and L.~L\"onnblad, A shoving model for collectivity in hadronic collisions,
[arXiv:1612.05132].


%\cite{ALICE:2019dfi}
\bibitem{ALICE:2019dfi}
S.~Acharya \textit{et al.} [ALICE], Charged-particle production as a function of multiplicity and transverse spherocity in $pp$ collisions at $\sqrt{s} =5.02$ and 13 TeV,
Eur. Phys. J. C \textbf{79}, 857 (2019).

\bibitem{ALICE:2022qxg}
S.~Acharya \textit{et al.} [ALICE], Study of charged-particle production at high $p_{\rm T}$ using event topology in $pp$, p--Pb and Pb--Pb collisions at $\sqrt{s_{\rm NN}}$ = 5.02 TeV,
Phys. Lett. B \textbf{843}, 137649 (2023).


%\cite{MovillaFernandez:1997fr}
\bibitem{MovillaFernandez:1997fr}
P.~A.~Movilla Fernandez \textit{et al.} [JADE],
A Study of event shapes and determinations of $\alpha_{s}$ using data of $e^{+}e^{-}$ annihilations at $\sqrt{s}$ = 22 to 44-GeV,
Eur. Phys. J. C \textbf{1}, 461 (1998).

%\cite{Donoghue:1979vi}
\bibitem{Donoghue:1979vi}
J.~F.~Donoghue, F.~E.~Low and S.~Y.~Pi, Tensor Analysis of Hadronic Jets in Quantum Chromodynamics,
Phys. Rev. D \textbf{20}, 2759 (1979).

%\cite{Dasgupta:2003iq}
\bibitem{Dasgupta:2003iq}
M.~Dasgupta and G.~P.~Salam, Event shapes in $e^{+}e^{-}$ annihilation and deep inelastic scattering,
J. Phys. G \textbf{30}, R143 (2004).

%\cite{Banfi:2004nk}
\bibitem{Banfi:2004nk}
A.~Banfi, G.~P.~Salam and G.~Zanderighi, Resummed event shapes at hadron - hadron colliders,
JHEP \textbf{08}, 062 (2004).

%\cite{Bethke:2009ehn}
\bibitem{Bethke:2009ehn}
S.~Bethke \textit{et al.} [JADE], Determination of the Strong Coupling $\alpha_{s}$ from hadronic Event Shapes with O($\alpha^{3}_{s}$) and resummed QCD predictions using JADE Data,
Eur. Phys. J. C \textbf{64}, 351 (2009).

%\cite{OPAL:1989fep}
\bibitem{OPAL:1989fep}
M.~Z.~Akrawy \textit{et al.} [OPAL], A Study of Jet Production Rates and a Test of QCD on the Z0 Resonance,
Phys. Lett. B \textbf{235}, 389 (1990).

%\cite{JADE:1982xwg}
\bibitem{JADE:1982xwg}
W.~Bartel \textit{et al.} [JADE], Observation of Four - Jet Structure in $e^+ e^-$ Annihilation at $\sqrt{s}=33$-{GeV},
Phys. Lett. B \textbf{115}, 338 (1982).


%\cite{Brandt:1964sa}
\bibitem{Brandt:1964sa}
S.~Brandt, C.~Peyrou, R.~Sosnowski and A.~Wroblewski, The Principal axis of jets. An Attempt to analyze high-energy collisions as two-body processes,
Phys. Lett. \textbf{12}, 57 (1964).

%\cite{ATLAS:2012tch}
\bibitem{ATLAS:2012tch}
G.~Aad \textit{et al.} [ATLAS], Measurement of event shapes at large momentum transfer with the ATLAS detector in $pp$ collisions at $\sqrt{s}=7$ TeV,
Eur. Phys. J. C \textbf{72}, 2211 (2012).

%\cite{CMS:2011usu}
\bibitem{CMS:2011usu}
V.~Khachatryan \textit{et al.} [CMS], First Measurement of Hadronic Event Shapes in $pp$ Collisions at $\sqrt {s}=7$ TeV,
Phys. Lett. B \textbf{699}, 48 (2011). 

%\cite{Banfi:2010xy}
\bibitem{Banfi:2010xy}
A.~Banfi, G.~P.~Salam and G.~Zanderighi, Phenomenology of event shapes at hadron colliders,
JHEP \textbf{06}, 038 (2010).


%\cite{STAR:2002hbo}
\bibitem{STAR:2002hbo}
C.~Adler \textit{et al.} [STAR], Elliptic flow from two and four particle correlations in Au+Au collisions at $\sqrt{s_{\rm NN}}$ = 130-GeV,
Phys. Rev. C \textbf{66}, 034904 (2002).

%\cite{Fox:1978vu}
\bibitem{Fox:1978vu}
G.~C.~Fox and S.~Wolfram, Observables for the Analysis of Event Shapes in e+ e- Annihilation and Other Processes,
Phys. Rev. Lett. \textbf{41}, 1581 (1978).


%\cite{Cesarotti:2020hwb}
\bibitem{Cesarotti:2020hwb}
C.~Cesarotti and J.~Thaler, A Robust Measure of Event Isotropy at Colliders,
JHEP \textbf{08}, 084 (2020).

%\cite{Cesarotti:2020ngq}
\bibitem{Cesarotti:2020ngq}
C.~Cesarotti, M.~Reece and M.~J.~Strassler, The efficacy of event isotropy as an event shape observable, JHEP \textbf{07}, 215 (2021).

%\cite{ATLAS:2023mny}
\bibitem{ATLAS:2023mny}
G.~Aad \textit{et al.} [ATLAS], Measurements of multijet event isotropies using optimal transport with the ATLAS detector,
JHEP \textbf{10}, 060 (2023).


\bibitem{Martin:2016igp}
T.~Martin, P.~Skands and S.~Farrington, Probing Collective Effects in Hadronisation with the Extremes of the Underlying Event,
Eur. Phys. J. C \textbf{76} 299 (2016).

%\cite{Bencedi:2021tst}
\bibitem{Bencedi:2021tst}
G.~Bencedi, A.~Ortiz and A.~Paz, Disentangling the hard gluon bremsstrahlung effects from the relative transverse activity classifier in $pp$ collisions,
Phys. Rev. D \textbf{104}, 016017 (2021).

%\cite{Ortiz:2017jaz}
\bibitem{Ortiz:2017jaz}
A.~Ortiz and L.~Valencia Palomo, Universality of the underlying event in $pp$ collisions,
Phys. Rev. D \textbf{96}, 114019 (2017).

%\cite{Ortiz:2024ndh}
\bibitem{Ortiz:2024ndh}
A.~Ortiz, G.~Bencedi and F.~Fan, Flattenicity as centrality estimator in p{\textendash}Pb collisions simulated with PYTHIA/Angantyr,
J. Phys. G \textbf{51}, 125003 (2024).

%\cite{Weber:2018ddv}
\bibitem{Weber:2018ddv}
S.~G.~Weber, A.~Dubla, A.~Andronic and A.~Morsch, Elucidating the multiplicity dependence of $\mathrm {J}/\psi $ production in proton\textendash{}proton collisions with PYTHIA8,
Eur. Phys. J. C \textbf{79}, 36 (2019).


%\cite{ALICE:2012cor}
\bibitem{ALICE:2012cor}
B.~Abelev \textit{et al.} [ALICE], Transverse sphericity of primary charged particles in minimum bias proton-proton collisions at $\sqrt{s}=0.9$, 2.76 and 7 TeV,
Eur. Phys. J. C \textbf{72}, 2124 (2012).

%\cite{CMS:2025sws}
\bibitem{CMS:2025sws}
V.~Chekhovsky \textit{et al.} [CMS],Measurement of event shapes in minimum-bias events from proton-proton collisions at s=13{\,}{\,}TeV, Phys. Rev. D \textbf{112}, 112006 (2025).
%0 citations counted in INSPIRE as of 01 Jun 2025

%\cite{ATLAS:2012uka}
\bibitem{ATLAS:2012uka}
G.~Aad \textit{et al.} [ATLAS], Measurement of charged-particle event shape variables in $\sqrt{s}=7$ TeV proton-proton interactions with the ATLAS detector,
Phys. Rev. D \textbf{88}, 032004 (2013).

%\cite{ALICE:2011ac}
\bibitem{ALICE:2011ac}
B.~Abelev \textit{et al.} [ALICE], Underlying Event measurements in $pp$ collisions at $\sqrt{s}=0.9$ and 7 TeV with the ALICE experiment at the LHC,
JHEP \textbf{07}, 116 (2012).

%\cite{ALICE:2019bdw}
\bibitem{ALICE:2019bdw}
S.~Acharya \textit{et al.} [ALICE], Event-shape and multiplicity dependence of freeze-out radii in $pp$ collisions at $ \sqrt{s} $ = 7 TeV,
JHEP \textbf{09}, 108 (2019).

%\cite{ALICE:2019mmy}
\bibitem{ALICE:2019mmy}
S.~Acharya \textit{et al.} [ALICE], Underlying Event properties in $pp$ collisions at $\sqrt{s}$ = 13 TeV,
JHEP \textbf{04}, 192 (2020).

%\cite{ALICE:2022fnb}
\bibitem{ALICE:2022fnb}
S.~Acharya \textit{et al.} [ALICE], Underlying-event properties in $pp$ and p\textendash{}Pb collisions at $ \sqrt{s_{\textrm{NN}}} $ = 5.02 TeV,
JHEP \textbf{06}, 023 (2023).

%\cite{ALICE:2023yuk}
\bibitem{ALICE:2023yuk}
S.~Acharya \textit{et al.} [ALICE], Production of pions, kaons, and protons as a function of the relative transverse activity classifier in $pp$ collisions at $ \sqrt{s} $ = 13 TeV,
JHEP \textbf{06}, 027 (2023).

%\cite{ALICE:2023csm}
\bibitem{ALICE:2023csm}
S.~Acharya \textit{et al.} [ALICE], Charged-particle production as a function of the relative transverse activity classifier in pp, p\textendash{}Pb, and Pb\textendash{}Pb collisions at the LHC,
JHEP \textbf{01}, 056 (2024).
%10 citations counted in INSPIRE as of 19 Feb 2025

%\cite{ALICE:2023bga}
\bibitem{ALICE:2023bga}
S.~Acharya \textit{et al.} [ALICE], Light-flavor particle production in high-multiplicity $pp$ collisions at $ \sqrt{\textrm{s}} $= 13 TeV as a function of transverse spherocity,
JHEP \textbf{05}, 184 (2024).

\bibitem{ALICE:2023dei}
S.~Acharya \textit{et al.} [ALICE], Modification of charged-particle jets in event-shape engineered Pb\textendash{}Pb collisions at $\sqrt{s_{\rm NN}}$ =5.02 TeV,
Phys. Lett. B \textbf{851}, 138584 (2024).

%\cite{ALICE:2024vaf}
\bibitem{ALICE:2024vaf}
S.~Acharya \textit{et al.} [ALICE], Particle production as a function of charged-particle flattenicity in $pp$ collisions at $\sqrt{s}$ = 13 TeV,
Phys. Rev. D \textbf{111}, 012010 (2025).
%\cite{ALICE:2023dei}


%\cite{Rath:2019izg}
\bibitem{Rath:2019izg}
R.~Rath, A.~Khuntia, S.~Tripathy and R.~Sahoo, A Baseline Study of the Event-Shape and Multiplicity Dependence of Chemical Freeze-Out Parameters in Proton-Proton Collisions at $\sqrt{s}$ = 13 TeV Using PYTHIA8,
Physics \textbf{2}, 679 (2020).

%\cite{Tripathy:2019blo}
\bibitem{Tripathy:2019blo}
S.~Tripathy, A.~Bisht, R.~Sahoo, A.~Khuntia and M.~P.~Salvan, Event Shape and Multiplicity Dependence of Freeze-Out Scenario and System Thermodynamics in Proton+Proton Collisions at $\sqrt {s}$ = 13 TeV Using PYTHIA8,
Adv. High Energy Phys. \textbf{2021}, 8822524 (2021).

%\cite{Khuntia:2018qox}
\bibitem{Khuntia:2018qox}
A.~Khuntia, S.~Tripathy, A.~Bisht and R.~Sahoo, Event shape engineering and multiplicity dependent study of identified particle production in proton + proton collisions at $\sqrt{s}$ = 13 TeV using PYTHIA8,
J. Phys. G \textbf{48}, 035102 (2021).
%19 citations counted in INSPIRE as of 24 Jul 2023

%\cite{R:2023lku}
\bibitem{R:2023lku}
A.~M.~K.~R, S.~Prasad, S.~Tripathy, N.~Mallick and R.~Sahoo, Investigating radial-flow-like effects via pseudorapidity and transverse spherocity dependence of particle production in $pp$ collisions at the LHC,
% [arXiv:2309.08336 [hep-ph]].
Eur. Phys. J. Plus \textbf{140}, 110 (2025).
%https://doi.org/10.1140/epjp/s13360-025-05996-9

%\cite{Prasad:2024gqq}
\bibitem{Prasad:2024gqq}
S.~Prasad, B.~Sahoo, S.~Tripathy, N.~Mallick and R.~Sahoo, Probing strangeness with event topology classifiers in $pp$ collisions at energies available at the CERN Large Hadron Collider with the rope hadronization mechanism in PYTHIA,
Phys. Rev. C \textbf{111}, 044902 (2025).
% doi:10.1103/PhysRevC.111.044902
% [arXiv:2409.05454 [hep-ph]].
%4 citations counted in INSPIRE as of 21 Apr 2025


%\cite{Deb:2020ige}
\bibitem{Deb:2020ige}
S.~Deb, R.~Sahoo, D.~Thakur, S.~Tripathy and A.~Khuntia, Investigating heavy-flavor vs light-flavor puzzle with event topology and multiplicity in proton + proton collisions at = 13 TeV using PYTHIA8,
J. Phys. G \textbf{48}, 095104 (2021).

%\cite{Khatun:2019dml}
\bibitem{Khatun:2019dml}
A.~Khatun, D.~Thakur, S.~Deb and R.~Sahoo, $J/\psi$  production dynamics: event shape, multiplicity and rapidity dependence in proton+proton collisions at LHC energies using PYTHIA8,
J. Phys. G \textbf{47}, 055110 (2020).

%\cite{Deb:2019yjo}
\bibitem{Deb:2019yjo}
S.~Deb, G.~Sarwar, R.~Sahoo and J.~e.~Alam, Study of QCD dynamics using small systems,
Eur. Phys. J. A \textbf{57}, 195 (2021).

%\cite{Deb:2020ezw}
\bibitem{Deb:2020ezw}
S.~Deb, S.~Tripathy, G.~Sarwar, R.~Sahoo and J.~e.~Alam, Deciphering QCD dynamics in small collision systems using event shape and final state multiplicity at the Large Hadron Collider,
Eur. Phys. J. A \textbf{56}, 252 (2020).

%\cite{Mallick:2020dzv}
\bibitem{Mallick:2020dzv}
N.~Mallick, S.~Tripathy, R.~Sahoo and A.~Ortiz, First implementation of transverse spherocity analysis for heavy-ion collisions at the Large Hadron Collider energies,
[arXiv:2001.06849].

%\cite{Tripathy:2025npe}
\bibitem{Tripathy:2025npe}
S.~Tripathy, S.~Prasad and R.~Sahoo,
Event-shape dependence of symmetry plane correlations using the Gaussian estimator in Pb-Pb collisions at the LHC using a multiphase transport model,
Phys. Rev. D \textbf{112}, 114012 (2025).
% doi:10.1103/cxx6-kf5y
% [arXiv:2504.09275 [nucl-ex]].
% %4 citations counted in INSPIRE as of 11 Dec 2025

%\cite{Prasad:2021bdq}
\bibitem{Prasad:2021bdq}
S.~Prasad, N.~Mallick, D.~Behera, R.~Sahoo and S.~Tripathy, Event topology and global observables in heavy-ion collisions at the Large Hadron Collider,
Sci. Rep. \textbf{12}, 3917 (2022).


%\cite{Mallick:2020ium}
\bibitem{Mallick:2020ium}
N.~Mallick, R.~Sahoo, S.~Tripathy and A.~Ortiz, Study of Transverse Spherocity and Azimuthal Anisotropy in Pb-Pb collisions at $\sqrt{s_{\rm{NN}}}$ = 5.02 TeV using A Multi-Phase Transport Model,
J. Phys. G \textbf{48}, 045104 (2021).

%\cite{Mallick:2021hcs}
\bibitem{Mallick:2021hcs}
N.~Mallick, S.~Tripathy and R.~Sahoo, Event topology and constituent-quark scaling of elliptic flow in heavy-ion collisions at the Large Hadron Collider using a multiphase transport model,
Eur. Phys. J. C \textbf{82}, 524 (2022).

%\cite{Prasad:2022zbr}
\bibitem{Prasad:2022zbr}
S.~Prasad, N.~Mallick, S.~Tripathy and R.~Sahoo, Probing initial geometrical anisotropy and final azimuthal anisotropy in heavy-ion collisions at Large Hadron Collider energies through event-shape engineering,
Phys. Rev. D \textbf{107}, 074011 (2023).

%\cite{Prasad:2025ezg}
\bibitem{Prasad:2025ezg}
S.~Prasad, A.~M.~Kavumpadikkal Radhakrishnan, R.~Sahoo and N.~Mallick,
Higher order flow coefficients {\textendash} a messenger of QCD medium formed in heavy-ion collisions at the Large Hadron Collider,
Phys. Lett. B \textbf{868}, 139753 (2025).
% doi:10.1016/j.physletb.2025.139753
% [arXiv:2505.01471 [nucl-th]].
%2 citations counted in INSPIRE as of 28 Oct 2025

%\cite{AxialFieldSpectrometer:1983drg}
\bibitem{AxialFieldSpectrometer:1983drg}
T.~Akesson \textit{et al.} [Axial Field Spectrometer], The Dominance of Jets at Large Transverse Energy in a Full Azimuth Hadron Calorimeter at {ISR} Energies,
Phys. Lett. B \textbf{128}, 354 (1983).

%\cite{UA1:1987esv}
\bibitem{UA1:1987esv}
C.~Albajar \textit{et al.} [UA1], Analysis of the Highest Transverse Energy Events Seen in the UA1 Detector at the S$p\bar{p}$S Collider,
Z. Phys. C \textbf{36}, 33 (1987).

%\cite{UA2:1987hpe}
\bibitem{UA2:1987hpe}
R.~Ansari \textit{et al.} [UA2], Jet Measures and Hadronic Event Shapes at the {CERN} $\bar{p} p$ Collider,
Z. Phys. C \textbf{36} (1987), 175

%\cite{CDF:2011yfm}
\bibitem{CDF:2011yfm}
T.~Aaltonen \textit{et al.} [CDF], Measurement of Event Shapes in Proton-Antiproton Collisions at Center-of-Mass Energy 1.96 TeV,
Phys. Rev. D \textbf{83}, 112007 (2011).

%\cite{MovillaFernandez:2001ed}
\bibitem{MovillaFernandez:2001ed}
P.~A.~Movilla Fernandez, S.~Bethke, O.~Biebel and S.~Kluth, Tests of power corrections for event shapes in $e^{+}e^{-}$ annihilation,
Eur. Phys. J. C \textbf{22}, 1 (2001).

%\cite{ZEUS:2002tyf}
\bibitem{ZEUS:2002tyf}
S.~Chekanov \textit{et al.} [ZEUS], Measurement of event shapes in deep inelastic scattering at HERA,
Eur. Phys. J. C \textbf{27}, 531 (2003).

%\cite{L3:2002oql}
\bibitem{L3:2002oql}
P.~Achard \textit{et al.} [L3], Determination of $\alpha_s$ from hadronic event shapes in $e^{+} e^{-}$ annihilation at 192 $\leq \sqrt{s} \leq$ 208 GeV,
Phys. Lett. B \textbf{536}, 217 (2002).

%\cite{ALEPH:2003obs}
\bibitem{ALEPH:2003obs}
A.~Heister \textit{et al.} [ALEPH], Studies of QCD at $e^{+}e^{-}$ centre-of-mass energies between 91-GeV and 209-GeV,
Eur. Phys. J. C \textbf{35}, 457 (2004).

%\cite{DELPHI:2004omy}
\bibitem{DELPHI:2004omy}
J.~Abdallah \textit{et al.} [DELPHI], The Measurement of $\alpha_{s}$ from event shapes with the DELPHI detector at the highest LEP energies,
Eur. Phys. J. C \textbf{37}, 1 (2004).

%\cite{OPAL:2004wof}
\bibitem{OPAL:2004wof}
G.~Abbiendi \textit{et al.} [OPAL], Measurement of event shape distributions and moments in $e^{+}e^{-} -> $ hadrons at 91-GeV - 209-GeV and a determination of $\alpha_{s}$,
Eur. Phys. J. C \textbf{40}, 287 (2005).

%\cite{H1:2005zsk}
\bibitem{H1:2005zsk}
A.~Aktas \textit{et al.} [H1], Measurement of event shape variables in deep-inelastic scattering at HERA,
Eur. Phys. J. C \textbf{46}, 343 (2006).


%\cite{Lin:2004en}
\bibitem{Lin:2004en}
Z.~W.~Lin, C.~M.~Ko, B.~A.~Li, B.~Zhang and S.~Pal, A Multi-phase transport model for relativistic heavy ion collisions,
Phys. Rev. C \textbf{72}, 064901 (2005).
%1408 citations counted in INSPIRE as of 13 Aug 2024

%\cite{He:2017tla}
\bibitem{He:2017tla}
Y.~He and Z.~W.~Lin, Improved Quark Coalescence for a Multi-Phase Transport Model,
Phys. Rev. C \textbf{96}, 014910 (2017).



%\cite{Andersson:1983ia}
\bibitem{Andersson:1983ia}
B.~Andersson, G.~Gustafson, G.~Ingelman and T.~Sjostrand, Parton Fragmentation and String Dynamics,
Phys. Rept. \textbf{97}, 31 (1983).
% doi:10.1016/0370-1573(83)90080-7
%3986 citations counted in INSPIRE as of 22 Aug 2023

%\cite{Corke:2010yf}
\bibitem{Corke:2010yf}
R.~Corke and T.~Sjostrand, Interleaved Parton Showers and Tuning Prospects,
JHEP \textbf{03}, 032 (2011).

%\cite{Sjostrand:2007gs}
\bibitem{Sjostrand:2007gs}
T.~Sjostrand, S.~Mrenna and P.~Z.~Skands, A Brief Introduction to PYTHIA 8.1,
Comput. Phys. Commun. \textbf{178}, 852 (2008).

\bibitem{manual}
Pythia8 online manual: \url{https://pythia.org/latest-manual/Welcome.html}.

%\cite{ALICE:2012fjm}
\bibitem{ALICE:2012fjm}
B.~Abelev \textit{et al.} [ALICE], Measurement of inelastic, single- and double-diffraction cross sections in proton--proton collisions at the LHC with ALICE,
Eur. Phys. J. C \textbf{73}, 2456 (2013).
% doi:10.1140/epjc/s10052-013-2456-0
% [arXiv:1208.4968 [hep-ex]].
%355 citations counted in INSPIRE as of 25 Apr 2025

\bibitem{Miller:2007ri}
M.~L.~Miller, K.~Reygers, S.~J.~Sanders and P.~Steinberg, Glauber modeling in high energy nuclear collisions,
Ann. Rev. Nucl. Part. Sci. \textbf{57}, 205 (2007).

\bibitem{Loizides:2016} C.~Loizides, Glauber modeling of high-energy nuclear collisions at sub-nucleon level,
Phys. Rev. C \textbf{94}, 024914 (2016). 

\bibitem{Glauber:1970}  R.~J.~Glauber and G.~Matthiae, High-energy scattering of protons by nuclei,
Nucl. Phys. B \textbf{21}, 135 (1970).

\bibitem{Wong:1994book} C.~Y.~Wong, Introduction to High-Energy Heavy-Ion Collisions (World Scientific, Singapore, 1994).

\bibitem{Kolb:2001} P.~F.~Kolb, U.~Heinz, P.~Huovinen, K.~J.~Eskola, K.~Tuominen, Centrality dependence of multiplicity, transverse energy, and elliptic flow from hydrodynamics,
Nucl. Phys. A \textbf{696}, 197 (2001).

\bibitem{Bysiak:FIT}
Sebastian Bysiak, The Fast Interaction Trigger Upgrade for ALICE”. In: Proceedings
%of The Eighth Annual Conference on Large Hadron Collider Physics - 
PoS (LHCP2020) 382, 251 (2020).

\bibitem{ALICE:MFTTDR}
“Technical Design Report for the Muon Forward Tracker,” tech. rep., (2015).
\url{https://cds.cern.ch/record/1981898}.
%\cite{Sjostrand:2004pf}

%\cite{CMS:2024ykx}
\bibitem{CMS:2024ykx}
A.~Hayrapetyan \textit{et al.} [CMS], Pseudorapidity distributions of charged hadrons in lead-lead collisions at$\sqrt{s_{ \rm NN}}$ = 5.36 TeV,
Phys. Lett. B \textbf{861}, 139279 (2025).
%\cite{CMS:2011aqh}
\bibitem{CMS:2011aqh}
S.~Chatrchyan \textit{et al.} [CMS], Dependence on pseudorapidity and centrality of charged hadron production in PbPb collisions at a nucleon-nucleon centre-of-mass energy of 2.76 TeV,
JHEP \textbf{08}, 141 (2011).

\bibitem{CMS:2010qvf}
V.~Khachatryan \textit{et al.} [CMS], Charged-particle Multiplicities in $pp$ Interactions at $\sqrt{s}=0.9$, 2.36, and 7 TeV,
JHEP \textbf{01}, 079 (2011).

%\cite{ATLAS:2010jvh}
\bibitem{ATLAS:2010jvh}
G.~Aad \textit{et al.} [ATLAS], Charged-particle multiplicities in pp interactions measured with the ATLAS detector at the LHC,
New J. Phys. \textbf{13}, 053033 (2011).

%\cite{ATLAS:2010zmc}
\bibitem{ATLAS:2010zmc}
G.~Aad \textit{et al.} [ATLAS], Charged-particle multiplicities in $pp$ interactions at $\sqrt{s}=900$ GeV measured with the ATLAS detector at the LHC,
Phys. Lett. B \textbf{688}, 21 (2010).

%\cite{ATLAS:2016zkp}
\bibitem{ATLAS:2016zkp}
G.~Aad \textit{et al.} [ATLAS], Charged-particle distributions in $\sqrt{s}$ = 13 TeV pp interactions measured with the ATLAS detector at the LHC,
Phys. Lett. B \textbf{758}, 67 (2016).

\bibitem{Sjostrand:2004pf}
T.~Sjostrand and P.~Z.~Skands, Multiple interactions and the structure of beam remnants,
JHEP \textbf{03}, 053 (2004).
% doi:10.1088/1126-6708/2004/03/053
% [arXiv:hep-ph/0402078 [hep-ph]].
%319 citations counted in INSPIRE as of 30 Mar 2906

%\cite{Bartalini:2011jp}
\bibitem{Bartalini:2011jp}
P.~Bartalini, E.~L.~Berger, B.~Blok, G.~Calucci, R.~Corke, M.~Diehl, Y.~Dokshitzer, L.~Fano, L.~Frankfurt and J.~R.~Gaunt, \textit{et al.}, Multi-Parton Interactions at the LHC,
[arXiv:1111.0469].
%139 citations counted in INSPIRE as of 30 Mar 2024

%\cite{Hanson:1975fe}
\bibitem{Hanson:1975fe}
G.~Hanson, G.~S.~Abrams, A.~Boyarski, M.~Breidenbach, F.~Bulos, W.~Chinowsky, G.~J.~Feldman, C.~E.~Friedberg, D.~Fryberger and G.~Goldhaber, \textit{et al.}, Evidence for Jet Structure in Hadron Production by $e^{+}e^{-}$ Annihilation,
Phys. Rev. Lett. \textbf{35}, 1609 (1975).


\bibitem{Cuautle:2014yda} 
E.~Cuautle, R.~Jimenez, I.~Maldonado, A.~Ortiz, G.~Paic and E.~Perez, Disentangling the soft and hard components of the $pp$ collisions using the sphero(i)city approach,
[arXiv:1404.2372].

\bibitem{Cuautle:2015kra} 
A.~Ortiz, G.~Paic and E.~Cuautle, Mid-rapidity charged hadron transverse spherocity in $pp$ collisions simulated with Pythia,
Nucl.\ Phys.\ A {\bf 941}, 78 (2015).	  

\bibitem{Ortiz:2017jho} A.~Ortiz, Experimental results on event shapes at hadron colliders, Adv. Ser. Dir. High Energy Phys. 29, 343 (2018).

%\cite{CMS:2019csb}
\bibitem{CMS:2019csb}
A.~M.~Sirunyan \textit{et al.} [CMS], Extraction and validation of a new set of CMS PYTHIA8 tunes from underlying-event measurements,
Eur. Phys. J. C \textbf{80}, 4 (2020).

%\cite{ATLAS:2016puo}
\bibitem{ATLAS:2016puo}
 [ATLAS], The Pythia 8 A3 tune description of ATLAS minimum bias and inelastic measurements incorporating the Donnachie-Landshoff diffractive model,
ATL-PHYS-PUB-2016-017.

%\cite{TheATLAScollaboration:2014rfk}
\bibitem{TheATLAScollaboration:2014rfk}
ATLAS Pythia 8 tunes to 7 TeV data, ATL-PHYS-PUB-2014-021.

%\cite{CMS:2020dqt}
\bibitem{CMS:2020dqt}
A.~M.~Sirunyan \textit{et al.} [CMS], Development and validation of HERWIG 7 tunes from CMS underlying-event measurements,
Eur. Phys. J. C \textbf{81}, 312 (2021).


%\cite{Ortiz:2022zqr}
\bibitem{Ortiz:2022zqr}
A.~Ortiz and G.~Paic, A look into the \textquotedblleft{}hedgehog\textquotedblright{} events in $pp$ collisions,
Rev. Mex. Fis. Suppl. \textbf{3}, 040911 (2022).
% doi:10.31349/SuplRevMexFis.3.040911
% [arXiv:2204.13733 [hep-ph]].
%4 citations counted in INSPIRE as of 25 Jan 2024


%\cite{Ortiz:2022mfv}
\bibitem{Ortiz:2022mfv}
A.~Ortiz, A.~Khuntia, O.~V\'azquez-Rueda, S.~Tripathy, G.~Bencedi, S.~Prasad and F.~Fan, Unveiling the effects of multiple soft partonic interactions in $pp$ collisions at $\sqrt{s}$ = 13.6 TeV using a new event classifier,
Phys. Rev. D \textbf{107}, 076012 (2023).

\bibitem{QM2025:flatPoster}
S. Prasad [ALICE], presented in Quark Matter 2025 conference, Available Online: \url{https://indico.cern.ch/event/1334113/contributions/6291974/}

%\cite{ATLAS:2015qwl}
\bibitem{ATLAS:2015qwl}
G.~Aad \textit{et al.} [ATLAS], Measurement of the correlation between flow harmonics of different order in lead-lead collisions at $\sqrt{s_{ \rm NN}}$ = 2.76 TeV with the ATLAS detector,
Phys. Rev. C \textbf{92}, 034903 (2015).


%\cite{CMS:2018mdd}
\bibitem{CMS:2018mdd}
A.~M.~Sirunyan \textit{et al.} [CMS], Study of the underlying event in top quark pair production in $pp$ collisions at 13 $~\text {Te}\text {V}$,
Eur. Phys. J. C \textbf{79}, 123 (2019).
% doi:10.1140/epjc/s10052-019-6620-z
% [arXiv:1807.02810 [hep-ex]].
%39 citations counted in INSPIRE as of 20 Feb 2025


%\cite{Bernaciak:2012nh}
\bibitem{Bernaciak:2012nh}
C.~Bernaciak, M.~S.~A.~Buschmann, A.~Butter and T.~Plehn, Fox-Wolfram Moments in Higgs Physics,
Phys. Rev. D \textbf{87}, 073014 (2013).


%\cite{Spiller:2015axa}
\bibitem{Spiller:2015axa}
L.~A.~Spiller, Modification of Fox-Wolfram Moments for Hadron Colliders,
JHEP \textbf{03}, 027 (2016).

%\cite{Kong:2024wdk}
\bibitem{Kong:2024wdk}
W.~X.~Kong and B.~W.~Zhang, Fox-Wolfram moment of jet production in relativistic heavy-ion collisions*, Chin. Phys. \textbf{49}, 094101 (2025).

%\cite{Barber:1979yr}
\bibitem{Barber:1979yr}
D.~P.~Barber, U.~Becker, H.~Benda, A.~Boehm, J.~G.~Branson, J.~Bron, D.~Buikman, J.~Burger, C.~C.~Chang and H.~S.~Chen, \textit{et al.}, Discovery of Three Jet Events and a Test of Quantum Chromodynamics at PETRA Energies,
Phys. Rev. Lett. \textbf{43}, 830 (1979).

%\cite{Clavelli:1981yh}
\bibitem{Clavelli:1981yh}
L.~Clavelli and D.~Wyler, Kinematical Bounds on Jet Variables and the Heavy Jet Mass Distribution,
Phys. Lett. B \textbf{103}, 383 (1981).

%\cite{CMS:2014tkl}
\bibitem{CMS:2014tkl}
V.~Khachatryan \textit{et al.} [CMS],
Study of Hadronic Event-Shape Variables in Multijet Final States in pp Collisions at $\sqrt{s}$ = 7 TeV,
JHEP \textbf{10}, 087 (2014).
% doi:10.1007/JHEP10(2014)087
% [arXiv:1407.2856 [hep-ex]].
%39 citations counted in INSPIRE as of 02 Nov 2025

%\cite{CMS:2018svp}
\bibitem{CMS:2018svp}
A.~M.~Sirunyan \textit{et al.} [CMS],
Event shape variables measured using multijet final states in proton-proton collisions at $ \sqrt{s}=13 $ TeV,
JHEP \textbf{12}, 117 (2018).
% doi:10.1007/JHEP12(2018)117
% [arXiv:1811.00588 [hep-ex]].
%33 citations counted in INSPIRE as of 28 Oct 2025



%\cite{Larkoski:2017jix}
\bibitem{Larkoski:2017jix}
A.~J.~Larkoski, I.~Moult and B.~Nachman,
Jet Substructure at the Large Hadron Collider: A Review of Recent Advances in Theory and Machine Learning,
Phys. Rept. \textbf{841}, 1 (2020).
% doi:10.1016/j.physrep.2019.11.001
% [arXiv:1709.04464 [hep-ph]].
%611 citations counted in INSPIRE as of 29 Nov 2025

%\cite{Kogler:2018hem}
\bibitem{Kogler:2018hem}
R.~Kogler, B.~Nachman, A.~Schmidt, L.~Asquith, M.~Campanelli, C.~Delitzsch, P.~Harris, A.~Hinzmann, D.~Kar and C.~McLean, \textit{et al.}
Jet Substructure at the Large Hadron Collider: Experimental Review,
Rev. Mod. Phys. \textbf{91}, 045003 (2019).
% doi:10.1103/RevModPhys.91.045003
% [arXiv:1803.06991 [hep-ex]].
%362 citations counted in INSPIRE as of 29 Nov 2025

%\cite{Krohn:2009th}
\bibitem{Krohn:2009th}
D.~Krohn, J.~Thaler and L.~T.~Wang,
Jet Trimming,
JHEP \textbf{02}, 084 (2010).
% doi:10.1007/JHEP02(2010)084
% [arXiv:0912.1342 [hep-ph]].
%1027 citations counted in INSPIRE as of 29 Nov 2025

%\cite{Ellis:2009su}
\bibitem{Ellis:2009su}
S.~D.~Ellis, C.~K.~Vermilion and J.~R.~Walsh,
Techniques for improved heavy particle searches with jet substructure,
Phys. Rev. D \textbf{80}, 051501 (2009).
% doi:10.1103/PhysRevD.80.051501
% [arXiv:0903.5081 [hep-ph]].
%449 citations counted in INSPIRE as of 29 Nov 2025

%\cite{Ellis:2009me}
\bibitem{Ellis:2009me}
S.~D.~Ellis, C.~K.~Vermilion and J.~R.~Walsh,
Recombination Algorithms and Jet Substructure: Pruning as a Tool for Heavy Particle Searches,
Phys. Rev. D \textbf{81}, 094023 (2010).
% doi:10.1103/PhysRevD.81.094023
% [arXiv:0912.0033 [hep-ph]].
%537 citations counted in INSPIRE as of 29 Nov 2025

%\cite{Butterworth:2008iy}
\bibitem{Butterworth:2008iy}
J.~M.~Butterworth, A.~R.~Davison, M.~Rubin and G.~P.~Salam,
Jet substructure as a new Higgs search channel at the LHC,
Phys. Rev. Lett. \textbf{100}, 242001 (2008).
% doi:10.1103/PhysRevLett.100.242001
% [arXiv:0802.2470 [hep-ph]].
%1305 citations counted in INSPIRE as of 29 Nov 2025

%\cite{Dasgupta:2013ihk}
\bibitem{Dasgupta:2013ihk}
M.~Dasgupta, A.~Fregoso, S.~Marzani and G.~P.~Salam,
Towards an understanding of jet substructure
JHEP \textbf{09}, 029 (2013).
% doi:10.1007/JHEP09(2013)029
% [arXiv:1307.0007 [hep-ph]].
%708 citations counted in INSPIRE as of 29 Nov 2025

%\cite{Larkoski:2014wba}
\bibitem{Larkoski:2014wba}
A.~J.~Larkoski, S.~Marzani, G.~Soyez and J.~Thaler,
Soft Drop,
JHEP \textbf{05}, 146 (2014).
% doi:10.1007/JHEP05(2014)146
% [arXiv:1402.2657 [hep-ph]].
%1195 citations counted in INSPIRE as of 29 Nov 2025


%\cite{L3:1992nwf}
\bibitem{L3:1992nwf}
B.~Adeva \textit{et al.} [L3], Studies of hadronic event structure and comparisons with QCD models at the Z0 resonance,
Z. Phys. C \textbf{55}, 39 (1992).

%\cite{L3:1992btq}
\bibitem{L3:1992btq}
O.~Adrian \textit{et al.} [L3], Determination of $\alpha_s$ from hadronic event shapes measured on the Z0 resonance,
Phys. Lett. B \textbf{284}, 471 (1992).

%\cite{L3:1995eyy}
\bibitem{L3:1995eyy}
M.~Acciarri \textit{et al.} [L3], Study of the structure of hadronic events and determination of $\alpha_{s}$ at $\sqrt{s}$ = 130 GeV and 136 GeV,
Phys. Lett. B \textbf{371} 137 (1996).

%\cite{L3:1997bxr}
\bibitem{L3:1997bxr}
M.~Acciarri \textit{et al.} [L3], QCD studies and determination of $\alpha_{s}$ in $e^{+}e^{-}$ collisions at $\sqrt{s}$ = 161 GeV and 172 GeV,
Phys. Lett. B \textbf{404}, 390 (1997).

%\cite{L3:1997dkv}
\bibitem{L3:1997dkv}
M.~Acciarri \textit{et al.} [L3], Study of hadronic events and measurements of $\alpha_{s}$ between 30 GeV and 91 GeV,
Phys. Lett. B \textbf{411}, 339 (1997).


%\cite{L3:1998ubl}
\bibitem{L3:1998ubl}
M.~Acciarri \textit{et al.} [L3], QCD results from studies of hadronic events produced in $e^{+}e^{-}$ annihilations at $\sqrt{s}$ = 183 GeV,
Phys. Lett. B \textbf{444}, 569 (1998).

%\cite{L3:2000shd}
\bibitem{L3:2000shd}
M.~Acciarri \textit{et al.} [L3], QCD studies in $e^{+} e^{-}$ annihilation from 30 GeV to 189 GeV,
Phys. Lett. B \textbf{489}, 65 (2000).

%\cite{L3:2004cdh}
\bibitem{L3:2004cdh}
P.~Achard \textit{et al.} [L3], Studies of hadronic event structure in $e^{+} e^{-}$ annihilation from 30 GeV to 209 GeV with the L3 detector,
Phys. Rept. \textbf{399}, 71 (2004).

%\cite{Kittel:2005fu}
\bibitem{Kittel:2005fu}
W.~Kittel and E.~A.~De Wolf, Soft multihadron dynamics, World Scientific (2005).

%\cite{Catani:1992jc}
\bibitem{Catani:1992jc}
S.~Catani, G.~Turnock and B.~R.~Webber, Jet broadening measures in $e^{+} e^{-}$ annihilation,
Phys. Lett. B \textbf{295}, 269 (1992).

%\cite{Catani:1992ua}
\bibitem{Catani:1992ua}
S.~Catani, L.~Trentadue, G.~Turnock and B.~R.~Webber, Resummation of large logarithms in $e^{+}e^{-}$ event shape distributions,
Nucl. Phys. B \textbf{407}, 3 (1993).

%\cite{Ellis:1980wv}
\bibitem{Ellis:1980wv}
R.~K.~Ellis, D.~A.~Ross and A.~E.~Terrano, The Perturbative Calculation of Jet Structure in e+ e- Annihilation,
Nucl. Phys. B \textbf{178}, 421 (1981).

%\cite{Soldner-Rembold:2004lzl}
\bibitem{Soldner-Rembold:2004lzl}
S.~Soldner-Rembold, QCD studies and alpha(s) measurements at LEP,
[arXiv:hep-ex/0411006 [hep-ex]].



%\cite{ZEUS:2006vwm}
\bibitem{ZEUS:2006vwm}
S.~Chekanov \textit{et al.} [ZEUS], Event shapes in deep inelastic scattering at HERA,
Nucl. Phys. B \textbf{767} (2007), 1-28.

%\cite{Field:2001aok}
\bibitem{Field:2001aok}
R.~D.~Field [CDF], The Underlying Event in Hard Scattering Processes,
eConf \textbf{C010630} (2001), P501
[arXiv:hep-ph/0201192 [hep-ph]].

%\cite{Banfi:2010zza}
\bibitem{Banfi:2010zza}
A.~Banfi, Event shapes at hadron colliders,
PoS \textbf{DIS2010} (2010), 099.

%\cite{Belle:2004yyi}
\bibitem{Belle:2004yyi}
K.~Abe \textit{et al.} [Belle], Measurement of branching fraction and CP asymmetry in B ---{\ensuremath{>}} eta h decays,
[arXiv:hep-ex/0408131 [hep-ex]].

%\cite{Ortiz:2020rwg}
\bibitem{Ortiz:2020rwg}
A.~Ortiz, A.~Paz, J.~D.~Romo, S.~Tripathy, E.~A.~Zepeda and I.~Bautista, Multiparton interactions in $pp$ collisions from machine learning-based regression,
Phys. Rev. D \textbf{102}, 076014 (2020).
% doi:10.1103/PhysRevD.102.076014
% [arXiv:2004.03800 [hep-ph]].
%26 citations counted in INSPIRE as of 09 Feb 2025

%\cite{ALICE:2020nkc}
\bibitem{ALICE:2020nkc}
S.~Acharya \textit{et al.} [ALICE], Multiplicity dependence of $\pi $, K, and p production in $pp$ collisions at $\sqrt{s} = 13$ TeV,
Eur. Phys. J. C \textbf{80}, 693 (2020).
%doi:10.1140/epjc/s10052-020-8125-1
%[arXiv:2003.02394 [nucl-ex]].
%98 citations counted in INSPIRE as of 07 Mar 2024

%\cite{ALICE:2022wpn}
\bibitem{ALICE:2022wpn}
S.~Acharya \textit{et al.} [ALICE], The ALICE experiment: a journey through QCD,
Eur. Phys. J. C \textbf{84}, 813 (2024).


%\cite{OrtizVelasquez:2013ofg}
\bibitem{OrtizVelasquez:2013ofg}
A.~Ortiz Velasquez, P.~Christiansen, E.~Cuautle Flores, I.~Maldonado Cervantes and G.~Pai\'c, Color Reconnection and Flowlike Patterns in $pp$ Collisions,
Phys. Rev. Lett. \textbf{111}, 042001 (2013).


%\cite{ALICE:2019hno}
\bibitem{ALICE:2019hno}
S.~Acharya \textit{et al.} [ALICE], Production of charged pions, kaons, and (anti-)protons in Pb-Pb and inelastic $pp$ collisions at $\sqrt {s_{\rm NN}}$ = 5.02 TeV,
Phys. Rev. C \textbf{101}, 044907 (2020).
% doi:10.1103/PhysRevC.101.044907
% [arXiv:1910.07678 [nucl-ex]].
%301 citations counted in INSPIRE as of 07 Feb 2025


%\cite{Ortiz:2016kpz}
\bibitem{Ortiz:2016kpz}
A.~Ortiz, G.~Bencedi and H.~Bello, Revealing the source of the radial flow patterns in proton\textendash{}proton collisions using hard probes,
J. Phys. G \textbf{44}, 065001 (2017).
% doi:10.1088/1361-6471/aa6594
% [arXiv:1608.04784 [hep-ph]].
%49 citations counted in INSPIRE as of 20 Feb 2025

%\cite{Tang:2008ud}
\bibitem{Tang:2008ud}
Z.~Tang, Y.~Xu, L.~Ruan, G.~van Buren, F.~Wang and Z.~Xu,
Spectra and radial flow at RHIC with Tsallis statistics in a Blast-Wave description,
Phys. Rev. C \textbf{79}, 051901 (2009).
% doi:10.1103/PhysRevC.79.051901
% [arXiv:0812.1609 [nucl-ex]].
%232 citations counted in INSPIRE as of 15 Nov 2025

%\cite{He:2025bnp}
\bibitem{He:2025bnp}
J.~He, X.~Peng, Z.~Yin and L.~Zheng,
Extracting the kinetic freeze-out properties of high energy pp collisions at the LHC with event shape classifiers,
Chin. Phys. C \textbf{50}, 014108 (2026).
% doi:10.1088/1674-1137/ae07ba
% [arXiv:2506.20612 [hep-ph]].
%0 citations counted in INSPIRE as of 15 Nov 2025


%\cite{Ollitrault:1992bk}
\bibitem{Ollitrault:1992bk}
J.~Y.~Ollitrault, Anisotropy as a signature of transverse collective flow,
Phys. Rev. D \textbf{46}, 229 (1992).
% doi:10.1103/PhysRevD.46.229
%1572 citations counted in INSPIRE as of 19 Feb 2025


%\cite{Jiang:2021foj}
\bibitem{Jiang:2021foj}
Z.~F.~Jiang, C.~B.~Yang and Q.~Peng, Directed flow of charged particles within idealized viscous hydrodynamics at energies available at the BNL Relativistic Heavy Ion Collider and at the CERN Large Hadron Collider,
Phys. Rev. C \textbf{104}, 064903 (2021).
% doi:10.1103/PhysRevC.104.064903
% [arXiv:2111.01994 [hep-ph]].
%14 citations counted in INSPIRE as of 19 Feb 2025


%\cite{ALICE:2024vzv}
\bibitem{ALICE:2024vzv}
S.~Acharya \textit{et al.} [ALICE], Observation of partonic flow in proton-proton and proton-nucleus collisions, Nature Commun. {\bf 17}, 2585 (2026).

%\cite{STAR:2003wqp}
\bibitem{STAR:2003wqp}
J.~Adams \textit{et al.} [STAR], Particle type dependence of azimuthal anisotropy and nuclear modification of particle production in Au + Au collisions at $\sqrt{s_{\rm NN}}$ = 200-GeV,
Phys. Rev. Lett. \textbf{92}, 052302 (2004).
% doi:10.1103/PhysRevLett.92.052302
% [arXiv:nucl-ex/0306007 [nucl-ex]].
% %744 citations counted in INSPIRE as of 19 Feb 2025

%\cite{CMS:2018loe}
\bibitem{CMS:2018loe}
A.~M.~Sirunyan \textit{et al.} [CMS],
Elliptic flow of charm and strange hadrons in high-multiplicity pPb collisions at $\sqrt{s_{_\mathrm{NN}}} =$ 8.16 TeV,
Phys. Rev. Lett. \textbf{121}, 082301 (2018).
% doi:10.1103/PhysRevLett.121.082301
% [arXiv:1804.09767 [hep-ex]].
% %145 citations counted in INSPIRE as of 19 Feb 2025

\bibitem{9160009}
T.~Bonnevier Wallstedt, Investigation of event-shape classifiers for proton-proton collisions with the ALICE experiment, 
Student Paper, Lund University, (2024). Available at: \url{http://lup.lub.lu.se/student-papers/record/9160009}.



%\cite{Chen:2023njr}
\bibitem{Chen:2023njr}
Y.~C.~Chen, Y.~Chen, A.~Badea, A.~Baty, G.~M.~Innocenti, M.~Maggi, C.~McGinn, M.~Peters, T.~A.~Sheng and J.~Thaler, \textit{et al.}, Long-range near-side correlation in $e^{+}e^{-}$ collisions at 183-209 GeV with ALEPH archived data,
Phys. Lett. B \textbf{856}, 138957 (2024).




%\cite{CMS:2013lua}
\bibitem{CMS:2013lua}
S.~Chatrchyan \textit{et al.} [CMS],
Event Shapes and Azimuthal Correlations in $Z$ + Jets Events in $pp$ Collisions at $\sqrt{s}=7$ TeV,
Phys. Lett. B \textbf{722}, 238 (2013).
% doi:10.1016/j.physletb.2013.04.025
% [arXiv:1301.1646 [hep-ex]].
%106 citations counted in INSPIRE as of 29 Oct 2025


%\cite{CMS:2020fae}
\bibitem{CMS:2020fae}
A.~M.~Sirunyan \textit{et al.} [CMS],
Investigation into the event-activity dependence of $\Upsilon$(nS) relative production in proton-proton collisions at $ \sqrt{s} $ = 7 TeV,
JHEP \textbf{11}, 001 (2020).
% doi:10.1007/JHEP11(2020)001
% [arXiv:2007.04277 [hep-ex]].
%41 citations counted in INSPIRE as of 28 Oct 2025

%\cite{CMS:2015wcf}
\bibitem{CMS:2015wcf}
V.~Khachatryan \textit{et al.} [CMS], Event generator tunes obtained from underlying event and multiparton scattering measurements,
Eur. Phys. J. C \textbf{76}, 155 (2016).


%\cite{Baty:2021ugw}
\bibitem{Baty:2021ugw}
A.~Baty, P.~Gardner and W.~Li,
Novel observables for exploring QCD collective evolution and quantum entanglement within individual jets,
Phys. Rev. C \textbf{107}, 064908 (2023).
% doi:10.1103/PhysRevC.107.064908
% [arXiv:2104.11735 [hep-ph]].
% %30 citations counted in INSPIRE as of 03 Nov 2025

%\cite{CMS:2023iam}
\bibitem{CMS:2023iam}
A.~Hayrapetyan \textit{et al.} [CMS],
Observation of Enhanced Long-Range Elliptic Anisotropies Inside High-Multiplicity Jets in pp Collisions at $\sqrt{s}=13${\,}{\,}TeV,
Phys. Rev. Lett. \textbf{133}, 142301 (2024).
% doi:10.1103/PhysRevLett.133.142301
% [arXiv:2312.17103 [hep-ex]].
%25 citations counted in INSPIRE as of 29 Oct 2025



%\cite{ALICE:2018vuu}
\bibitem{ALICE:2018vuu}
S.~Acharya \textit{et al.} [ALICE], Transverse momentum spectra and nuclear modification factors of charged particles in $pp$, p-Pb and Pb-Pb collisions at the LHC,
JHEP \textbf{11}, 013 (2018).

%\cite{ALICE:2017svf}
\bibitem{ALICE:2017svf}
S.~Acharya \textit{et al.} [ALICE], Constraints on jet quenching in p-Pb collisions at $\sqrt{s_{\rm NN}}$ = 5.02 TeV measured by the event-activity dependence of semi-inclusive hadron-jet distributions,
Phys. Lett. B \textbf{783}, 95 (2018).

%\cite{ALICE:2023jye}
\bibitem{ALICE:2023jye}
S.~Acharya \textit{et al.} [ALICE], Measurements of jet quenching using semi-inclusive hadron+jet distributions in $pp$ and central Pb-Pb collisions at $\sqrt{s_{\rm NN}}$ = 5.02 TeV,
Phys. Rev. C \textbf{110}, 014906 (2024).


%\cite{STAR:2006vcp}
\bibitem{Adams:2006yt}
J.~Adams \textit{et al.} [STAR], Direct observation of dijets in central Au+Au collisions at $\sqrt{s_{\rm NN}}$ = 200-GeV,
Phys. Rev. Lett. \textbf{97}, 162301 (2006).

%\cite{ALICE:2011gpa}
\bibitem{ALICE:2011gpa}
K.~Aamodt \textit{et al.} [ALICE], Particle-yield modification in jet-like azimuthal di-hadron correlations in Pb-Pb collisions at $\sqrt{s_{NN}} = 2.76$ TeV,
Phys. Rev. Lett. \textbf{108}, 092301 (2012).
%186 citations counted in INSPIRE as of 19 Feb 2025

%\cite{ALICE:2016gso}
\bibitem{ALICE:2016gso}
J.~Adam \textit{et al.} [ALICE],
Jet-like correlations with neutral pion triggers in pp and central Pb\textendash{}Pb collisions at 2.76 TeV,
Phys. Lett. B \textbf{763}, 238 (2016).
%49 citations counted in INSPIRE as of 19 Feb 2025

%\cite{Qin:2015srf}
\bibitem{Qin:2015srf}
G.~Y.~Qin and X.~N.~Wang, Jet quenching in high-energy heavy-ion collisions,
Int. J. Mod. Phys. E \textbf{24}, 1530014 (2015).
%366 citations counted in INSPIRE as of 19 Feb 2025

%\cite{Mendez:2025dqz}
\bibitem{Mendez:2025dqz}
J.~E.~M.~M{\'e}ndez and A.~Ortiz, Investigating the most active pp collisions (top 0.1{\%}) using the tools developed by experiments at the LHC,
J. Phys. G \textbf{52}, 095001 (2025).


%\cite{Cacciari:2011ma}
\bibitem{Cacciari:2011ma}
M.~Cacciari, G.~P.~Salam and G.~Soyez, FastJet User Manual,
Eur. Phys. J. C \textbf{72}, 1896 (2012).


%\cite{Sahu:2025mla}
\bibitem{Sahu:2025mla}
D.~Sahu, A.~Ortiz and V.~Vazquez Campos, Understanding the medium-like effects in the jet-like yield in pp and p{\textendash}Pb collisions using event generators,
J. Phys. G \textbf{52}, 125104 (2025).
%0 citations counted in INSPIRE as of 15 Nov 2025

%\cite{Peng:2025mpf}
\bibitem{Peng:2025mpf}
Y.~Peng, Y.~Wu, X.~Peng, Z.~Yin and L.~Zheng, Investigating jet-induced identified hadron production from the relative transverse activity classifier in pp collisions at the LHC,''
Phys. Rev. D \textbf{112}, 11 (2025).

%\cite{ALICE:2015lib}
\bibitem{ALICE:2015lib}
J.~Adam \textit{et al.} [ALICE], Event shape engineering for inclusive spectra and elliptic flow in Pb-Pb collisions at $\sqrt{s_{\rm NN}}=2.76$ TeV,
Phys. Rev. C \textbf{93}, 034916 (2016).
% doi:10.1103/PhysRevC.93.034916
% [arXiv:1507.06194 [nucl-ex]].
%62 citations counted in INSPIRE as of 28 May 2025



%\cite{Sambataro:2022sns}
\bibitem{Sambataro:2022sns}
M.~L.~Sambataro, Y.~Sun, V.~Minissale, S.~Plumari and V.~Greco, Event-shape engineering analysis of D meson in ultrarelativistic heavy-ion collisions, Eur. Phys. J. C \textbf{82}, 833 (2022).

%\cite{ALICE:2023wdn}
\bibitem{ALICE:2023wdn}
S.~Acharya \textit{et al.} [ALICE], Symmetry plane correlations in Pb-Pb collisions at $\sqrt{s_{\rm NN}} = 2.76$TeV,
Eur. Phys. J. C \textbf{83}, 576 (2023).
%doi:10.1140/epjc/s10052-023-11658-w
%[arXiv:2302.01234 [nucl-ex]].
%4 citations counted in INSPIRE as of 20 Feb 2025

%\cite{Bhalerao:2011yg}
\bibitem{Bhalerao:2011yg}
R.~S.~Bhalerao, M.~Luzum and J.~Y.~Ollitrault, Determining initial-state fluctuations from flow measurements in heavy-ion collisions,
Phys. Rev. C \textbf{84}, 034910 (2011).
%doi:10.1103/PhysRevC.84.034910
%[arXiv:1104.4740 [nucl-th]].
%163 citations counted in INSPIRE as of 23 Oct 2024


%\cite{ALICE:2024fus}
\bibitem{ALICE:2024fus}
S.~Acharya \textit{et al.} [ALICE],
``Higher-order symmetry plane correlations in Pb-Pb collisions at $\sqrt{s_{\rm NN}}$ = 5.02 TeV,''
Phys. Rev. C \textbf{111}, 064913 (2025).


%\cite{Jia:2014jca}
\bibitem{Jia:2014jca}
J.~Jia, Event-shape fluctuations and flow correlations in ultra-relativistic heavy-ion collisions,
J. Phys. G \textbf{41}, 124003 (2014).
% doi:10.1088/0954-3899/41/12/124003
% [arXiv:1407.6057 [nucl-ex]].
%72 citations counted in INSPIRE as of 07 Apr 2025

%\cite{Niemi:2015qia}
\bibitem{Niemi:2015qia}
H.~Niemi, K.~J.~Eskola and R.~Paatelainen,Event-by-event fluctuations in a perturbative QCD + saturation + hydrodynamics model: Determining QCD matter shear viscosity in ultrarelativistic heavy-ion collisions,
Phys. Rev. C \textbf{93}, 024907 (2016).
% doi:10.1103/PhysRevC.93.024907
% [arXiv:1505.02677 [hep-ph]].
%347 citations counted in INSPIRE as of 07 Apr 2025

%\cite{Qian:2016pau}
\bibitem{Qian:2016pau}
J.~Qian and U.~Heinz, Hydrodynamic flow amplitude correlations in event-by-event fluctuating heavy-ion collisions,
Phys. Rev. C \textbf{94}, 024910 (2016).
% doi:10.1103/PhysRevC.94.024910
% [arXiv:1607.01732 [nucl-th]].
%36 citations counted in INSPIRE as of 07 Apr 2025

%\cite{ALICE:2017kwu}
\bibitem{ALICE:2017kwu}
S.~Acharya \textit{et al.} [ALICE],
Systematic studies of correlations between different order flow harmonics in Pb-Pb collisions at $\sqrt{s_{\rm NN}}$ = 2.76 TeV,
Phys. Rev. C \textbf{97}, 024906 (2018).
% doi:10.1103/PhysRevC.97.024906
% [arXiv:1709.01127 [nucl-ex]].
%57 citations counted in INSPIRE as of 07 Apr 2025

%\cite{ALICE:2016kpq}
\bibitem{ALICE:2016kpq}
J.~Adam \textit{et al.} [ALICE],
Correlated event-by-event fluctuations of flow harmonics in Pb-Pb collisions at $\sqrt{s_{_{\rm NN}}}=2.76$ TeV,
Phys. Rev. Lett. \textbf{117}, 182301 (2016).
% doi:10.1103/PhysRevLett.117.182301
% [arXiv:1604.07663 [nucl-ex]].
%226 citations counted in INSPIRE as of 07 Apr 2025

%\cite{Niemi:2012aj}
\bibitem{Niemi:2012aj}
H.~Niemi, G.~S.~Denicol, H.~Holopainen and P.~Huovinen, Event-by-event distributions of azimuthal asymmetries in ultrarelativistic heavy-ion collisions,
Phys. Rev. C \textbf{87}, 054901 (2013).
% doi:10.1103/PhysRevC.87.054901
% [arXiv:1212.1008 [nucl-th]].
%326 citations counted in INSPIRE as of 07 Apr 2025


%\cite{ALICE:2017sss}
\bibitem{ALICE:2017sss}
S.~Acharya \textit{et al.} [ALICE], Constraining the magnitude of the Chiral Magnetic Effect with Event Shape Engineering in Pb-Pb collisions at $\sqrt{s_\mathrm{NN}}$ = 2.76 TeV,
Phys. Lett. B \textbf{777}, 151 (2018).


%\cite{CMS:2017lrw}
\bibitem{CMS:2017lrw}
A.~M.~Sirunyan \textit{et al.} [CMS], Constraints on the chiral magnetic effect using charge-dependent azimuthal correlations in $p\mathrm{Pb}$ and PbPb collisions at the CERN Large Hadron Collider,
Phys. Rev. C \textbf{97}, 044912 (2018).

%\cite{Milton:2021wku}
\bibitem{Milton:2021wku}
R.~Milton, G.~Wang, M.~Sergeeva, S.~Shi, J.~Liao and H.~Z.~Huang, Utilization of event shape in search of the chiral magnetic effect in heavy-ion collisions,
Phys. Rev. C \textbf{104}, 064906 (2021).

%\cite{Xu:2023elq}
\bibitem{Xu:2023elq}
Z.~Xu, B.~Chan, G.~Wang, A.~Tang and H.~Z.~Huang, Event shape selection method in search of the chiral magnetic effect in heavy-ion collisions,
Phys. Lett. B \textbf{848}, 138367 (2024).


%\cite{Li:2024gdz}
\bibitem{Li:2024gdz}
H.~S.~Li, Y.~Feng and F.~Wang, Investigating event-shape methods in the search for the chiral magnetic effect in relativistic heavy ion collisions, Phys. Rev. C \textbf{112}, 054901 (2025).

%\cite{Schukraft:2012ah}
\bibitem{Schukraft:2012ah}
J.~Schukraft, A.~Timmins and S.~A.~Voloshin, Ultra-relativistic nuclear collisions: event shape engineering,
Phys. Lett. B \textbf{719}, 394  (2013).


\bibitem{Komiske:2017aww}
P.~T.~Komiske, E.~M.~Metodiev and J.~Thaler,
Energy flow polynomials: A complete linear basis for jet substructure,
JHEP \textbf{04} (2018), 013.

\bibitem{Komiske:2018cqr}
P.~T.~Komiske, E.~M.~Metodiev and J.~Thaler,
Energy Flow Networks: Deep Sets for Particle Jets,
JHEP \textbf{01} (2019), 121.


%\cite{Andrews:2018nwy}
\bibitem{Andrews:2018nwy}
M.~Andrews, M.~Paulini, S.~Gleyzer and B.~Poczos,
End-to-End Physics Event Classification with CMS Open Data: Applying Image-Based Deep Learning to Detector Data for the Direct Classification of Collision Events at the LHC,
Comput. Softw. Big Sci. \textbf{4} (2020), 6.


%\cite{Jiang:2024vvn}
\bibitem{Jiang:2024vvn}
Z.~Jiang, B.~Carlson, A.~Deiana, J.~Eastlack, S.~Hauck, S.~C.~Hsu, R.~Narayan, S.~Parajuli, D.~Yin and B.~Zuo,
Machine learning evaluation in the Global Event Processor FPGA for the ATLAS trigger upgrade,
JINST \textbf{19} (2024), P05031.

%\cite{Lee:2022kdn}
\bibitem{Lee:2022kdn}
K.~Lee, J.~Mulligan, M.~P{\l}osko{\'n}, F.~Ringer and F.~Yuan,
Machine learning-based jet and event classification at the Electron-Ion Collider with applications to hadron structure and spin physics,
JHEP \textbf{03} (2023), 085.


\bibitem{Ortiz:2021peu}
A.~Ortiz and E.~A.~Zepeda,
``Extraction of the multiplicity dependence of multiparton interactions from LHC pp data using machine learning techniques,''
J. Phys. G \textbf{48}, 085014 (2021).


%\cite{Mallick:2021wop}
\bibitem{Mallick:2021wop}
N.~Mallick, S.~Tripathy, A.~N.~Mishra, S.~Deb and R.~Sahoo, Estimation of Impact Parameter and Transverse Spherocity in heavy-ion collisions at the LHC energies using Machine Learning,
Phys. Rev. D \textbf{103}, 094031 (2021).

%\cite{Basak:2025idd}
\bibitem{Basak:2025idd}
D.~Basak, H.~Hushnud and K.~Dey,
Harnessing data-driven methods for precise model independent event shape estimation in relativistic heavy-ion collisions,
J. Phys. G {\bf 53}, 015102 (2026).



%\cite{Skands:2014pea}
\bibitem{Skands:2014pea}
P.~Skands, S.~Carrazza and J.~Rojo, Tuning PYTHIA 8.1: the Monash 2013 Tune,
Eur. Phys. J. C \textbf{74}, 3024 (2014).



% %\cite{Lai:2021ckt}
% \bibitem{Lai:2021ckt}
% Y.~S.~Lai, J.~Mulligan, M.~P{\l}osko{\'n} and F.~Ringer,
% The information content of jet quenching and machine learning assisted observable design,
% JHEP \textbf{10} (2022), 011

%\cite{Andersson:1983jt}
\bibitem{Andersson:1983jt}
B.~Andersson, G.~Gustafson and B.~Soderberg, A General Model for Jet Fragmentation,
Z. Phys. C \textbf{20}, 317 (1983).
% doi:10.1007/BF01407824
%632 citations counted in INSPIRE as of 10 Feb 2025

%\cite{Argyropoulos:2014zoa}
\bibitem{Argyropoulos:2014zoa}
S.~Argyropoulos and T.~Sj\"ostrand, Effects of color reconnection on $t\bar{t}$ final states at the LHC,
JHEP \textbf{11}, 043 (2014).


%\cite{ALICE:2015qqj}
\bibitem{ALICE:2015qqj}
J.~Adam \textit{et al.} [ALICE],
Pseudorapidity and transverse-momentum distributions of charged particles in proton{\textendash}proton collisions at $\sqrt s=$ 13 TeV,
Phys. Lett. B \textbf{753}, 319 (2016).
% doi:10.1016/j.physletb.2015.12.030
% [arXiv:1509.08734 [nucl-ex]].
%201 citations counted in INSPIRE as of 16 Dec 2025

%\cite{Fries:2003kq}
\bibitem{Fries:2003kq}
R.~J.~Fries, B.~Muller, C.~Nonaka and S.~A.~Bass, Hadron production in heavy ion collisions: Fragmentation and recombination from a dense parton phase,
Phys. Rev. C \textbf{68}, 044902 (2003).
%736 citations counted in INSPIRE as of 13 Aug 2024

%\cite{Fries:2003vb}
\bibitem{Fries:2003vb}
R.~J.~Fries, B.~Muller, C.~Nonaka and S.~A.~Bass, Hadronization in heavy ion collisions: Recombination and fragmentation of partons,
Phys. Rev. Lett. \textbf{90}, 202303 (2003).
%863 citations counted in INSPIRE as of 13 Aug 2024


%\cite{Wang:1991hta}
\bibitem{Wang:1991hta}
X.~N.~Wang and M.~Gyulassy, HIJING: A Monte Carlo model for multiple jet production in p p, p A and A A collisions,
Phys. Rev. D \textbf{44}, 3501 (1991).
%2042 citations counted in INSPIRE as of 13 Aug 2024

%\cite{Zhang:1997ej}
\bibitem{Zhang:1997ej}
B.~Zhang, ZPC 1.0.1: A Parton cascade for ultrarelativistic heavy ion collisions,
Comput. Phys. Commun. \textbf{109}, 193 (1998).
%478 citations counted in INSPIRE as of 13 Aug 2024


%\cite{Lin:2001zk}
\bibitem{Lin:2001zk}
Z.~w.~Lin and C.~M.~Ko, Partonic effects on the elliptic flow at RHIC,
Phys. Rev. C \textbf{65}, 034904 (2002).

%\cite{Li:2001xh}
\bibitem{Li:2001xh}
B.~Li, A.~T.~Sustich, B.~Zhang and C.~M.~Ko, Studies of superdense hadronic matter in a relativistic transport model,
Int. J. Mod. Phys. E \textbf{10}, 267 (2001).
%78 citations counted in INSPIRE as of 13 Aug 2024

%\cite{Greco:2003mm}
\bibitem{Greco:2003mm}
V.~Greco, C.~M.~Ko and P.~Levai, Parton coalescence at RHIC,
Phys. Rev. C \textbf{68}, 034904 (2003).
%556 citations counted in INSPIRE as of 13 Aug 2024

%\cite{Tripathy:2018bib}
\bibitem{Tripathy:2018bib}
S.~Tripathy, S.~De, M.~Younus and R.~Sahoo, Predictions for azimuthal anisotropy in Xe+Xe collisions at $\sqrt{s_{NN}}=$ 5.44 TeV using a multiphase transport model,
Phys. Rev. C \textbf{98}, 064904 (2018).
%19 citations counted in INSPIRE as of 13 Aug 2024

\bibitem{Pierog:2013ria}
T.~Pierog,  {\it et. al.}, EPOS LHC: Test of collective hadronization with data measured at the CERN Large Hadron Collider,
Phys. Rev. C \textbf{92}  034906 (2015).

\bibitem{Pierog:2009zt}
T.~Pierog and K.~Werner, EPOS Model and Ultra High Energy Cosmic Rays,
Nucl. Phys. B Proc. Suppl. \textbf{196}, 102 (2009).
% doi:10.1016/j.nuclphysbps.2009.09.017
% [arXiv:0905.1198 [hep-ph]].
%280 citations counted in INSPIRE as of 28 Dec 2025

%\cite{Drescher:2000ha}
\bibitem{Drescher:2000ha}
H.~J.~Drescher, M.~Hladik, S.~Ostapchenko, T.~Pierog and K.~Werner, Parton based Gribov-Regge theory,
Phys. Rept. \textbf{350}, 93 (2001).
% doi:10.1016/S0370-1573(00)00122-8
% [arXiv:hep-ph/0007198 [hep-ph]].
%553 citations counted in INSPIRE as of 28 Dec 2025

%\cite{Sjostrand:1993yb}
\bibitem{Sjostrand:1993yb}
T.~Sjostrand,
High-energy physics event generation with PYTHIA 5.7 and JETSET 7.4,
Comput. Phys. Commun. \textbf{82}, 74 (1994).
% doi:10.1016/0010-4655(94)90132-5
%4398 citations counted in INSPIRE as of 28 Dec 2025

%\cite{Lonnblad:1992tz}
\bibitem{Lonnblad:1992tz}
L.~Lonnblad,
ARIADNE version 4: A Program for simulation of QCD cascades implementing the color dipole model,
Comput. Phys. Commun. \textbf{71}, 15 (1992).
% doi:10.1016/0010-4655(92)90068-A
%1377 citations counted in INSPIRE as of 28 Dec 2025

%\cite{Odorico:1989bi}
\bibitem{Odorico:1989bi}
R.~Odorico, 
Cojets 5.12: A Monte Carlo Simulation Program for $\bar{p} p$ and $p p$ Collisions,
Comput. Phys. Commun. \textbf{59}, 527 (1990).
% doi:10.1016/0010-4655(90)90094-H
%42 citations counted in INSPIRE as of 28 Dec 2025

%\cite{Engel:1994vs}
\bibitem{Engel:1994vs}
R.~Engel,
Photoproduction within the two component dual parton model. 1. Amplitudes and cross sections,
Z. Phys. C \textbf{66}, 203 (1995).
% doi:10.1007/BF01496594
%603 citations counted in INSPIRE as of 28 Dec 2025

%\cite{Engel:1995yda}
\bibitem{Engel:1995yda}
R.~Engel and J.~Ranft,
Hadronic photon-photon interactions at high-energies,
Phys. Rev. D \textbf{54}, 4244 (1996).
% doi:10.1103/PhysRevD.54.4244
% [arXiv:hep-ph/9509373 [hep-ph]].
%526 citations counted in INSPIRE as of 28 Dec 2025

%\cite{Stelzer:1994ta}
\bibitem{Stelzer:1994ta}
T.~Stelzer and W.~F.~Long,
Automatic generation of tree level helicity amplitudes,
Comput. Phys. Commun. \textbf{81}, 357 (1994).
% doi:10.1016/0010-4655(94)90084-1
% [arXiv:hep-ph/9401258 [hep-ph]].
%1040 citations counted in INSPIRE as of 29 Dec 2025

%\cite{Nason:2004rx}
\bibitem{Nason:2004rx}
P.~Nason,
A New method for combining NLO QCD with shower Monte Carlo algorithms,
JHEP \textbf{11}, 040 (2004).
% doi:10.1088/1126-6708/2004/11/040
% [arXiv:hep-ph/0409146 [hep-ph]].
%4654 citations counted in INSPIRE as of 29 Dec 2025

%\cite{Frixione:2007vw}
\bibitem{Frixione:2007vw}
S.~Frixione, P.~Nason and C.~Oleari,
Matching NLO QCD computations with Parton Shower simulations: the POWHEG method,
JHEP \textbf{11}, 070 (2007).
% doi:10.1088/1126-6708/2007/11/070
% [arXiv:0709.2092 [hep-ph]].
%5351 citations counted in INSPIRE as of 29 Dec 2025

\bibitem{SHERPA:gitlab}
Sherpa Manual 3.0.0 documentation, 
Available Online: \url{https://sherpa-team.gitlab.io/sherpa/v3.0.0alpha1/manual/introduction.html}

%\cite{Eskola:1999fc}
\bibitem{Eskola:1999fc}
K.~J.~Eskola, K.~Kajantie, P.~V.~Ruuskanen and K.~Tuominen,
Scaling of transverse energies and multiplicities with atomic number and energy in ultrarelativistic nuclear collisions,
Nucl. Phys. B \textbf{570}, 379 (2000).
% doi:10.1016/S0550-3213(99)00720-8
% [arXiv:hep-ph/9909456 [hep-ph]].
%404 citations counted in INSPIRE as of 29 Dec 2025

%\cite{Kharzeev:2004if}
\bibitem{Kharzeev:2004if}
D.~Kharzeev, E.~Levin and M.~Nardi,
Color glass condensate at the LHC: Hadron multiplicities in pp, pA and AA collisions,
Nucl. Phys. A \textbf{747}, 609 (2005).
% doi:10.1016/j.nuclphysa.2004.10.018
% [arXiv:hep-ph/0408050 [hep-ph]].
%325 citations counted in INSPIRE as of 29 Dec 2025

%\cite{Loizides:2017ack}
\bibitem{Loizides:2017ack}
C.~Loizides, J.~Kamin and D.~d'Enterria,
Improved Monte Carlo Glauber predictions at present and future nuclear colliders,
Phys. Rev. C \textbf{97}, 054910 (2018)
[erratum: Phys. Rev. C \textbf{99}, 019901 (2019)].
% doi:10.1103/PhysRevC.97.054910
% [arXiv:1710.07098 [nucl-ex]].
%350 citations counted in INSPIRE as of 29 Dec 2025

%%%%%%%%%%%%%% Unused References %%%%%%%%%%%%%%%%%

%e+e- collisions
%\cite{Palni:2020shu}
%\bibitem{Palni:2020shu}
%P.~Palni, A.~Khuntia and P.~Bartalini,
%``Evolution of strange and multi-strange hadron production with relative transverse multiplicity activity in underlying event,''
%Eur. Phys. J. C \textbf{80}, 919. (2020).

%\cite{Greco:2003xt}
%\bibitem{Greco:2003xt}
%V.~Greco, C.~M.~Ko and P.~Levai,
%``Parton coalescence and anti-proton / pion anomaly at RHIC,''
%Phys. Rev. Lett. \textbf{90}, 202302 (2003).
%754 citations counted in INSPIRE as of 13 Aug 2024

%\cite{Li:1995pra}
%\bibitem{Li:1995pra}
%B.~A.~Li and C.~M.~Ko,
%``Formation of superdense hadronic matter in high-energy heavy ion collisions,''
%Phys. Rev. C \textbf{52}, 2037 (1995).
%487 citations counted in INSPIRE as of 13 Aug 2024

%\cite{Loizides:2017ack}
%\bibitem{Loizides:2017ack}
%C.~Loizides, J.~Kamin and D.~d'Enterria,
%``Improved Monte Carlo Glauber predictions at present and future nuclear colliders,''
%Phys. Rev. C \textbf{97}, 054910 (2018)
%[erratum: Phys. Rev. C \textbf{99}, 019901 (2019)].
%284 citations counted in INSPIRE as of 13 Aug 2024

\end{thebibliography}
\end{document}